%% file: Manuscript.tex
\documentclass[12pt]{article}

\usepackage{geometry}
\usepackage[T1]{fontenc}        

\usepackage[utf8x]{inputenc}    
\usepackage[comma,authoryear]{natbib}
\usepackage{ucs}                
\usepackage{amsmath}            
\usepackage{amsfonts}
\usepackage{amssymb}
\usepackage{amsthm} 
\usepackage{thmtools}
\usepackage{dsfont}
\usepackage[shortlabels]{enumitem}
\usepackage{ifthen}
\usepackage{color}
\usepackage{graphicx} 
\usepackage{textcomp}            
\usepackage{palatino}            
\usepackage{eulervm}            
\usepackage{varioref}            
\usepackage[colorlinks=false,plainpages=false]{hyperref}
\usepackage{url}                
\usepackage{cleveref}             
\usepackage{fancyhdr}
\usepackage{float}
\usepackage{multirow}
\usepackage{rotating}
\usepackage{subfig}
\usepackage{xr} 
\usepackage{xcite}
\usepackage{setspace} 
\usepackage{tabularx}

\usepackage{xr}

\newcommand{\R}{{\mathbb{R}}}         

 \newcommand{\N}{{\mathbb{N}}}         

\usepackage[colorinlistoftodos, textwidth=65mm, shadow]{todonotes}

\newcommand{\bqan}{\begin{eqnarray}}
\newcommand{\eqan}{\end{eqnarray}}
\newcommand{\bit}{\begin{itemize}}
\newcommand{\eit}{\end{itemize}}

\newtheoremstyle{Test1}
  {2 \baselineskip}
  {1.5 \baselineskip}
  {\itshape}
  {-0.0ex}
  {\fontfamily{ppl}\fontseries{l}\fontshape{n}}
  {:}
  {\newline}
   {}

\theoremstyle{Test1}



\relax 
\begin{document}
\def\spacingset#1{\renewcommand{\baselinestretch}%
{#1}\small\normalsize} \spacingset{1}
\spacingset{1.5}

\externaldocument[I-]{SupplementaryMaterial.tex}

\vspace{-5cm}
\title{\Large \bf Goodness (of fit) of Imputation Accuracy: The GoodImpact Analysis}

\author{{Maria Thurow}$^1$\thanks{
}, Florian Dumpert$^2$, Burim Ramosaj$^1$ and Markus Pauly$^1$\\
   }

\vspace{-8ex}
  \date{}
\maketitle 
\vspace{-1.25cm}
\begin{center}\noindent${}^{1}$ {TU Dortmund University, Department of Statistics, Germany\\\mbox{ }\hspace{1 ex}Mail: maria.thurow@tu-dortmund.de}\\
 \noindent${}^{2}$ {Federal Statistical Office of Germany (DESTATIS)}\\
 \end{center}
\begin{abstract}
\noindent In statistical survey analysis, (partial) non-responders  are integral elements during data acquisition. Treating missing values during data preparation and data analysis is therefore a non-trivial underpinning. Focusing on different data sets from the Federal Statistical Office of Germany (DESTATIS), we investigate various imputation methods regarding their imputation accuracy. Since the latter is not uniquely determined in theory and practice, we study different measures for assessing imputation accuracy: Beyond the most common measures, the normalized-root mean squared error (NRMSE) and the proportion of false classification (PFC), we put a special focus on (distribution) distance- and association measures for assessing imputation accuracy.
The aim is to deliver guidelines for correctly assessing distributional accuracy after imputation. Our empirical findings indicate a discrepancy between the NRMSE resp. PFC and distance measures. While the latter measure distributional similarities, NRMSE and PFC focus on data reproducibility. We realize that a low NRMSE or PFC seem not to imply lower distributional discrepancies. Although several measures for assessing distributional discrepancies exist, our results indicate that not all of them are suitable for evaluating imputation-induced differences.  
\end{abstract}
\noindent{\textbf{Keywords:}} Missing Values, Multiple Imputation, Distributional Similarities, Kolmogorov-Smirnov-Test, Random Forest, MICE



\section{Introduction}\label{sec:int}

Official statistics process various kind of surveys and data sets for manifold issues. %
%
Often, the goal 
is to estimate key figures of an underlying population on the basis of a sample. For example one likes to estimate the median earnings of employees in certain industries, broken down by employee education and company categories. Thereby, data returned from the sample can be incomplete, e.g. due to nonresponse, and official statistics has to decide how to handle these cases.
%
There exist many different ways to deal with missing data, where one can distinguish between three main approaches: {\it partial deletion of missing information, imputation methods} and {\it data adjusting methods}. We refer to the latter as all statistical methods that incorporate missingness directly. 
 For example, one could adopt maximum-likelihood based estimators by integrating out missing information similarly to the Expectation-Maximization algorithm (\citealp{dempster1977maximum}) or use the missing structure of the data to obtain test statistics that make use of the full data  (\citealp{amro2017permuting,amro2019multiplication,amro2019asymptotic}). Such {\it data adjusting methods} require deep statistical knowledge and are restricted to the problem-specific analysis. 
 In contrast, {\it partial deletion methods} may lead to the loss of valuable information, especially when the presence of missing values itself delivers important insights. For the Federal Statistical Office as a federal authority, it is important to deliver information in terms of data to the general audience. Therefore {\it imputing missing values} during data preparation can be considered as a universal approach for applying various statistical methods without restricting the attention to some sub-models preliminary. This, however, imposes some conditions on the imputation method itself: it should be as general as possible to allow various subsequent data analyses 
 (\citealp{meng1994multiple}). 
In the following, we therefore only consider imputation approaches which are 
{\it "methods that allow incomplete cases to be included in the analysis"} \citep{littlequote2011}.

There exists an abundance of approaches to impute missing values. 
Quick fixes for missing values are {\it naive imputation} methods such as mean or mode imputation. 
While the idea seems to be simple and fast, several authors have demonstrated their heavy disadvantage  \citep{schafer1997,rubin2004multiple,van_Buuren_flexible_2018}. 
More sophisticated methods therefore rely on regression and classification models and predict missing values using the corresponding model \citep{missforest,missranger}. However, in \cite{rubin2004multiple}, it has been criticised that any imputation method that singly imputes missing values imposes the belief that the information was given preliminary and subsequent analysis treats them as known and fixed. Hence, incorporating uncertainty arising from the imputation itself is a crucial point. This was proposed to be done using {\it multiple imputations} and corresponding aggregation structures for obtaining final test statistics, see e.g. \cite{rubin2004multiple}.

In any case, there exists no uniformly best imputation method. Indeed, the 
choice of a good imputation method heavily depends on the particular research question at hand. A fact that is often forgotten in practice.
To exemplify this, let us consider the problem of imputing missing covariates in classical regression problems. There exist several solutions for this task. 
For example, \cite{jones1996indicator} and \cite{jiang2019logistic} developed solutions that put a special focus on residual variance estimators and potential bias while \cite{van2005accurate} proposed methods that have advantages in predicting new outcomes. The difference between those examples is the statistical emphasis: while the first two examples aim to cover later statistical inference under missing values, the last two are more interested in point predictors.

The discrepancies in research targets additionally hampers the choice of the 'best' imputation strategy and raises the question {\it how to compare imputation methods?} Typical evaluation measures are:
\begin{itemize}
 \item Summary statistics like measures for location, dispersion or shape, 
 as well as measures of association. 
 \item NRMSE (cont.) or PFC (discrete) for reflecting data reproducibility. 
\end{itemize}
While summary statistics only cover special distributional aspects, NRMSE and PFC are among the most common measures to evaluate imputation accuracy, see e.g. \cite{waljee2013comparison},  \cite{missforest}, \cite{van_Buuren_flexible_2018}, \cite{Ramosaj_Pauly_2019} and \cite{ramosaj2020cautionary}. However, they can also have serious limitations. For example a recent study of \citealp{ramosaj2020cautionary} showed that procedures with the  'best' predictive accuracy resulted in the most serious type I error rate inflation in subsequent statistical testing.

Choosing the correct imputation strategy in official statistics is additionally hampered by another complicating issue: there is often no pre-defined target evaluation. Moreover, 
there are several, not always complementary, sub-goals of imputation in official statistics \citep{chambers2001} :
\begin{itemize}
    \item Predictive accuracy: preservation of true values, e.g. low NRMSE and PFC,
    \item Ranking accuracy: preservation of order in the imputed values,
    \item Distributional accuracy: preserving the distribution of the true data values,
    \item Estimation accuracy: reproducing the lower order moments of the distributions of the true values
\end{itemize}
and, where required, an overall goal in imputation tasks: plausibility. Here, the number of broken pre-defined plausibility constraints (e.\,g., negative yearly income might be implausible for employees) is an important figure in such cases. The less constraints are broken the better the imputation.

However, as mentioned above, very often the challenge for official statistics is not primarily to 
predict the missing values as well as possible (predictive accuracy). Instead, the main goal is 
useful inference in spite of incomplete data material (known as inferential accuracy). 
In fact, following \cite{littlequote2011} 
\textit{"the main reason for imputation is not to recover the information in the missing values 
($\dots$) but rather to allow the information in observed values in the incomplete cases to be retained."}

The evaluation according to inferential accuracy often presupposes that it is already known which (few) key figures of the population are of interest. In this case, the imputation can be optimized accordingly. But what happens if completed data sets are to be released in a suitable format for science or, for example, passed on to supranational statistical authorities such as Eurostat? Then, it is often not known at the time of imputation which indicators are to be calculated from the data. In this case, it would be desirable that the characteristics in a completed data set had at least the same low (also mixed) moments as the hypothetical complete data set. This would restore the location, variance, and correlation among the variables by imputation and would provide useful results for population estimates based on those (estimation accuracy). Thinking ahead, %
it is obvious to consider not only the lower moments, but also higher moments, quantiles and extreme values; eventually the distribution. Ideally, a completed data set has the same distribution as the hypothetical complete data set (distributional accuracy). 
This is an ambitious goal which raises the question {\it how to measure distributional accuracy in a simulation study?}

Preliminary considerations can be found in \cite{dumpert2020mlimp}. Therein the distributional accuracy of simple regression imputation of metric variables 
is investigated by summary statistics and the p-value of the Kolmogorov-Smirnov test. 
A visual approach in this direction was recently given by the Imputation Assessment and Comparison Tool 
\citep[ImpACT]{gray2019,gray2020}. ImpACT allows for univariate distribution analysis of imputation methods 
by comparing kernel density curves, fringe plots, histograms, and jitter-and-box plot combinations. These visual outputs are very intuitive, comprehensive, and allow for exploratory analysis. However, these should be only a first part of a more general tool box as assessing distributional aspects of imputation approaches only based on visual analysis may be misleading. The present paper therefore complements this tool box by studying how good imputation methods can reflect the features' true marginal distributions. 

It is structured as follows: In Section~\ref{sec:dat}, we give a brief introduction of the considered data set from the DESTATIS and also recapture the different missing mechanism. 
Section \ref{sec:imp} gives a thorough introduction to the imputation methods under study, covering state-of-the-art methods such as multiple imputation using chained equation (mice) 
and Machine Learning (ML) based methods from Random Forest models. Different evaluation measures for imputation methods are given in Section \ref{sec:evm}. An extensive simulation study based on the employee data of the DESTATIS is presented and discussed in Section \ref{sec:sim}.

\section{Data Set and Missing Settings}\label{sec:dat}

Our empirical analysis regarding distributional discrepancies after imputation is based on the DESTATIS employee data and hospital data. Due to similarity of results, we only present our analyses for the employee data set. The latter can nevertheless be requested from the authors from interested readers. The next sub-sections will cover a brief description of the data and the generation of missing values in our simulation study. 

\subsection{Data Set}\label{sec:dataset}
\input{dataset}

\subsection{Missing Settings}\label{sec:missingSetup}
\input{missing_mechanisms}

\section{Imputation Methods}\label{sec:imp}
\input{imputation}

\section{How to Judge Imputation Quality? -- Evaluation Measures}\label{sec:evm}
\input{measures}

\section{Simulation Study and Evaluation}\label{sec:sim}

In this section, our simulation results based on the DESTATIS employee data set are presented and discussed, focussing on the evaluation measures from the previous section. 
This is possible as we only insert missing values artificially as explained in Section~\ref{sec:missingSetup} and thus have access to the true (unobserved) data. For ease of presentation some results are moved to the Supplement.

\subsection{Simulation Setup}
\label{sec:simsetup}
The data set used for the simulation is described in Section \ref{sec:dataset} and is designed through the introduction of (artificial) missing values in a Monte-Carlo iterated fashion. Therefore, we first insert missing values into selected variables of the data set. Then the incomplete data set is imputed with the methods presented in Section \ref{sec:imp}. The distances between the distributions of the original data and the imputed data are calculated for each variable of each imputed data set using suitable distance measures introduced in Section \ref{sec:evm}. In addition, the measures $NRMSE$ and $PFC$ are calculated for each imputation and 
 permutation-based p-values are calculated.

Missing values have not been inserted to every variable, but to only $24$ potential variables in the employee data. A detailed description can be found in  Table $5$ and $6$ of the Supplement indicating the implemented missing mechanism for each considered variable.

We used three different values $r \in \{ 0.01, 0.05, 0.1 \}$ for the overall missing rate in \texttt{prodNA} under both, the MCAR and the MAR framework. Note that the MAR-mechanism as prescribed in Section~\ref{sec:missingSetup} has been applied on three variables, while the rest of the treated variables have a MCAR mechanism. Under the missing procedure described in Section \ref{sec:missingSetup}, this will result into an overall MAR mechanism. To be more precise, various dependence structures have been used for the three variables in the MAR mechanism. As explained in the second Step of Section $2$ in the Supplement regarding the size-allocation of the probabilities $\{p_i\}_i$, largest or smallest values of the latter have been allocated according to the following logic:
\begin{itemize}
\item The higher the age, the less likely it is to have missing values in the variable (normalized) gross annual earning, which results from the variable-aggregation explained in Section \ref{sec:dat}. 
\item The lower the (standardized) weekly working time (\texttt{ef53}), the higher the probability that the value of the bonus for special working hours (\texttt{ef23}) is missing.
\item The probability of a missing value in the performance group (for employees with payment according to individual agreement; variable \texttt{ef9}) differs for the different types of education (according to the ISCED scale, \texttt{ef43}). The probability of not specifying the performance group is highest when the employee's education is in the 'lower secondary education' category. The second highest probability of a missing value is for employees with education from the category 'upper secondary education' and the lowest probability of a missing value occurs for employees with a tertiary education.
\end{itemize}
A summarizing table for the implemented missing mechanism is given in Table $5$ and $6$ in the Supplement. 
We used $100$ Monte-Carlo iterates for every simulation set-up, which consists of a fixed missing mechanism and an overall missing rate $r$. For the estimation of the p-values as introduced in Section \ref{sec:evm}, we made use of $perm = 999$ permutation cycles with $m = 5$ imputations.  Regarding the MICE imputation procedure, we distinguished between metric and categorical variables. Specifically, the options \texttt{pmm}, \texttt{norm} and \texttt{rf} have been used for metric outcomes, while the option \texttt{rf} was used for categorical outcomes only. In the subsequent evaluation, the imputations are labeled by \texttt{Mice.Pmm}, \texttt{Mice.Norm} and \texttt{Mice.RF}. Strong dependencies between the variables can result in computational difficulties when applying, e.g., the MICE algorithm. Therefore, we restricted the selection of potential covariates in the imputation algorithm to those metric variables, which have a pairwise Pearson-correlation smaller than $0.8$. 
For categorical variables such restrictions have not been made, i.e. all were considered within each MICE procedure. 

To keep the imputation time at a minimal level, we made use of parallel computing for both, the MICE and the Amelia procedure. This was conducted using the \texttt{micemd} package (\citealp{paralmice}) and the \texttt{parallel} option in \texttt{Amelia}. Regarding the multiple imputation logic, we used $m = 5$ repetitions and the aggregation of the corresponding distance measures obtained from the $m = 5$ imputed data sets was conducted through averaging (\citealp{rubin2004multiple}).

\subsection{Predictive Imputation Accuracy}
\label{subsec:Impgueten}

\begin{figure}[H]
		\centering
		\subfloat[ $NRMSE$ ]{{\includegraphics[width=7cm]{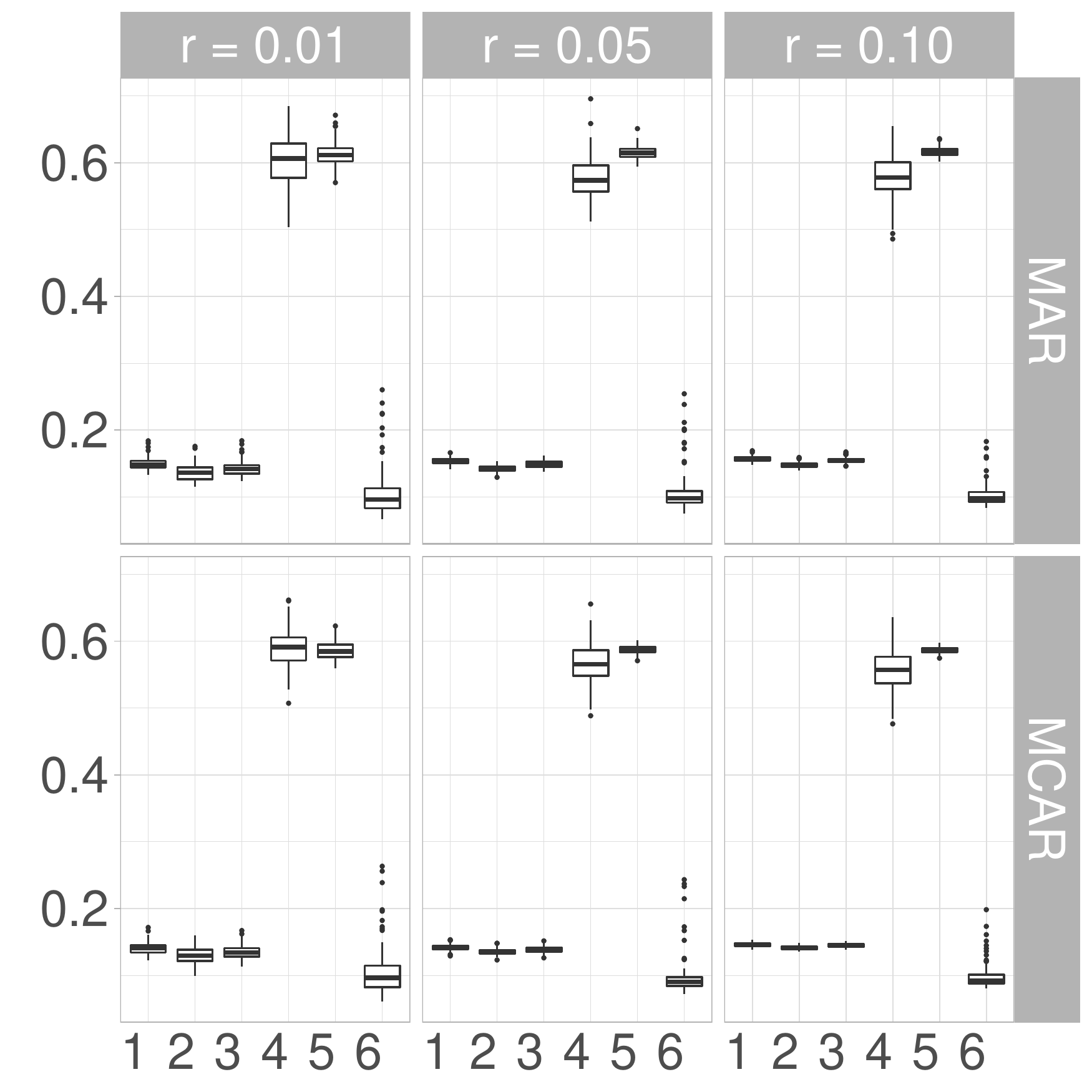} }}%
		\qquad
		\subfloat[$PFC$]{{\includegraphics[width=7cm]{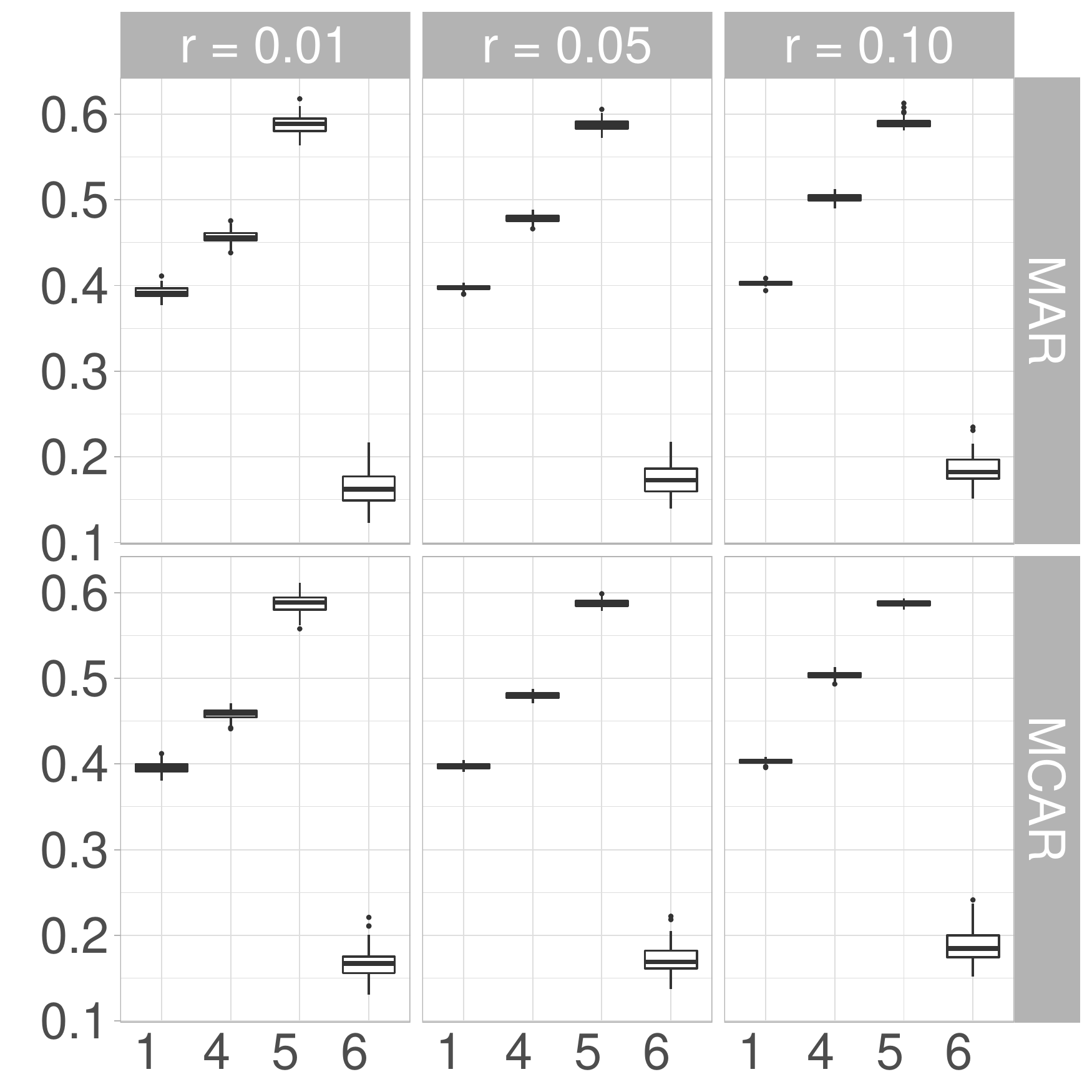} }}%
		\caption{Boxplot for the imputation accuracy using $MC = 100$ iterations. \\ 1: \texttt{Amelia}, 2: \texttt{Mice.Norm}, 3: \texttt{Mice.Pmm}, 4: \texttt{Mice.RF}, 5: \texttt{Naive}, 6: \texttt{missRanger}} \label{BoxPlot_NRMSE_PFC_Employee}
	\end{figure}

We use the NRMSE and the PFC 
for partly assessing imputation accuracy. 
Their behavior 
was examined on the DESTATIS employee data set through summarizing boxplots. They are given in Figure~\ref{BoxPlot_NRMSE_PFC_Employee}. Having a closer look at the NRMSE- and PFC -median of all imputation methods, we observe an almost unchanged behaviour across different missing rates $r \in \{ 0.01, 0.05, 0.1 \}$ and missing mechanisms. Regarding NRMSE, the \texttt{Mice.RF} imputation method and the naive imputation lead to the highest NRMSE values with median values around $0.6$. With increasing missing rate the naive method turned slightly worse than \texttt{Mice.RF}. 
In contrast, the \texttt{missRanger} method showed the most preferable behaviour with median NRMSE around $0.1$ for both missing mechanisms. \texttt{Amelia}, \texttt{Mice.Norm} and \texttt{Mice.Pmm} turned out to behave rather similarly across the different missing rates and missing mechanisms. They are close to the \texttt{missRanger} method, but show less variation in imputation accuracy than \texttt{missRanger} while being slightly worse than the latter. Regarding imputation accuracy for categorical variables, we observed a different pattern. 
The \texttt{missRanger} method yielded by far the lowest PFC values across all missing rates and missing mechanisms, while the naive imputation turned out to be the worst. Moreover,  \texttt{Amelia} turned out to perform better than its MICE counterpart and the \texttt{Mice.RF} procedure was better than before. Note that 
all three MICE procedures make use of a Random Forest based imputation scheme. 
Thus, only 
\texttt{Mice.RF} is shown.


\subsection{Distance Measures}
\label{subsec:Abstaende}

Beside predictive accuracy 
we also computed several measures to 
assess distributional discrepancies. Here the aim is to check the extend to which the proposed imputation methods are able to reflect 
the true marginal feature 
distributions. 

\subsubsection*{$\chi^2$-Association and Cramérs $V$}

In the DESTATIS employee data, the different categorical variables do not share the same domain having thus an impact on the range of the classical $\chi^2$ test statistic. To make the association measure comparable between the different variables, we therefore compute Cramérs $V$ statistic standardizing the range to the unit interval $[0,1]$. This will ease the interpretation making comparison 
more accessible. As the $\chi^2$ test statistic and Cramérs $V$ are used for testing independence, an increased realization of both statistics indicates a high association between the original and imputed variables. 
For our purposes, we also modified  Cramérs $V$ slightly to \textit{turn it more in form of a usual distance measure}. That is, we used 
\begin{align}
 \kappa &= \frac{1}{|\boldsymbol{F}_{sim}|\cdot 100} \sum\limits_{i = 1}^{100} \sum\limits_{j \in \boldsymbol{F}_{sim}} |V_{i,j}^2 - 1|,
\end{align}
where $V_{i,j}$ is the Cramérs $V$ statistic for the $j$-th categorical variable in the $i$-th simulation run. $\boldsymbol{F}_{sim}$ is the index set of all categorical variables with  missing values. The usage of $\kappa$ allows us to define two categorical variables \textit{distributional similar}, if $\kappa$ is close to zero. The simulation results for $\kappa$ can be found in Table~\ref{Kappa_Association_VRE}. 
\begin{table}[h!]
\centering
\begin{tabular}[t]{|l|c|c|c|c|c|c|}

\hline

	& \multicolumn{6}{c|}{Missing mechanism and rate}\\
	\cline{2-7}
	& \multicolumn{3}{c|}{MAR} & \multicolumn{3}{c|}{MCAR}\\
	\cline{2-7}
	& 1\,\% & 5\,\% & 10\,\% & 1\,\% & 5\,\% & 10\,\% \\
\hline

	\texttt{Amelia} & 0.015 & 0.070 & 0.131 & 0.015 & 0.070 & 0.131 \\


		
	\texttt{Mice.RF} & 0.013 & 0.068 & 0.139 & 0.013 & 0.068 &0.139\\

	\texttt{Naive}  & 0.013 & 0.064& 0.123 & 0.013 & 0.064 & 0.123\\
		
	 \texttt{missRanger}  & 0.005 & 0.025 & 0.051 & 0.005 & 0.025 & 0.052\\

\hline
	
\hline
\end{tabular}
\caption{Realizations of $\kappa$ using $MC = 100$ iterations.}\label{Kappa_Association_VRE}
\end{table}
It is apparent that $\kappa$ increases with increases missing rates from $1 \%$ to $10\%$ for all  
imputation methods. No difference between MAR and MCAR situation were observed. 
Again, the \textit{missRanger} resulted into the lowest $\kappa$ values across the different imputation methods indicating a high dependence structure between the original and imputed variables. 
Again, only \texttt{Mice.RF} is shown. 
Regarding \texttt{Amelia}, the imputation procedure resulted in almost all scenarios with the highest value of $\kappa$, but was often similar to the MICE procedures. A similar effect could be observed for Cramérs $V$ measure, see Figure 1 in the Supplement.

\subsubsection*{Two-sample Kolmogorov-Smirnov Test}

\begin{figure}[H]
	\centering
	\includegraphics[width=15cm]{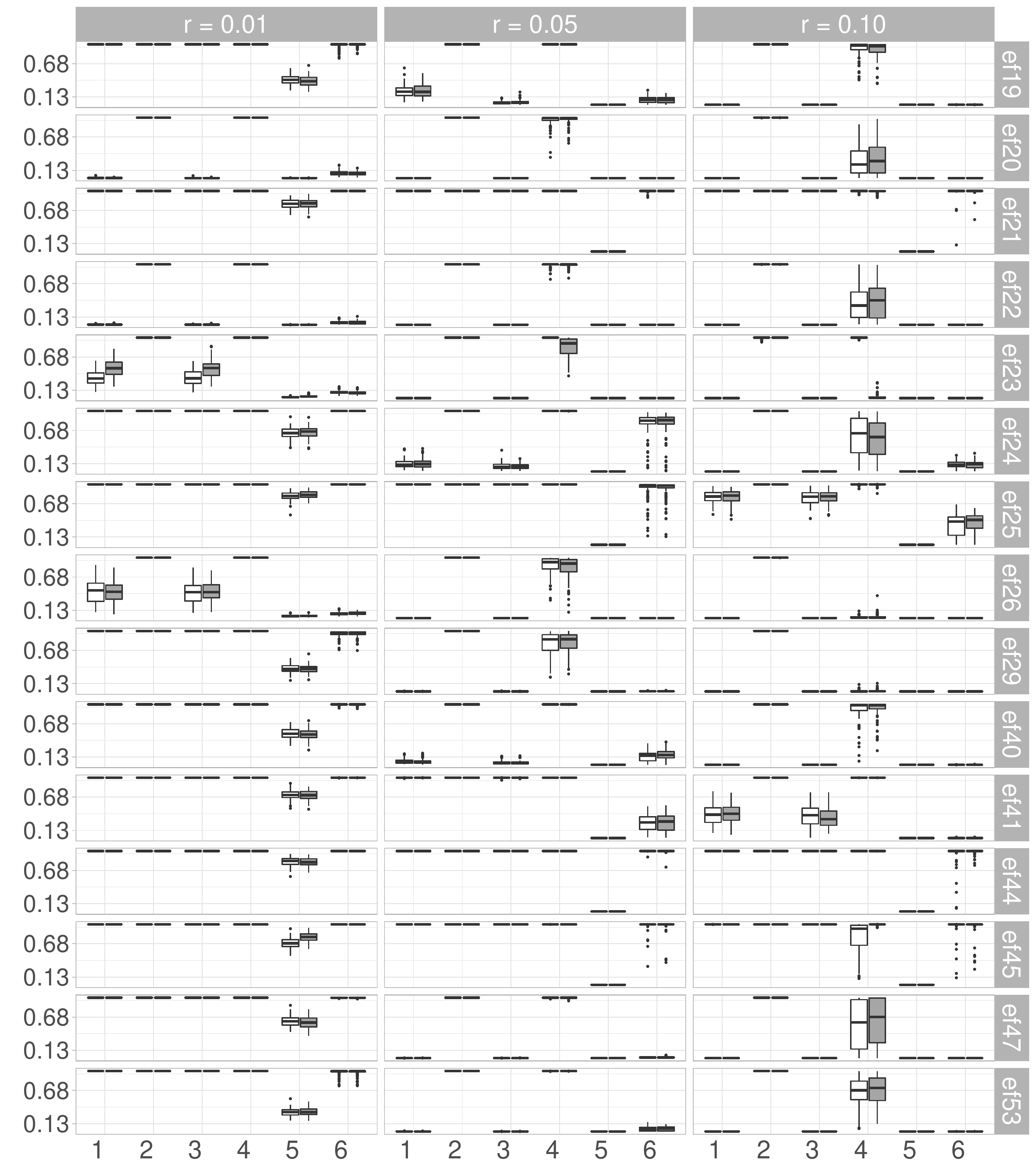}
	\caption{Pair of boxplot for the \textbf{p-values} of the \textbf{Kolmogorov-Smirnov statistic} using $MC = 100$ iterations and $perm = 999$ permutations. Each boxplot pair corresponds to the following missing mechanism: the left one to the MCAR, and the right one to the MAR mechanism. \\1: \texttt{Amelia}, 2: \texttt{Mice.Norm}, 3: \texttt{Mice.Pmm}, 4: \texttt{Mice.RF}, 5: \texttt{Naive}, 6: \texttt{missRanger}}\label{KSStat_pVal}
\end{figure}

In order to assess the distributional accuracy of metric and ordinal variables, we used the two-sample Kolmogorov-Smirnov statistic (KS-statistic). Therein, we first estimated the empirical distribution functions for every variable. Then, the corresponding KS-statistic is computed between the empirical distribution function of the original and imputed data. Permutation-based p-values are obtained as described in Section \ref{sec:evm}. The results of the KS-statistic is presented in the Supplement, in Figure 2. Here, we give a brief overview on the obtained p-value results. In Figure \ref{KSStat_pVal} it is noticeable, that the missing mechanism does not seem to have an impact on the computed p-values throughout the variables. While most of the imputation methods indicated a similar p-value trend for missing rates at $1\%$, the \texttt{Naive} method indicated a considerably different KS-statistic for every variable. The effect strengthens, as the missing rate increases indicating a considerable distributional difference as given in the null-hypothesis $(\ref{generalNULL})$. While most of the imputation methods flipped from a non-considerable result for low missing rates to a considerable one for higher missing rates, the \texttt{Mice.Norm} approach remained stable delivering no empirical evidence for distributional discrepancies under the KS-statistic. Hence, the method seems to be more robust towards an increased missing rate, preserving distributional properties more similar to the original data set. Compared with the other MICE approaches, the \texttt{Mice.RF} method indicated a more appropriate behaviour than the \texttt{Mice.Pmm} method. They both lost in distributional similarities, especially when the missing rate increased, but this effect was less distinct for \texttt{Mice.RF}. Regarding \texttt{Amelia}, the imputation method indicated considerably different imputations for missing rates greater than $1\%$. \texttt{missRanger} behaved worse than its mice counterpart  \texttt{Mice.RF} regarding \textit{distributional recovery}. Although it showed quite good results for the smallest missing rate of $1 \%$, 
it exhibit a rather bad {distributional fit} for larger missing rates.

\subsubsection*{Two-sample Cramer-von-Mises Statistic}

\begin{figure}[H]
	\centering
	\includegraphics[width=15cm]{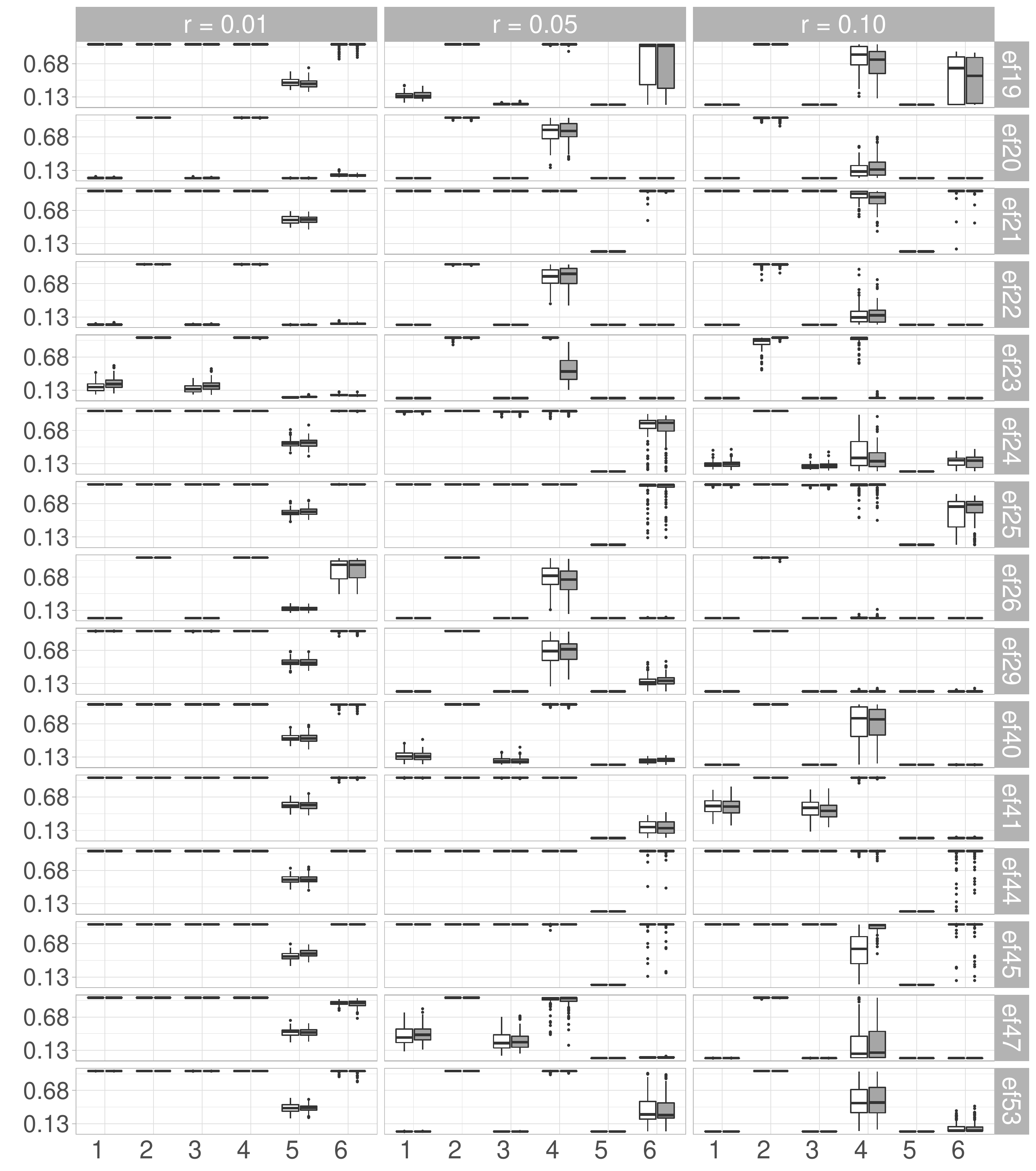}
	\caption{Pair of boxplot for the \textbf{p-values} of the \textbf{Cramer-von-Mises statistic} using $MC = 100$ iterations and $perm = 999$ permutations. Each boxplot pair corresponds to the following missing mechanism: the left one to the MCAR, and the right one to the MAR mechanism. \\1: \texttt{Amelia}, 2: \texttt{Mice.Norm}, 3: \texttt{Mice.Pmm}, 4: \texttt{Mice.RF}, 5: \texttt{Naive}, 6: \texttt{missRanger}}\label{CSStat_pVal}
\end{figure}

Similarly to the KS-statistic, we computed the Cramer-von-Mises statistic (CM-statistic) for every variable based on the estimated empirical distribution function of the original and imputed data. The difference of the CS-statistic to the KS-statistic lies in the penalization of differences between two distribution functions. While the KS-statistic considers the maximal absolute difference, the CM-statistics aggregates dissimilarities through squared penalization. The realizations of the CM-statistic can be found in the Supplement, Figure $3$. Here, we present the permutation based p-value results described in Section \ref{sec:evm}. They are given in Figure \ref{CSStat_pVal}. Similarly to the KS-statistic, the \texttt{Naive} method performed worse indicating considerable distributional differences measured by the CM-statistic between the imputed and original data set. Again, the \texttt{Mice.Norm} approach resulted into the best results delivering no evidence for distributional differences between the imputed and original data set. The performance of \texttt{Mice.RF} was also strong, but more sensitive towards an increased missing rate. \texttt{Mice.Pmm} performed worse being more affected by higher missing rates. \texttt{Amelia} resulted into the worst results. For missing rates larger than $1\%$, the imputation method indicated considerably different results regarding distributional differences measured by the CM-statitic. The Random-Forest based method \texttt{missForest} was slightly better, but suffered also when missing rates turned larger. Hence, both method did not show appropriate results under the CM statistic.

\subsubsection*{Kullback-Leibler Divergence}

\begin{figure}[H]
	\centering
	\includegraphics[width=15cm]{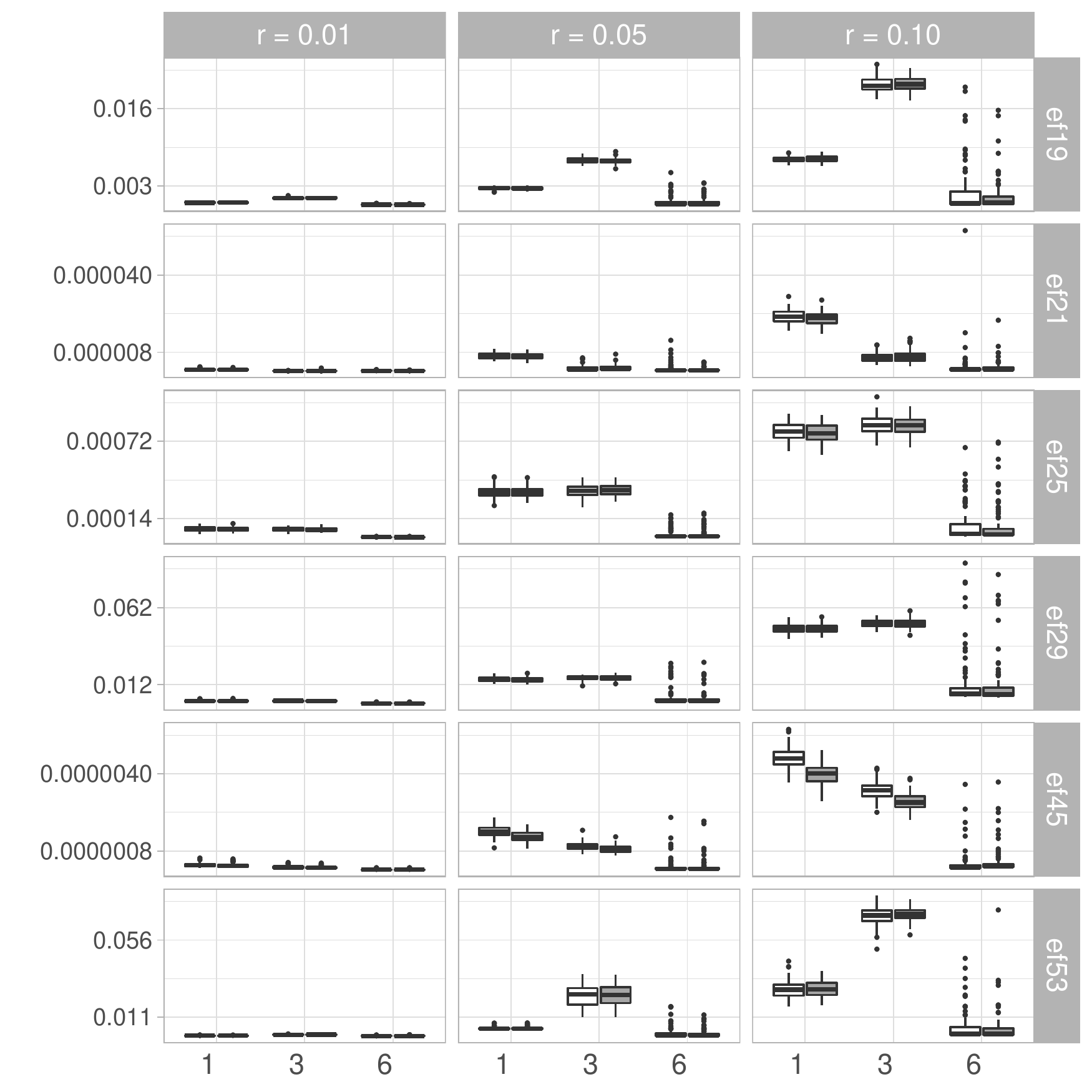}
	\caption{Pair of boxplot for the \textbf{Kullback-Leibler divergence} using $MC = 100$ iterations. Each boxplot pair corresponds to the following missing mechanism: the left one to the MCAR, and the right one to the MAR mechanism. \\1: \texttt{Amelia}, 3: \texttt{Mice.Pmm}, 6: \texttt{missRanger}}\label{KLStat_VREdata}
\end{figure}

In addition to the previous results, we make use of the Kullback-Leibler divergence (KL-divergence) as an information theoretical measure for assessing distributional discrepancies. Differently to the KS- and CM-statistic, it is not based on the empirical distribution function, but on probability density functions. Their estimation has been conducted within the \textsf{R} computing software using the default method for an automated bandwidth selection with \textit{Silverman's rule of thumb}. Critical values for the estimation of the KL-divergence are values of zero for one of the kernel density estimators. This was the case for the variables \texttt{ef22} (total income for additional working hours), \texttt{ef23} (allowance for special working hours), \texttt{ef24} (income tax), \texttt{ef26} (working days subject to social security contributions), \texttt{ef40} (company affiliation), \texttt{ef44} (monthly net income) and \texttt{ef47} (normalized special payments) attaining values of zero for the kernel density estimator based on the imputed data set. This led to infinite values for the KL-divergence. Therefore, Figure \ref{KLStat_VREdata} excludes these variables. Furthermore, the imputation procedures \texttt{Mice.Norm} and \texttt{Mice.RF} led more often to such critical values, too. Therefore, these procedures have also been excluded from our simulation results. Regarding the \texttt{Naive} approach, high realizations of the KL-divergence made the comparison with the other imputation techniques in one boxplot impossible. Therefore, the results of this approach is dropped as well. Similarly to the results for the KS- and CM-statistic, the KL-divergence increased on average and in volatility for an increased missing rate $r$. Under this measure, the \texttt{missRanger} method performed best, followed by the \texttt{Amelia} procedure. A clear difference between the missing mechanisms was not observable. The \texttt{Mice.Pmm} imputation performed worse under this measure. 

\subsubsection*{Mallows $L^2$-Distance}
\begin{figure}[H]
	\centering
	\includegraphics[width=15cm]{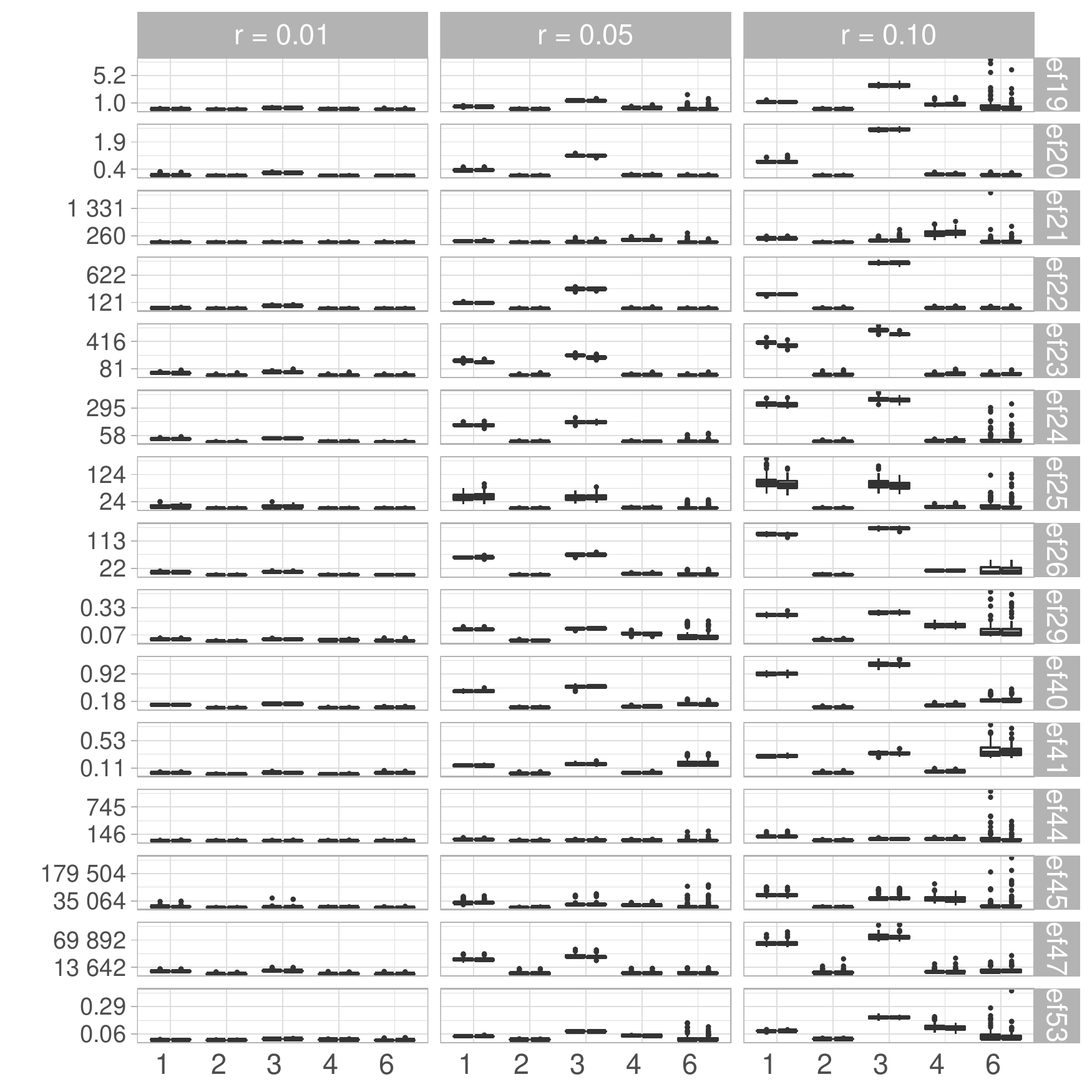}
	\caption{Pair of boxplot for the \textbf{Mallows $\boldsymbol{L^2}$-Distance} using $MC = 100$ iterations. Each boxplot pair corresponds to the following missing mechanism: the left one to the MCAR, and the right one to the MAR mechanism. \\1: \texttt{Amelia}, 2: \texttt{Mice.Norm}, 3: \texttt{Mice.Pmm}, 4: \texttt{Mice.RF}, 6: \texttt{missRanger}}\label{MallowsDistance_VREdata}
\end{figure}

Mallows $L^2$-distance measure (ML2-statistic) is based on the quadratic penalization of the inverse distribution functions of two comparing distributions. This is different to the previous cases, since the latter make use of empirical distribution function estimators and kernel density estimators for distributional discrepancies. We estimate the quantile function similarly to \cite{estim_Mallows} using the empirical squared error of the order statistics. The simulation results for the \texttt{Naive} imputation method were clearly beyond the range of the ML2-statistic for the other imputation method. For some variables, for example, \texttt{naive} resulted into $700$ times larger ML2-statistic than \texttt{Mice.RF} making it impossible to include in a joint boxplot. Therefore, we dropped the results for this imputation technique in order to make the boxplot in Figure \ref{MallowsDistance_VREdata} comparable with the other methods. Similarly to the previous results, an increased missing rate led to an increased ML2-statistic, both on average and in volatility. Again, the \texttt{Mice.Norm} imputation procedure performed comparably well, leading to the lowest realizations of the ML2-statistic. In addition, the \texttt{Mice.Norm} imputation approach was more robust towards an increased missing rate. \texttt{Amelia} and \texttt{Mice.Pmm} performed comparable, while the latter imputation method resulted into larger realizations of the ML2-statistic for some variables. \texttt{missRanger} indicated on average a similar behavior to the \texttt{Mice.Norm} procedure, but was more volatile during the simulation runs.  \\
 
All considered measures were also investigated whether they show mutual same directions. Therefore, in Table $45$ of the Supplement, we additionally computed Pearson's correlation coefficient across the different imputation methods, missing mechanism and Monte-Carlo iteration. The results indicate that all considered distance measures for metric variables behave similarly showing positive correlation among each other. 

\subsection{Computational Complexity}

Beside the analysis of distributional accuracy through imputation and data recovery potential, we were interested in the computational time costs for the various methods we used. Although several imputation techniques such as \texttt{missRagner} or \texttt{Mice.RF} are based on Machine Learning methods, that usually require hyperparameter tuning in prediction tasks, we did not incorporate tuning strategies in imputation. This, because computational time would drastically increase since tunig has to happen on a variable-by-variable basis. We rather focus on the default parameters proposed for \texttt{missRanger} resp. \texttt{missForest} and \texttt{Mice.RF}. Averaging over $MC = 100$ Monte-Carlo iterations, the computation time  for  various missing rates and imputation methods is reported in Table~\ref{TimeComplexity}.

Recall that the employee data set under study consists of $25,974$ observations measured on $27$ variables. We did not distinguish between the different missing mechanism, since the latter should not have a considerable impact on the time complexity, due to the nature of the considered imputation methods. Therefore, we restrict our attention to the MAR mechanism.

\begin{table}[h!]
\centering
\begin{tabularx}{.99\textwidth}{lXXX}
\hline
	& \multicolumn{3}{c}{Time complexity in seconds}\\
	\cline{1-4}
	$r = $ & 1\,\% & 5\,\% & 10\,\% \\
\hline

	\texttt{Amelia}$^\star$ & 150.77 & 971.44 &  3203.53  \\

	\texttt{Mice.Norm}$^\star$ & 467.55 & 898.52 & 1297.73 \\

	\texttt{Mice.Pmm}$^\star$ & 477.22 & 937.98 & 1371.69 \\
		
	\texttt{Mice.RF}$^\star$ & 3104.12 & 3634.12 & 4030.20 \\

	\texttt{missRanger}  & 479.81 & 605.35 & 637.16 \\
		
	 \texttt{Naive}  & 0.0046 & 0.048 & 0.048 \\

\hline
\end{tabularx}
\caption{Time complexity in seconds of the various imputation methods on the employee data set averaged over $100$ Monte-Carlo runs. Methods applying the multiple imputation logic are marked with $\star$ using $m = 5$ imputations.}\label{TimeComplexity}
\end{table}

It is noticeable that the \texttt{missRanger} procedure is less affected by an increased missing rate as the computational time increases only slightly. Furthermore, the method seemed to result into the lowest time complexity of all non naive methods if $r> 1 \%$. However, a direct comparison to the other methods has to be interpreted carefully, since \texttt{Amelia} and the MICE procedures are applied multiple times $(m = 5)$. Among the MICE methods, \texttt{Mice.Norm} resulted into the lowest computational efforts on average while being less sensitive to an increased missing rate than \texttt{Mice.RF} or \texttt{Mice.Pmm}. Compared with all other multiple imputation procedures, \texttt{Amelia} turned out to be the fastest method for missing rates up to $1\%$, but the result quickly changed indicating a very sensitive behaviour of \texttt{Amelia} to increased missing rates. Taking into account the previous results on distributional preserving measures as well, the \texttt{Mice.Norm} procedure remains competitive.

\subsection{Summary of the Simulation Results}

We compared the performance of various imputation techniques for different types of variables using the PFC, NRMSE and distributional distance based measures. Noticeable is the difference between the NRMSE and PFC results and the distributional distance based measures. While  \texttt{missRanger} indicated the lowest PFC and NRMSE results, it resulted into larger realizations of some of the other measures such as the KS- or CM-statistic. This indicates that the \texttt{missRanger} approach is very suitable for data reproducibility, but not so much for distribution-preserving data recovery. Different are the results when changing the Random Forest based imputation technique to \texttt{Mice.RF}. Therein, the method indicated comparably well distribution preserving measures showing low KS- , CM- and ML2-statistics. Regarding categorical variables, the $\kappa$ measure used for assessing distributional discrepancy resulted into similar recommendations as the NRMSE and PFC measures. This, because the proposed measure have not to be understood directly as a distance measure, but as a \textit{dependence measure} being therefore in line with the imputation accuracy measures. 
The best results leading to the lowest realizations of the KS-, CM- and ML2-statistic could be obtained from the \texttt{Mice.Norm} approach. In addition, the latter imputation method also delivered stable p-value estimations for testing the hypothesis in $(\ref{generalNULL})$, when the missing rate increased. In general, there was not enough evidence to conclude that the latter method imputed missing values distributionally different to the original data set. This was different for the other methods, were statistical test decisions flipped depending on the missing rate. Simultaneously, the \texttt{Mice.Norm} imputation method performed worse when evaluating it with the NRMSE or PFC measure.

\section{Discussion and Outlook}

We conducted an extensive simulation study on the evaluation of various imputation methods on data sets obtained from the Federal Statistical Office of Germany (DESTATIS). Therein, several measures have been proposed for assessing the accuracy of imputation methods under single and multiple imputation frameworks. 
For first evaluation we made use of the common data preserving measures PFC and NRMSE. In addition, we compared the same imputation techniques with measures reflecting the potential of distributional recovery. In particular, for categorical outcomes, we proposed a slight modification of Cramér's $V$, which is based on the $\chi^2$ statistic. This measure indicates potential dependencies between two variables and is not a typical (distribution based) distance measure. For metric outcomes, we took into consideration several measures that are based on empirical distribution functions, kernel density estimators or quantile functions such as the Kolmogorov-Smirnov statistic, the Cramer-von-Mises statistic, the Kullback-Leibler divergence and Mallow's $L^2$ distance.  Our simulation study indicates that both classes of accuracy measures, the NRMSE resp. PFC results and the distributional based distance measure recommend different imputation procedures for its usage in missing value imputation. While the \texttt{missRanger} approach yielded low NRMSE and PFC results across different missing rates, the same method performed worse when evaluating it with the Kolmogorov-Smirnov statistic or the Cramer-von-Mises statistic. The results may indicate that an imputation scheme might be suitable for data reproducibility, but not for distribution preserving imputations. This effect is supported by the simulation results obtained for the \texttt{Mice.Norm} approach. This MICE method is based on a Bayesian linear regression model with normality assumptions and indicated high NRMSE and PFC values while also yielding low Kolmogorov-Smirnov and Cramer-von-Mises statistics. Under the \texttt{Mice.Norm} approach, the p-values obtained for the edf-based measures were robust towards an increased missing rate delivering in almost all variables not enough evidence for distributional differences. Furthermore, the Random Forest based imputation method \texttt{Mice.RF} performed better than its counterpart \texttt{missRanger} with regard to distributional recovery. 

Our results can be used as a preliminary step for evaluating imputation schemes from a more flexible statistical inference perspective. While the NRMSE and PFC measures seem to be suitable for evaluating data recovery potential and therefore point predictions, distribution preserving imputation schemes are required to obtain correct statistical inference procedures. Therefore, solely selecting an imputation scheme based on the NRMSE and PFC measure is not the correct way, especially when the data analyst's approach is focused on statistical decisions, testing procedures or uncertainty quantification. Future work will be concerned with extensions from univariate to multivariate investigations to cover dependencies among the variables, e.g. by Copulas, which so far have been mostly neglected. 

\subsection*{Acknowledgements}
The work of Burim Ramosaj and Markus Pauly was supported by the German Research Foundation (DFG).

\newpage
\bibliographystyle{apalike}
\bibliography{Manuscript}
\end{document}


%
%
%
%
%

\title{\Large \bf Supplementary Material to\\ 
Goodness (of fit) of Imputation Accuracy: The GoodImpact Analysis
}

\author{{Maria Thurow}$^1$\footnote{Email address: maria.thurow@tu-dortmund.de}, Florian Dumpert$^2$, Burim Ramosaj$^1$ and Markus Pauly$^1$\\[1ex]
   }

\vspace{-8ex}
  \date{}
\maketitle \vspace*{1ex}

\noindent${}^{1}$ {TU Dortmund University, Department of Statistics, Germany  }\\
 \noindent${}^{2}$ {Federal Statistical Office of Germany (DESTATIS)}\\

Additional simulation results for the employee data set can be found in this supplement with a brief description. Furthermore, simulation results for the hospital data set of DESTATIS are presented here as well. 

\newpage
\tableofcontents


\vfill
\vfill

\section{Data Set Description}\label{sec:data}
\subsection{Employee Data}
\begin{table}[H]
\caption{Variabale description of the \textbf{employee data set}. Observed values in German language are kept. } \label{variables_vse_employees_1}
\begin{center}
\begin{tabular}{|l|l|l|c|}
\hline
Variable & Name & Scale of& Observed Values\\
&& Measurement &\\
\hline
  \texttt{ef1}&Company Number& nominal & \{1,...,881\}\\[0.7em]
 \texttt{ef2}& Type of Sheet & nominal & \{ \textit{Arbeitnehmerdatensatz} \}\\[0.7em]
 \texttt{ef3}&Employee Number&nominal&\{1,...,25974\}\\[0.7em]
 \texttt{ef4}& Economic Sector & nominal &\{B \& C, D \& E, F, G, H,\\
 &&& I, J, K, L \& M, N, O \& P,\\
 &&& Q \& WZ75, R, S\} \\[0.7em]
 
 \texttt{region}& Regional & nominal & \{ \textit{Westdeutschland (und Berlin)},\\
 &Information&& \textit{Ostdeutschland}\}\\[0.7em]
 
 \texttt{ef9}& Group with  &nominal& \{  \textit{keine Angabe},\\
 & gratifications  && \textit{Leitender Arbeitnehmer (AN)},\\
 &according to && \textit{AN mit besonderen Erfahrungen},\\
 &free agreement&& \textit{AN mit mehrjähriger Berufs-}\\
 &&& \textit{erfahrung, AN ohne}\\
 &&& \textit{Entscheidungsbefugnis,}\\
 &&& \textit{AN in einfacher Tätigkeit}\}\\[0.7em]
 \texttt{ef10}& Sex &nominal&\{\textit{männlich, weiblich}\}\\[0.7em]
 \texttt{ef15}&Classification of&nominal&\{01-06, 07-24, 25-32, 33-43,\\
 &the profession&&44-51, 52-53, 54, 60-61,\\
 &(KldB)&&62-63, 68, 69-70 \& 752 \& 753,\\
 &&&71-74, 751 \& 76, 77, 78, 82-83,\\
 &&&84-86, 87-88, 90-93, übrige\}\\[0.7em]
 \hline
\end{tabular}
\end{center}
\end{table}

\begin{table}[H]
\caption*{\label{variables_vse_employees_2}Variabale description of the \textbf{employee data set}. Observed values in German language are kept. (continued).}
\begin{center}
\begin{tabular}{|l|l|l|c|}
\hline
Variable & Name & Scale of& Observed Values\\
&& Measurement &\\
\hline
  \texttt{ef16u1}& Position in & nominal & \{ \textit{Auszubildender,}\\
  &Profession&& \textit{Arbeiter (nicht Facharbeiter),}\\
  &&& \textit{Facharbeiter, Meister,}\\
  &&& \textit{Angestellter, Beamter,}\\
  &&& \textit{Heimarbeiter, Beamter (Teilzeit),}\\
  &&& \textit{Teilzeit weniger 18 Std.,}\\
  &&& \textit{Teilzeit mindestens 18 Std.}\}\\[0.7em]
  \texttt{ef16u2}& Education & nominal &\{ \textit{Hauptschulabschl. (ohne BA),}\\
  &BA: Vacational && \textit{Hauptschulabschl. (mit BA),}\\
  &training && \textit{Abitur (ohne BA),}\\
  &&&\textit{Abitur (mit BA),}\\
  &&& \textit{Fachhochschulabschluss,} \\
  &&& \textit{Hochschul-/Universitäts-,} \\
  &&& \textit{abschluss, unbekannt}\}\\[0.7em]
  
  \texttt{ef17}& Type of &nominal&\{ \textit{unbefristet, befristet,}\\
  & Employment && \textit{Azubis, Altersteilzeit,}\\
  &Contract&& \textit{geringfügig beschäftigt} \}\\[0.7em]
  
  \texttt{ef18}& Regular Weekly & cardinal&\{ 0,...,45 \}
  \\
  & Working Time (in h) &&\\[0.7em] 
  
  \texttt{ef19}& Paid Hours & cardinal&\{0,...,180\}
  \\
  & Without Overtime && \\[0.7em]
  
    \texttt{ef20}& Paid Overtime & cardinal&\{0,...,50\}
    \\[0.7em]
    
    \texttt{ef21}& Gross Monthly &cardinal&[0, 7000]
    \\
    & Income in Euro &&\\[0.7em]
    
    \texttt{ef22} & Total Income & cardinal & [0, 2692]
    \\
  &for Overtime&&\\[0.7em]
  \hline
\end{tabular}
\end{center}
\end{table}

\begin{table}[H]
\caption*{Variabale description of the \textbf{employee data set}. Observed values in German language are kept. (continued).}
\begin{center}
\begin{tabular}{|l|l|l|c|}
\hline
Variable & Name & Scale of& Observed Values\\
&& Measurement &\\
\hline
  \texttt{ef23}& Bonus for Special & cardinal & [0, 3253]
  \\
  & Working Hours &&\\[0.7em]
  
    \texttt{ef24}& Income Tax & cardinal & [0, 2925]
    \\[0.7em]
    
	\texttt{ef25}& Social Insurance & cardinal & [0, 1299]
	\\
  & Contributions & & \\[0.7em]
    
    \texttt{ef26} & Days Subject to Social & cardinal & \{0,...,360\}\\
    & Insurance Contributions &&\\
    & in the Reporting Year &&\\[0.7em]
    
    \texttt{ef27}& Gross Annual Earnings & cardinal&[0,84000]
    \\[0.7em]
    
    \texttt{ef28}& Special Payments & cardinal & [0, 48308]
    \\[0.7em]
    
    \texttt{ef29}& Vacation Entitlement 2010 &cardinal&\{0,...,40\}
    \\[0.7em]
    
    \texttt{ef38}& Extrapolation Factor & cardinal&[57.05, 5975.89]\\
    &for Employees&&\\[0.7em]
    
    \texttt{ef40}& Company Affiliation & cardinal &\{0,..., 45\}
    \\
    &(in years)&&\\[0.7em]
    
    \texttt{ef41}& Age (in years) & cardinal &\{ \textit{16 oder jünger, 17,...,}\\
    &&&\textit{65, 66 und älter} \}\\[0.7em]
    
    \texttt{ef42}&Job according to & nominal& \{ 1,...,9, \textit{keine Angabe} \} \\
    &the ISCO-key&&\\[0.7em]
    
    \texttt{ef43} & Education According to & nominal & \{\textit{Sekundarbildung I,}\\
    &the ISCED-Key&& \textit{Sekundarbildung II,}\\
    &&& \textit{Tertiäre Bildung}\}\\[0.7em]
    \hline
\end{tabular}
\end{center}
\end{table}

\begin{table}[H]
\caption*{Variabale description of the \textbf{employee data set}. Observed values in German language are kept. (continued).}
\begin{center}
\begin{tabular}{|l|l|l|c|}
\hline
Variable & Name & Scale of& Observed Values\\
&& Measurement &\\
\hline
  \texttt{ef44}& Net Monthly Earnings & cardinal & [10, 6321]
  \\[0.7em]
  
  \texttt{ef45}
  & Standardized Gross & cardinal & [204, 84000]
  \\
  & Annual Earnings &&\\[0.7em]
  
  \texttt{ef47}
  & Standardized Special Payment & cardinal & [0, 42960]\\
  &$\texttt{ef47}=\cfrac{\texttt{ef27}\cdot 360}{\texttt{ef26}}$&&[0, 21090]\\[0.7em]
  
  \texttt{ef50}& Number of Working Weeks & cardinal & [0, 51.43]\\
  &$\texttt{ef50}=\cfrac{\texttt{ef26}}{7}$&&\\[0.7em]
  
  \texttt{ef53}& Weekly Working Hours of a &cardinal&[0, 40.05]\\
  & Marginally Employed Person &&\\
  &$\texttt{ef53}=\cfrac{\texttt{ef19}}{4.345}$&&\\[0.7em]
  \hline
\end{tabular}
\vspace{7cm}
\end{center}
    
    
    
\end{table}
\newpage
\subsection{Hospital Data}

The $2010$ hospital data set of the DESTATIS consists of $40$ variables measured in $1,701,542$  observations. It consists of observations from patients randomly selected from $50 \%$ of the hospitals in Germany, while only $20 \%$ of the patients have been randomly selected among the ones treated at these hospitals. They are described in Table \ref{Variablen_drg_1}. In some of the variables of the hospital data, secondary diagnosis and surgeries resp. medical procedures are recorded. However, most of these variables are missing leading to the substitution of these through dummy variables. For the secondary diagnosis, the variables measured in \texttt{icd\_nd1} -\texttt{icd\_nd10} are transformed to dummy variables indicating whether patient had the secondary diagnosis ($= 1$) or not $(=0)$. Using the international, statistical grouping code of diseases, $21$ categories have been identified for potential secondary diseases leading to $21$ new dummy variables. They are described in Table \ref{Tab:ICD1}. For surgeries and medical procedures, a similar substitution has been conducted leading to the introduction of six dummy variables as well. They are described in Table \ref{Tab:OPS}.
 
\newpage
\begin{table}[H]
\caption{Variabale description of the \textbf{hospital data set}. Observed values in German language are kept.}\label{Variablen_drg_1}
\begin{center}
\begin{tabular}{|l|l|l|c|}
\hline
Variable & Name & Scale of& Observed Values\\
&& Measurement &\\
\hline
\texttt{fall\_nr} & Case Number & nominal & $\{1,...,1701542\}$\\[0.7em]
\texttt{kh\_region} & Region of the & nominal & \{ \textit{Nord, Süd, Ost} \}\\
&Institute&&\\[0.7em]
\texttt{kh\_typ\_gem} &  Type of Structural & nominal & \{ \textit{Agglomerationsräume,} \\
& Housing Aarea of &&	\textit{Verstädterte Räume,}\\
& the Institute && \textit{Ländliche Räume}\} \\[0.7em]

\texttt{pat\_region} & Region of the Patient & nominal & \{ \textit{Nord, Süd, Ost} \}\\[0.7em]

\texttt{pat\_typ\_gem} & Type of Structural & nominal & \{ \textit{Agglomerationsräume,} \\

&Housing Area of&&	\textit{Verstädterte Räume,}\\

&the Patient&& \textit{Ländliche Räume}\} \\[0.7em]

\texttt{sex}& Sex & nominal & \{\textit{weiblich, männlich}\}\\ [0.7em]

\texttt{typ\_alter}& Group of Age &ordinal&\{ \textit{$<1$ Jahr,}\\

&&& \textit{$1 \leq$ und $<10$ Jahre,}\\

&&& \textit{$10 \leq$ und $<20$ Jahre,}\\

&&& \textit{$20 \leq$ und $<30$ Jahre,}\\

&&& \textit{$30  \leq$ und $<40$ Jahre, }\\

&&& \textit{$40 \leq$ und $<50$ Jahre, }\\

&&& \textit{$50 \leq$ und $<60$ Jahre,}\\

&&& \textit{$60 \leq$ und $<70$ Jahre,}\\

&&& \textit{$70 \leq$ und $<80$ Jahre,}\\

&&& \textit{$80$ Jahre $\leq$} \}\\[0.7em]
\hline
\end{tabular}
\vspace{0.7em}
\end{center}
\end{table}

\begin{table}[H]
\caption*{Variable description of the \textbf{hospital data set}. Observed values in German language are kept (continued).}
\begin{center}
\begin{tabular}{|l|l|l|c|}
\hline
Variable & Name & Scale of& Observed Values\\
&& Measurement &\\
\hline

\texttt{aufn\_anl}& Reason for Admission & nominal &\{ \textit{Einweisung durch Arzt,}\\

&&& \textit{Notfall, Verlegung,}\\

&&& \textit{Geburt}\}\\[0.7em]

 \texttt{aufn\_gew} & Weight at Admission & ordinal & \{\textit{<3000g,}\\

 & (Children until the &&\textit{ 3000g $\leq$ und < 4500g,}\\

 & Age of 1 Year)&& \textit{4500g $\leq$ und < 7500g,}\\

 &&& \textit{7500g $\leq$}\}\\[0.7em]

\texttt{entl\_grd} & Reason for Dismissal & nominal & \{ \textit{Behandlung regulär beendet,}\\

&&& \textit{Entlassung gegen ärztl. Rat,}\\

&&& \textit{Verlegung, Entlassung,}\\

&&& \textit{Tod, Sonstiger Grund} \}\\[0.7em]

\texttt{typ\_vwd} & Type of Length of Stay & ordinal & \{ \textit{Stundenfall, 1-3 Tage,}\\

&&& \textit{4-7 Tage, 8-10 Tage,}\\

&&& \textit{11-14 Tage, 15-21 Tage,}\\

&&& \textit{22 Tage und länger}\}\\[0.7em]

\texttt{icd\_hd3} &  ICD-Code Main & nominal & \{A00-A09,...,Z80-Z99\}\\

&Diagnosis&&\\
&(Group of Diagnosis)&&\\[0.7em]
	
\texttt{icd\_nd1-} & ICD-Code from up to&nominal&\{A00-A09,...,Z80-Z99\}\\

\texttt{icd\_nd10}& 10 Secondary &&\\
	&Diagnosis&&\\
	&(Group of Diagnosis)&&\\[0.7em]

\texttt{ops\_ko1-} & OPS-Code from up to & nominal&\{110-110,...,999-999\}\\

\texttt{ops\_ko10} & 10 Procedures &&\\[0.7em]

\texttt{typ\_op} & Surgeries & nominal & \{\textit{Ja, Nein}\}\\[0.7em]
\hline
\end{tabular}
\vspace{0.7em}
\end{center}
\end{table}


\begin{table}[H]
\caption*{Variable description of the \textbf{hospital data set}. Observed values in German language are kept (continued).}
\begin{center}
\begin{tabular}{|l|l|l|c|}
\hline
Variable & Name & Scale of& Observed Values\\
&& Measurement &\\
\hline

\texttt{fab\_max} & Specialist & nominal & \{ \textit{01: Innere Medizin,...,}\\
& Department &&	\textit{37: Sonstige} \\
& longest Stay &&\textit{Fachabteilungen} \}\\[0.7em]

\texttt{tage\_max}& Maximal Number & ordinal & \{ \textit{Stundenfall, 1-3 Tage,}\\

& of Days Stood&& \textit{4-7 Tage, 8-10 Tage,}\\

&in the Department&& \textit{11-14 Tage, 15-21 Tage,}\\

&&& \textit{22 Tage und länger}\}\\[0.7em]

 \texttt{drgh}&DRG-Code&nominal&\{901A,...,Z66Z\}\\[0.7em]

\texttt{partition}&DRG-Partition&nominal&\{medizinische\\
&&&Fallpauschalen,\\
&&&operative Fallpauschalen,\\

&&&andere Fallpauschalen\}\\[0.7em]

\texttt{cm\_vol}&Case-Mix-Volume&metrisch&[716, 99999259703]\\[0.7em]

\texttt{beatm}&Time of Mechanical &ordinal&\{ \textit{nicht beatmet,}\\

& Ventilation && \textit{bis 12 Stunden beatmet,}\\

&&& \textit{13 bis 72 Stunden beatmet,}\\

&&& \textit{73 bis 168 Stunden beatmet,}\\

&&& \textit{über 168 Stunden beatmet}\}\\[0.7em]

\texttt{split}&Partition of Basis-&nominal&\{A,...,Z\}\\
& DRG according to&&\\
&Level of Severity&&\\[0.7em]
\hline
\end{tabular}
\vspace{0.7em}
\end{center}
\end{table}

\begin{table}[H]
\caption{New Dummy Variables introduced in the hospital data and allocation to Diagnosis Codes.}\label{Tab:ICD1}
\begin{center}
\begin{tabular}{|l|l|l|}
\hline
New Variable & Description & Diagnosis Code \\
\hline

 \texttt{nd1}& Certain infectious and parasitic disease. &A00,...,B99\\[0.7em]
 
 \texttt{nd2}& Renewals. &C00,...,D48\\[0.7em]
 
\texttt{nd3}& Diseases of the blood and hematopoietic&D50,...,D90\\
 &Organs and certain disorders with &\\

 & the involvement of the immune system.&\\[0.7em]

 \texttt{nd4}&Endocrines, nutritional and metabolic &E00,...,E90\\
 &disorder. &\\[0.7em]
 
 \texttt{nd5}&Psychological problems and  behavioural&F00,...,F99\\
 &disorder. &\\[0.7em]
 
 \texttt{nd6}& Diseases of the nervous system. &G00,...,G99\\[0.7em]

 \texttt{nd7}& Diseases of the eye and its appendages.&H00,...,H59\\[0.7em]

 \texttt{nd8}&Diseases of the ear and the mastoid process. &H60,...,H95\\[0.7em]

 \texttt{nd9}& Diseases of the circulatory system. &I00,...,I99\\[0.7em]

 \texttt{nd10}&Diseases of the respiratory system. &J00,...,J99\\[0.7em]

 \texttt{nd11}&Diseases of the digestive system.&K00,...,K93\\[0.7em]

 \texttt{nd12}&Diseases of the skin and the hypodermis. &L00,...,L99\\[0.7em]

 \texttt{nd13}&Diseases of the muscle-sceleton system and &M00,...,M99\\

 &the connective tissue.&\\[0.7em]
 
 \texttt{nd14}&Diseases of the genitourinary system. &N00,...,N99\\[0.7em]

 \texttt{nd15}&Pregnancy, birth and puerperium. &O00,...,O99\\[0.7em]
\hline
\end{tabular}
\vspace{0.7em}
\end{center}
\end{table}

\begin{table}[H]
\caption*{New Dummy Variables introduced in the hospital data and allocation to Diagnosis Codes (continued).}
\begin{center}
\begin{tabular}{|l|l|l|}
\hline
New Varibale& Description & Diagnosis Code\\
\hline

 \texttt{nd16}& Certain states with source in the &P00,...,P96\\

 &perinatal period. &\\[0.7em]

 \texttt{nd17}&Congenital malformations, deformations &Q00,...,Q99\\

 &and chromosome anomalies.&\\[0.7em]
 
 \texttt{nd18}&Symptoms and abnormal clinical &R00,...,R99\\

 &laboratory findings, which have not been &\\
 & classified elsewhere. &\\[0.7em]

 \texttt{nd19}&Injuries, toxication and certain &S00,...,T98\\

 &implications of external causes. &\\[0.7em]
 
 \texttt{nd20}&External causes of morbidity and mortality. &V01,...,Y84\\[0.7em]

 \texttt{nd21}& Factors influencing health condition &Z00,...,Z99\\

 & yielding to the utilization of&\\

 &the health system.&\\[0.7em]
 \hline
\end{tabular}
\vspace{0.7em}
\end{center}
\end{table}

\begin{table}[H]
\caption{New dummy variables in the hospital data and allocation to the procedural code.}\label{Tab:OPS}
\begin{center}
\begin{tabular}{|l|l|l|}
\hline
New Variable & Description & Procedural Code\\
\hline

 \texttt{ops1}&Diagnostic Measures &1-10,...,1-99\\[0.7em]

 \texttt{ops2}&Image-induced diagnostic.&3-03,...,3-99\\[0.7em]

 \texttt{ops3}& Surgeries. &5-01,...,5-99\\[0.7em]

 \texttt{ops4}&Drugs. &6-00\\[0.7em]

 \texttt{ops5}& Non-surgical and therapeutic measures. &8-01,...,8-99\\[0.7em]

 \texttt{ops6}& Complementary measures. &9-20,...,9-99\\[0.7em]

\hline
\end{tabular}
\vspace{0.7em}
\end{center}
\end{table}

\newpage

\section{Missing Mechanism}\label{Supp:MissMech}

We describe the generation of missing values, especially for the considered MAR mechanism, more into detail. Note that the latter was only considered for the employee data set. For the hospital data set, just the MCAR mechanism was used. If we denote with $\mathbf{X} = [\mathbf{X}_1, \dots, \mathbf{X}_k] \in \R^{n \times k}$ the data matrix and recall the definition of the \textit{variable missing rate} and the \textit{overall missing rate}, then our procedure for generating missing values according to the MAR is conducted iteratively using the following steps:

\begin{enumerate}
    \item We start with the first variable $\mathbf{X}_1$ for which we generate missing values using the \texttt{prodNA} function in \textsf{R} with an overall missing rate $r$. 
    Note that in this stage, the number of missing values per variable is given by $n_{mis}^{(j)} = n \cdot r_j$ and will be used for keeping the overall missing rate equal to $r$.  
    \item \label{SecondStep} Starting with the next variable $\mathbf{X}_2$, we generate iid random numbers $\{p_i\}_{i = 1}^{p^{(1)}}\sim Unif(0,1)$ , where $1 \le p^{(1)}\le n$ refers to the number of distinct observed characteristics $\{ a_i^{(1)} \}_{i = 1}^{p^{(1)}}$ of variable $\mathbf{X}_1$ in the data matrix $\mathbf{X}$. Note that missing values in $\mathbf{X}_1$ are dropped at this stage. The allocation of the probabilities $\{ p_i\}_i$ is conducted based on the largest resp. smallest value of $\mathbf{X}_2$, depending on the pre-defined logic. That is, the largest value of $\mathbf{X}_1$ denoted by $a_{(p)}^{(1)}$ is linked with the smallest or largest realized value of $\{p_i\}_i$. The allocated probabilities will be denoted as $\{ p_i^{(alloc)}\}_{i = 1}^{p^{(1)}}$. Furthermore, we denote with $\{ H_i^{(1)} \}_{i = 1}^{p^{(1)}}$ the absolute frequencies of $\{ a_i^{(1)} \}_{i = 1}^{p^{(1)}}$ in $\mathbf{X}_1$ and finally compute the probability of obtaining a missing value in $\mathbf{X}_2$, given $a_{r}^{(1)}$, for $r = 1, \dots, p^{(1)}$ the following way:
    \begin{align}
        \widehat{\mathbb{P}}[ m_{i2} = 1 | X_{i1} = a_r^{(1)} ] := \pi_r &= \frac{ p_r^{(alloc)} \cdot H_r^{(1)}  }{ \sum\limits_{r = 1}^{p^{(1)} } p_r^{(alloc)} H_r^{(1)} }
    \end{align}
    \item \label{ThirdStep} Considering the set $I_r = \{ 1 \le i \le n : X_{i1} =  a_r^{(1)} \}$, we randomly select $n_{mis}^{(j)} \cdot \pi_r$ indices among $I_r$ and set the corresponding values in $\mathbf{X}_2$ to $m_{i2} = 1$. This is conducted for every characteristic $r = 1, \dots, p^{(1)}$. This way, we obtain the overall missing rate $r$, since $\sum_{r=1}^{p^{(1)}} n_{mis}^{(j)} \pi_r = n_{mis}^{(j)}$.
    \item Steps \ref{SecondStep}. and \ref{ThirdStep}. are repeated, until all variables in $\mathbf{X}$ have been treated.
 \end{enumerate}
 
 For the MAR mechanism in the employee data set, three (context based) variables have been selected for the insertion of MAR-based missing values, while the rest of the $24$ variables have MCAR-based missing values, see Section 5 in the main article. The following tables describe, in which of the variables the MCAR or MAR mechanism was implemented.
 
 \begin{table}[H]
\caption{Variables of the \textbf{employee} data set, in which missing values have been inserted. The bracketed variables after the MAR indicates on which variable the dependence for the MAR mechanism has been implemented. }\label{InsertedNA_Employee}
\begin{minipage}[t]{\textwidth}
\centering
\label{Tab:fehlend_sim_vse}
\begin{tabular}[t]{|l|l|l|}
\hline
	{Variable} & {Scale of Measurement} & {Implemented Mechanism}\\
\hline

	\texttt{ef4} & nominal & MCAR \\

	\texttt{ef9}& nominal & MCAR \& MAR (\texttt{ef43})\\

	\texttt{ef10} & nominal & MCAR\\
	
	\texttt{ef15}& nominal & MCAR\\

	\texttt{ef16u1} & nominal im Beruf & MCAR \\
	
	\texttt{ef16u2}& nominal & MCAR\\

	\texttt{ef17}& nominal & MCAR\\
	

	\texttt{ef19} & metric & MCAR \\
	
	\texttt{ef20}& metric & MCAR\\
	
	\texttt{ef21} & metric & MCAR\\

	\texttt{ef22}&  metric & MCAR \\

	\texttt{ef23} & metric & MCAR \& MAR (\texttt{ef53})\\

	\texttt{ef25} & metric & MCAR\\
	
	\texttt{ef26}& metric & MCAR \\

	\texttt{ef29} & metric & MCAR \\
	
	\texttt{ef40}& metric & MCAR \\

	\texttt{ef41} & metric & MCAR \\
	
	\texttt{ef42}& nominal & MCAR\\
	
	\texttt{ef43} & nominal & MCAR \\
	
	\texttt{ef44}& metric & MCAR \\

	\texttt{ef45} & metric & MCAR \& MAR (\texttt{ef41})\\
	
	\texttt{ef47}& metric & MCAR\\
	
	\texttt{ef53}& metric & MCAR\\

\hline
\end{tabular}
\end{minipage}
\end{table}

\begin{table}[H]
\caption{Variables of the \textbf{hospital} data set, in which missing values have been inserted.}\label{InsertedNA_Hospital}
\begin{minipage}[t]{\textwidth}
\centering
\label{Tab:fehlend_sim_drg}
\begin{tabular}[t]{|l|l|l|}
\hline
	{Variable} & {Scale of Measurement} & {Implemented Mechanism}\\
\hline

	\texttt{sex} & nominal & MCAR\\

	\texttt{typ\_alter}& ordinal & MCAR\\

	\texttt{aufn\_anl} & nominal & MCAR\\
	
	\texttt{entl\_grd}& nominal & MCAR\\

	\texttt{typ\_vwd} & ordinal & MCAR\\
	
	\texttt{typ\_op}& nominal & MCAR\\

	\texttt{fab\_max} & nominal & MCAR\\
	
	\texttt{cm\_vol}& metric & MCAR\\

	\texttt{beatm} & ordinal & MCAR\\

\hline
\end{tabular}
\end{minipage}
\end{table}

\newpage
\section{More on the Simulations}\label{sec:msim}
In this section, we present additional simulation results for the employee data and extend the simulation to the hospital data as well.

\subsection{Employee Data}

\subsubsection{Nominal and Ordinal Variables}
\begin{figure}[H]
	\centering
	\includegraphics[width=15cm]{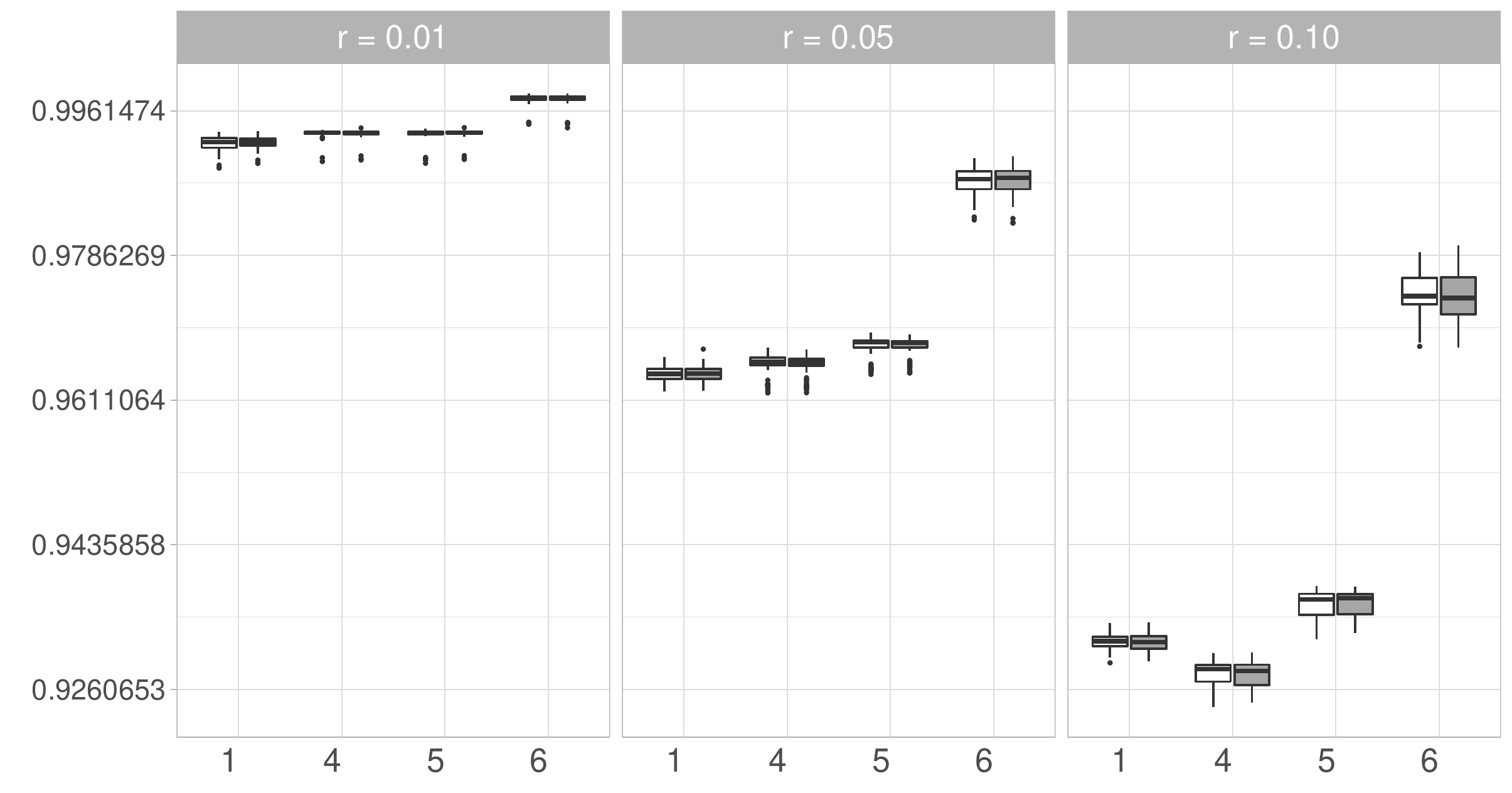}
	\caption{Pair of boxplot of \textbf{Cramérs V} on the employee data set using $MC = 100$ iterations. Each boxplot pair corresponds to the following missing mechanism: the left one to the MCAR, and the right one to the MAR mechanism. \\1: \texttt{Amelia}, 2: \texttt{Mice.Norm}, 3: \texttt{Mice.Pmm}, 4: \texttt{Mice.RF}, 5: \texttt{Naive}, 6: \texttt{missRanger}}\label{Boxplot_CramersVal}
\end{figure}

We present summary statistics of the $\chi^2$ statistics and Cramérs $V$ statistics in addition to the results given in the main article. In addition, the boxplot of Cramérs $V$ statistic is given in Figure \ref{Boxplot_CramersVal}. Note that a value close to $1$ indicates a high dependence between two random variables. Comparing the different imputation methods, it is noticeable that \texttt{missRanger} was less affected by an increased missing rate yielding in addition the highest Cramérs V measures. Note that the MICE procedures yielded similar results since all methods made use of a Random-Forest based imputation scheme for categorical outcomes. Similar to the main result \texttt{Amelia} performed worse.

In the upcoming part, you can find summary statistics of the $\chi^2$ statistic and Cramérs V measure.  
\subsubsubsection{Summary Statistics}
\input{table_fivepoint_Chisq_vse.txt}
\input{table_fivepoint_CramersV_vse.txt}

\begin{table}[H]
\caption{minimum ($q_0$), first quartile ($q_{0.25}$),
               median ($q_{0.5}$), third quartile ($q_{0.75}$),
               maximum ($q_{1}$) and standard deviation (sd) of the observed values of the PFC for the different
               imputation methods and missing rates (r) for the employee data}
\centering
\label{Tab:vse_pfc_summary_1}
\begin{tabular}[t]{|l|l|c|c|c|c|c|c|c|}
\hline
\multirow{2}{*}{method}&missing&\multirow{2}{*}{r}&\multirow{2}{*}{$q_{0}$}&\multirow{2}{*}{$q_{0.25}$}&\multirow{2}{*}{$q_{0.5}$}&\multirow{2}{*}{$q_{0.75}$}&\multirow{2}{*}{$q_{1}$}&\multirow{2}{*}{sd}\\
&mechanism&&&&&&&\\
\hline
\hline	 
 
	 Amelia&MAR&0.01& 0.38 & 0.39 & 0.39 & 0.40 & 0.41 & 0.01 \\
	 &&0.05 & 0.39 & 0.40 & 0.40 & 0.40 & 0.40 & <0.01\\
	 &&0.10& 0.39 & 0.40 & 0.40 & 0.40 & 0.41 & <0.01\\
	 \cline{2-9}
	 &MCAR&0.01& 0.38 & 0.39 & 0.40 & 0.40 & 0.41 & 0.01\\
	 &&0.05& 0.39 & 0.40 & 0.40 & 0.40 & 0.40 & <0.01\\
	 &&0.10&  0.40 & 0.40 & 0.40 & 0.40 & 0.41 & <0.01\\
	 \hline
	 Mice.Norm&MAR&10.01& 0.45 & 0.47 & 0.47 & 0.48 & 0.49 & 0.01 \\
	 &&0.05 & 0.47 & 0.48 & 0.49 & 0.49 & 0.50 & 0.01\\
	 &&0.10& 0.49 & 0.50 & 0.50 & 0.51 & 0.52 & <0.01\\
	 \cline{2-9}
	 &MCAR&0.01& 0.45 & 0.47 & 0.47 & 0.48 & 0.49 & 0.01\\
	 &&0.05& 0.48 & 0.49 & 0.49 & 0.50 & 0.50 & 0.01\\
	 &&0.10&  0.50 & 0.50 & 0.51 & 0.51 & 0.52 & <0.01\\ 
\hline
\end{tabular}
\end{table}

\begin{table}[H]
\caption{minimum ($q_0$), first quartile ($q_{0.25}$),
               median ($q_{0.5}$), third quartile ($q_{0.75}$),
               maximum ($q_{1}$) and standard deviation (sd) of the observed values of the PFC for the different
               imputation methods and missing rates (r) for the employee data}
\centering
\label{Tab:vse_pfc_summary_2}
\begin{tabular}[t]{|l|l|c|c|c|c|c|c|c|}
\hline
\multirow{2}{*}{method}&missing&\multirow{2}{*}{r}&\multirow{2}{*}{$q_{0}$}&\multirow{2}{*}{$q_{0.25}$}&\multirow{2}{*}{$q_{0.5}$}&\multirow{2}{*}{$q_{0.75}$}&\multirow{2}{*}{$q_{1}$}&\multirow{2}{*}{sd}\\
&mechanism&&&&&&&\\
\hline
\hline	 
 
 	 Mice.PMM&MAR&0.01& 0.45 & 0.47 & 0.47 & 0.48 & 0.49 & 0.01 \\
	 &&0.05& 0.47 & 0.48 & 0.48 & 0.48 & 0.50 & 0.01\\
	 &&0.10& 0.49 & 0.50 & 0.50 & 0.50 & 0.51 & <0.01\\
	 \cline{2-9}
	 &MCAR&0.01& 0.46 & 0.47 & 0.47 & 0.48 & 0.49 & 0.01\\
	 &&0.05& 0.47 & 0.48 & 0.48 & 0.49 & 0.50 & 0.01\\
	 &&0.10&  0.50 & 0.50 & 0.50 & 0.51 & 0.51 & <0.01\\
	 \hline
	 Mice.RF&MAR&0.01& 0.44 & 0.45 & 0.46 & 0.46 & 0.48 & 0.01 \\
	 &&0.05& 0.47 & 0.48 & 0.48 & 0.48 & 0.49 & <0.01\\
	 &&0.10& 0.49 & 0.50 & 0.50 & 0.51 & 0.51 & <0.01\\
	 \cline{2-9}
	 &MCAR&0.01& 0.44 & 0.45 & 0.46 & 0.46 & 0.47 & 0.01\\
	 &&0.05& 0.47 & 0.48 & 0.48 & 0.48 & 0.49 & <0.01\\
	 &&0.10&  0.49 & 0.50 & 0.50 & 0.51 & 0.51 & <0.01\\
	 \hline
	 Naive Imputation&MAR&0.01& 0.56 & 0.58 & 0.59 & 0.59 & 0.62 & 0.01 \\
	 &&0.05& 0.57 & 0.58 & 0.59 & 0.59 & 0.61 & 0.01\\
	 &&0.10& 0.58 & 0.59 & 0.59 & 0.59 & 0.61 & 0.01\\
	 \cline{2-9}
	 &MCAR&0.01& 0.56 & 0.58 & 0.59 & 0.59 & 0.61 & 0.01\\
	 &&0.05& 0.58 & 0.58 & 0.59 & 0.59 & 0.60 & <0.01\\
	 &&0.10&  0.58 & 0.59 & 0.59 & 0.59 & 0.59 & <0.01\\
	 \hline
	 MissRanger&MAR&0.01& 0.12 & 0.15 & 0.16 & 0.18 & 0.22 & 0.02 \\
	 &&0.05& 0.14 & 0.16 & 0.17 & 0.19 & 0.22 & 0.02\\
	 &&0.10& 0.15 & 0.17 & 0.18 & 0.20 & 0.23 & 0.02\\
	 \cline{2-9}
	 &MCAR&0.01& 0.13 & 0.16 & 0.17 & 0.17 & 0.22 & 0.02\\
	 &&0.05& 0.14 & 0.16 & 0.17 & 0.18 & 0.22 & 0.02\\
	 &&0.10&  0.15 & 0.17 & 0.18 & 0.20 & 0.24 & 0.02\\
\hline
\end{tabular}
\end{table}

\subsubsection{Continuous Variables}

This Sub-Section covers the results for the continuous outcomes of the employee data set. We first present boxplots of the Kolmogorov-Smirnov statistic and Cramer-von-Mises statistic. At the end of this Sub-Section, a detailed summary statistics for all variables and imputation measures is given. 

\subsubsubsection{Boxplots of the Kolmogorov-Smirnov Statistic}

\begin{figure}[H]
	\centering
	\includegraphics[width=15cm]{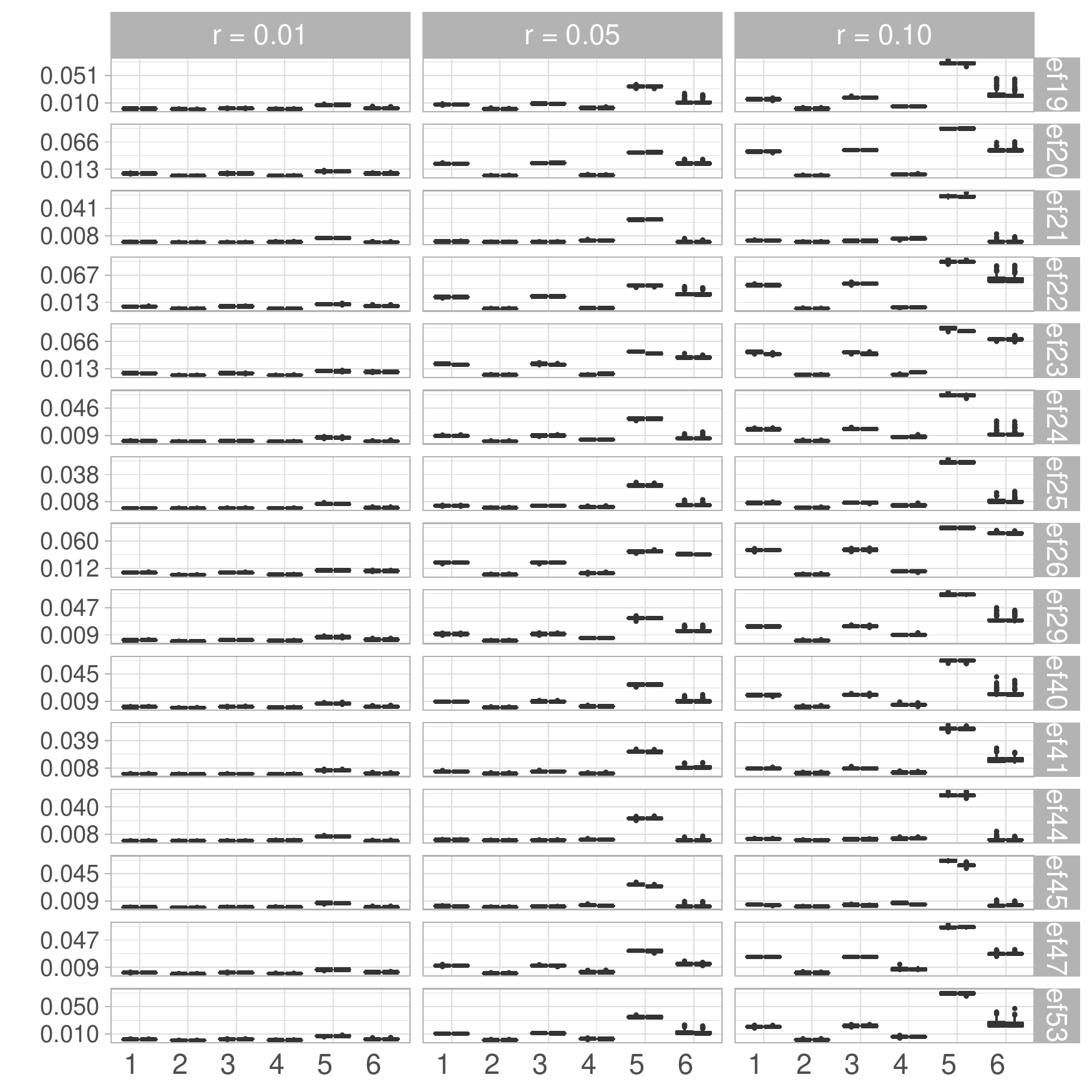}
	\caption{Pair of boxplot for the \textbf{Kolmogorov-Smirnov statistic} using $MC = 100$ iterations. Each boxplot pair corresponds to the following missing mechanism: the left one to the MCAR, and the right one to the MAR mechanism. \\1: \texttt{Amelia}, 2: \texttt{Mice.Norm}, 3: \texttt{Mice.Pmm}, 4: \texttt{Mice.RF}, 5: \texttt{Naive}, 6: \texttt{missRanger}}\label{Supp:KSStat_VREdata}
\end{figure}

For the Kolmogorov-Smirnov statistic (KS-statistic),we first estimated the empirical distribution functions for every variable. Then, the corresponding KS-statistic is computed between the empirical distribution function of the original and imputed data. The results are given in Figure~\ref{Supp:KSStat_VREdata}. It is surprising, that the KS-statistic does not vary much between the different Monte-Carlo iterates. For the different missing mechanisms, no clear difference could be observed and a large impact on the KS-statistic is not deducible. Furthermore, for the variables \texttt{ef21} (gross monthly income), \texttt{ef24} (income tax), \texttt{ef25} (social security contributions), \texttt{ef44} (net monthly income) and \texttt{ef45} (normalized gross income per year), all imputation method indicated a KS -statistic close to zero, except for the \texttt{Naive} method. The reason for this effect is that all mentioned variables are directly related to the \textit{income} variable and can therefore be well-estimated. The regression based approaches in \texttt{Amelia}, the MICE procedures and in \texttt{missRanger} benefit from this effect. In contrast, the \texttt{Naive} method performed for these variables correspondingly worse, since dependence structures are ignored by simply averaging the observed results for missing values. For an increased missing rate, the KS-statistic realized a slight inflation for every variable and imputation method. The \texttt{Mice.Norm} approach resulted into the lowest KS-statistic on average, whereas the \texttt{Naive} method indicated the highest KS-statistic. The Random Forest based MICE procedure \texttt{Mice.RF} was close to the \texttt{Mice.Norm} approach. Moreover, 
differences between the three MICE approaches for the metric case could be observed. \texttt{Mice.Pmm}, performed worse than \texttt{Mice.Norm} behaving similarly to the \texttt{Amelia} procedure. The \texttt{missRanger} method indicated the highest variation of the KS-statistic, but performed similar to \texttt{Mice.Pmm} and \texttt{Amelia}. 

\subsubsubsection{Boxplot of the Cramer-von-Mises Statistic}

\begin{figure}[H]
	\centering
	\includegraphics[width=15cm]{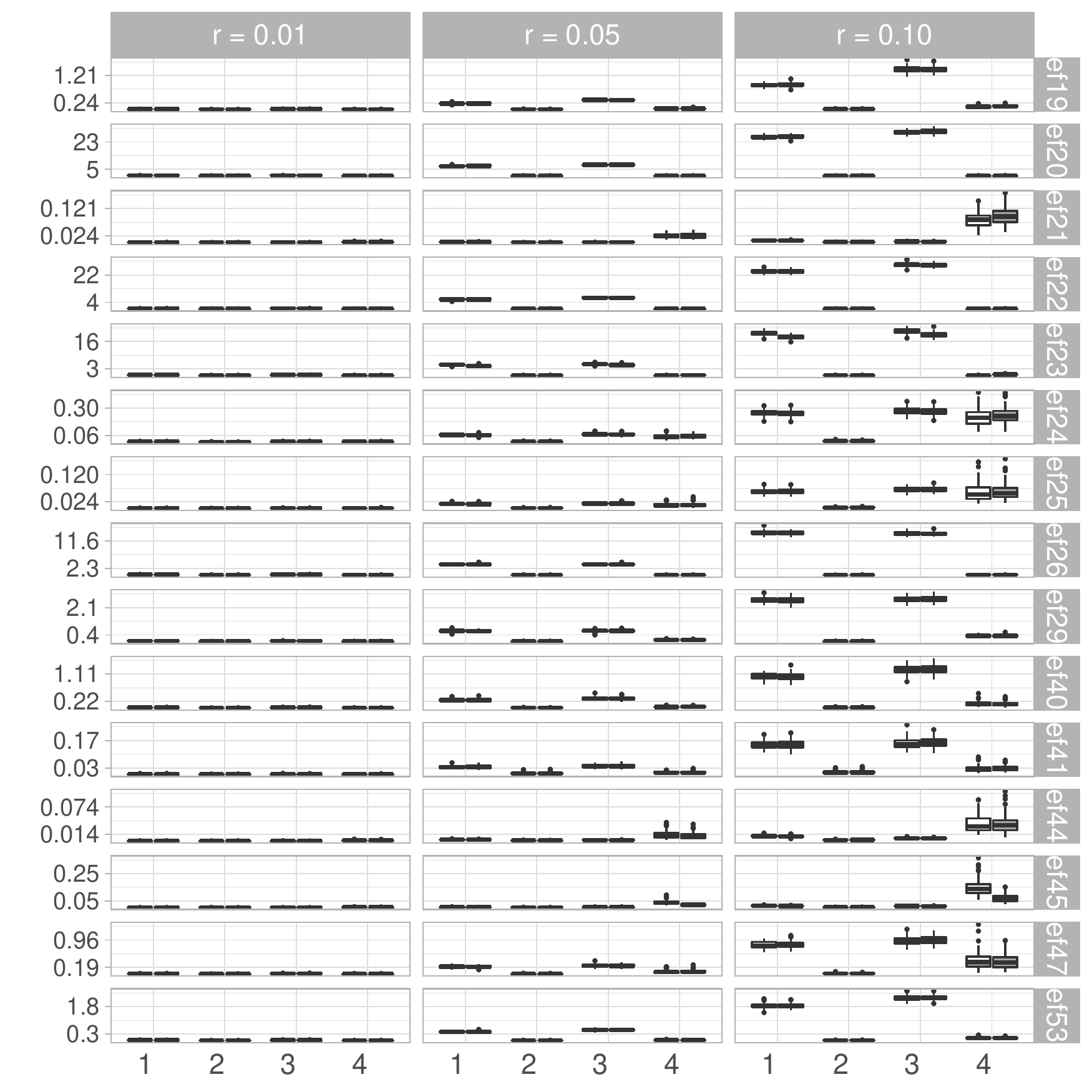}
	\caption{Pair of boxplot for the \textbf{Cramer-von-Mises statistic} using $MC = 100$ iterations. Each boxplot pair corresponds to the following missing mechanism: the left one to the MCAR, and the right one to the MAR mechanism. \\1: \texttt{Amelia}, 2: \texttt{Mice.Norm}, 3: \texttt{Mice.Pmm}, 4: \texttt{Mice.RF}}\label{CSStat_VREdata}
\end{figure}

 The results for the Cramer-von-Mises statistic (CM-statistic) can be found in Figure \ref{CSStat_VREdata}. Due to relatively high realizations of the CM-statistic for the \texttt{Naive} and \texttt{missRanger} imputation method, only the boxplot for the MICE procedures and \texttt{Amelia} is presented. The \texttt{Naive} method indicated a CM-statistic of $0.74, 18.10$ and $71.40$ on average for each missing rate $r \in \{0.01, 0.05, 0.1 \}$ and the MCAR mechanism, while the \texttt{missRanger} approach resulted into CM-statistic values of around $0.54$, $12.91$ and $51.47$. The difference to the MAR mechanism was small. Small, but neglectable differences to the MAR mechanism were observable. The results indicated that both approaches, \texttt{missRanger} and the \texttt{Naive} approach performed worse compared to the other imputation methods. Similarly to the KS-statistic, the CM-statistic inflated on average for an increased missing rate together, while the standard deviation between the different Monte-Carlo runs also increased. Regarding the MICE procedures, the difference between the MCAR and MAR mechanism was slightly greater compared to the KS-statistic. Again, the \texttt{Mice.Norm} approach resulted into the lowest CM-statistics across the different missing rates. Furthermore, it seems that this method is more robust towards an increased missing rate. \texttt{Amelia} and \texttt{Mice.RF} performed similar on average, but the Random Forest based imputation procedure revealed more volatile realizations of the CM statistic than \texttt{Amelia}. However, this is not the case for all variables, since for some of the variables, \texttt{Mice.RF} was also similar to \texttt{Mice.Norm}. Compared to these method, the \texttt{Mice.Pmm} approach performed worse for all variables.

\subsubsubsection{Association of Distance Measures}

Additionally to the obtained results, we were interested whether the different distributional discrepancy measures behave similarly, i.e show the same directional behaviour. Therefore, we computed correlation matrices for the Kolmogorov-Smirnov statistic (KS), the Cramer-von-Mises statistic (CM), the Kullback-Leibler divergence (KL) and Mallows $L^2$ distance (ML2). The result is given in Table \ref{Tab:Correlation_VSE}. Note that the latter has been computed on all imputation methods, missing mechanisms and Monte-Carlo iterations. The result indicates that all measures behave similarly, that is, if a method performs worse in terms of one of the given measures, the other measures tend to indicate a similar effect.

\begin{table}[H]
\caption{Pearson's correlation over all imputation methods and missing mechanisms as well as the Monte-Carlo iteration based on the employee data set and its metric variables.}
\centering
\label{Tab:Correlation_VSE}
\begin{tabular}[t]{|l|c|c|c|c|}
\hline
 & CS & KS & KL & ML2 \\
\hline
\hline	 
	 CS  & 1 & 0.938 & 0.903 &  0.891  \\
	 KS  &  0.938 & 1 &  0.883 &  0.888  \\
	 KL  & 0.903 & 0.883 & 1 & 0.935 \\
	 ML2  & 0.892 & 0.888 &  0.935 & 1  \\
\hline
\end{tabular}
\end{table}

\subsubsubsection{Summary Statistics}
\input{table_fivepoint_KolmogorowSmirnow_vse.txt}
\input{table_fivepoint_CramerVonMises_vse.txt}
\input{table_fivepoint_KullbackLeibler_vse.txt}
\input{table_fivepoint_Mallows_vse.txt}

\begin{table}[H]
\caption{minimum ($q_0$), first quartile ($q_{0.25}$),
               median ($q_{0.5}$), third quartile ($q_{0.75}$),
               maximum ($q_{1}$) and standard deviation (sd) of the observed values of the NRMSE for the different
               imputation methods and missing rates (r) for the employee data}
\centering
\label{Tab:vse_nrmse_summary_1}
\begin{tabular}[t]{|l|l|c|c|c|c|c|c|c|}
\hline
\multirow{2}{*}{method}&missing&\multirow{2}{*}{r}&\multirow{2}{*}{$q_{0}$}&\multirow{2}{*}{$q_{0.25}$}&\multirow{2}{*}{$q_{0.5}$}&\multirow{2}{*}{$q_{0.75}$}&\multirow{2}{*}{$q_{1}$}&\multirow{2}{*}{sd}\\
&mechanism&&&&&&&\\
\hline
\hline	 
	 Amelia&MAR&0.01& 0.13 & 0.14 & 0.15 & 0.15 & 0.18 & 0.01 \\
	 &&0.05 & 0.14 & 0.15 & 0.15 & 0.16 & 0.17 &  <0.01\\
	 &&0.10 & 0.14 & 0.15 & 0.15 &  0.16 & 0.17 & <0.01\\
	 \cline{2-9}
	 &MCAR&0.01& 0.12 & 0.13 & 0.14 & 0.15 & 0.17 & 0.01\\
	 &&0.05& 0.13 & 0.14 & 0.14 & 0.14 & 0.15 & <0.01\\
	 &&0.10 &  0.14 & 0.14 & 0.15 & 0.15 & 0.15 & <0.01\\
	 \hline
	 Mice.Norm&MAR&0.01&  0.11 & 0.13 & 0.14 & 0.14 & 0.18 & 0.01 \\
	 &&0.05& 0.13 & 0.14 & 0.14 & 0.15 & 0.15 & <0.01\\
	 &&0.10& 0.14 & 0.15 & 0.15 & 0.15 & 0.16 & <0.01\\
	 \cline{2-9}
	 &MCAR&0.01& 0.10 & 0.12 & 0.13 & 0.14 & 0.16 & 0.01\\
	 &&0.05& 0.12 & 0.13 & 0.13 & 0.14 & 0.15 & <0.01\\
	 &&0.10&  0.14 & 0.14 & 0.14 & 0.14 & 0.15 & <0.01\\
	 \hline
	 Mice.PMM&MAR&0.01& 0.12 & 0.13 & 0.14 & 0.15 & 0.18 & 0.01 \\
	 &&0.05&0.14 & 0.14 & 0.15 & 0.15 & 0.16 & <0.01\\
	 &&0.10& 0.15 & 0.15 & 0.15 &  0.16 & 0.17 & <0.01\\
	 \cline{2-9}
	 &MCAR&0.01& 0.11 & 0.13 & 0.13 & 0.14 & 0.17 & 0.01\\
	 &&0.05& 0.13 & 0.14 & 0.14 & 0.14 & 0.15 & <0.01\\
	 &&0.10&  0.14 & 0.14 & 0.14 & 0.15 & 0.15 & <0.01\\
	 \hline
	 Mice.RF&MAR&0.01& 0.50 & 0.58 & 0.61 & 0.63 & 0.68 & 0.04 \\
	 &&0.05& 0.51 & 0.56 & 0.57 & 0.60 & 0.70 & 0.03\\
	 &&0.10& 0.49 & 0.56 & 0.58 &  0.60 & 0.65 & 0.03\\
	 \cline{2-9}
	 &MCAR&0.01& 0.51 & 0.57 & 0.59 & 0.61 & 0.66 & 0.03\\
	 &&0.05& 0.49 & 0.55 & 0.57 & 0.59 & 0.66 & 0.03\\
	 &&0.10&  0.48 & 0.54 & 0.56 & 0.58 & 0.64 & 0.03\\
\hline
\end{tabular}
\end{table}

\begin{table}[H]
\caption{minimum ($q_0$), first quartile ($q_{0.25}$),
               median ($q_{0.5}$), third quartile ($q_{0.75}$),
               maximum ($q_{1}$) and standard deviation (sd) of the observed values of the NRMSE for the different
               imputation methods and missing rates (r) for the employee data}
\centering
\label{Tab:vse_nrmse_summary_2}
\begin{tabular}[t]{|l|l|c|c|c|c|c|c|c|}
\hline
\multirow{2}{*}{method}&missing&\multirow{2}{*}{r}&\multirow{2}{*}{$q_{0}$}&\multirow{2}{*}{$q_{0.25}$}&\multirow{2}{*}{$q_{0.5}$}&\multirow{2}{*}{$q_{0.75}$}&\multirow{2}{*}{$q_{1}$}&\multirow{2}{*}{sd}\\
&mechanism&&&&&&&\\
\hline
\hline	 
	 Naive Imputation&MAR&0.01& 0.57 & 0.60 & 0.61 & 0.62 & 0.67 & 0.02 \\
	 &&0.05& 0.59 & 0.61 & 0.61 & 0.62 & 0.65 & 0.01\\
	 &&0.10& 0.60 & 0.61 & 0.62 & 0.62 & 0.64 & 0.01\\
	 \cline{2-9}
	 &MCAR&0.01& 0.56 & 0.58 & 0.58 & 0.59 & 0.62 & 0.01\\
	 &&0.05& 0.57 & 0.58 & 0.59 & 0.59 & 0.60 & 0.01\\
	 &&0.10 &  0.57 & 0.58 & 0.59 & 0.59 & 0.60 & <0.01\\
	 \hline
	 MissRanger&MAR&0.01& 0.07 & 0.08 & 0.10 & 0.11 & 0.26 & 0.04 \\
	 &&0.05& 0.08 & 0.09 & 0.10 & 0.11 & 0.25 & 0.03\\
	 &&0.10& 0.08 & 0.09 & 0.10 &  0.11 & 0.18 & 0.02\\
	 \cline{2-9}
	 &MCAR&0.01& 0.06 & 0.08 & 0.10 & 0.11 & 0.26 & 0.04\\
	 &&0.05& 0.07 & 0.08 & 0.09 & 0.10 & 0.24 & 0.03\\
	 &&0.10 &  0.08 & 0.09 & 0.09 & 0.10 & 0.20 & 0.02\\
\hline
\end{tabular}

\end{table}
\newpage

\subsection{Hospital Data}
For the simulation on the hospital data, only random subsets per Monte-Carlo iteration are considered. This, because the treatment of all observations would be too time consuming. Therefore, at every Monte-Carlo iteration, we randomly selected $10,000$ observations, while the $40$ variables have been modified as described in the first section replacing secondary diagnosis and surgery/medical procedures variable by corresponding dummy variables. The introduction of artificial missing values is not conducted on all variables, but only to the following nine variables: \texttt{sex}, \texttt{type\_alter}, \texttt{aufn\_anl}, \texttt{entl\_grd}, \texttt{typ\_vwd}, \texttt{typ\_op}, \texttt{fab\_max}, \texttt{cm\_vol} and \texttt{beatm}. Furhtermore, we did not introduce missing values under both, the MCAR and MAR mechanism, but focused only on the MCAR mechanism. The upcoming results are therefore restricted to the MCAR mechanism for the hospital data set. 

\subsubsection{Imputation Accuracy: NRMSE \& PFC}

\begin{figure}[H]
	\centering
	\subfloat[ $NRMSE$ ]{{\includegraphics[width=7cm]{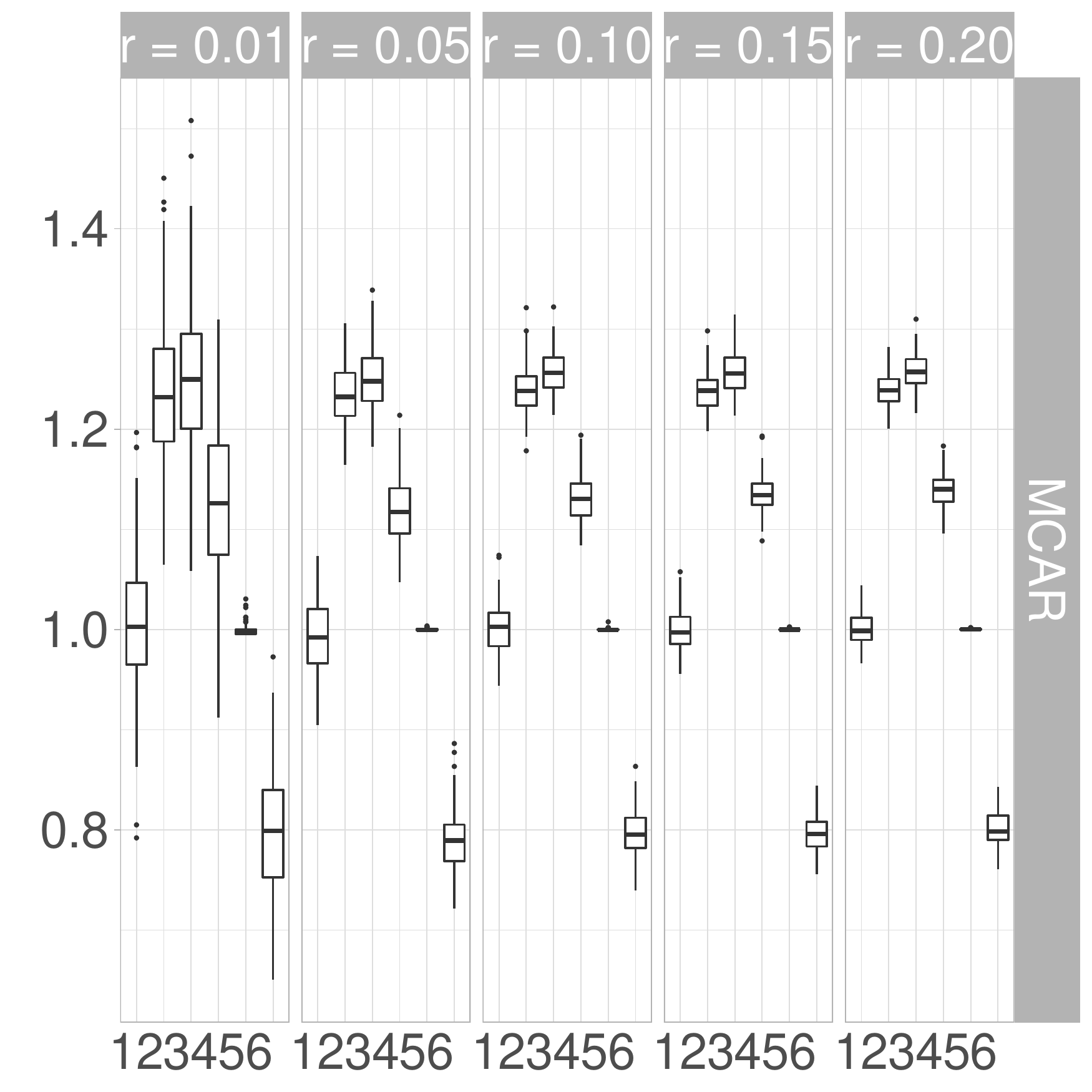} }}%
	\qquad
	\subfloat[$PFC$]{{\includegraphics[width=7cm]{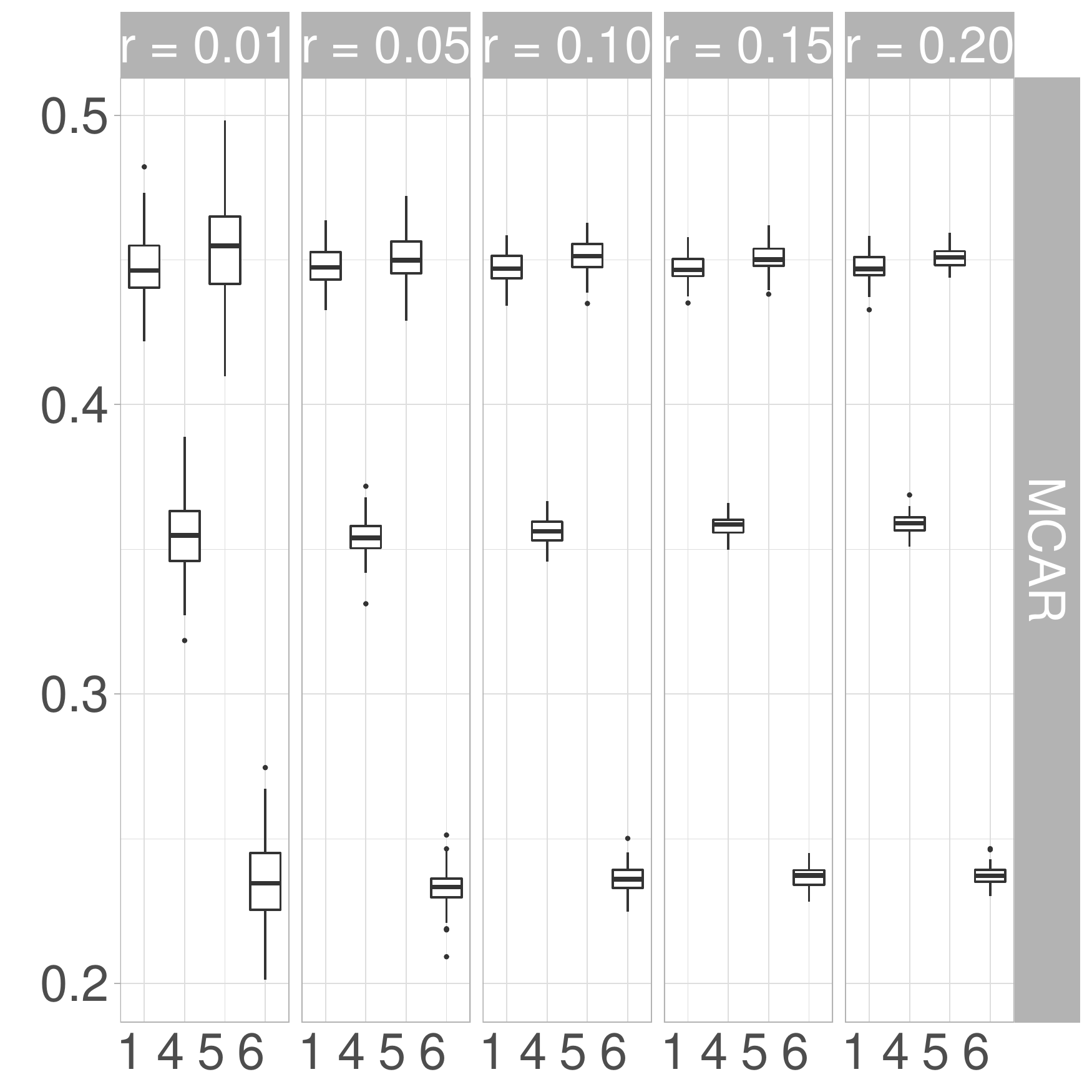} }}%
	\caption{Boxplot for the imputation accuracy using $MC = 100$ iterations on the \textbf{hospital data set}. 1: \texttt{Amelia}, 2: \texttt{Mice.Norm}, 3: \texttt{Mice.Pmm}, 4: \texttt{Mice.RF}, 5: \texttt{Naive}, 6: \texttt{missRanger}} \label{BoxPlot_NRMSE_PFC_Hospital}
\end{figure}

When considering the NRMSE and PFC as an imputation quality measure on the hospital data, one quickly realize the comparably low results of the \texttt{missRanger} procedure given in Figure \ref{BoxPlot_NRMSE_PFC_Hospital} for both, continuous and categorical outcomes. The error stabilizes with an increased missing rate being close to $0.8$ for the NRMSE or $0.24$ for the PFC. In addition, the method was more robust towards varying missing rates. The MICE procedures indicated higher PFC values compared to the \texttt{missRanger} approach and were also more robust towards increasing missing rates. \texttt{Amelia} and the \texttt{Naive} method performed for categorical outcomes comparable with PFC values around $0.45$. For continuous outcomes, \texttt{Amelia} indicated on average lower NRMSE values than the MICE procedures. Especially the predictive mean matching approach resulted into the highest NRMSE values being comparable to the \texttt{Mice.Norm} method.

\begin{table}[H]
\caption{minimum ($q_0$), first quartile ($q_{0.25}$),
               median ($q_{0.5}$), third quartile ($q_{0.75}$),
               maximum ($q_{1}$) and standard deviation (sd) of the observed values of the NRMSE for the different
               imputation methods and missing rates (r) for the hospital data}
\centering
\label{Tab:drg_nrmse_summary}
\begin{tabular}[t]{|l|c|c|c|c|c|c|c|}
\hline
method&r&$q_{0}$& $q_{0.25}$&$q_{0.5}$&$q_{0.75}$& $q_{1}$&sd\\
\hline
	 
	 \hline
	 Amelia&0.01& 0.79 & 0.97 & 1.00 & 1.05 & 1.20 & 0.07 \\
	 &0.05& 0.90 & 0.97 & 0.99 & 1.02 & 1.07 & 0.04\\
	 &0.10& 0.94 & 0.98 & 1.00 & 1.02 & 1.07 & 0.03\\
	 &0.15& 0.96 & 0.99 & 1.00 & 1.01 & 1.06 & 0.02\\
	 &0.20& 0.98 & 0.99 & 1.00 & 1.01 & 1.04 & 0.02\\
	 
	 \hline
	 Mice.Norm&0.01& 1.06 & 1.19 & 1.23 & 1.28 & 1.45 & 0.08\\
	 &0.05& 1.16 & 1.21 & 1.23 & 1.26 & 1.31 & 0.03\\
	 &0.10& 1.18 & 1.22 & 1.24 & 1.25 & 1.32 & 0.02\\
	 &0.15& 1.20 & 1.22 & 1.24 & 1.25 & 1.30 & 0.02\\
	 &0.20& 1.20 & 1.23 & 1.24 & 1.25 & 1.28 & 0.02 \\
	 
	 \hline
	 Mice.PMM&0.01& 1.06 & 1.20 & 1.25 & 1.30 & 1.51 & 0.07\\
	 &0.05& 1.18 & 1.23 & 1.25 & 1.27 & 1.34 & 0.03\\
	 &0.10& 1.21 & 1.24 & 1.26 & 1.27 & 1.32 & 0.02\\
	 &0.15& 1.21 & 1.24 & 1.26 & 1.27 & 1.31 & 0.02\\
	 &0.20& 1.22 & 1.25 & 1.26 & 1.27 & 1.31 & 0.02\\
	 
	 \hline
	 Mice.RF&0.01&  0.91 & 1.07 & 1.13 & 1.18 & 1.31 & 0.08\\
	 &0.05& 1.05 & 1.10 & 1.12 & 1.14 & 1.21 & 0.04\\
	 &0.10& 1.08 & 1.11 & 1.13 & 1.15 & 1.19 & 0.02\\
	 &0.15& 1.09 & 1.12 & 1.13 & 1.15 & 1.19 & 0.02\\
	 &0.20& 1.10 & 1.13 & 1.14 & 1.15 & 1.18 & 0.02\\
	 
	 \hline
	 Naive Imputation&0.01& 0.99 & 1.00 & 1.00 & 1.00 & 1.03 & 0.01 \\
	 &0.05& 1.00 & 1.00 & 1.00 & 1.00 & 1.00 & <0.01\\
	 &0.10& 1.00 & 1.00 & 1.00 & 1.00 & 1.01 & <0.01\\
	 &0.15& 1.00 & 1.00 & 1.00 & 1.00 & 1.00 & <0.01\\
	 &0.20&  1.00 & 1.00 & 1.00 & 1.00 & 1.00 & <0.01\\
	 
	 \hline
	 MissRanger&0.01& 0.65 & 0.75 & 0.80 & 0.84 & 0.97 & 0.07\\
	 &0.05&  0.72 & 0.77 & 0.79 & 0.81 & 0.89 & 0.03\\
	 &0.10& 0.74 & 0.78 & 0.80 & 0.81 & 0.86 & 0.02\\
	 &0.15& 0.76 & 0.78 & 0.80 & 0.81 & 0.84 & 0.02\\
	 &0.20& 0.77 & 0.79 & 0.80 & 0.82 & 0.84 & 0.02 \\
\hline
\end{tabular}
\end{table}

\newpage
\begin{table}[H]
\caption{minimum ($q_0$), first quartile ($q_{0.25}$),
               median ($q_{0.5}$), third quartile ($q_{0.75}$),
               maximum ($q_{1}$) and standard deviation (sd) of the observed values of the PFC for the different
               imputation methods and missing rates (r) for the hospital data}
\centering
\label{Tab:drg_pfc_summary}
\begin{tabular}[t]{|l|c|c|c|c|c|c|c|}
\hline
method&r&$q_{0}$& $q_{0.25}$&$q_{0.5}$&$q_{0.75}$& $q_{1}$&sd\\
\hline
 
	 \hline
	 Amelia&0.01& 0.42 & 0.44 & 0.45 & 0.45 & 0.48 & 0.01\\
	 &0.05& 0.43 & 0.44 & 0.45 & 0.45 & 0.46 & 0.01\\
	 &0.10& 0.43 & 0.44 & 0.45 & 0.45 & 0.46 & 0.01\\
	 &0.15& 0.44 & 0.44 & 0.45 & 0.45 & 0.46 & <0.01\\
	 &0.20& 0.44 & 0.44 & 0.45 & 0.45 & 0.46 & <0.01\\
	 
	 \hline
	 Mice.Norm&0.01& 0.32 & 0.35 & 0.35 & 0.37 & 0.40 & 0.01\\
	 &0.05& 0.34 & 0.35 & 0.35 & 0.36 & 0.37 & 0.01\\
	 &0.10& 0.35 & 0.35 & 0.36 & 0.36 & 0.37 & <0.01\\
	 &0.15& 0.35 & 0.36 & 0.36 & 0.36 & 0.37 & <0.01\\
	 &0.20& 0.35 & 0.36 & 0.36 & 0.36 & 0.37 & <0.01 \\
	 
	 \hline
	 Mice.PMM&0.01& 0.32 & 0.34 & 0.35 & 0.36 & 0.39 & 0.01\\
	 &0.05& 0.34 & 0.35 & 0.35 & 0.36 & 0.37 & 0.01\\
	 &0.10& 0.35 & 0.35 & 0.36 & 0.36 & 0.37 & <0.01\\
	 &0.15&  0.35 & 0.36 & 0.36 & 0.36 & 0.37 & <0.01 \\
	 &0.20& 0.35 & 0.36 & 0.36 & 0.36 & 0.37 & <0.01\\
	 
	 \hline
	 Mice.RF&0.01& 0.32 & 0.35 & 0.35 & 0.36 & 0.39 & 0.01\\
	 &0.05& 0.33 & 0.35 & 0.35 & 0.36 & 0.37 & 0.01\\
	 &0.10& 0.35 & 0.35 & 0.36 & 0.36 & 0.37 & <0.01\\
	 &0.15& 0.35 & 0.36 & 0.36 & 0.36 & 0.37 & <0.01\\
	 &0.20& 0.35 & 0.36 & 0.36 & 0.36 & 0.37 & <0.01\\
	 
	 \hline
	 Naive Imputation&0.01& 0.41 & 0.44 & 0.45 & 0.47 & 0.50 & 0.02\\
	 &0.05& 0.43 & 0.45 & 0.45 & 0.46 & 0.47 & 0.01\\
	 &0.10& 0.43 & 0.45 & 0.45 & 0.46 & 0.46 & 0.01\\
	 &0.15 & 0.44 & 0.45 & 0.45 & 0.45 & 0.46 & <0.01\\
	 &0.20 &  0.45 & 0.45 & 0.45 & 0.45 & 0.46 & <0.01\\
	 
	 \hline
	 MissRanger&0.01& 0.20 & 0.23 & 0.23 & 0.25 & 0.27 & 0.01\\
	 &0.05& 0.21 & 0.23 & 0.23 & 0.24 & 0.25 & 0.01\\
	 &0.10& 0.22 & 0.23 & 0.24 & 0.24 & 0.25 & <0.01\\
	 &0.15& 0.23 & 0.23 & 0.24 & 0.24 & 0.25 & <0.01\\
	 &0.20& 0.23 & 0.24 & 0.24 & 0.24 & 0.25 & <0.01\\
\hline
\end{tabular}
\end{table}

\subsubsection{Nominal and Ordinal Variables}

Distributional differences of the categorical variables of the hospital data set are measured through $\kappa$, which has been defined in the main article. The results can be found in Table \ref{KappaVal_Hospital}. Since the MICE procedures use all the Random Forest method for imputing categorical outcomes, the results are the same. It is noticable that $\kappa$ increased, when the missing rate increased for all considered imputation methods. However, the increments where much smaller under the \texttt{missRanger} method indicating a stronger dependence between the original and imputed data set. The lowest values could be observed for the \texttt{missRanger} method. 

\begin{table}[H]
\caption{Realizations of $\kappa$ on the \textbf{hospital data set} using  $MC = 100$ iterations.}\label{KappaVal_Hospital}
\begin{minipage}[t]{\textwidth}
\centering
\label{Tab:drg_chisq_diff_max}
\begin{tabular}[t]{|l|c|c|c|c|c|}
\hline
	& \multicolumn{5}{c|}{Missing rate $r$}\\
	\cline{2-6}
	& 1\,\% & 5\,\% & 10\,\% & 15\,\% & 20\,\% \\
    \hline

	\texttt{Amelia} & $0.025$ & $0.110$ & $0.196$ & $0.269$ & $0.333$\\

	\texttt{Mice.Norm} & $0.013$ & $0.063$ & $0.124$ & $0.185$ & $0.243$\\

	\texttt{Mice.Pmm} & $0.013$ & $0.063$ & $0.125$ &  $0.185$& $0.243$\\
	
	\texttt{Mice.RF}& $0.013$ & $0.063$ & $0.124$ & $0.185$ & $0.243$ \\
	
	\texttt{Naive}  & $0.013$ & $0.064$ & $0.125$ & $0.182$ & $ 0.236 $ \\

	\texttt{missRanger} & $0.008$ & $0.041$ & $0.081$ & $0.121$ & $0.160$\\
	
\hline
\end{tabular}
\end{minipage}
\end{table}

\begin{figure}[H]
	\centering
	\includegraphics[width=15cm]{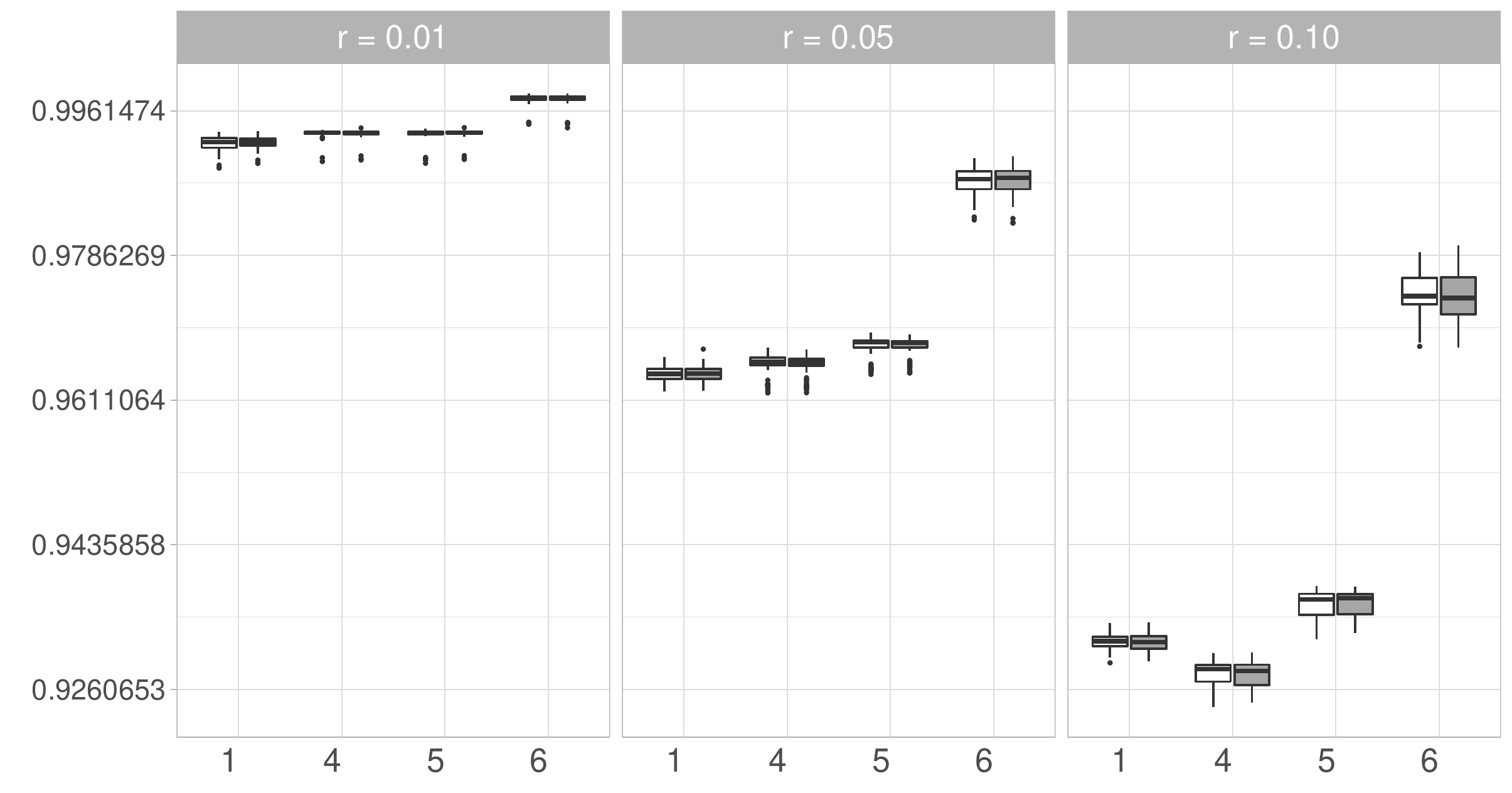}
	\caption{Pair of boxplot of \textbf{Cramérs V} on the hospital data set using $MC = 100$ iterations. Each boxplot pair corresponds to the following missing mechanism: the left one to the MCAR, and the right one to the MAR mechanism. \\1: \texttt{Amelia}, 2: \texttt{Mice.Norm}, 3: \texttt{Mice.Pmm}, 4: \texttt{Mice.RF}, 5: \texttt{Naive}, 6: \texttt{missRanger}}\label{Boxplot_CramersVal_Hospital}
\end{figure}

In addition to the $\kappa$ values presented above, we give a brief overview on Cramérs V measure in form of a boxplot in Figure \ref{Boxplot_CramersVal_Hospital}. Since the MICE methods all make use of a Random-Forest based imputation scheme, we only report the \texttt{Mice.RF} method at this step. It is noticable that the \texttt{missRanger} method resulted into the highest values indicating stronger dependencies than the other methods. \texttt{Amelia} performed worse, while the \texttt{Mice.RF} procedure yielded only for $r = 0.1$ the lowest Cramérs V measure on average. Note that an increased missing rate have a negative effect resulting on lower Cramérs V measures across the imputation methods on average. 

\subsubsubsection{Association of Distance Measures}

Additionally to the obtained results, we were interested whether the different distributional discrepancy measures behave similarly, i.e show the same directional behaviour. We computed correlation matrices for distance measures of ordinal variables. Hence, we took into consideration the Cramérs V measure (CV), the Cramer-von-Mises statistic (CM) and the Kolmogorov-Smirnov statistic (KS). The result is given in Table \ref{Tab:Correlation_Ordinal_DRG} and indicates that all measures behave similarly, that is, if a method performs worse in terms of one of the given measures, the other measures tend to indicate a similar effect. The negative correlation results from the adverse scale of Cramérs V, where higher values indicate stronger dependencies between imputed and original data set.

\begin{table}[H]
\caption{Pearson's correlation over all imputation methods  as well as the Monte-Carlo iteration based on the hospital data set and its ordinal variables.}
\centering
\label{Tab:Correlation_Ordinal_DRG}
\begin{tabular}[t]{|l|c|c|c|}
\hline
 & CV & CM & KL \\
\hline
\hline	 
	 CV  & 1 & -0.887 & -0.953   \\
	 CM  &  -0.887 & 1 &  0.952  \\
	 KL  & -0.954 &  0.952 & 1  \\
\hline
\end{tabular}
\end{table}

\subsubsubsection{Summary Statistics}
\pagestyle{empty}
\input{table_fivepoint_Chisq_drg.txt}
\input{table_fivepoint_CramersV_drg.txt}
\pagestyle{plain}

\subsubsection{Continuous Variables}

We consider four variables in the hospital data set for the computation of the Kolmogorov-Smirnov statistic, the Cramer-von-Mises statistic, the Kullback-Leibler divergence and Mallows $L^2$ distance.

\subsubsubsection{Two-sample Kolmogorov-Smirnov Test}

\begin{figure}[H]
	\centering
	\includegraphics[width=15cm]{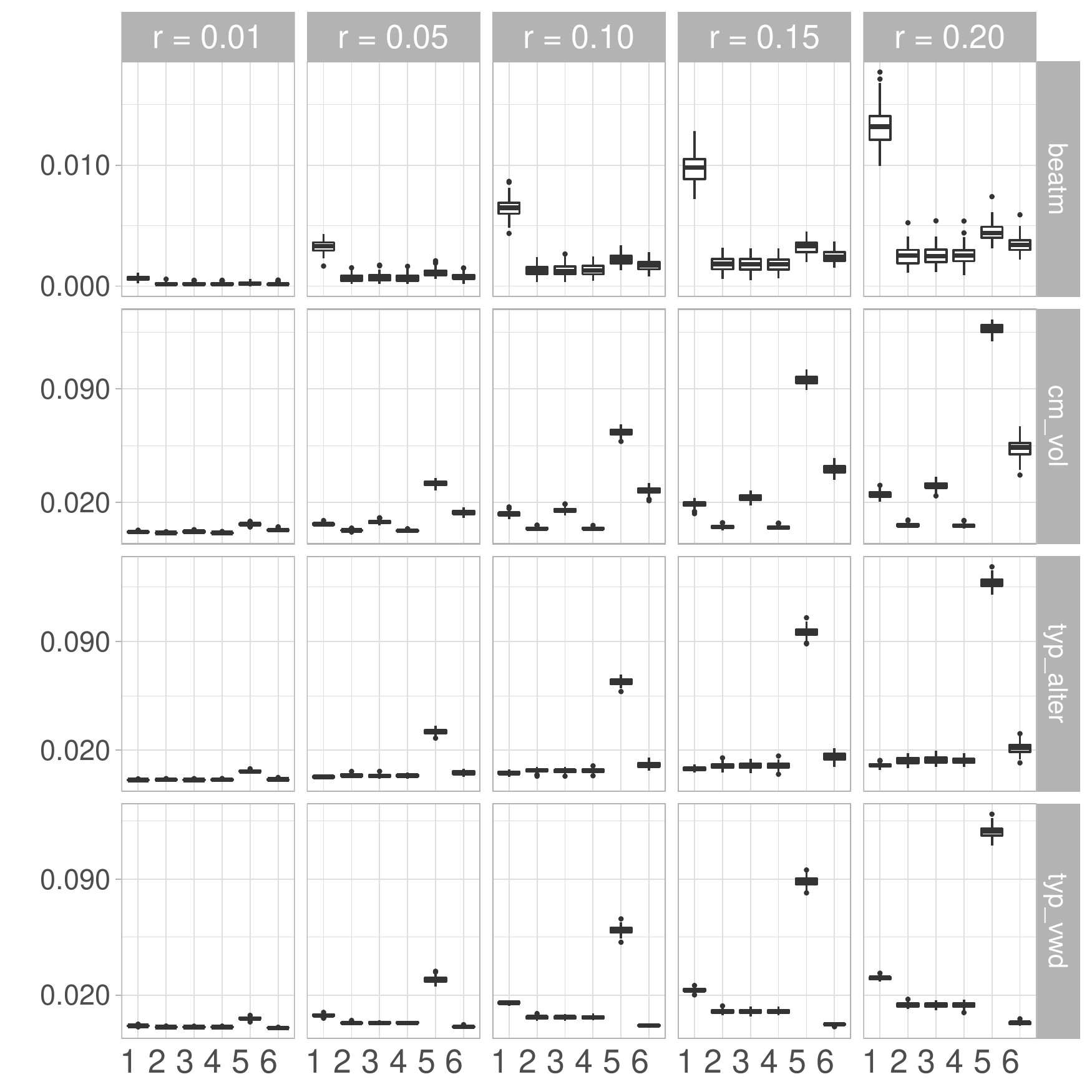}
	\caption{Boxplot of the \textbf{Kolmogorov-Smirnov statistic} using $MC = 100$ iterations on the \textbf{hospital data}. 1: \texttt{Amelia}, 2: \texttt{Mice.Norm}, 3: \texttt{Mice.Pmm}, 4: \texttt{Mice.RF}, 5: \texttt{Naive}, 6: \texttt{missRanger}}\label{KSStat_hospital}
\end{figure}

Realizations of the Kolmogorov-Smirnov (KS) statistic can be found in Figure \ref{KSStat_hospital}. Therein, it can be seen that the \texttt{Amelia} procedure resulted into the highest realizations of the KS statistic, especially for variable \texttt{beatm}. For the other variables, the \texttt{Naive} method indicated the highest KS statistic, while \texttt{missRanger} or \texttt{Amelia} resulted into the second highest realization alternately. The best results could be achieved when using the MICE procedures. Therein, all three methods performed comparably well. It is noticeable that an increased missing rate resulted into higher KS statistics throughout the different methods. 

\subsubsubsection{Two-sample Cramer-von-Mises Statistic}

\begin{figure}[H]
	\centering
	\includegraphics[width=15cm]{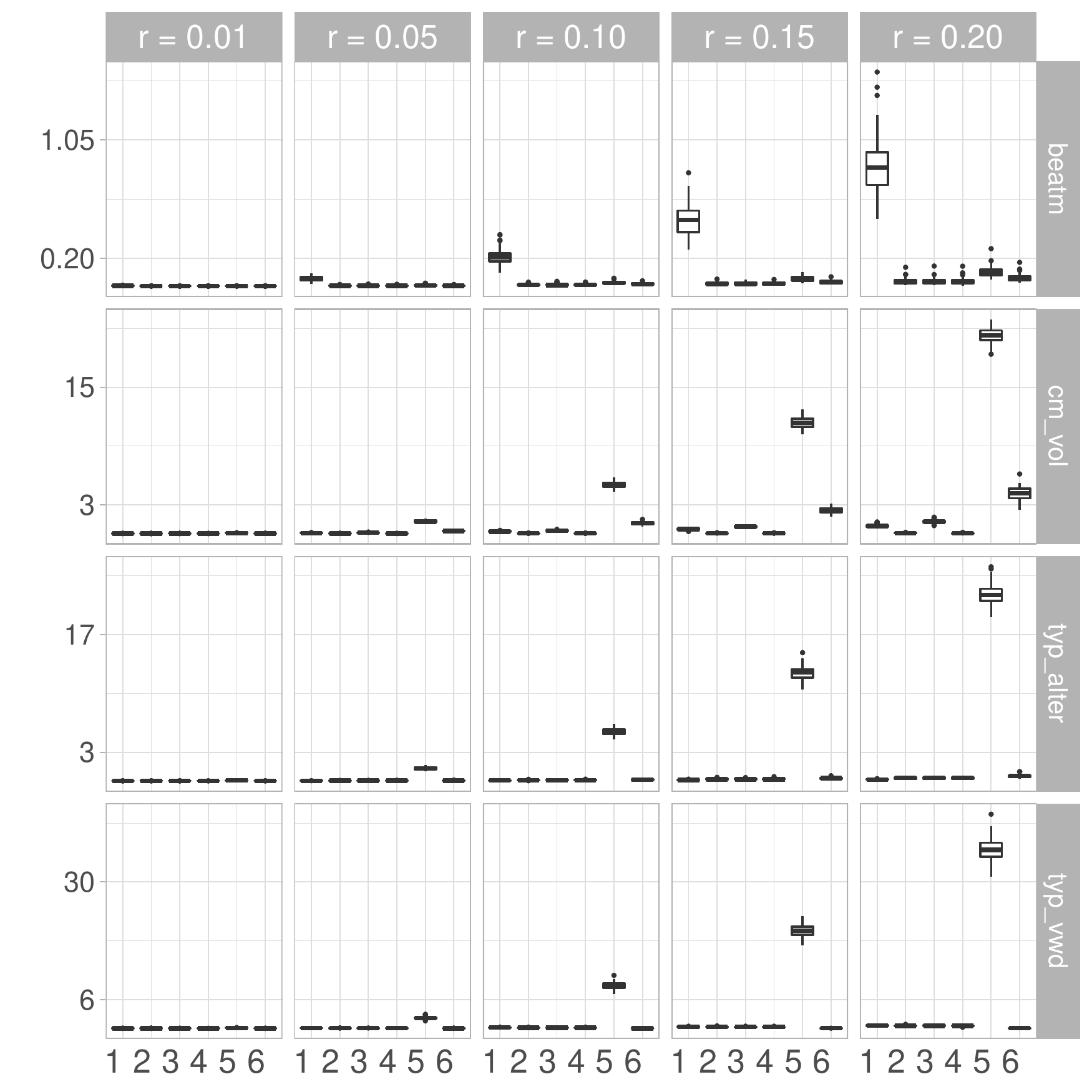}
	\caption{Boxplot of the \textbf{Cramer-von-Mises statistic} using $MC = 100$ iterations on the \textbf{hospital data}. 1: \texttt{Amelia}, 2: \texttt{Mice.Norm}, 3: \texttt{Mice.Pmm}, 4: \texttt{Mice.RF}, 5: \texttt{Naive}, 6: \texttt{missRanger}}\label{CMStat_hospital}
\end{figure}

The Cramer-von-Mises (CM) statistic can be found in Figure \ref{CMStat_hospital}. The results are similar to the KS statistic, but realized larger realizations for the CM statistic in general. \texttt{Amelia}, the \texttt{Naive} method and \texttt{missRanger} performed worse alternately, while all three MICE procedures indicated the lowest CM statistics.

\subsubsubsection{Kullback-Leibler Divergence}

\begin{figure}[H]
	\centering
	\includegraphics[width=15cm]{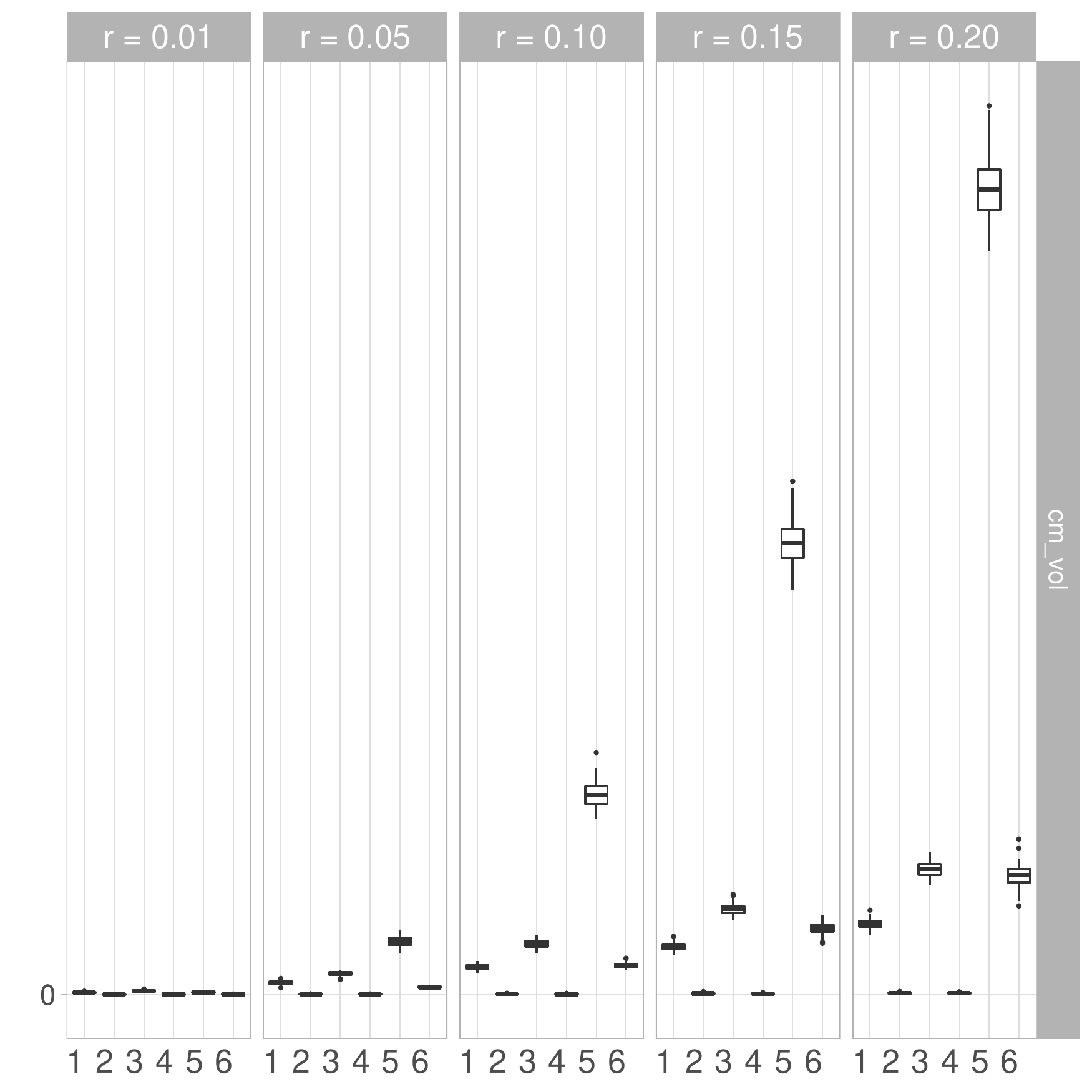}
	\caption{Boxplot of the \textbf{Kullback-Leiber Divergence} using $MC = 100$ iterations on the \textbf{hospital data}. 1: \texttt{Amelia}, 2: \texttt{Mice.Norm}, 3: \texttt{Mice.Pmm}, 4: \texttt{Mice.RF}, 5: \texttt{Naive}, 6: \texttt{missRanger}}\label{KLDiv_hospital}
\end{figure}

\begin{figure}[H]
	\centering
	\includegraphics[width=15cm]{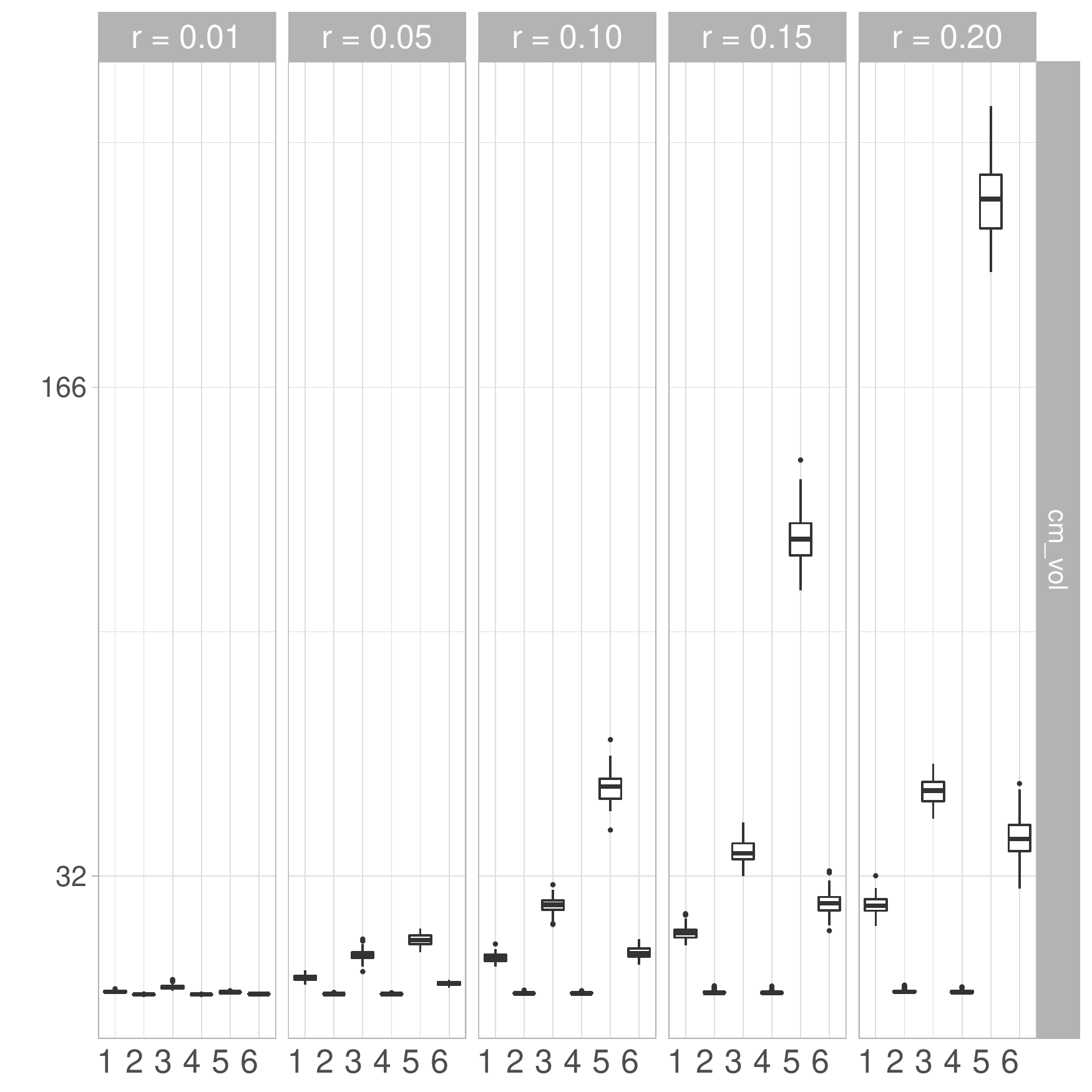}
	\caption{Boxplot of the \textbf{Mallows $\mathbf{L^2}$ statistic} using $MC = 100$ iterations on the \textbf{hospital data}. 1: \texttt{Amelia}, 2: \texttt{Mice.Norm}, 3: \texttt{Mice.Pmm}, 4: \texttt{Mice.RF}, 5: \texttt{Naive}, 6: \texttt{missRanger}}\label{Mallow_hospital}
\end{figure}

Different to the previous results, the Kullback-Leibler (KL) divergence and Mallows $L^2$ (ML2) distance can only be computed for realizations of continuous random variables. Since only one variable in the hospital data set is continuous, Figures \ref{KLDiv_hospital} and \ref{Mallow_hospital}. The results for the KL divergence and ML2 distance are similar. While the \texttt{Naive} method indicated the highest results and therefore the greatest distributional difference under these measures, the \texttt{Mice.Norm} approach and the \texttt{Mice.RF} method yielded favourable results. Different to the previous measures, the \texttt{Mice.Pmm} method performed worse compared to the other MICE method and \texttt{missRanger}. \texttt{Amelia} was only slightly better than \texttt{Mice.Pmm}. Similarly, all imputation methods were negatively affected by an increased missing rate in terms of distributional discrepancies under both measures.

\subsubsubsection{Mallows $L^2$-Distance}

\subsubsubsection{Association of Distance Measures}

We were also interested whether the different distributional discrepancy measures behave similarly, i.e show the same directional behaviour. Therefore, we computed correlation matrices for the Kolmogorov-Smirnov statistic (KS), the Cramer-von-Mises statistic (CM), the Kullback-Leibler divergence (KL) and Mallows $L^2$ distance (ML2). The result is given in Table \ref{Tab:Correlation_metric_hospital}. Note that the latter has been computed on all imputation methods, missing mechanisms and Monte-Carlo iterations. The result indicate that all measures behave similarly, that is, if a method performs worse in terms of one of the given measures, the other measures tend to indicate a similar effect.

\begin{table}[H]
\caption{Pearson's correlation over all imputation methods  as well as the Monte-Carlo iteration based on the hospital data set and its metric variables.}
\centering
\label{Tab:Correlation_metric_hospital}
\begin{tabular}[t]{|l|c|c|c|c|}
\hline
 & CS & KS & KL & ML2 \\
\hline
\hline	 
	 CS  & 1 & 0.945 & 0.933 &  0.940  \\
	 KS  &  0.945 & 1 & 0.948 &  0.923  \\
	 KL  & 0.933 & 0.948 & 1 & 0.954\\
	 ML2  &  0.940 & 0.923 &  0.954 & 1  \\
\hline
\end{tabular}
\end{table}

\subsubsubsection{Summary Statistics}
\pagestyle{empty}
\input{table_fivepoint_KolmogorowSmirnow_drg.txt}
\input{table_fivepoint_CramerVonMises_drg.txt}
\input{table_fivepoint_KullbackLeibler_drg.txt}
\input{table_fivepoint_Mallows_drg.txt}
\pagestyle{plain}

%% file: dataset.tex
We make use of anonymized data sets from the research centers of the DESTATIS (so-called \emph{campus files}). The considered employee data set results from the \textit{Structure of Earnings Survey 2010} and consists of 25,974 observations with 33 variables. A detailed variable-by-variable description can be found in the Supplement. \\
Redundant variables for later analysis such as the ID number and the type of survey have been removed. In the original data set, missing values are present in three variables. This is the case for the variables \textit{normalized gross annual earnings} (\texttt{ef45}) and for \textit{normalized special payments} (\texttt{ef47}) as well as for the variable \textit{special payments} (\texttt{ef28}) itself. The first two variables consist of around $81 \%$ missing cases, while \texttt{ef28} indicated only $0.3 \%$ missing instances. The reason for this discrepancy lies at the definition of the variables itself. While some of the employees did not report earnings and special payments for the whole year, the normalization is 
the theoretical projection of earnings one would achieve with the same employment working throughout the year. This, because some employees might start working not at the beginning of the year, but later on. This was actually the case for $19\%$ of the recorded employees yielding to $81 \%$ cases with full-year earnings and hence, $81\%$ missing instances. Since our analysis will be focused on distributional discrepancies after imputation, we assigned for every missing value in these three variables a constant value of $0$, since all of them are metric and non-negative. This should not have any impacts on our final results, even if distributional distortions might occur through the assignment since we make a distributional comparison before and after inserting artificial missing values and imputing them. In addition, the non-normalized counterparts of the variables \texttt{ef45} and \texttt{ef47} are listed in the data set as the variables \textit{gross annual earnings} (\texttt{ef27}) and \textit{special payments} (\texttt{ef28}). All four of them reflect almost the same information. Therefore an aggregation of the form 
$
    \max\{ \texttt{ef27}, \texttt{ef45} \} \text{ and } \max\{ \texttt{ef28}, \texttt{ef47} \}
$ 
was conducted reducing the number of considered variables. 
After this aggregation 
there are 73 employees showing no information in \texttt{ef28} and \texttt{ef47} leading to the constant assignment of $0$ as mentioned above. The weekly working times of marginally employed people are given in variable \texttt{ef53} and show a constant value of $0$ for all employees, who are not marginally employed. This information is, however, given in variable \texttt{ef18} such that a junction of these two variables was conducted filling all instances in \texttt{ef53} with the needed information in \texttt{ef18}. Furthermore, it should be noted that the variables \textit{working days throughout the year} (\texttt{ef26})  and \textit{working weeks in one year} (\texttt{ef50}) are in a one-to-one relation leading to the deletion of variable \texttt{ef26}. 
After these preparations our data set contains $27$ variables, with no missing entries for further analysis. 
Beside the conducted data preparation steps, highly correlated variables among the $27$ variables are still present. Some of the considered imputation methods such as MICE can suffer from the presence of such correlations. A detailed description how this will be overcome is discussed in Section \ref{sec:simsetup}.

%% file: missing_mechanisms.tex
We denote with $\mathbf{X} = [\mathbf{X}_1, \dots, \mathbf{X}_k] \in \R^{n \times k}$ the data matrix containing $n$ observations in $k$ variables. Furthermore, the corresponding missing matrix $\mathbf{M} = (m_{ij})_{i, j} \in  \{0,1\}^{n \times k}$ indicates whether the observations in $X_{i j}$ are missing ($m_{i, j} = 1$) or not ($m_{ij} = 0$). Different to \cite{rubin2004multiple}, we focus on only two missing mechanisms: missing completely at random (MCAR) and missing at random (MAR). Assuming that the missing matrix $\mathbf{M}$ originates from a probability distribution with parameter $\xi$, while $f$ denotes a corresponding density or probability mass function, the considered missing mechanisms can be defined as follows \citep{rubin2004multiple}:

\begin{enumerate}
    \item The missing mechanism is said to be MCAR, if $f(\mathbf{M} | \mathbf{X}, \xi) =f(\mathbf{M}| \xi)$ for all $\mathbf{X}$ and $\xi$. That is, the missing values do occur in a completely random fashion. 
    \item The missing mechanism is said to be MAR, if $f(\mathbf{M} |  \mathbf{X}, \xi) = f(\mathbf{M} | \mathbf{X}_{obs}, \xi)$, for all $\mathbf{X}_{mis}$ and $\xi$. In this case, $\mathbf{X}_{obs}$ and $\mathbf{X}_{mis}$ denote the observed and missing parts of $\mathbf{X}$. Hence, the occurance of missing values only depends on the observed parts and the missing data generating parameter $\xi$. Unobserved values do not impact the missing instances. 
\end{enumerate}

We use the \textsf{R}-function \texttt{prodNA} given in the \texttt{missForest}-package to artificially introducing missing values according to the MCAR framework.
The corresponding algorithm randomly selects $\lceil r \cdot (n \cdot k) \rceil$ entries from $\mathbf{X}$ and sets the corresponding entries in $\mathbf{M}$ to $m_{ij} = 1$. Hence, we have to distinguish between two quantities:
\begin{enumerate}
\item 
The \textit{variable missing rate} $r_j$, $j = 1,\dots, k$, representing the relative amount of missing instances in variable $j$ given by $$ r_j = \frac{1}{n} \sum\limits_{i = 1}^n \mathds{1}\{ m_{ij} = 1 \}$$
\item and the \textit{overall missing rate} $r$ defined as the relative amount of missing instances based on every entry in the data matrix, i.e.  $$ r = \frac{1}{n\cdot k} \sum\limits_{i = 1}^n \sum\limits_{j = 1}^k  \mathds{1}\{ m_{ij} = 1 \}  = \frac{1}{k} \sum\limits_{j = 1}^k r_j.$$
\end{enumerate}
For the MAR mechanism we use an iterative approach taking into account two variables
: one, where missing values are inserted and the other one for introducing dependencies among the observed variables. The concrete procedure is described in detail in the supplement.


%% file: imputation.tex
There exist plenty of possible ways to impute missing values in practice. These can, for example, be roughly subsumed
into the following (non-disjunct) groups:
\begin{itemize}
\item[(1)] single variable-based imputation techniques such as mean, median or carryover imputations, 
\item[(2)] data augmentation, i.e. regression based modelling, including multivariate response regression, 
\item[(3)] multiple imputation using chained equations (MICE), i.e. iteratively applying univariate response regression resp. classification.
\end{itemize}

The last two are usually used within the multiple imputation logic, i.e. generating multiple values for missing instances as draws from the trained regression resp. classification method. 
\cite{rubin2004multiple} introduced the logic of imputing missing values 
from a Bayesian perspective. This makes the understanding of potential implications on the frequentist's application perspective difficult. Therefore, multiple imputation and its invention has to be understood as the bridge between the Bayesian imputation world and the frequentist's analysis world. Recent research (\citealp{rahman2011decision}; \citealp{silva2011missing}; \citealp{missforest} or \citealp{Ramosaj_Pauly_2019}) has been focused on machine learning based imputation techniques while some of these methods are even combined with Bayesian elements (\citealp{doove2014recursive}).  

Nevertheless, there is in general no optimal procedure and not all methods are applicable to both, categorical and metric data. We thus only compare a few procedures for our investigations. Our selections were based on the following points: (i) easy to use, (ii) implementation in statistical software, (iii) recommendations from previous 
simulation results in the literature and (iv) wide spread in statistical practice. This resulted in a total of seven different imputation methods:
\texttt{naive imputation}, \texttt{Amelia}, \texttt{missForest}, \texttt{missRanger}, multiple imputation by
chained equations (MICE) based on a nonparametric Random Forest prediction model (\texttt{Mice.RF}) and MICE with fully conditional specification using either predictive mean matching (\texttt{Mice.Pmm}) or a normal model (\texttt{Mice.Norm}). 
Let us briefly describe the key aspects of these methods and the used implementation:

We use \texttt{naive imputation} as described in \citet[p.\,12]{van_Buuren_flexible_2018} as a simple benchmark. That is, missing categorical variables are imputed by the mode while  missing metric outcomes are replaced by the mean of the observed outcomes of the respective variable.

As second method we use \texttt{Amelia}, a bootstrap-based multiple imputation approach implemented in the \texttt{R}-package \texttt{Amelia II} \citep{amelia}. It is based on an iterative application of the EM algorithm on multiple bootstrap samples of the data. Its initial implementation assumes multivariate normality and observations being MAR. To make the algorithm applicable to categorical variables, these are transformed, e.g. by using ranks in case of ordinal, and dummy variables for nominal data. In case of many categorical variables and large data sets this may cause a longer runtime which we counteract by choosing a larger tolerance of $0.005$ in the EM algorithm. To further stabilize the algorithm we follow \citet[Section 4.6]{amelia} and choose a ridge prior of $1\,\%$.

Random Forest based imputation methods have recently been recommended in several papers for imputing mixed-type data \citep{missforest,waljee2013comparison,starkweather2014new,missranger}. The most famous one is the \texttt{missForest} proposed by \cite{missforest} with around $1,450$ Google Scholar citations as of January 2021. It is implemented in the \texttt{R} package \texttt{missForest}. The key idea is to train random forest models for each variable (starting with the variable with the fewest missing values) using the observed data as training set. These models then predict the missing observations and the process is repeated in an iterative fashion until a stopping criterion is reached (which ever comes first: distance between the old and the newly imputed data matrix is negligible or the maximum number of iterations (default: \texttt{maxiter}=10) is reached). Due to its nonparametric nature, the \texttt{missForest} does in principle not rely on stringent distributional assumptions and is applicable for imputing both, continuous and categorical variables.
Recently, a faster variant of the \texttt{missForest} algorithm was proposed by \cite{missranger}. It is based upon the \texttt{R} package \texttt{ranger} \citep{wright2017ranger} and additionally offers the possibility for introducing predictive mean matching principles in the imputation. 
Apart from a possibly better reproduction of the variability, this has the advantage that the imputations lie in the range of the observed values. This is why we have chosen only this option for \texttt{missRanger} 
in our study. For the implementation we used the default number of trees in a random forest ( ${\texttt{num.trees} = 500}$).

As, e.g., stressed by \cite{erler2016dealing}, the current gold standard to impute missing values is multiple imputation with chained equations. This is also reflected by the around  $6,400$ Google Scholar citations (as of January 2021) of the paper by \citet{mice}, which describes its implementation in the \texttt{R} package \texttt{mice}. It offers a flexible and general class of methods for multiple imputation. Similar to \texttt{missForest} and  \texttt{missRanger} prediction models are learned to impute the missing observations. However, different to the two Random Forest implementations, MICE utilizes multiple imputations. For our purposes, we thereby studied three options: Predictive Mean Matching (\texttt{method}=\texttt{pmm}), Bayesian imputation under the normal linear model (\texttt{method}=\texttt{norm}) and Random Forest (\texttt{method}=\texttt{rf}). These 
 are abbreviated as \texttt{Mice.Pmm}, \texttt{Mice.Norm} and \texttt{Mice.RF}, respectively. 
A detailed explanation can be found in \citet{van_Buuren_flexible_2018}. We have used 
\texttt{Mice.PMM} and \texttt{Mice.Norm} only for continuous variables. In this case they are both based on a linear model. In contrast, \texttt{Mice.RF} uses classification and regression trees and is applied for the imputation of continuous and categorical variables. Beyond the number of trees in each Random Forest ($\texttt{ntree}=10$) the major difference to \texttt{missForest} and \texttt{missRanger} is that multiple imputations are performed to better account for the inherent uncertainty of the data \citep{van_Buuren_flexible_2018, ramosaj2020cautionary}.

%% file: measures.tex
There are several possibilities to evaluate imputation quality 
in simulations. Thereby, the choice should depend on the envisioned statistical analysis tasks. For example, if only the point estimation of a single parameter (mean, variance, coefficient of variation etc.) is of interest, a distance of the corresponding (sample) estimators/statistics calculated on the observed data sets (i.\,e., the complete data sets for our study) and on the imputed data sets can serve as quality measure. 
However, this is rarely the case in practice, where rather sophisticated analyses are carried out. In fact, in official statistics data must often be provided in a way that allows manifold applications. Thus, we advocate that evaluation measures describing the distributional distance should come more to fore. 

We nevertheless start this section with two of the most common measures for predictive imputation accuracy: the percentage of falsely classified/imputed entries ($PFC$) for categorical variables and the normalised root mean squared error ($NRMSE$) for continuous outcomes. The latter is based on the quadratic imputation errors, see, e.g., \cite{missforest}, \cite{NRMSE_PFC} or \cite{Ramosaj_Pauly_2019} for their concrete definition. If one is only interested in accuracy results measured by NRMSE and PFC, values close to zero should reflect a \textit{good} (predictive) imputation procedure. This way, the imputed value is very close to the true, unobserved value. Note that in practice, NRMSE and PFC can only be estimated, and rarely computed exactly. 

In addition to these two common measures 
we also study several distance measures for distributions in order to measure distributional accuracy.  For categorical variables we utilize Cramérs $V$ which is based upon Pearson's chi-squared statistic but gives easy-to-interpret values between $0$ and $1$ \citep{cohen1988statistical}.

For metric variables, we use three different types of (pseudo-)distances that are either based upon empirical distribution functions (edf), empirical quantile functions (eqf) or kernel density estimators.

From the edf-class, we consider the following two-sample goodness-of-fit statistics: 
(i) the Kolmogorov-Smirnow-statistic (KS) and (ii) 
the Cramer-von-Mises-statistic (CM)
, see \cite{shorack2009empirical} for their definition and \cite{janssen2000global} for more details on goodness-of-fit tests.

As an eqf-based distance measure we use the Mallows $L^2$-distance \citep{estim_Mallows}. It is also known as (second) Wasserstein or earth mover metric and was used in early consistency proofs for the bootstrap \citep{bickel1981some}. 

Finally, we also consider the Kullback-Leibler divergence between the kernel density estimators obtained from the imputed and the complete data set, respectively. To this end we have chosen a Gaussian kernel and calculated the estimators with the \texttt{R}-function 
\texttt{density} that exhibits an automatic bandwidth selection. \\

In addition to the computation of the various distance measures for continuous and categorical outcomes, we conducted permutation tests (being aware of potential non-exchangeable situations under $H_0$) based on the two edf-statistics to 
test the general hypothesis of equal distributions
\begin{align}\label{generalNULL}
    H_0: F_j^{(true)} = F_j^{(imp)} \quad vs. \quad H_1: F_j^{(true)} \neq F_j^{(imp)}.
\end{align}
Here, $F_j^{(true)}$ and $F_j^{(imp)}$ denote the cumulative distribution function of the $j$-th variable in the original and the imputed data set, respectively. In order to incorporate the multiple imputation logic as prescribed in \cite{rubin2004multiple}, the following logic is applied for estimating p-values under $(\ref{generalNULL})$ for a fixed variable $j \in \{1, \dots, p\}$.
\begin{enumerate}
    \item Compute all considered edf-based statistic KS and CM 
    for both, 
    the original data set and the different imputed data sets. Regarding the latter, all edf-based statistics are aggregated through averaging over the $m$ obtained statistics. We denote the obtained statistics by $KS_{n, m}^{(0)}$ and $CM_{n,m}^{(0)}$. \label{TestStep1}
    \item For every imputed data set, separately permute the observations in variable $j$ between the imputed data set and the original data set \citep{manly2006randomization}. Roughly speaking, this should not distort the test statistics' distribution under the null given in $(\ref{generalNULL})$.  \label{TestStep2}
    \item For every imputed and permuted data set, compute the corresponding edf-statistic and aggregate the result through averaging. We denote the obtained statistics at this stage by $KS_{n,m}^{(\ell)}$ and $CM_{n,m}^{(\ell)}$. 
    \label{TestStep3}
    \item Repeat steps \ref{TestStep2}. -- \ref{TestStep3}. $perm \in \N$ times and finally estimate the following permutation-based p-values: 
    \begin{align}
    p_{n, m}^{*KS} &= \frac{1}{perm+1}\left(\sum\limits_{\ell = 1}^{perm} \mathds{1}\{ KS_{n, m}^{(\ell)} > |KS_{n,m}^{(0)}|  \} +1 \right), \\
    p_{n, m}^{*CM} &= \frac{1}{perm+1}\left(\sum\limits_{\ell = 1}^{perm} \mathds{1}\{ CM_{n, m}^{(\ell)} > |CM_{n,m}^{(0)}|  \} +1 \right). 
    \end{align}
\end{enumerate}

%% file: table_fivepoint_Chisq_vse.txt
\begin{table}[H] 
\caption{minimum ($q_0$), first quartile ($q_{0.25}$),
               median ($q_{0.5}$), mean ($\bar{x}$), third quartile ($q_{0.75}$),
               maximum ($q_{1}$) and standard deviation (sd) of the observed values of the realization of Pearson's chi-squared statistic $\left(\text{with factor } \frac{1}{1000}\right)$ for the variable \texttt{ef10} for the different
               imputation methods and missing rates (r)
               creating missing values using the MAR-mechanism for the employee data (VSE, $\chi^2$, \texttt{ef10}, MAR)}\label{Tab:vse.vals.chisq.MAR.ef10 }\begin{minipage}[t]{\textwidth} 

\centering 
\begin{tabular}[t]{|l|c|c|c|c|c|c|c|c|}\hline
 method&r&$q_{0}$& $q_{0.25}$&$q_{0.5}$&$\bar{x}$&$q_{0.75}$& $q_{1}$&sd\\ \hline 
\hline  &0.01&25.60&25.64&25.66&25.66&25.68&25.72&0.02\\
  &0.05&24.29&24.39&24.42&24.41&24.44&24.57&0.05\\
\multirow{-5}{*}{Amelia}&0.10&22.61&22.87&22.91&22.92&22.97&23.10&0.08\\

\hline  &0.01&25.42&25.48&25.50&25.50&25.53&25.58&0.03\\
  &0.05&23.51&23.62&23.66&23.66&23.69&23.85&0.06\\
\multirow{-5}{*}{Mice.Norm}&0.10&21.24&21.43&21.50&21.49&21.57&21.80&0.10\\

\hline  &0.01&25.43&25.48&25.51&25.51&25.53&25.60&0.03\\
  &0.05&23.52&23.64&23.68&23.68&23.72&23.85&0.07\\
\multirow{-5}{*}{Mice.PMM}&0.10&21.23&21.45&21.51&21.51&21.57&21.74&0.10\\

\hline  &0.01&25.44&25.49&25.52&25.52&25.54&25.59&0.03\\
  &0.05&23.53&23.66&23.71&23.71&23.76&23.92&0.07\\
\multirow{-5}{*}{Mice.RF}&0.10&21.30&21.45&21.53&21.52&21.58&21.74&0.10\\

\hline Naive &0.01&25.45&25.51&25.54&25.54&25.57&25.65&0.04\\
  Imputation&0.05&23.63&23.80&23.85&23.85&23.89&24.05&0.08\\
 &0.10&21.62&21.80&21.88&21.88&21.96&22.18&0.11\\

\hline  &0.01&25.64&25.76&25.79&25.78&25.80&25.86&0.04\\
  &0.05&24.47&24.97&25.06&25.02&25.13&25.24&0.16\\
\multirow{-5}{*}{MissRanger}&0.10&23.11&23.76&24.13&23.98&24.21&24.40&0.33\\

\hline 
\end{tabular} 
\end{minipage} 
\end{table}

\newpage 
\begin{table}[H] 
\caption{minimum ($q_0$), first quartile ($q_{0.25}$),
               median ($q_{0.5}$), mean ($\bar{x}$), third quartile ($q_{0.75}$),
               maximum ($q_{1}$) and standard deviation (sd) of the observed values of the realization of Pearson's chi-squared statistic $\left(\text{with factor } \frac{1}{1000}\right)$ for the variable \texttt{ef10} for the different
               imputation methods and missing rates (r)
               creating missing values using the MCAR-mechanism for the employee data (VSE, $\chi^2$, \texttt{ef10}, MCAR)}\label{Tab:vse.vals.chisq.MCAR.ef10 }\begin{minipage}[t]{\textwidth} 

\centering 
\begin{tabular}[t]{|l|c|c|c|c|c|c|c|c|}\hline
 method&r&$q_{0}$& $q_{0.25}$&$q_{0.5}$&$\bar{x}$&$q_{0.75}$& $q_{1}$&sd\\ \hline 
\hline  &0.01&25.59&25.64&25.66&25.66&25.68&25.71&0.02\\
  &0.05&24.19&24.37&24.42&24.41&24.45&24.55&0.06\\
\multirow{-5}{*}{Amelia}&0.10&22.64&22.86&22.93&22.91&22.97&23.07&0.08\\

\hline  &0.01&25.41&25.48&25.51&25.50&25.52&25.57&0.03\\
  &0.05&23.47&23.61&23.65&23.65&23.70&23.84&0.08\\
\multirow{-5}{*}{Mice.Norm}&0.10&21.33&21.44&21.51&21.51&21.57&21.71&0.09\\

\hline  &0.01&25.40&25.49&25.51&25.50&25.52&25.56&0.03\\
  &0.05&23.47&23.62&23.66&23.67&23.72&23.86&0.08\\
\multirow{-5}{*}{Mice.PMM}&0.10&21.32&21.45&21.53&21.52&21.60&21.78&0.09\\

\hline  &0.01&25.44&25.49&25.52&25.52&25.54&25.57&0.03\\
  &0.05&23.53&23.65&23.69&23.70&23.74&23.85&0.07\\
\multirow{-5}{*}{Mice.RF}&0.10&21.33&21.48&21.53&21.54&21.61&21.85&0.10\\

\hline Naive &0.01&25.40&25.52&25.54&25.54&25.56&25.65&0.04\\
  Imputation&0.05&23.61&23.79&23.85&23.84&23.90&24.09&0.09\\
 &0.10&21.69&21.83&21.89&21.89&21.95&22.18&0.09\\

\hline  &0.01&25.67&25.76&25.79&25.78&25.81&25.85&0.04\\
  &0.05&24.54&24.96&25.06&25.02&25.14&25.22&0.17\\
\multirow{-5}{*}{MissRanger}&0.10&22.99&23.82&24.02&23.95&24.20&24.45&0.31\\

\hline 
\end{tabular} 
\end{minipage} 
\end{table}

\newpage 
\begin{table}[H] 
\caption{minimum ($q_0$), first quartile ($q_{0.25}$),
               median ($q_{0.5}$), mean ($\bar{x}$), third quartile ($q_{0.75}$),
               maximum ($q_{1}$) and standard deviation (sd) of the observed values of the realization of Pearson's chi-squared statistic $\left(\text{with factor } \frac{1}{1000}\right)$ for the variable \texttt{ef15} for the different
               imputation methods and missing rates (r)
               creating missing values using the MAR-mechanism for the employee data (VSE, $\chi^2$, \texttt{ef15}, MAR)}\label{Tab:vse.vals.chisq.MAR.ef15 }\begin{minipage}[t]{\textwidth} 

\centering 
\begin{tabular}[t]{|l|c|c|c|c|c|c|c|c|}\hline
 method&r&$q_{0}$& $q_{0.25}$&$q_{0.5}$&$\bar{x}$&$q_{0.75}$& $q_{1}$&sd\\ \hline 
\hline  &0.01&483.33&484.64&485.18&485.05&485.52&486.42&0.64\\
  &0.05&449.05&451.52&452.42&452.30&453.14&454.64&1.24\\
\multirow{-5}{*}{Amelia}&0.10&409.70&412.42&413.44&413.59&414.60&417.47&1.76\\

\hline  &0.01&484.55&485.69&486.14&486.16&486.56&487.45&0.58\\
  &0.05&452.20&455.23&456.02&456.03&456.83&458.85&1.30\\
\multirow{-5}{*}{Mice.Norm}&0.10&414.30&417.15&418.28&418.31&419.40&423.28&1.66\\

\hline  &0.01&484.79&485.74&486.11&486.14&486.55&487.32&0.58\\
  &0.05&452.75&455.36&456.48&456.32&457.29&459.19&1.30\\
\multirow{-5}{*}{Mice.PMM}&0.10&414.67&417.30&418.44&418.56&419.67&422.94&1.68\\

\hline  &0.01&484.95&485.86&486.26&486.25&486.66&487.39&0.56\\
  &0.05&453.45&455.84&456.58&456.57&457.37&459.22&1.20\\
\multirow{-5}{*}{Mice.RF}&0.10&414.48&417.93&418.90&418.97&419.90&423.63&1.70\\

\hline Naive &0.01&486.27&487.00&487.37&487.39&487.80&488.44&0.51\\
  Imputation&0.05&460.86&463.48&464.10&464.08&464.82&466.87&1.04\\
 &0.10&434.04&436.26&437.11&437.03&437.68&440.54&1.30\\

\hline  &0.01&488.36&489.89&490.37&490.36&490.93&491.79&0.72\\
  &0.05&467.49&474.66&477.38&477.15&480.27&482.31&3.58\\
\multirow{-5}{*}{MissRanger}&0.10&442.40&454.06&460.43&460.13&466.17&470.69&6.72\\

\hline 
\end{tabular} 
\end{minipage} 
\end{table}

\newpage 
\begin{table}[H] 
\caption{minimum ($q_0$), first quartile ($q_{0.25}$),
               median ($q_{0.5}$), mean ($\bar{x}$), third quartile ($q_{0.75}$),
               maximum ($q_{1}$) and standard deviation (sd) of the observed values of the realization of Pearson's chi-squared statistic $\left(\text{with factor } \frac{1}{1000}\right)$ for the variable \texttt{ef15} for the different
               imputation methods and missing rates (r)
               creating missing values using the MCAR-mechanism for the employee data (VSE, $\chi^2$, \texttt{ef15}, MCAR)}\label{Tab:vse.vals.chisq.MCAR.ef15 }\begin{minipage}[t]{\textwidth} 

\centering 
\begin{tabular}[t]{|l|c|c|c|c|c|c|c|c|}\hline
 method&r&$q_{0}$& $q_{0.25}$&$q_{0.5}$&$\bar{x}$&$q_{0.75}$& $q_{1}$&sd\\ \hline 
\hline  &0.01&483.26&484.41&484.83&484.81&485.19&486.26&0.63\\
  &0.05&448.39&451.35&452.34&452.19&453.06&455.10&1.20\\
\multirow{-5}{*}{Amelia}&0.10&409.44&412.11&413.17&413.24&414.22&417.26&1.61\\

\hline  &0.01&484.28&485.55&485.87&485.91&486.34&487.28&0.57\\
  &0.05&452.84&455.05&456.15&455.95&456.83&459.62&1.31\\
\multirow{-5}{*}{Mice.Norm}&0.10&413.69&416.55&417.66&417.80&419.00&422.68&1.88\\

\hline  &0.01&484.53&485.47&485.95&485.91&486.32&487.58&0.58\\
  &0.05&453.22&455.41&456.25&456.20&457.00&459.11&1.20\\
\multirow{-5}{*}{Mice.PMM}&0.10&412.53&416.85&418.25&418.15&419.26&423.01&1.83\\

\hline  &0.01&484.73&485.67&486.06&486.03&486.43&487.65&0.58\\
  &0.05&453.55&455.79&456.56&456.51&457.51&459.48&1.30\\
\multirow{-5}{*}{Mice.RF}&0.10&413.36&417.22&418.19&418.43&419.55&423.68&1.85\\

\hline Naive &0.01&485.89&486.84&487.24&487.23&487.63&488.38&0.52\\
  Imputation&0.05&461.26&463.40&464.05&464.03&464.71&466.68&1.10\\
 &0.10&433.19&435.78&436.51&436.82&437.69&440.76&1.44\\

\hline  &0.01&488.99&489.86&490.62&490.49&491.03&492.03&0.72\\
  &0.05&468.96&474.79&476.95&477.37&480.62&483.29&3.56\\
\multirow{-5}{*}{MissRanger}&0.10&443.85&455.69&459.86&460.37&466.50&470.46&6.34\\

\hline 
\end{tabular} 
\end{minipage} 
\end{table}

\newpage 
\begin{table}[H] 
\caption{minimum ($q_0$), first quartile ($q_{0.25}$),
               median ($q_{0.5}$), mean ($\bar{x}$), third quartile ($q_{0.75}$),
               maximum ($q_{1}$) and standard deviation (sd) of the observed values of the realization of Pearson's chi-squared statistic $\left(\text{with factor } \frac{1}{1000}\right)$ for the variable \texttt{ef16u1} for the different
               imputation methods and missing rates (r)
               creating missing values using the MAR-mechanism for the employee data (VSE, $\chi^2$, \texttt{ef16u1}, MAR)}\label{Tab:vse.vals.chisq.MAR.ef16u1 }\begin{minipage}[t]{\textwidth} 

\centering 
\begin{tabular}[t]{|l|c|c|c|c|c|c|c|c|}\hline
 method&r&$q_{0}$& $q_{0.25}$&$q_{0.5}$&$\bar{x}$&$q_{0.75}$& $q_{1}$&sd\\ \hline 
\hline  &0.01&211.24&221.66&224.45&223.91&226.34&229.75&3.63\\
  &0.05&191.06&196.76&198.51&198.72&200.48&206.13&3.23\\
\multirow{-5}{*}{Amelia}&0.10&172.62&175.80&177.25&177.74&179.75&184.28&2.58\\

\hline  &0.01&217.36&230.44&230.79&230.18&231.00&231.65&2.64\\
  &0.05&202.77&216.83&218.24&216.59&219.23&221.63&4.72\\
\multirow{-5}{*}{Mice.Norm}&0.10&187.72&198.74&202.85&200.52&204.74&206.95&5.72\\

\hline  &0.01&217.07&230.46&230.79&230.13&230.97&231.71&2.66\\
  &0.05&204.35&217.18&218.96&217.11&219.65&222.22&4.69\\
\multirow{-5}{*}{Mice.PMM}&0.10&189.08&198.15&202.72&200.45&204.17&207.17&5.38\\

\hline  &0.01&217.50&230.68&230.97&230.34&231.16&231.82&2.64\\
  &0.05&202.53&217.29&218.45&216.95&219.67&222.76&4.62\\
\multirow{-5}{*}{Mice.RF}&0.10&185.83&198.51&202.58&200.28&204.13&207.37&5.56\\

\hline Naive &0.01&217.74&231.03&231.21&230.68&231.39&232.09&2.64\\
  Imputation&0.05&206.13&220.35&221.11&219.27&221.43&223.38&4.68\\
 &0.10&193.73&207.11&208.92&206.07&209.82&211.34&5.79\\

\hline  &0.01&219.98&232.53&232.96&232.32&233.21&233.60&2.55\\
  &0.05&211.78&226.77&229.64&227.50&230.95&231.98&5.08\\
\multirow{-5}{*}{MissRanger}&0.10&199.73&215.04&225.73&221.99&227.95&229.22&7.27\\

\hline 
\end{tabular} 
\end{minipage} 
\end{table}

\newpage 
\begin{table}[H] 
\caption{minimum ($q_0$), first quartile ($q_{0.25}$),
               median ($q_{0.5}$), mean ($\bar{x}$), third quartile ($q_{0.75}$),
               maximum ($q_{1}$) and standard deviation (sd) of the observed values of the realization of Pearson's chi-squared statistic $\left(\text{with factor } \frac{1}{1000}\right)$ for the variable \texttt{ef16u1} for the different
               imputation methods and missing rates (r)
               creating missing values using the MCAR-mechanism for the employee data (VSE, $\chi^2$, \texttt{ef16u1}, MCAR)}\label{Tab:vse.vals.chisq.MCAR.ef16u1 }\begin{minipage}[t]{\textwidth} 

\centering 
\begin{tabular}[t]{|l|c|c|c|c|c|c|c|c|}\hline
 method&r&$q_{0}$& $q_{0.25}$&$q_{0.5}$&$\bar{x}$&$q_{0.75}$& $q_{1}$&sd\\ \hline 
\hline  &0.01&213.33&222.87&224.76&224.30&226.20&229.82&3.13\\
  &0.05&191.20&196.48&198.94&198.58&200.76&210.64&3.24\\
\multirow{-5}{*}{Amelia}&0.10&171.92&175.55&177.10&177.52&179.49&185.27&2.72\\

\hline  &0.01&216.19&230.41&230.73&230.06&230.96&231.58&2.68\\
  &0.05&204.00&217.29&218.45&216.56&219.28&221.04&4.80\\
\multirow{-5}{*}{Mice.Norm}&0.10&187.05&192.85&203.13&200.20&204.64&207.31&5.91\\

\hline  &0.01&217.31&230.38&230.75&230.04&230.92&231.65&2.62\\
  &0.05&203.91&216.47&218.23&216.49&219.28&221.31&4.64\\
\multirow{-5}{*}{Mice.PMM}&0.10&188.13&193.54&202.87&200.33&204.43&207.08&5.68\\

\hline  &0.01&217.62&230.62&230.95&230.32&231.15&231.63&2.62\\
  &0.05&204.58&217.00&218.50&216.75&219.62&220.77&4.83\\
\multirow{-5}{*}{Mice.RF}&0.10&188.00&193.59&202.30&199.86&203.89&207.42&5.78\\

\hline Naive &0.01&217.81&230.95&231.22&230.69&231.36&231.93&2.58\\
  Imputation&0.05&207.68&220.42&221.01&219.14&221.49&222.54&4.74\\
 &0.10&194.54&197.71&208.91&205.83&209.88&212.21&5.98\\

\hline  &0.01&218.24&232.78&233.11&232.43&233.33&233.67&2.70\\
  &0.05&208.73&226.11&230.08&227.31&230.89&231.97&5.52\\
\multirow{-5}{*}{MissRanger}&0.10&196.52&215.26&223.27&220.73&227.61&230.27&7.64\\

\hline 
\end{tabular} 
\end{minipage} 
\end{table}

\newpage 
\begin{table}[H] 
\caption{minimum ($q_0$), first quartile ($q_{0.25}$),
               median ($q_{0.5}$), mean ($\bar{x}$), third quartile ($q_{0.75}$),
               maximum ($q_{1}$) and standard deviation (sd) of the observed values of the realization of Pearson's chi-squared statistic $\left(\text{with factor } \frac{1}{1000}\right)$ for the variable \texttt{ef16u2} for the different
               imputation methods and missing rates (r)
               creating missing values using the MAR-mechanism for the employee data (VSE, $\chi^2$, \texttt{ef16u2}, MAR)}\label{Tab:vse.vals.chisq.MAR.ef16u2 }\begin{minipage}[t]{\textwidth} 

\centering 
\begin{tabular}[t]{|l|c|c|c|c|c|c|c|c|}\hline
 method&r&$q_{0}$& $q_{0.25}$&$q_{0.5}$&$\bar{x}$&$q_{0.75}$& $q_{1}$&sd\\ \hline 
\hline  &0.01&152.76&153.18&153.30&153.30&153.45&153.64&0.18\\
  &0.05&142.26&143.14&143.44&143.40&143.66&144.39&0.40\\
\multirow{-5}{*}{Amelia}&0.10&130.45&131.38&131.75&131.80&132.15&133.28&0.60\\

\hline  &0.01&153.19&153.54&153.68&153.67&153.81&153.98&0.18\\
  &0.05&143.59&144.49&144.81&144.77&145.11&146.09&0.45\\
\multirow{-5}{*}{Mice.Norm}&0.10&131.52&133.20&133.54&133.61&133.97&135.42&0.61\\

\hline  &0.01&153.20&153.58&153.70&153.67&153.79&154.02&0.17\\
  &0.05&143.73&144.68&144.98&144.94&145.23&145.82&0.44\\
\multirow{-5}{*}{Mice.PMM}&0.10&132.16&133.28&133.72&133.76&134.24&136.37&0.70\\

\hline  &0.01&153.16&153.60&153.71&153.70&153.81&154.00&0.17\\
  &0.05&144.11&144.65&144.97&144.94&145.20&145.91&0.41\\
\multirow{-5}{*}{Mice.RF}&0.10&132.39&133.36&133.85&133.82&134.32&135.87&0.70\\

\hline Naive &0.01&153.49&153.88&153.98&153.98&154.12&154.34&0.17\\
  Imputation&0.05&145.70&146.38&146.67&146.64&146.91&147.73&0.39\\
 &0.10&136.47&137.47&137.75&137.80&138.13&139.34&0.53\\

\hline  &0.01&154.42&154.86&154.98&154.97&155.10&155.38&0.19\\
  &0.05&149.14&150.95&151.39&151.34&151.96&152.73&0.76\\
\multirow{-5}{*}{MissRanger}&0.10&142.86&145.38&146.74&146.46&147.87&148.62&1.54\\

\hline 
\end{tabular} 
\end{minipage} 
\end{table}

\newpage 
\begin{table}[H] 
\caption{minimum ($q_0$), first quartile ($q_{0.25}$),
               median ($q_{0.5}$), mean ($\bar{x}$), third quartile ($q_{0.75}$),
               maximum ($q_{1}$) and standard deviation (sd) of the observed values of the realization of Pearson's chi-squared statistic $\left(\text{with factor } \frac{1}{1000}\right)$ for the variable \texttt{ef16u2} for the different
               imputation methods and missing rates (r)
               creating missing values using the MCAR-mechanism for the employee data (VSE, $\chi^2$, \texttt{ef16u2}, MCAR)}\label{Tab:vse.vals.chisq.MCAR.ef16u2 }\begin{minipage}[t]{\textwidth} 

\centering 
\begin{tabular}[t]{|l|c|c|c|c|c|c|c|c|}\hline
 method&r&$q_{0}$& $q_{0.25}$&$q_{0.5}$&$\bar{x}$&$q_{0.75}$& $q_{1}$&sd\\ \hline 
\hline  &0.01&152.95&153.19&153.30&153.30&153.42&153.73&0.17\\
  &0.05&142.43&143.12&143.41&143.40&143.69&144.23&0.40\\
\multirow{-5}{*}{Amelia}&0.10&130.19&131.51&131.69&131.77&132.02&133.42&0.59\\

\hline  &0.01&153.22&153.57&153.70&153.67&153.77&154.07&0.15\\
  &0.05&143.56&144.41&144.67&144.69&144.99&145.49&0.43\\
\multirow{-5}{*}{Mice.Norm}&0.10&131.39&133.08&133.46&133.45&133.89&134.70&0.60\\

\hline  &0.01&153.22&153.60&153.70&153.68&153.76&154.07&0.17\\
  &0.05&143.96&144.52&144.91&144.86&145.15&145.88&0.43\\
\multirow{-5}{*}{Mice.PMM}&0.10&132.43&133.20&133.61&133.61&134.01&135.27&0.56\\

\hline  &0.01&153.28&153.61&153.74&153.72&153.83&154.04&0.16\\
  &0.05&143.68&144.65&144.97&144.93&145.25&145.94&0.46\\
\multirow{-5}{*}{Mice.RF}&0.10&131.98&133.41&133.68&133.70&134.10&134.95&0.57\\

\hline Naive &0.01&153.46&153.92&154.02&154.00&154.08&154.33&0.16\\
  Imputation&0.05&145.64&146.35&146.65&146.60&146.86&147.52&0.39\\
 &0.10&136.04&137.48&137.75&137.78&138.08&138.68&0.48\\

\hline  &0.01&154.48&154.85&154.99&154.96&155.11&155.35&0.19\\
  &0.05&149.07&150.69&151.39&151.30&151.96&152.45&0.81\\
\multirow{-5}{*}{MissRanger}&0.10&142.52&145.77&146.95&146.55&147.82&148.69&1.62\\

\hline 
\end{tabular} 
\end{minipage} 
\end{table}

\newpage 
\begin{table}[H] 
\caption{minimum ($q_0$), first quartile ($q_{0.25}$),
               median ($q_{0.5}$), mean ($\bar{x}$), third quartile ($q_{0.75}$),
               maximum ($q_{1}$) and standard deviation (sd) of the observed values of the realization of Pearson's chi-squared statistic $\left(\text{with factor } \frac{1}{1000}\right)$ for the variable \texttt{ef17} for the different
               imputation methods and missing rates (r)
               creating missing values using the MAR-mechanism for the employee data (VSE, $\chi^2$, \texttt{ef17}, MAR)}\label{Tab:vse.vals.chisq.MAR.ef17 }\begin{minipage}[t]{\textwidth} 

\centering 
\begin{tabular}[t]{|l|c|c|c|c|c|c|c|c|}\hline
 method&r&$q_{0}$& $q_{0.25}$&$q_{0.5}$&$\bar{x}$&$q_{0.75}$& $q_{1}$&sd\\ \hline 
\hline  &0.01&101.80&102.22&102.31&102.30&102.40&102.59&0.13\\
  &0.05&95.77&96.13&96.36&96.32&96.50&96.92&0.26\\
\multirow{-5}{*}{Amelia}&0.10&88.35&88.96&89.26&89.21&89.43&89.91&0.33\\

\hline  &0.01&102.00&102.39&102.46&102.47&102.58&102.75&0.15\\
  &0.05&95.64&96.80&97.07&97.02&97.26&98.03&0.42\\
\multirow{-5}{*}{Mice.Norm}&0.10&89.18&90.21&90.61&90.60&90.90&92.20&0.56\\

\hline  &0.01&101.93&102.37&102.44&102.45&102.55&102.79&0.15\\
  &0.05&95.94&96.93&97.32&97.27&97.56&98.10&0.44\\
\multirow{-5}{*}{Mice.PMM}&0.10&89.61&90.26&90.74&90.73&91.08&92.21&0.54\\

\hline  &0.01&102.15&102.47&102.57&102.56&102.64&102.86&0.14\\
  &0.05&96.34&96.96&97.20&97.22&97.49&98.07&0.39\\
\multirow{-5}{*}{Mice.RF}&0.10&88.97&89.97&90.26&90.26&90.54&92.15&0.50\\

\hline Naive &0.01&102.35&102.64&102.72&102.73&102.83&103.06&0.14\\
  Imputation&0.05&97.34&98.02&98.30&98.29&98.56&99.00&0.35\\
 &0.10&91.77&92.49&92.73&92.77&93.05&94.16&0.40\\

\hline  &0.01&103.17&103.57&103.64&103.62&103.69&103.79&0.12\\
  &0.05&99.92&101.83&102.54&102.27&102.77&103.04&0.69\\
\multirow{-5}{*}{MissRanger}&0.10&95.11&100.33&101.11&100.54&101.51&102.09&1.46\\

\hline 
\end{tabular} 
\end{minipage} 
\end{table}

\newpage 
\begin{table}[H] 
\caption{minimum ($q_0$), first quartile ($q_{0.25}$),
               median ($q_{0.5}$), mean ($\bar{x}$), third quartile ($q_{0.75}$),
               maximum ($q_{1}$) and standard deviation (sd) of the observed values of the realization of Pearson's chi-squared statistic $\left(\text{with factor } \frac{1}{1000}\right)$ for the variable \texttt{ef17} for the different
               imputation methods and missing rates (r)
               creating missing values using the MCAR-mechanism for the employee data (VSE, $\chi^2$, \texttt{ef17}, MCAR)}\label{Tab:vse.vals.chisq.MCAR.ef17 }\begin{minipage}[t]{\textwidth} 

\centering 
\begin{tabular}[t]{|l|c|c|c|c|c|c|c|c|}\hline
 method&r&$q_{0}$& $q_{0.25}$&$q_{0.5}$&$\bar{x}$&$q_{0.75}$& $q_{1}$&sd\\ \hline 
\hline  &0.01&102.04&102.24&102.35&102.35&102.46&102.71&0.14\\
  &0.05&95.73&96.18&96.37&96.37&96.57&97.04&0.28\\
\multirow{-5}{*}{Amelia}&0.10&88.28&89.02&89.22&89.25&89.52&89.96&0.38\\

\hline  &0.01&102.02&102.40&102.51&102.51&102.61&102.89&0.17\\
  &0.05&96.08&96.75&97.09&97.04&97.30&97.96&0.40\\
\multirow{-5}{*}{Mice.Norm}&0.10&88.95&90.28&90.70&90.59&90.99&91.73&0.59\\

\hline  &0.01&101.96&102.40&102.49&102.50&102.60&102.82&0.17\\
  &0.05&95.73&97.08&97.34&97.29&97.54&97.99&0.38\\
\multirow{-5}{*}{Mice.PMM}&0.10&89.20&90.31&90.64&90.65&90.95&91.87&0.54\\

\hline  &0.01&102.08&102.50&102.60&102.59&102.70&103.01&0.17\\
  &0.05&96.29&97.02&97.30&97.27&97.56&98.05&0.37\\
\multirow{-5}{*}{Mice.RF}&0.10&88.83&89.93&90.32&90.32&90.71&92.25&0.58\\

\hline Naive &0.01&102.26&102.67&102.79&102.77&102.87&103.14&0.17\\
  Imputation&0.05&97.47&98.10&98.33&98.31&98.60&99.04&0.34\\
 &0.10&91.43&92.42&92.79&92.77&93.09&93.92&0.50\\

\hline  &0.01&103.15&103.54&103.63&103.60&103.68&103.79&0.14\\
  &0.05&100.04&102.02&102.53&102.30&102.75&103.05&0.65\\
\multirow{-5}{*}{MissRanger}&0.10&96.28&100.40&101.10&100.62&101.45&101.86&1.30\\

\hline 
\end{tabular} 
\end{minipage} 
\end{table}

\newpage 
\begin{table}[H] 
\caption{minimum ($q_0$), first quartile ($q_{0.25}$),
               median ($q_{0.5}$), mean ($\bar{x}$), third quartile ($q_{0.75}$),
               maximum ($q_{1}$) and standard deviation (sd) of the observed values of the realization of Pearson's chi-squared statistic $\left(\text{with factor } \frac{1}{1000}\right)$ for the variable \texttt{ef4} for the different
               imputation methods and missing rates (r)
               creating missing values using the MAR-mechanism for the employee data (VSE, $\chi^2$, \texttt{ef4}, MAR)}\label{Tab:vse.vals.chisq.MAR.ef4 }\begin{minipage}[t]{\textwidth} 

\centering 
\begin{tabular}[t]{|l|c|c|c|c|c|c|c|c|}\hline
 method&r&$q_{0}$& $q_{0.25}$&$q_{0.5}$&$\bar{x}$&$q_{0.75}$& $q_{1}$&sd\\ \hline 
\hline  &0.01&336.18&336.39&336.46&336.45&336.51&336.63&0.10\\
  &0.05&330.95&331.37&331.55&331.53&331.68&332.18&0.23\\
\multirow{-5}{*}{Amelia}&0.10&324.03&324.94&325.12&325.18&325.45&326.12&0.40\\

\hline  &0.01&333.48&334.02&334.20&334.18&334.38&334.73&0.26\\
  &0.05&317.30&318.82&319.20&319.19&319.60&320.86&0.64\\
\multirow{-5}{*}{Mice.Norm}&0.10&296.15&298.05&298.67&298.68&299.29&300.70&0.89\\

\hline  &0.01&333.45&333.99&334.18&334.18&334.36&334.71&0.26\\
  &0.05&317.47&318.76&319.11&319.18&319.58&321.15&0.63\\
\multirow{-5}{*}{Mice.PMM}&0.10&296.52&297.77&298.28&298.39&299.01&300.91&0.92\\

\hline  &0.01&333.49&333.98&334.15&334.16&334.35&334.69&0.25\\
  &0.05&317.07&318.66&318.99&319.01&319.36&320.80&0.65\\
\multirow{-5}{*}{Mice.RF}&0.10&295.70&297.58&298.11&298.14&298.81&300.40&0.98\\

\hline Naive &0.01&332.80&333.31&333.52&333.49&333.72&334.19&0.29\\
  Imputation&0.05&315.68&316.95&317.35&317.38&317.84&319.14&0.65\\
 &0.10&296.39&297.93&298.37&298.44&298.93&300.67&0.83\\

\hline  &0.01&335.78&336.71&336.93&336.86&337.05&337.44&0.28\\
  &0.05&327.75&333.15&333.71&333.31&334.07&335.05&1.37\\
\multirow{-5}{*}{MissRanger}&0.10&318.30&327.78&328.99&328.73&329.89&333.47&2.25\\

\hline 
\end{tabular} 
\end{minipage} 
\end{table}

\newpage 
\begin{table}[H] 
\caption{minimum ($q_0$), first quartile ($q_{0.25}$),
               median ($q_{0.5}$), mean ($\bar{x}$), third quartile ($q_{0.75}$),
               maximum ($q_{1}$) and standard deviation (sd) of the observed values of the realization of Pearson's chi-squared statistic $\left(\text{with factor } \frac{1}{1000}\right)$ for the variable \texttt{ef4} for the different
               imputation methods and missing rates (r)
               creating missing values using the MCAR-mechanism for the employee data (VSE, $\chi^2$, \texttt{ef4}, MCAR)}\label{Tab:vse.vals.chisq.MCAR.ef4 }\begin{minipage}[t]{\textwidth} 

\centering 
\begin{tabular}[t]{|l|c|c|c|c|c|c|c|c|}\hline
 method&r&$q_{0}$& $q_{0.25}$&$q_{0.5}$&$\bar{x}$&$q_{0.75}$& $q_{1}$&sd\\ \hline 
\hline  &0.01&336.18&336.37&336.44&336.44&336.51&336.73&0.11\\
  &0.05&330.87&331.38&331.56&331.54&331.69&332.10&0.25\\
\multirow{-5}{*}{Amelia}&0.10&324.26&324.98&325.23&325.21&325.50&326.03&0.39\\

\hline  &0.01&333.60&334.00&334.15&334.17&334.36&334.81&0.28\\
  &0.05&317.55&318.82&319.19&319.15&319.56&320.66&0.58\\
\multirow{-5}{*}{Mice.Norm}&0.10&296.15&298.15&298.78&298.75&299.38&300.60&0.91\\

\hline  &0.01&333.52&333.98&334.19&334.17&334.40&334.77&0.29\\
  &0.05&317.39&318.81&319.22&319.11&319.47&320.20&0.56\\
\multirow{-5}{*}{Mice.PMM}&0.10&295.63&297.91&298.50&298.48&299.14&300.50&0.90\\

\hline  &0.01&333.43&333.89&334.17&334.14&334.36&334.74&0.30\\
  &0.05&317.12&318.74&319.06&318.98&319.32&320.40&0.59\\
\multirow{-5}{*}{Mice.RF}&0.10&295.77&297.48&298.24&298.16&298.81&300.13&0.93\\

\hline Naive &0.01&332.86&333.24&333.40&333.45&333.65&334.08&0.31\\
  Imputation&0.05&315.86&316.97&317.36&317.34&317.74&318.84&0.58\\
 &0.10&296.33&297.99&298.59&298.52&299.10&300.13&0.84\\

\hline  &0.01&336.04&336.74&336.95&336.90&337.07&337.29&0.27\\
  &0.05&328.79&333.22&333.77&333.62&334.18&335.26&1.00\\
\multirow{-5}{*}{MissRanger}&0.10&316.70&327.76&328.54&328.09&329.35&333.40&2.83\\

\hline 
\end{tabular} 
\end{minipage} 
\end{table}

\newpage 
\begin{table}[H] 
\caption{minimum ($q_0$), first quartile ($q_{0.25}$),
               median ($q_{0.5}$), mean ($\bar{x}$), third quartile ($q_{0.75}$),
               maximum ($q_{1}$) and standard deviation (sd) of the observed values of the realization of Pearson's chi-squared statistic $\left(\text{with factor } \frac{1}{1000}\right)$ for the variable \texttt{ef42} for the different
               imputation methods and missing rates (r)
               creating missing values using the MAR-mechanism for the employee data (VSE, $\chi^2$, \texttt{ef42}, MAR)}\label{Tab:vse.vals.chisq.MAR.ef42 }\begin{minipage}[t]{\textwidth} 

\centering 
\begin{tabular}[t]{|l|c|c|c|c|c|c|c|c|}\hline
 method&r&$q_{0}$& $q_{0.25}$&$q_{0.5}$&$\bar{x}$&$q_{0.75}$& $q_{1}$&sd\\ \hline 
\hline  &0.01&229.30&230.03&230.33&230.31&230.57&231.33&0.37\\
  &0.05&214.65&216.25&216.62&216.68&217.09&218.11&0.66\\
\multirow{-5}{*}{Amelia}&0.10&198.23&199.94&200.49&200.52&201.12&202.46&0.87\\

\hline  &0.01&229.43&230.54&230.84&230.78&231.06&231.52&0.39\\
  &0.05&216.39&217.67&218.50&218.32&218.99&219.56&0.78\\
\multirow{-5}{*}{Mice.Norm}&0.10&199.17&200.90&201.92&201.78&202.64&204.35&1.09\\

\hline  &0.01&229.71&230.56&230.78&230.78&231.08&231.52&0.39\\
  &0.05&216.97&218.04&218.51&218.56&219.09&220.65&0.74\\
\multirow{-5}{*}{Mice.PMM}&0.10&198.53&201.27&201.97&202.09&202.84&204.47&1.10\\

\hline  &0.01&229.69&230.63&230.87&230.83&231.13&231.60&0.38\\
  &0.05&216.66&218.06&218.77&218.74&219.34&220.45&0.77\\
\multirow{-5}{*}{Mice.RF}&0.10&199.90&201.48&202.33&202.25&203.01&204.49&1.11\\

\hline Naive &0.01&229.38&230.15&230.43&230.39&230.66&231.10&0.37\\
  Imputation&0.05&216.17&217.35&217.93&217.96&218.58&219.33&0.72\\
 &0.10&201.57&203.18&203.95&203.93&204.66&206.75&1.05\\

\hline  &0.01&231.32&232.35&232.70&232.63&233.03&233.40&0.48\\
  &0.05&221.81&225.67&227.94&227.27&229.40&230.62&2.53\\
\multirow{-5}{*}{MissRanger}&0.10&209.67&217.61&221.46&220.57&224.28&227.06&4.40\\

\hline 
\end{tabular} 
\end{minipage} 
\end{table}

\newpage 
\begin{table}[H] 
\caption{minimum ($q_0$), first quartile ($q_{0.25}$),
               median ($q_{0.5}$), mean ($\bar{x}$), third quartile ($q_{0.75}$),
               maximum ($q_{1}$) and standard deviation (sd) of the observed values of the realization of Pearson's chi-squared statistic $\left(\text{with factor } \frac{1}{1000}\right)$ for the variable \texttt{ef42} for the different
               imputation methods and missing rates (r)
               creating missing values using the MCAR-mechanism for the employee data (VSE, $\chi^2$, \texttt{ef42}, MCAR)}\label{Tab:vse.vals.chisq.MCAR.ef42 }\begin{minipage}[t]{\textwidth} 

\centering 
\begin{tabular}[t]{|l|c|c|c|c|c|c|c|c|}\hline
 method&r&$q_{0}$& $q_{0.25}$&$q_{0.5}$&$\bar{x}$&$q_{0.75}$& $q_{1}$&sd\\ \hline 
\hline  &0.01&229.23&230.05&230.28&230.26&230.45&230.96&0.34\\
  &0.05&214.46&216.27&216.76&216.70&217.12&217.99&0.64\\
\multirow{-5}{*}{Amelia}&0.10&198.88&200.02&200.46&200.49&201.03&202.33&0.80\\

\hline  &0.01&229.55&230.63&230.83&230.82&231.09&231.51&0.36\\
  &0.05&216.06&217.69&218.26&218.20&218.65&219.85&0.71\\
\multirow{-5}{*}{Mice.Norm}&0.10&199.29&200.91&201.65&201.63&202.28&203.77&0.99\\

\hline  &0.01&229.76&230.60&230.87&230.81&231.07&231.53&0.37\\
  &0.05&216.64&218.08&218.45&218.43&218.83&219.96&0.67\\
\multirow{-5}{*}{Mice.PMM}&0.10&199.22&201.21&201.80&201.82&202.58&204.27&1.02\\

\hline  &0.01&229.65&230.66&230.92&230.87&231.14&231.61&0.38\\
  &0.05&216.13&218.21&218.58&218.63&219.16&220.21&0.72\\
\multirow{-5}{*}{Mice.RF}&0.10&199.64&201.52&202.01&202.07&202.67&204.26&0.93\\

\hline Naive &0.01&229.18&230.17&230.46&230.40&230.68&231.11&0.38\\
  Imputation&0.05&215.81&217.45&217.96&217.86&218.34&219.26&0.67\\
 &0.10&201.53&203.21&203.78&203.80&204.35&205.84&0.89\\

\hline  &0.01&231.20&232.13&232.44&232.45&232.85&233.32&0.53\\
  &0.05&221.88&225.88&228.64&227.74&229.64&231.02&2.22\\
\multirow{-5}{*}{MissRanger}&0.10&211.17&217.60&221.19&220.88&224.92&226.83&4.24\\

\hline 
\end{tabular} 
\end{minipage} 
\end{table}

\newpage 
\begin{table}[H] 
\caption{minimum ($q_0$), first quartile ($q_{0.25}$),
               median ($q_{0.5}$), mean ($\bar{x}$), third quartile ($q_{0.75}$),
               maximum ($q_{1}$) and standard deviation (sd) of the observed values of the realization of Pearson's chi-squared statistic $\left(\text{with factor } \frac{1}{1000}\right)$ for the variable \texttt{ef43} for the different
               imputation methods and missing rates (r)
               creating missing values using the MAR-mechanism for the employee data (VSE, $\chi^2$, \texttt{ef43}, MAR)}\label{Tab:vse.vals.chisq.MAR.ef43 }\begin{minipage}[t]{\textwidth} 

\centering 
\begin{tabular}[t]{|l|c|c|c|c|c|c|c|c|}\hline
 method&r&$q_{0}$& $q_{0.25}$&$q_{0.5}$&$\bar{x}$&$q_{0.75}$& $q_{1}$&sd\\ \hline 
\hline  &0.01&51.62&51.68&51.69&51.69&51.71&51.76&0.03\\
  &0.05&50.44&50.54&50.58&50.58&50.61&50.73&0.06\\
\multirow{-5}{*}{Amelia}&0.10&48.71&48.95&49.01&49.01&49.08&49.32&0.11\\

\hline  &0.01&51.37&51.47&51.50&51.50&51.53&51.61&0.04\\
  &0.05&48.95&49.17&49.30&49.29&49.41&49.58&0.16\\
\multirow{-5}{*}{Mice.Norm}&0.10&45.52&46.02&46.19&46.18&46.34&46.77&0.25\\

\hline  &0.01&51.37&51.46&51.50&51.50&51.53&51.59&0.05\\
  &0.05&49.06&49.31&49.43&49.41&49.50&49.77&0.14\\
\multirow{-5}{*}{Mice.PMM}&0.10&45.62&46.12&46.33&46.29&46.47&46.91&0.28\\

\hline  &0.01&51.39&51.48&51.51&51.51&51.54&51.63&0.05\\
  &0.05&49.11&49.37&49.46&49.45&49.54&49.72&0.13\\
\multirow{-5}{*}{Mice.RF}&0.10&45.75&46.31&46.47&46.43&46.58&46.90&0.25\\

\hline Naive &0.01&51.16&51.25&51.30&51.30&51.35&51.42&0.07\\
  Imputation&0.05&48.46&48.63&48.73&48.73&48.82&49.00&0.13\\
 &0.10&45.15&45.50&45.62&45.64&45.76&46.14&0.21\\

\hline  &0.01&51.51&51.73&51.90&51.82&51.93&51.94&0.12\\
  &0.05&50.02&50.81&51.06&51.18&51.73&51.81&0.55\\
\multirow{-5}{*}{MissRanger}&0.10&46.32&49.27&49.96&50.07&51.33&51.54&1.35\\

\hline 
\end{tabular} 
\end{minipage} 
\end{table}

\newpage 
\begin{table}[H] 
\caption{minimum ($q_0$), first quartile ($q_{0.25}$),
               median ($q_{0.5}$), mean ($\bar{x}$), third quartile ($q_{0.75}$),
               maximum ($q_{1}$) and standard deviation (sd) of the observed values of the realization of Pearson's chi-squared statistic $\left(\text{with factor } \frac{1}{1000}\right)$ for the variable \texttt{ef43} for the different
               imputation methods and missing rates (r)
               creating missing values using the MCAR-mechanism for the employee data (VSE, $\chi^2$, \texttt{ef43}, MCAR)}\label{Tab:vse.vals.chisq.MCAR.ef43 }\begin{minipage}[t]{\textwidth} 

\centering 
\begin{tabular}[t]{|l|c|c|c|c|c|c|c|c|}\hline
 method&r&$q_{0}$& $q_{0.25}$&$q_{0.5}$&$\bar{x}$&$q_{0.75}$& $q_{1}$&sd\\ \hline 
\hline  &0.01&51.63&51.67&51.69&51.69&51.71&51.76&0.03\\
  &0.05&50.40&50.54&50.57&50.58&50.62&50.77&0.06\\
\multirow{-5}{*}{Amelia}&0.10&48.81&48.92&49.01&49.00&49.07&49.27&0.10\\

\hline  &0.01&51.38&51.46&51.49&51.49&51.52&51.61&0.05\\
  &0.05&48.85&49.15&49.28&49.26&49.37&49.59&0.15\\
\multirow{-5}{*}{Mice.Norm}&0.10&45.34&46.01&46.18&46.16&46.33&46.67&0.26\\

\hline  &0.01&51.37&51.46&51.49&51.49&51.52&51.59&0.04\\
  &0.05&48.90&49.27&49.41&49.39&49.49&49.72&0.16\\
\multirow{-5}{*}{Mice.PMM}&0.10&45.75&46.11&46.30&46.30&46.47&46.94&0.25\\

\hline  &0.01&51.36&51.48&51.51&51.51&51.54&51.63&0.05\\
  &0.05&48.96&49.37&49.49&49.48&49.59&49.87&0.16\\
\multirow{-5}{*}{Mice.RF}&0.10&45.82&46.19&46.38&46.40&46.59&47.00&0.26\\

\hline Naive &0.01&51.13&51.25&51.28&51.29&51.33&51.44&0.07\\
  Imputation&0.05&48.34&48.62&48.73&48.73&48.82&49.09&0.15\\
 &0.10&45.18&45.48&45.67&45.64&45.79&46.07&0.20\\

\hline  &0.01&51.53&51.71&51.89&51.82&51.92&51.95&0.12\\
  &0.05&49.76&50.66&51.08&51.15&51.74&51.85&0.63\\
\multirow{-5}{*}{MissRanger}&0.10&46.42&49.29&49.93&50.08&51.31&51.57&1.30\\

\hline 
\end{tabular} 
\end{minipage} 
\end{table}

\newpage 
\begin{table}[H] 
\caption{minimum ($q_0$), first quartile ($q_{0.25}$),
               median ($q_{0.5}$), mean ($\bar{x}$), third quartile ($q_{0.75}$),
               maximum ($q_{1}$) and standard deviation (sd) of the observed values of the realization of Pearson's chi-squared statistic $\left(\text{with factor } \frac{1}{1000}\right)$ for the variable \texttt{ef9} for the different
               imputation methods and missing rates (r)
               creating missing values using the MAR-mechanism for the employee data (VSE, $\chi^2$, \texttt{ef9}, MAR)}\label{Tab:vse.vals.chisq.MAR.ef9 }\begin{minipage}[t]{\textwidth} 

\centering 
\begin{tabular}[t]{|l|c|c|c|c|c|c|c|c|}\hline
 method&r&$q_{0}$& $q_{0.25}$&$q_{0.5}$&$\bar{x}$&$q_{0.75}$& $q_{1}$&sd\\ \hline 
\hline  &0.01&127.93&128.13&128.21&128.20&128.27&128.51&0.11\\
  &0.05&121.07&121.40&121.54&121.55&121.72&122.18&0.21\\
\multirow{-5}{*}{Amelia}&0.10&112.75&113.30&113.54&113.49&113.74&114.25&0.33\\

\hline  &0.01&127.52&127.78&127.85&127.85&127.94&128.26&0.13\\
  &0.05&118.69&119.60&119.86&119.85&120.13&120.82&0.40\\
\multirow{-5}{*}{Mice.Norm}&0.10&108.58&109.77&110.19&110.19&110.63&111.63&0.67\\

\hline  &0.01&127.48&127.77&127.84&127.84&127.93&128.24&0.14\\
  &0.05&118.80&119.84&120.08&120.09&120.33&121.22&0.39\\
\multirow{-5}{*}{Mice.PMM}&0.10&108.30&109.78&110.28&110.21&110.65&111.61&0.68\\

\hline  &0.01&127.58&127.89&127.98&127.97&128.06&128.36&0.13\\
  &0.05&119.29&120.03&120.20&120.24&120.50&120.94&0.34\\
\multirow{-5}{*}{Mice.RF}&0.10&107.96&109.76&110.28&110.21&110.68&111.43&0.62\\

\hline Naive &0.01&127.52&127.91&128.07&128.03&128.14&128.41&0.19\\
  Imputation&0.05&119.32&120.95&121.25&121.19&121.52&122.23&0.51\\
 &0.10&108.81&112.73&113.30&113.04&113.72&114.79&1.10\\

\hline  &0.01&128.49&129.15&129.27&129.21&129.35&129.59&0.22\\
  &0.05&122.43&126.26&126.76&126.52&127.19&127.55&0.94\\
\multirow{-5}{*}{MissRanger}&0.10&115.43&121.60&123.61&122.57&124.38&124.90&2.46\\

\hline 
\end{tabular} 
\end{minipage} 
\end{table}

\newpage 
\begin{table}[H] 
\caption{minimum ($q_0$), first quartile ($q_{0.25}$),
               median ($q_{0.5}$), mean ($\bar{x}$), third quartile ($q_{0.75}$),
               maximum ($q_{1}$) and standard deviation (sd) of the observed values of the realization of Pearson's chi-squared statistic $\left(\text{with factor } \frac{1}{1000}\right)$ for the variable \texttt{ef9} for the different
               imputation methods and missing rates (r)
               creating missing values using the MCAR-mechanism for the employee data (VSE, $\chi^2$, \texttt{ef9}, MCAR)}\label{Tab:vse.vals.chisq.MCAR.ef9 }\begin{minipage}[t]{\textwidth} 

\centering 
\begin{tabular}[t]{|l|c|c|c|c|c|c|c|c|}\hline
 method&r&$q_{0}$& $q_{0.25}$&$q_{0.5}$&$\bar{x}$&$q_{0.75}$& $q_{1}$&sd\\ \hline 
\hline  &0.01&127.85&128.10&128.19&128.18&128.25&128.47&0.12\\
  &0.05&120.77&121.37&121.56&121.55&121.74&122.15&0.25\\
\multirow{-5}{*}{Amelia}&0.10&112.78&113.23&113.52&113.50&113.71&114.34&0.33\\

\hline  &0.01&127.55&127.72&127.82&127.82&127.90&128.17&0.14\\
  &0.05&118.60&119.45&119.66&119.68&119.94&120.52&0.38\\
\multirow{-5}{*}{Mice.Norm}&0.10&108.71&109.72&110.12&110.13&110.53&111.57&0.62\\

\hline  &0.01&127.40&127.74&127.82&127.82&127.92&128.23&0.15\\
  &0.05&118.94&119.62&119.90&119.90&120.20&120.85&0.41\\
\multirow{-5}{*}{Mice.PMM}&0.10&107.94&109.65&109.99&110.01&110.40&111.45&0.60\\

\hline  &0.01&127.61&127.86&127.95&127.95&128.04&128.25&0.13\\
  &0.05&119.19&119.87&120.10&120.09&120.30&120.80&0.33\\
\multirow{-5}{*}{Mice.RF}&0.10&108.58&109.85&110.20&110.23&110.63&111.75&0.56\\

\hline Naive &0.01&127.73&127.99&128.10&128.10&128.20&128.53&0.15\\
  Imputation&0.05&120.40&121.11&121.29&121.28&121.48&122.20&0.30\\
 &0.10&112.47&113.18&113.40&113.41&113.67&114.18&0.37\\

\hline  &0.01&128.48&129.08&129.23&129.17&129.34&129.49&0.23\\
  &0.05&122.81&126.19&126.80&126.48&127.15&127.52&0.98\\
\multirow{-5}{*}{MissRanger}&0.10&116.26&121.49&122.88&122.30&123.98&124.72&2.18\\

\hline 
\end{tabular} 
\end{minipage} 
\end{table}

\newpage 

%% file: table_fivepoint_CramersV_vse.txt
\begin{table}[H] 
\caption{minimum ($q_0$), first quartile ($q_{0.25}$),
               median ($q_{0.5}$), mean ($\bar{x}$), third quartile ($q_{0.75}$),
               maximum ($q_{1}$) and standard deviation (sd) of the observed values of Cram\'{e}rs $V$ for the variable \texttt{ef10} for the different
               imputation methods and missing rates (r)
               creating missing values using the MAR-mechanism for the employee data (VSE, Cram\'{e}rs $V$, \texttt{ef10}, MAR)}\label{Tab:vse.vals.cramersv.MAR.ef10 }\begin{minipage}[t]{\textwidth} 

\centering 
\begin{tabular}[t]{|l|c|c|c|c|c|c|c|c|}\hline
 method&r&$q_{0}$& $q_{0.25}$&$q_{0.5}$&$\bar{x}$&$q_{0.75}$& $q_{1}$&sd\\ \hline 
\hline  &0.01&0.99&0.99&0.99&0.99&0.99&1.00&0.00\\
  &0.05&0.97&0.97&0.97&0.97&0.97&0.97&0.00\\
\multirow{-5}{*}{Amelia}&0.10&0.93&0.94&0.94&0.94&0.94&0.94&0.00\\

\hline  &0.01&0.99&0.99&0.99&0.99&0.99&0.99&0.00\\
  &0.05&0.95&0.95&0.95&0.95&0.96&0.96&0.00\\
\multirow{-5}{*}{Mice.Norm}&0.10&0.90&0.91&0.91&0.91&0.91&0.92&0.00\\

\hline  &0.01&0.99&0.99&0.99&0.99&0.99&0.99&0.00\\
  &0.05&0.95&0.95&0.95&0.95&0.96&0.96&0.00\\
\multirow{-5}{*}{Mice.PMM}&0.10&0.90&0.91&0.91&0.91&0.91&0.91&0.00\\

\hline  &0.01&0.99&0.99&0.99&0.99&0.99&0.99&0.00\\
  &0.05&0.95&0.95&0.96&0.96&0.96&0.96&0.00\\
\multirow{-5}{*}{Mice.RF}&0.10&0.91&0.91&0.91&0.91&0.91&0.91&0.00\\

\hline Naive &0.01&0.99&0.99&0.99&0.99&0.99&0.99&0.00\\
  Imputation&0.05&0.95&0.96&0.96&0.96&0.96&0.96&0.00\\
 &0.10&0.91&0.92&0.92&0.92&0.92&0.92&0.00\\

\hline  &0.01&0.99&1.00&1.00&1.00&1.00&1.00&0.00\\
  &0.05&0.97&0.98&0.98&0.98&0.98&0.99&0.00\\
\multirow{-5}{*}{MissRanger}&0.10&0.94&0.96&0.96&0.96&0.97&0.97&0.01\\

\hline 
\end{tabular} 
\end{minipage} 
\end{table}

\newpage 
\begin{table}[H] 
\caption{minimum ($q_0$), first quartile ($q_{0.25}$),
               median ($q_{0.5}$), mean ($\bar{x}$), third quartile ($q_{0.75}$),
               maximum ($q_{1}$) and standard deviation (sd) of the observed values of Cram\'{e}rs $V$ for the variable \texttt{ef10} for the different
               imputation methods and missing rates (r)
               creating missing values using the MCAR-mechanism for the employee data (VSE, Cram\'{e}rs $V$, \texttt{ef10}, MCAR)}\label{Tab:vse.vals.cramersv.MCAR.ef10 }\begin{minipage}[t]{\textwidth} 

\centering 
\begin{tabular}[t]{|l|c|c|c|c|c|c|c|c|}\hline
 method&r&$q_{0}$& $q_{0.25}$&$q_{0.5}$&$\bar{x}$&$q_{0.75}$& $q_{1}$&sd\\ \hline 
\hline  &0.01&0.99&0.99&0.99&0.99&0.99&0.99&0.00\\
  &0.05&0.97&0.97&0.97&0.97&0.97&0.97&0.00\\
\multirow{-5}{*}{Amelia}&0.10&0.93&0.94&0.94&0.94&0.94&0.94&0.00\\

\hline  &0.01&0.99&0.99&0.99&0.99&0.99&0.99&0.00\\
  &0.05&0.95&0.95&0.95&0.95&0.96&0.96&0.00\\
\multirow{-5}{*}{Mice.Norm}&0.10&0.91&0.91&0.91&0.91&0.91&0.91&0.00\\

\hline  &0.01&0.99&0.99&0.99&0.99&0.99&0.99&0.00\\
  &0.05&0.95&0.95&0.95&0.95&0.96&0.96&0.00\\
\multirow{-5}{*}{Mice.PMM}&0.10&0.91&0.91&0.91&0.91&0.91&0.92&0.00\\

\hline  &0.01&0.99&0.99&0.99&0.99&0.99&0.99&0.00\\
  &0.05&0.95&0.95&0.96&0.96&0.96&0.96&0.00\\
\multirow{-5}{*}{Mice.RF}&0.10&0.91&0.91&0.91&0.91&0.91&0.92&0.00\\

\hline Naive &0.01&0.99&0.99&0.99&0.99&0.99&0.99&0.00\\
  Imputation&0.05&0.95&0.96&0.96&0.96&0.96&0.96&0.00\\
 &0.10&0.91&0.92&0.92&0.92&0.92&0.92&0.00\\

\hline  &0.01&0.99&1.00&1.00&1.00&1.00&1.00&0.00\\
  &0.05&0.97&0.98&0.98&0.98&0.98&0.99&0.00\\
\multirow{-5}{*}{MissRanger}&0.10&0.94&0.96&0.96&0.96&0.97&0.97&0.01\\

\hline 
\end{tabular} 
\end{minipage} 
\end{table}

\newpage 
\begin{table}[H] 
\caption{minimum ($q_0$), first quartile ($q_{0.25}$),
               median ($q_{0.5}$), mean ($\bar{x}$), third quartile ($q_{0.75}$),
               maximum ($q_{1}$) and standard deviation (sd) of the observed values of Cram\'{e}rs $V$ for the variable \texttt{ef15} for the different
               imputation methods and missing rates (r)
               creating missing values using the MAR-mechanism for the employee data (VSE, Cram\'{e}rs $V$, \texttt{ef15}, MAR)}\label{Tab:vse.vals.cramersv.MAR.ef15 }\begin{minipage}[t]{\textwidth} 

\centering 
\begin{tabular}[t]{|l|c|c|c|c|c|c|c|c|}\hline
 method&r&$q_{0}$& $q_{0.25}$&$q_{0.5}$&$\bar{x}$&$q_{0.75}$& $q_{1}$&sd\\ \hline 
\hline  &0.01&0.99&0.99&0.99&0.99&0.99&0.99&0.00\\
  &0.05&0.95&0.96&0.96&0.96&0.96&0.96&0.00\\
\multirow{-5}{*}{Amelia}&0.10&0.91&0.91&0.92&0.92&0.92&0.92&0.00\\

\hline  &0.01&0.99&0.99&0.99&0.99&0.99&0.99&0.00\\
  &0.05&0.96&0.96&0.96&0.96&0.96&0.96&0.00\\
\multirow{-5}{*}{Mice.Norm}&0.10&0.92&0.92&0.92&0.92&0.92&0.93&0.00\\

\hline  &0.01&0.99&0.99&0.99&0.99&0.99&0.99&0.00\\
  &0.05&0.96&0.96&0.96&0.96&0.96&0.96&0.00\\
\multirow{-5}{*}{Mice.PMM}&0.10&0.92&0.92&0.92&0.92&0.92&0.93&0.00\\

\hline  &0.01&0.99&0.99&0.99&0.99&0.99&0.99&0.00\\
  &0.05&0.96&0.96&0.96&0.96&0.96&0.96&0.00\\
\multirow{-5}{*}{Mice.RF}&0.10&0.92&0.92&0.92&0.92&0.92&0.93&0.00\\

\hline Naive &0.01&0.99&0.99&0.99&0.99&0.99&0.99&0.00\\
  Imputation&0.05&0.97&0.97&0.97&0.97&0.97&0.97&0.00\\
 &0.10&0.94&0.94&0.94&0.94&0.94&0.94&0.00\\

\hline  &0.01&0.99&1.00&1.00&1.00&1.00&1.00&0.00\\
  &0.05&0.97&0.98&0.98&0.98&0.99&0.99&0.00\\
\multirow{-5}{*}{MissRanger}&0.10&0.95&0.96&0.97&0.97&0.97&0.98&0.01\\

\hline 
\end{tabular} 
\end{minipage} 
\end{table}

\newpage 
\begin{table}[H] 
\caption{minimum ($q_0$), first quartile ($q_{0.25}$),
               median ($q_{0.5}$), mean ($\bar{x}$), third quartile ($q_{0.75}$),
               maximum ($q_{1}$) and standard deviation (sd) of the observed values of Cram\'{e}rs $V$ for the variable \texttt{ef15} for the different
               imputation methods and missing rates (r)
               creating missing values using the MCAR-mechanism for the employee data (VSE, Cram\'{e}rs $V$, \texttt{ef15}, MCAR)}\label{Tab:vse.vals.cramersv.MCAR.ef15 }\begin{minipage}[t]{\textwidth} 

\centering 
\begin{tabular}[t]{|l|c|c|c|c|c|c|c|c|}\hline
 method&r&$q_{0}$& $q_{0.25}$&$q_{0.5}$&$\bar{x}$&$q_{0.75}$& $q_{1}$&sd\\ \hline 
\hline  &0.01&0.99&0.99&0.99&0.99&0.99&0.99&0.00\\
  &0.05&0.95&0.96&0.96&0.96&0.96&0.96&0.00\\
\multirow{-5}{*}{Amelia}&0.10&0.91&0.91&0.91&0.92&0.92&0.92&0.00\\

\hline  &0.01&0.99&0.99&0.99&0.99&0.99&0.99&0.00\\
  &0.05&0.96&0.96&0.96&0.96&0.96&0.97&0.00\\
\multirow{-5}{*}{Mice.Norm}&0.10&0.92&0.92&0.92&0.92&0.92&0.93&0.00\\

\hline  &0.01&0.99&0.99&0.99&0.99&0.99&0.99&0.00\\
  &0.05&0.96&0.96&0.96&0.96&0.96&0.96&0.00\\
\multirow{-5}{*}{Mice.PMM}&0.10&0.91&0.92&0.92&0.92&0.92&0.93&0.00\\

\hline  &0.01&0.99&0.99&0.99&0.99&0.99&0.99&0.00\\
  &0.05&0.96&0.96&0.96&0.96&0.96&0.96&0.00\\
\multirow{-5}{*}{Mice.RF}&0.10&0.92&0.92&0.92&0.92&0.92&0.93&0.00\\

\hline Naive &0.01&0.99&0.99&0.99&0.99&0.99&0.99&0.00\\
  Imputation&0.05&0.97&0.97&0.97&0.97&0.97&0.97&0.00\\
 &0.10&0.94&0.94&0.94&0.94&0.94&0.95&0.00\\

\hline  &0.01&1.00&1.00&1.00&1.00&1.00&1.00&0.00\\
  &0.05&0.97&0.98&0.98&0.98&0.99&0.99&0.00\\
\multirow{-5}{*}{MissRanger}&0.10&0.95&0.96&0.97&0.97&0.97&0.98&0.01\\

\hline 
\end{tabular} 
\end{minipage} 
\end{table}

\newpage 
\begin{table}[H] 
\caption{minimum ($q_0$), first quartile ($q_{0.25}$),
               median ($q_{0.5}$), mean ($\bar{x}$), third quartile ($q_{0.75}$),
               maximum ($q_{1}$) and standard deviation (sd) of the observed values of Cram\'{e}rs $V$ for the variable \texttt{ef16u1} for the different
               imputation methods and missing rates (r)
               creating missing values using the MAR-mechanism for the employee data (VSE, Cram\'{e}rs $V$, \texttt{ef16u1}, MAR)}\label{Tab:vse.vals.cramersv.MAR.ef16u1 }\begin{minipage}[t]{\textwidth} 

\centering 
\begin{tabular}[t]{|l|c|c|c|c|c|c|c|c|}\hline
 method&r&$q_{0}$& $q_{0.25}$&$q_{0.5}$&$\bar{x}$&$q_{0.75}$& $q_{1}$&sd\\ \hline 
\hline  &0.01&0.95&0.97&0.98&0.98&0.98&0.99&0.01\\
  &0.05&0.90&0.92&0.92&0.92&0.93&0.94&0.01\\
\multirow{-5}{*}{Amelia}&0.10&0.86&0.87&0.87&0.87&0.88&0.89&0.01\\

\hline  &0.01&0.96&0.99&0.99&0.99&0.99&1.00&0.01\\
  &0.05&0.93&0.96&0.97&0.96&0.97&0.97&0.01\\
\multirow{-5}{*}{Mice.Norm}&0.10&0.90&0.92&0.93&0.93&0.94&0.94&0.01\\

\hline  &0.01&0.96&0.99&0.99&0.99&0.99&1.00&0.01\\
  &0.05&0.93&0.96&0.97&0.96&0.97&0.97&0.01\\
\multirow{-5}{*}{Mice.PMM}&0.10&0.90&0.92&0.93&0.93&0.93&0.94&0.01\\

\hline  &0.01&0.96&0.99&0.99&0.99&0.99&1.00&0.01\\
  &0.05&0.93&0.96&0.97&0.96&0.97&0.98&0.01\\
\multirow{-5}{*}{Mice.RF}&0.10&0.89&0.92&0.93&0.93&0.93&0.94&0.01\\

\hline Naive &0.01&0.97&0.99&0.99&0.99&0.99&1.00&0.01\\
  Imputation&0.05&0.94&0.97&0.97&0.97&0.97&0.98&0.01\\
 &0.10&0.91&0.94&0.95&0.94&0.95&0.95&0.01\\

\hline  &0.01&0.97&1.00&1.00&1.00&1.00&1.00&0.01\\
  &0.05&0.95&0.98&0.99&0.99&0.99&1.00&0.01\\
\multirow{-5}{*}{MissRanger}&0.10&0.92&0.96&0.98&0.97&0.99&0.99&0.02\\

\hline 
\end{tabular} 
\end{minipage} 
\end{table}

\newpage 
\begin{table}[H] 
\caption{minimum ($q_0$), first quartile ($q_{0.25}$),
               median ($q_{0.5}$), mean ($\bar{x}$), third quartile ($q_{0.75}$),
               maximum ($q_{1}$) and standard deviation (sd) of the observed values of Cram\'{e}rs $V$ for the variable \texttt{ef16u1} for the different
               imputation methods and missing rates (r)
               creating missing values using the MCAR-mechanism for the employee data (VSE, Cram\'{e}rs $V$, \texttt{ef16u1}, MCAR)}\label{Tab:vse.vals.cramersv.MCAR.ef16u1 }\begin{minipage}[t]{\textwidth} 

\centering 
\begin{tabular}[t]{|l|c|c|c|c|c|c|c|c|}\hline
 method&r&$q_{0}$& $q_{0.25}$&$q_{0.5}$&$\bar{x}$&$q_{0.75}$& $q_{1}$&sd\\ \hline 
\hline  &0.01&0.96&0.98&0.98&0.98&0.98&0.99&0.01\\
  &0.05&0.90&0.92&0.92&0.92&0.93&0.95&0.01\\
\multirow{-5}{*}{Amelia}&0.10&0.86&0.87&0.87&0.87&0.88&0.89&0.01\\

\hline  &0.01&0.96&0.99&0.99&0.99&0.99&1.00&0.01\\
  &0.05&0.93&0.96&0.97&0.96&0.97&0.97&0.01\\
\multirow{-5}{*}{Mice.Norm}&0.10&0.89&0.91&0.93&0.93&0.94&0.94&0.01\\

\hline  &0.01&0.96&0.99&0.99&0.99&0.99&1.00&0.01\\
  &0.05&0.93&0.96&0.97&0.96&0.97&0.97&0.01\\
\multirow{-5}{*}{Mice.PMM}&0.10&0.90&0.91&0.93&0.93&0.94&0.94&0.01\\

\hline  &0.01&0.96&0.99&0.99&0.99&0.99&1.00&0.01\\
  &0.05&0.94&0.96&0.97&0.96&0.97&0.97&0.01\\
\multirow{-5}{*}{Mice.RF}&0.10&0.90&0.91&0.93&0.92&0.93&0.94&0.01\\

\hline Naive &0.01&0.97&0.99&0.99&0.99&0.99&1.00&0.01\\
  Imputation&0.05&0.94&0.97&0.97&0.97&0.97&0.98&0.01\\
 &0.10&0.91&0.92&0.95&0.94&0.95&0.95&0.01\\

\hline  &0.01&0.97&1.00&1.00&1.00&1.00&1.00&0.01\\
  &0.05&0.94&0.98&0.99&0.99&0.99&1.00&0.01\\
\multirow{-5}{*}{MissRanger}&0.10&0.92&0.96&0.98&0.97&0.99&0.99&0.02\\

\hline 
\end{tabular} 
\end{minipage} 
\end{table}

\newpage 
\begin{table}[H] 
\caption{minimum ($q_0$), first quartile ($q_{0.25}$),
               median ($q_{0.5}$), mean ($\bar{x}$), third quartile ($q_{0.75}$),
               maximum ($q_{1}$) and standard deviation (sd) of the observed values of Cram\'{e}rs $V$ for the variable \texttt{ef16u2} for the different
               imputation methods and missing rates (r)
               creating missing values using the MAR-mechanism for the employee data (VSE, Cram\'{e}rs $V$, \texttt{ef16u2}, MAR)}\label{Tab:vse.vals.cramersv.MAR.ef16u2 }\begin{minipage}[t]{\textwidth} 

\centering 
\begin{tabular}[t]{|l|c|c|c|c|c|c|c|c|}\hline
 method&r&$q_{0}$& $q_{0.25}$&$q_{0.5}$&$\bar{x}$&$q_{0.75}$& $q_{1}$&sd\\ \hline 
\hline  &0.01&0.99&0.99&0.99&0.99&0.99&0.99&0.00\\
  &0.05&0.96&0.96&0.96&0.96&0.96&0.96&0.00\\
\multirow{-5}{*}{Amelia}&0.10&0.91&0.92&0.92&0.92&0.92&0.92&0.00\\

\hline  &0.01&0.99&0.99&0.99&0.99&0.99&0.99&0.00\\
  &0.05&0.96&0.96&0.96&0.96&0.96&0.97&0.00\\
\multirow{-5}{*}{Mice.Norm}&0.10&0.92&0.92&0.93&0.93&0.93&0.93&0.00\\

\hline  &0.01&0.99&0.99&0.99&0.99&0.99&0.99&0.00\\
  &0.05&0.96&0.96&0.96&0.96&0.97&0.97&0.00\\
\multirow{-5}{*}{Mice.PMM}&0.10&0.92&0.92&0.93&0.93&0.93&0.94&0.00\\

\hline  &0.01&0.99&0.99&0.99&0.99&0.99&0.99&0.00\\
  &0.05&0.96&0.96&0.96&0.96&0.97&0.97&0.00\\
\multirow{-5}{*}{Mice.RF}&0.10&0.92&0.93&0.93&0.93&0.93&0.93&0.00\\

\hline Naive &0.01&0.99&0.99&0.99&0.99&0.99&1.00&0.00\\
  Imputation&0.05&0.97&0.97&0.97&0.97&0.97&0.97&0.00\\
 &0.10&0.94&0.94&0.94&0.94&0.94&0.95&0.00\\

\hline  &0.01&1.00&1.00&1.00&1.00&1.00&1.00&0.00\\
  &0.05&0.98&0.98&0.99&0.99&0.99&0.99&0.00\\
\multirow{-5}{*}{MissRanger}&0.10&0.96&0.97&0.97&0.97&0.97&0.98&0.01\\

\hline 
\end{tabular} 
\end{minipage} 
\end{table}

\newpage 
\begin{table}[H] 
\caption{minimum ($q_0$), first quartile ($q_{0.25}$),
               median ($q_{0.5}$), mean ($\bar{x}$), third quartile ($q_{0.75}$),
               maximum ($q_{1}$) and standard deviation (sd) of the observed values of Cram\'{e}rs $V$ for the variable \texttt{ef16u2} for the different
               imputation methods and missing rates (r)
               creating missing values using the MCAR-mechanism for the employee data (VSE, Cram\'{e}rs $V$, \texttt{ef16u2}, MCAR)}\label{Tab:vse.vals.cramersv.MCAR.ef16u2 }\begin{minipage}[t]{\textwidth} 

\centering 
\begin{tabular}[t]{|l|c|c|c|c|c|c|c|c|}\hline
 method&r&$q_{0}$& $q_{0.25}$&$q_{0.5}$&$\bar{x}$&$q_{0.75}$& $q_{1}$&sd\\ \hline 
\hline  &0.01&0.99&0.99&0.99&0.99&0.99&0.99&0.00\\
  &0.05&0.96&0.96&0.96&0.96&0.96&0.96&0.00\\
\multirow{-5}{*}{Amelia}&0.10&0.91&0.92&0.92&0.92&0.92&0.93&0.00\\

\hline  &0.01&0.99&0.99&0.99&0.99&0.99&0.99&0.00\\
  &0.05&0.96&0.96&0.96&0.96&0.96&0.97&0.00\\
\multirow{-5}{*}{Mice.Norm}&0.10&0.92&0.92&0.93&0.93&0.93&0.93&0.00\\

\hline  &0.01&0.99&0.99&0.99&0.99&0.99&0.99&0.00\\
  &0.05&0.96&0.96&0.96&0.96&0.97&0.97&0.00\\
\multirow{-5}{*}{Mice.PMM}&0.10&0.92&0.92&0.93&0.93&0.93&0.93&0.00\\

\hline  &0.01&0.99&0.99&0.99&0.99&0.99&0.99&0.00\\
  &0.05&0.96&0.96&0.96&0.96&0.97&0.97&0.00\\
\multirow{-5}{*}{Mice.RF}&0.10&0.92&0.93&0.93&0.93&0.93&0.93&0.00\\

\hline Naive &0.01&0.99&0.99&0.99&0.99&0.99&1.00&0.00\\
  Imputation&0.05&0.97&0.97&0.97&0.97&0.97&0.97&0.00\\
 &0.10&0.93&0.94&0.94&0.94&0.94&0.94&0.00\\

\hline  &0.01&1.00&1.00&1.00&1.00&1.00&1.00&0.00\\
  &0.05&0.98&0.98&0.99&0.99&0.99&0.99&0.00\\
\multirow{-5}{*}{MissRanger}&0.10&0.96&0.97&0.97&0.97&0.97&0.98&0.01\\

\hline 
\end{tabular} 
\end{minipage} 
\end{table}

\newpage 
\begin{table}[H] 
\caption{minimum ($q_0$), first quartile ($q_{0.25}$),
               median ($q_{0.5}$), mean ($\bar{x}$), third quartile ($q_{0.75}$),
               maximum ($q_{1}$) and standard deviation (sd) of the observed values of Cram\'{e}rs $V$ for the variable \texttt{ef17} for the different
               imputation methods and missing rates (r)
               creating missing values using the MAR-mechanism for the employee data (VSE, Cram\'{e}rs $V$, \texttt{ef17}, MAR)}\label{Tab:vse.vals.cramersv.MAR.ef17 }\begin{minipage}[t]{\textwidth} 

\centering 
\begin{tabular}[t]{|l|c|c|c|c|c|c|c|c|}\hline
 method&r&$q_{0}$& $q_{0.25}$&$q_{0.5}$&$\bar{x}$&$q_{0.75}$& $q_{1}$&sd\\ \hline 
\hline  &0.01&0.99&0.99&0.99&0.99&0.99&0.99&0.00\\
  &0.05&0.96&0.96&0.96&0.96&0.96&0.97&0.00\\
\multirow{-5}{*}{Amelia}&0.10&0.92&0.93&0.93&0.93&0.93&0.93&0.00\\

\hline  &0.01&0.99&0.99&0.99&0.99&0.99&0.99&0.00\\
  &0.05&0.96&0.97&0.97&0.97&0.97&0.97&0.00\\
\multirow{-5}{*}{Mice.Norm}&0.10&0.93&0.93&0.93&0.93&0.94&0.94&0.00\\

\hline  &0.01&0.99&0.99&0.99&0.99&0.99&0.99&0.00\\
  &0.05&0.96&0.97&0.97&0.97&0.97&0.97&0.00\\
\multirow{-5}{*}{Mice.PMM}&0.10&0.93&0.93&0.93&0.93&0.94&0.94&0.00\\

\hline  &0.01&0.99&0.99&0.99&0.99&0.99&0.99&0.00\\
  &0.05&0.96&0.97&0.97&0.97&0.97&0.97&0.00\\
\multirow{-5}{*}{Mice.RF}&0.10&0.93&0.93&0.93&0.93&0.93&0.94&0.00\\

\hline Naive &0.01&0.99&0.99&0.99&0.99&0.99&1.00&0.00\\
  Imputation&0.05&0.97&0.97&0.97&0.97&0.97&0.98&0.00\\
 &0.10&0.94&0.94&0.94&0.94&0.95&0.95&0.00\\

\hline  &0.01&1.00&1.00&1.00&1.00&1.00&1.00&0.00\\
  &0.05&0.98&0.99&0.99&0.99&0.99&1.00&0.00\\
\multirow{-5}{*}{MissRanger}&0.10&0.96&0.98&0.99&0.98&0.99&0.99&0.01\\

\hline 
\end{tabular} 
\end{minipage} 
\end{table}

\newpage 
\begin{table}[H] 
\caption{minimum ($q_0$), first quartile ($q_{0.25}$),
               median ($q_{0.5}$), mean ($\bar{x}$), third quartile ($q_{0.75}$),
               maximum ($q_{1}$) and standard deviation (sd) of the observed values of Cram\'{e}rs $V$ for the variable \texttt{ef17} for the different
               imputation methods and missing rates (r)
               creating missing values using the MCAR-mechanism for the employee data (VSE, Cram\'{e}rs $V$, \texttt{ef17}, MCAR)}\label{Tab:vse.vals.cramersv.MCAR.ef17 }\begin{minipage}[t]{\textwidth} 

\centering 
\begin{tabular}[t]{|l|c|c|c|c|c|c|c|c|}\hline
 method&r&$q_{0}$& $q_{0.25}$&$q_{0.5}$&$\bar{x}$&$q_{0.75}$& $q_{1}$&sd\\ \hline 
\hline  &0.01&0.99&0.99&0.99&0.99&0.99&0.99&0.00\\
  &0.05&0.96&0.96&0.96&0.96&0.96&0.97&0.00\\
\multirow{-5}{*}{Amelia}&0.10&0.92&0.93&0.93&0.93&0.93&0.93&0.00\\

\hline  &0.01&0.99&0.99&0.99&0.99&0.99&1.00&0.00\\
  &0.05&0.96&0.97&0.97&0.97&0.97&0.97&0.00\\
\multirow{-5}{*}{Mice.Norm}&0.10&0.93&0.93&0.93&0.93&0.94&0.94&0.00\\

\hline  &0.01&0.99&0.99&0.99&0.99&0.99&0.99&0.00\\
  &0.05&0.96&0.97&0.97&0.97&0.97&0.97&0.00\\
\multirow{-5}{*}{Mice.PMM}&0.10&0.93&0.93&0.93&0.93&0.94&0.94&0.00\\

\hline  &0.01&0.99&0.99&0.99&0.99&0.99&1.00&0.00\\
  &0.05&0.96&0.97&0.97&0.97&0.97&0.97&0.00\\
\multirow{-5}{*}{Mice.RF}&0.10&0.92&0.93&0.93&0.93&0.93&0.94&0.00\\

\hline Naive &0.01&0.99&0.99&0.99&0.99&1.00&1.00&0.00\\
  Imputation&0.05&0.97&0.97&0.97&0.97&0.97&0.98&0.00\\
 &0.10&0.94&0.94&0.95&0.94&0.95&0.95&0.00\\

\hline  &0.01&1.00&1.00&1.00&1.00&1.00&1.00&0.00\\
  &0.05&0.98&0.99&0.99&0.99&0.99&1.00&0.00\\
\multirow{-5}{*}{MissRanger}&0.10&0.96&0.98&0.99&0.98&0.99&0.99&0.01\\

\hline 
\end{tabular} 
\end{minipage} 
\end{table}

\newpage 
\begin{table}[H] 
\caption{minimum ($q_0$), first quartile ($q_{0.25}$),
               median ($q_{0.5}$), mean ($\bar{x}$), third quartile ($q_{0.75}$),
               maximum ($q_{1}$) and standard deviation (sd) of the observed values of Cram\'{e}rs $V$ for the variable \texttt{ef4} for the different
               imputation methods and missing rates (r)
               creating missing values using the MAR-mechanism for the employee data (VSE, Cram\'{e}rs $V$, \texttt{ef4}, MAR)}\label{Tab:vse.vals.cramersv.MAR.ef4 }\begin{minipage}[t]{\textwidth} 

\centering 
\begin{tabular}[t]{|l|c|c|c|c|c|c|c|c|}\hline
 method&r&$q_{0}$& $q_{0.25}$&$q_{0.5}$&$\bar{x}$&$q_{0.75}$& $q_{1}$&sd\\ \hline 
\hline  &0.01&1.00&1.00&1.00&1.00&1.00&1.00&0.00\\
  &0.05&0.99&0.99&0.99&0.99&0.99&0.99&0.00\\
\multirow{-5}{*}{Amelia}&0.10&0.98&0.98&0.98&0.98&0.98&0.98&0.00\\

\hline  &0.01&0.99&0.99&0.99&0.99&1.00&1.00&0.00\\
  &0.05&0.97&0.97&0.97&0.97&0.97&0.97&0.00\\
\multirow{-5}{*}{Mice.Norm}&0.10&0.94&0.94&0.94&0.94&0.94&0.94&0.00\\

\hline  &0.01&0.99&0.99&0.99&0.99&1.00&1.00&0.00\\
  &0.05&0.97&0.97&0.97&0.97&0.97&0.98&0.00\\
\multirow{-5}{*}{Mice.PMM}&0.10&0.94&0.94&0.94&0.94&0.94&0.94&0.00\\

\hline  &0.01&0.99&0.99&0.99&0.99&1.00&1.00&0.00\\
  &0.05&0.97&0.97&0.97&0.97&0.97&0.97&0.00\\
\multirow{-5}{*}{Mice.RF}&0.10&0.94&0.94&0.94&0.94&0.94&0.94&0.00\\

\hline Naive &0.01&0.99&0.99&0.99&0.99&0.99&0.99&0.00\\
  Imputation&0.05&0.97&0.97&0.97&0.97&0.97&0.97&0.00\\
 &0.10&0.94&0.94&0.94&0.94&0.94&0.94&0.00\\

\hline  &0.01&1.00&1.00&1.00&1.00&1.00&1.00&0.00\\
  &0.05&0.99&0.99&0.99&0.99&0.99&1.00&0.00\\
\multirow{-5}{*}{MissRanger}&0.10&0.97&0.99&0.99&0.99&0.99&0.99&0.00\\

\hline 
\end{tabular} 
\end{minipage} 
\end{table}

\newpage 
\begin{table}[H] 
\caption{minimum ($q_0$), first quartile ($q_{0.25}$),
               median ($q_{0.5}$), mean ($\bar{x}$), third quartile ($q_{0.75}$),
               maximum ($q_{1}$) and standard deviation (sd) of the observed values of Cram\'{e}rs $V$ for the variable \texttt{ef4} for the different
               imputation methods and missing rates (r)
               creating missing values using the MCAR-mechanism for the employee data (VSE, Cram\'{e}rs $V$, \texttt{ef4}, MCAR)}\label{Tab:vse.vals.cramersv.MCAR.ef4 }\begin{minipage}[t]{\textwidth} 

\centering 
\begin{tabular}[t]{|l|c|c|c|c|c|c|c|c|}\hline
 method&r&$q_{0}$& $q_{0.25}$&$q_{0.5}$&$\bar{x}$&$q_{0.75}$& $q_{1}$&sd\\ \hline 
\hline  &0.01&1.00&1.00&1.00&1.00&1.00&1.00&0.00\\
  &0.05&0.99&0.99&0.99&0.99&0.99&0.99&0.00\\
\multirow{-5}{*}{Amelia}&0.10&0.98&0.98&0.98&0.98&0.98&0.98&0.00\\

\hline  &0.01&0.99&0.99&0.99&0.99&1.00&1.00&0.00\\
  &0.05&0.97&0.97&0.97&0.97&0.97&0.97&0.00\\
\multirow{-5}{*}{Mice.Norm}&0.10&0.94&0.94&0.94&0.94&0.94&0.94&0.00\\

\hline  &0.01&0.99&0.99&0.99&0.99&1.00&1.00&0.00\\
  &0.05&0.97&0.97&0.97&0.97&0.97&0.97&0.00\\
\multirow{-5}{*}{Mice.PMM}&0.10&0.94&0.94&0.94&0.94&0.94&0.94&0.00\\

\hline  &0.01&0.99&0.99&0.99&0.99&1.00&1.00&0.00\\
  &0.05&0.97&0.97&0.97&0.97&0.97&0.97&0.00\\
\multirow{-5}{*}{Mice.RF}&0.10&0.94&0.94&0.94&0.94&0.94&0.94&0.00\\

\hline Naive &0.01&0.99&0.99&0.99&0.99&0.99&0.99&0.00\\
  Imputation&0.05&0.97&0.97&0.97&0.97&0.97&0.97&0.00\\
 &0.10&0.94&0.94&0.94&0.94&0.94&0.94&0.00\\

\hline  &0.01&1.00&1.00&1.00&1.00&1.00&1.00&0.00\\
  &0.05&0.99&0.99&0.99&0.99&0.99&1.00&0.00\\
\multirow{-5}{*}{MissRanger}&0.10&0.97&0.99&0.99&0.99&0.99&0.99&0.00\\

\hline 
\end{tabular} 
\end{minipage} 
\end{table}

\newpage 
\begin{table}[H] 
\caption{minimum ($q_0$), first quartile ($q_{0.25}$),
               median ($q_{0.5}$), mean ($\bar{x}$), third quartile ($q_{0.75}$),
               maximum ($q_{1}$) and standard deviation (sd) of the observed values of Cram\'{e}rs $V$ for the variable \texttt{ef42} for the different
               imputation methods and missing rates (r)
               creating missing values using the MAR-mechanism for the employee data (VSE, Cram\'{e}rs $V$, \texttt{ef42}, MAR)}\label{Tab:vse.vals.cramersv.MAR.ef42 }\begin{minipage}[t]{\textwidth} 

\centering 
\begin{tabular}[t]{|l|c|c|c|c|c|c|c|c|}\hline
 method&r&$q_{0}$& $q_{0.25}$&$q_{0.5}$&$\bar{x}$&$q_{0.75}$& $q_{1}$&sd\\ \hline 
\hline  &0.01&0.99&0.99&0.99&0.99&0.99&0.99&0.00\\
  &0.05&0.96&0.96&0.96&0.96&0.96&0.97&0.00\\
\multirow{-5}{*}{Amelia}&0.10&0.92&0.92&0.93&0.93&0.93&0.93&0.00\\

\hline  &0.01&0.99&0.99&0.99&0.99&0.99&1.00&0.00\\
  &0.05&0.96&0.96&0.97&0.97&0.97&0.97&0.00\\
\multirow{-5}{*}{Mice.Norm}&0.10&0.92&0.93&0.93&0.93&0.93&0.93&0.00\\

\hline  &0.01&0.99&0.99&0.99&0.99&0.99&1.00&0.00\\
  &0.05&0.96&0.97&0.97&0.97&0.97&0.97&0.00\\
\multirow{-5}{*}{Mice.PMM}&0.10&0.92&0.93&0.93&0.93&0.93&0.94&0.00\\

\hline  &0.01&0.99&0.99&0.99&0.99&0.99&1.00&0.00\\
  &0.05&0.96&0.97&0.97&0.97&0.97&0.97&0.00\\
\multirow{-5}{*}{Mice.RF}&0.10&0.92&0.93&0.93&0.93&0.93&0.94&0.00\\

\hline Naive &0.01&0.99&0.99&0.99&0.99&0.99&0.99&0.00\\
  Imputation&0.05&0.96&0.96&0.97&0.97&0.97&0.97&0.00\\
 &0.10&0.93&0.93&0.93&0.93&0.94&0.94&0.00\\

\hline  &0.01&0.99&1.00&1.00&1.00&1.00&1.00&0.00\\
  &0.05&0.97&0.98&0.99&0.99&0.99&0.99&0.01\\
\multirow{-5}{*}{MissRanger}&0.10&0.95&0.96&0.97&0.97&0.98&0.99&0.01\\

\hline 
\end{tabular} 
\end{minipage} 
\end{table}

\newpage 
\begin{table}[H] 
\caption{minimum ($q_0$), first quartile ($q_{0.25}$),
               median ($q_{0.5}$), mean ($\bar{x}$), third quartile ($q_{0.75}$),
               maximum ($q_{1}$) and standard deviation (sd) of the observed values of Cram\'{e}rs $V$ for the variable \texttt{ef42} for the different
               imputation methods and missing rates (r)
               creating missing values using the MCAR-mechanism for the employee data (VSE, Cram\'{e}rs $V$, \texttt{ef42}, MCAR)}\label{Tab:vse.vals.cramersv.MCAR.ef42 }\begin{minipage}[t]{\textwidth} 

\centering 
\begin{tabular}[t]{|l|c|c|c|c|c|c|c|c|}\hline
 method&r&$q_{0}$& $q_{0.25}$&$q_{0.5}$&$\bar{x}$&$q_{0.75}$& $q_{1}$&sd\\ \hline 
\hline  &0.01&0.99&0.99&0.99&0.99&0.99&0.99&0.00\\
  &0.05&0.96&0.96&0.96&0.96&0.96&0.97&0.00\\
\multirow{-5}{*}{Amelia}&0.10&0.92&0.93&0.93&0.93&0.93&0.93&0.00\\

\hline  &0.01&0.99&0.99&0.99&0.99&0.99&1.00&0.00\\
  &0.05&0.96&0.97&0.97&0.97&0.97&0.97&0.00\\
\multirow{-5}{*}{Mice.Norm}&0.10&0.92&0.93&0.93&0.93&0.93&0.93&0.00\\

\hline  &0.01&0.99&0.99&0.99&0.99&0.99&1.00&0.00\\
  &0.05&0.96&0.97&0.97&0.97&0.97&0.97&0.00\\
\multirow{-5}{*}{Mice.PMM}&0.10&0.92&0.93&0.93&0.93&0.93&0.93&0.00\\

\hline  &0.01&0.99&0.99&0.99&0.99&0.99&1.00&0.00\\
  &0.05&0.96&0.97&0.97&0.97&0.97&0.97&0.00\\
\multirow{-5}{*}{Mice.RF}&0.10&0.92&0.93&0.93&0.93&0.93&0.93&0.00\\

\hline Naive &0.01&0.99&0.99&0.99&0.99&0.99&0.99&0.00\\
  Imputation&0.05&0.96&0.96&0.97&0.97&0.97&0.97&0.00\\
 &0.10&0.93&0.93&0.93&0.93&0.93&0.94&0.00\\

\hline  &0.01&0.99&1.00&1.00&1.00&1.00&1.00&0.00\\
  &0.05&0.97&0.98&0.99&0.99&0.99&0.99&0.00\\
\multirow{-5}{*}{MissRanger}&0.10&0.95&0.96&0.97&0.97&0.98&0.99&0.01\\

\hline 
\end{tabular} 
\end{minipage} 
\end{table}

\newpage 
\begin{table}[H] 
\caption{minimum ($q_0$), first quartile ($q_{0.25}$),
               median ($q_{0.5}$), mean ($\bar{x}$), third quartile ($q_{0.75}$),
               maximum ($q_{1}$) and standard deviation (sd) of the observed values of Cram\'{e}rs $V$ for the variable \texttt{ef43} for the different
               imputation methods and missing rates (r)
               creating missing values using the MAR-mechanism for the employee data (VSE, Cram\'{e}rs $V$, \texttt{ef43}, MAR)}\label{Tab:vse.vals.cramersv.MAR.ef43 }\begin{minipage}[t]{\textwidth} 

\centering 
\begin{tabular}[t]{|l|c|c|c|c|c|c|c|c|}\hline
 method&r&$q_{0}$& $q_{0.25}$&$q_{0.5}$&$\bar{x}$&$q_{0.75}$& $q_{1}$&sd\\ \hline 
\hline  &0.01&1.00&1.00&1.00&1.00&1.00&1.00&0.00\\
  &0.05&0.99&0.99&0.99&0.99&0.99&0.99&0.00\\
\multirow{-5}{*}{Amelia}&0.10&0.97&0.97&0.97&0.97&0.97&0.97&0.00\\

\hline  &0.01&0.99&1.00&1.00&1.00&1.00&1.00&0.00\\
  &0.05&0.97&0.97&0.97&0.97&0.98&0.98&0.00\\
\multirow{-5}{*}{Mice.Norm}&0.10&0.94&0.94&0.94&0.94&0.94&0.95&0.00\\

\hline  &0.01&0.99&1.00&1.00&1.00&1.00&1.00&0.00\\
  &0.05&0.97&0.97&0.98&0.98&0.98&0.98&0.00\\
\multirow{-5}{*}{Mice.PMM}&0.10&0.94&0.94&0.94&0.94&0.95&0.95&0.00\\

\hline  &0.01&0.99&1.00&1.00&1.00&1.00&1.00&0.00\\
  &0.05&0.97&0.97&0.98&0.98&0.98&0.98&0.00\\
\multirow{-5}{*}{Mice.RF}&0.10&0.94&0.94&0.95&0.95&0.95&0.95&0.00\\

\hline Naive &0.01&0.99&0.99&0.99&0.99&0.99&0.99&0.00\\
  Imputation&0.05&0.97&0.97&0.97&0.97&0.97&0.97&0.00\\
 &0.10&0.93&0.94&0.94&0.94&0.94&0.94&0.00\\

\hline  &0.01&1.00&1.00&1.00&1.00&1.00&1.00&0.00\\
  &0.05&0.98&0.99&0.99&0.99&1.00&1.00&0.01\\
\multirow{-5}{*}{MissRanger}&0.10&0.94&0.97&0.98&0.98&0.99&1.00&0.01\\

\hline 
\end{tabular} 
\end{minipage} 
\end{table}

\newpage 
\begin{table}[H] 
\caption{minimum ($q_0$), first quartile ($q_{0.25}$),
               median ($q_{0.5}$), mean ($\bar{x}$), third quartile ($q_{0.75}$),
               maximum ($q_{1}$) and standard deviation (sd) of the observed values of Cram\'{e}rs $V$ for the variable \texttt{ef43} for the different
               imputation methods and missing rates (r)
               creating missing values using the MCAR-mechanism for the employee data (VSE, Cram\'{e}rs $V$, \texttt{ef43}, MCAR)}\label{Tab:vse.vals.cramersv.MCAR.ef43 }\begin{minipage}[t]{\textwidth} 

\centering 
\begin{tabular}[t]{|l|c|c|c|c|c|c|c|c|}\hline
 method&r&$q_{0}$& $q_{0.25}$&$q_{0.5}$&$\bar{x}$&$q_{0.75}$& $q_{1}$&sd\\ \hline 
\hline  &0.01&1.00&1.00&1.00&1.00&1.00&1.00&0.00\\
  &0.05&0.98&0.99&0.99&0.99&0.99&0.99&0.00\\
\multirow{-5}{*}{Amelia}&0.10&0.97&0.97&0.97&0.97&0.97&0.97&0.00\\

\hline  &0.01&0.99&1.00&1.00&1.00&1.00&1.00&0.00\\
  &0.05&0.97&0.97&0.97&0.97&0.97&0.98&0.00\\
\multirow{-5}{*}{Mice.Norm}&0.10&0.93&0.94&0.94&0.94&0.94&0.95&0.00\\

\hline  &0.01&0.99&1.00&1.00&1.00&1.00&1.00&0.00\\
  &0.05&0.97&0.97&0.98&0.98&0.98&0.98&0.00\\
\multirow{-5}{*}{Mice.PMM}&0.10&0.94&0.94&0.94&0.94&0.95&0.95&0.00\\

\hline  &0.01&0.99&1.00&1.00&1.00&1.00&1.00&0.00\\
  &0.05&0.97&0.97&0.98&0.98&0.98&0.98&0.00\\
\multirow{-5}{*}{Mice.RF}&0.10&0.94&0.94&0.94&0.95&0.95&0.95&0.00\\

\hline Naive &0.01&0.99&0.99&0.99&0.99&0.99&1.00&0.00\\
  Imputation&0.05&0.96&0.97&0.97&0.97&0.97&0.97&0.00\\
 &0.10&0.93&0.94&0.94&0.94&0.94&0.94&0.00\\

\hline  &0.01&1.00&1.00&1.00&1.00&1.00&1.00&0.00\\
  &0.05&0.98&0.99&0.99&0.99&1.00&1.00&0.01\\
\multirow{-5}{*}{MissRanger}&0.10&0.95&0.97&0.98&0.98&0.99&1.00&0.01\\

\hline 
\end{tabular} 
\end{minipage} 
\end{table}

\newpage 
\begin{table}[H] 
\caption{minimum ($q_0$), first quartile ($q_{0.25}$),
               median ($q_{0.5}$), mean ($\bar{x}$), third quartile ($q_{0.75}$),
               maximum ($q_{1}$) and standard deviation (sd) of the observed values of Cram\'{e}rs $V$ for the variable \texttt{ef9} for the different
               imputation methods and missing rates (r)
               creating missing values using the MAR-mechanism for the employee data (VSE, Cram\'{e}rs $V$, \texttt{ef9}, MAR)}\label{Tab:vse.vals.cramersv.MAR.ef9 }\begin{minipage}[t]{\textwidth} 

\centering 
\begin{tabular}[t]{|l|c|c|c|c|c|c|c|c|}\hline
 method&r&$q_{0}$& $q_{0.25}$&$q_{0.5}$&$\bar{x}$&$q_{0.75}$& $q_{1}$&sd\\ \hline 
\hline  &0.01&0.99&0.99&0.99&0.99&0.99&0.99&0.00\\
  &0.05&0.97&0.97&0.97&0.97&0.97&0.97&0.00\\
\multirow{-5}{*}{Amelia}&0.10&0.93&0.93&0.93&0.93&0.94&0.94&0.00\\

\hline  &0.01&0.99&0.99&0.99&0.99&0.99&0.99&0.00\\
  &0.05&0.96&0.96&0.96&0.96&0.96&0.96&0.00\\
\multirow{-5}{*}{Mice.Norm}&0.10&0.91&0.92&0.92&0.92&0.92&0.93&0.00\\

\hline  &0.01&0.99&0.99&0.99&0.99&0.99&0.99&0.00\\
  &0.05&0.96&0.96&0.96&0.96&0.96&0.97&0.00\\
\multirow{-5}{*}{Mice.PMM}&0.10&0.91&0.92&0.92&0.92&0.92&0.93&0.00\\

\hline  &0.01&0.99&0.99&0.99&0.99&0.99&0.99&0.00\\
  &0.05&0.96&0.96&0.96&0.96&0.96&0.96&0.00\\
\multirow{-5}{*}{Mice.RF}&0.10&0.91&0.92&0.92&0.92&0.92&0.93&0.00\\

\hline Naive &0.01&0.99&0.99&0.99&0.99&0.99&0.99&0.00\\
  Imputation&0.05&0.96&0.97&0.97&0.97&0.97&0.97&0.00\\
 &0.10&0.92&0.93&0.93&0.93&0.94&0.94&0.00\\

\hline  &0.01&0.99&1.00&1.00&1.00&1.00&1.00&0.00\\
  &0.05&0.97&0.99&0.99&0.99&0.99&0.99&0.00\\
\multirow{-5}{*}{MissRanger}&0.10&0.94&0.97&0.98&0.97&0.98&0.98&0.01\\

\hline 
\end{tabular} 
\end{minipage} 
\end{table}

\newpage 
\begin{table}[H] 
\caption{minimum ($q_0$), first quartile ($q_{0.25}$),
               median ($q_{0.5}$), mean ($\bar{x}$), third quartile ($q_{0.75}$),
               maximum ($q_{1}$) and standard deviation (sd) of the observed values of Cram\'{e}rs $V$ for the variable \texttt{ef9} for the different
               imputation methods and missing rates (r)
               creating missing values using the MCAR-mechanism for the employee data (VSE, Cram\'{e}rs $V$, \texttt{ef9}, MCAR)}\label{Tab:vse.vals.cramersv.MCAR.ef9 }\begin{minipage}[t]{\textwidth} 

\centering 
\begin{tabular}[t]{|l|c|c|c|c|c|c|c|c|}\hline
 method&r&$q_{0}$& $q_{0.25}$&$q_{0.5}$&$\bar{x}$&$q_{0.75}$& $q_{1}$&sd\\ \hline 
\hline  &0.01&0.99&0.99&0.99&0.99&0.99&0.99&0.00\\
  &0.05&0.96&0.97&0.97&0.97&0.97&0.97&0.00\\
\multirow{-5}{*}{Amelia}&0.10&0.93&0.93&0.93&0.93&0.94&0.94&0.00\\

\hline  &0.01&0.99&0.99&0.99&0.99&0.99&0.99&0.00\\
  &0.05&0.96&0.96&0.96&0.96&0.96&0.96&0.00\\
\multirow{-5}{*}{Mice.Norm}&0.10&0.91&0.92&0.92&0.92&0.92&0.93&0.00\\

\hline  &0.01&0.99&0.99&0.99&0.99&0.99&0.99&0.00\\
  &0.05&0.96&0.96&0.96&0.96&0.96&0.96&0.00\\
\multirow{-5}{*}{Mice.PMM}&0.10&0.91&0.92&0.92&0.92&0.92&0.93&0.00\\

\hline  &0.01&0.99&0.99&0.99&0.99&0.99&0.99&0.00\\
  &0.05&0.96&0.96&0.96&0.96&0.96&0.96&0.00\\
\multirow{-5}{*}{Mice.RF}&0.10&0.91&0.92&0.92&0.92&0.92&0.93&0.00\\

\hline Naive &0.01&0.99&0.99&0.99&0.99&0.99&0.99&0.00\\
  Imputation&0.05&0.96&0.97&0.97&0.97&0.97&0.97&0.00\\
 &0.10&0.93&0.93&0.93&0.93&0.94&0.94&0.00\\

\hline  &0.01&0.99&1.00&1.00&1.00&1.00&1.00&0.00\\
  &0.05&0.97&0.99&0.99&0.99&0.99&0.99&0.00\\
\multirow{-5}{*}{MissRanger}&0.10&0.95&0.97&0.97&0.97&0.98&0.98&0.01\\

\hline 
\end{tabular} 
\end{minipage} 
\end{table}

\newpage

%% file: table_fivepoint_KolmogorowSmirnow_vse.txt
\begin{table}[H] 
\caption{minimum ($q_0$), first quartile ($q_{0.25}$),
               median ($q_{0.5}$), mean ($\bar{x}$), third quartile ($q_{0.75}$),
               maximum ($q_{1}$) and standard deviation (sd) of the observed values of the realization of the Kolmogorow-Smirnow-statistic for the variable \texttt{ef19} for the different
               imputation methods and missing rates (r)
               creating missing values using the MAR-mechanism for the employee data (VSE, Kolmogorow-Smirnow, \texttt{ef19}, MAR)}\label{Tab:vse.vals.ksmirn.MAR.ef19 }\begin{minipage}[t]{\textwidth} 

\centering 
\begin{tabular}[t]{|l|c|c|c|c|c|c|c|c|}\hline
 method&r&$q_{0}$& $q_{0.25}$&$q_{0.5}$&$\bar{x}$&$q_{0.75}$& $q_{1}$&sd\\ \hline 
\hline  &0.01&0.00&0.00&0.00&0.00&0.00&0.00&0.00\\
  &0.05&0.01&0.01&0.01&0.01&0.01&0.01&0.00\\
\multirow{-5}{*}{Amelia}&0.10&0.01&0.02&0.02&0.02&0.02&0.02&0.00\\

\hline  &0.01&0.00&0.00&0.00&0.00&0.00&0.00&0.00\\
  &0.05&0.00&0.00&0.00&0.00&0.00&0.00&0.00\\
\multirow{-5}{*}{Mice.Norm}&0.10&0.00&0.00&0.00&0.00&0.00&0.00&0.00\\

\hline  &0.01&0.00&0.00&0.00&0.00&0.00&0.00&0.00\\
  &0.05&0.01&0.01&0.01&0.01&0.01&0.01&0.00\\
\multirow{-5}{*}{Mice.PMM}&0.10&0.02&0.02&0.02&0.02&0.02&0.02&0.00\\

\hline  &0.01&0.00&0.00&0.00&0.00&0.00&0.00&0.00\\
  &0.05&0.00&0.00&0.00&0.00&0.00&0.00&0.00\\
\multirow{-5}{*}{Mice.RF}&0.10&0.00&0.00&0.00&0.00&0.01&0.01&0.00\\

\hline Naive &0.01&0.01&0.01&0.01&0.01&0.01&0.01&0.00\\
  Imputation&0.05&0.03&0.03&0.03&0.03&0.04&0.04&0.00\\
 &0.10&0.07&0.07&0.07&0.07&0.07&0.08&0.00\\

\hline  &0.01&0.00&0.00&0.00&0.00&0.00&0.00&0.00\\
  &0.05&0.01&0.01&0.01&0.01&0.01&0.02&0.00\\
\multirow{-5}{*}{MissRanger}&0.10&0.02&0.02&0.02&0.02&0.02&0.05&0.01\\

\hline 
\end{tabular} 
\end{minipage} 
\end{table}

\newpage 
\begin{table}[H] 
\caption{minimum ($q_0$), first quartile ($q_{0.25}$),
               median ($q_{0.5}$), mean ($\bar{x}$), third quartile ($q_{0.75}$),
               maximum ($q_{1}$) and standard deviation (sd) of the observed values of the realization of the Kolmogorow-Smirnow-statistic for the variable \texttt{ef19} for the different
               imputation methods and missing rates (r)
               creating missing values using the MCAR-mechanism for the employee data (VSE, Kolmogorow-Smirnow, \texttt{ef19}, MCAR)}\label{Tab:vse.vals.ksmirn.MCAR.ef19 }\begin{minipage}[t]{\textwidth} 

\centering 
\begin{tabular}[t]{|l|c|c|c|c|c|c|c|c|}\hline
 method&r&$q_{0}$& $q_{0.25}$&$q_{0.5}$&$\bar{x}$&$q_{0.75}$& $q_{1}$&sd\\ \hline 
\hline  &0.01&0.00&0.00&0.00&0.00&0.00&0.00&0.00\\
  &0.05&0.01&0.01&0.01&0.01&0.01&0.01&0.00\\
\multirow{-5}{*}{Amelia}&0.10&0.01&0.02&0.02&0.02&0.02&0.02&0.00\\

\hline  &0.01&0.00&0.00&0.00&0.00&0.00&0.00&0.00\\
  &0.05&0.00&0.00&0.00&0.00&0.00&0.00&0.00\\
\multirow{-5}{*}{Mice.Norm}&0.10&0.00&0.00&0.00&0.00&0.00&0.00&0.00\\

\hline  &0.01&0.00&0.00&0.00&0.00&0.00&0.00&0.00\\
  &0.05&0.01&0.01&0.01&0.01&0.01&0.01&0.00\\
\multirow{-5}{*}{Mice.PMM}&0.10&0.02&0.02&0.02&0.02&0.02&0.02&0.00\\

\hline  &0.01&0.00&0.00&0.00&0.00&0.00&0.00&0.00\\
  &0.05&0.00&0.00&0.00&0.00&0.00&0.00&0.00\\
\multirow{-5}{*}{Mice.RF}&0.10&0.00&0.00&0.00&0.01&0.01&0.01&0.00\\

\hline Naive &0.01&0.01&0.01&0.01&0.01&0.01&0.01&0.00\\
  Imputation&0.05&0.03&0.03&0.03&0.03&0.04&0.04&0.00\\
 &0.10&0.06&0.07&0.07&0.07&0.07&0.07&0.00\\

\hline  &0.01&0.00&0.00&0.00&0.00&0.00&0.00&0.00\\
  &0.05&0.01&0.01&0.01&0.01&0.01&0.02&0.00\\
\multirow{-5}{*}{MissRanger}&0.10&0.02&0.02&0.02&0.02&0.02&0.05&0.01\\

\hline 
\end{tabular} 
\end{minipage} 
\end{table}

\newpage 
\begin{table}[H] 
\caption{minimum ($q_0$), first quartile ($q_{0.25}$),
               median ($q_{0.5}$), mean ($\bar{x}$), third quartile ($q_{0.75}$),
               maximum ($q_{1}$) and standard deviation (sd) of the observed values of the realization of the Kolmogorow-Smirnow-statistic for the variable \texttt{ef20} for the different
               imputation methods and missing rates (r)
               creating missing values using the MAR-mechanism for the employee data (VSE, Kolmogorow-Smirnow, \texttt{ef20}, MAR)}\label{Tab:vse.vals.ksmirn.MAR.ef20 }\begin{minipage}[t]{\textwidth} 

\centering 
\begin{tabular}[t]{|l|c|c|c|c|c|c|c|c|}\hline
 method&r&$q_{0}$& $q_{0.25}$&$q_{0.5}$&$\bar{x}$&$q_{0.75}$& $q_{1}$&sd\\ \hline 
\hline  &0.01&0.00&0.00&0.00&0.00&0.01&0.01&0.00\\
  &0.05&0.02&0.02&0.02&0.02&0.02&0.03&0.00\\
\multirow{-5}{*}{Amelia}&0.10&0.05&0.05&0.05&0.05&0.05&0.05&0.00\\

\hline  &0.01&0.00&0.00&0.00&0.00&0.00&0.00&0.00\\
  &0.05&0.00&0.00&0.00&0.00&0.00&0.00&0.00\\
\multirow{-5}{*}{Mice.Norm}&0.10&0.00&0.00&0.00&0.00&0.00&0.00&0.00\\

\hline  &0.01&0.00&0.00&0.01&0.01&0.01&0.01&0.00\\
  &0.05&0.02&0.02&0.03&0.03&0.03&0.03&0.00\\
\multirow{-5}{*}{Mice.PMM}&0.10&0.05&0.05&0.05&0.05&0.05&0.05&0.00\\

\hline  &0.01&0.00&0.00&0.00&0.00&0.00&0.00&0.00\\
  &0.05&0.00&0.00&0.00&0.00&0.00&0.00&0.00\\
\multirow{-5}{*}{Mice.RF}&0.10&0.00&0.00&0.00&0.00&0.00&0.00&0.00\\

\hline Naive &0.01&0.01&0.01&0.01&0.01&0.01&0.01&0.00\\
  Imputation&0.05&0.04&0.05&0.05&0.05&0.05&0.05&0.00\\
 &0.10&0.09&0.09&0.09&0.09&0.09&0.10&0.00\\

\hline  &0.01&0.00&0.00&0.00&0.00&0.01&0.01&0.00\\
  &0.05&0.02&0.02&0.02&0.03&0.03&0.03&0.00\\
\multirow{-5}{*}{MissRanger}&0.10&0.05&0.05&0.05&0.05&0.05&0.07&0.00\\

\hline 
\end{tabular} 
\end{minipage} 
\end{table}

\newpage 
\begin{table}[H] 
\caption{minimum ($q_0$), first quartile ($q_{0.25}$),
               median ($q_{0.5}$), mean ($\bar{x}$), third quartile ($q_{0.75}$),
               maximum ($q_{1}$) and standard deviation (sd) of the observed values of the realization of the Kolmogorow-Smirnow-statistic for the variable \texttt{ef20} for the different
               imputation methods and missing rates (r)
               creating missing values using the MCAR-mechanism for the employee data (VSE, Kolmogorow-Smirnow, \texttt{ef20}, MCAR)}\label{Tab:vse.vals.ksmirn.MCAR.ef20 }\begin{minipage}[t]{\textwidth} 

\centering 
\begin{tabular}[t]{|l|c|c|c|c|c|c|c|c|}\hline
 method&r&$q_{0}$& $q_{0.25}$&$q_{0.5}$&$\bar{x}$&$q_{0.75}$& $q_{1}$&sd\\ \hline 
\hline  &0.01&0.00&0.00&0.00&0.00&0.01&0.01&0.00\\
  &0.05&0.02&0.02&0.02&0.02&0.02&0.03&0.00\\
\multirow{-5}{*}{Amelia}&0.10&0.05&0.05&0.05&0.05&0.05&0.05&0.00\\

\hline  &0.01&0.00&0.00&0.00&0.00&0.00&0.00&0.00\\
  &0.05&0.00&0.00&0.00&0.00&0.00&0.00&0.00\\
\multirow{-5}{*}{Mice.Norm}&0.10&0.00&0.00&0.00&0.00&0.00&0.00&0.00\\

\hline  &0.01&0.00&0.00&0.01&0.01&0.01&0.01&0.00\\
  &0.05&0.02&0.02&0.03&0.03&0.03&0.03&0.00\\
\multirow{-5}{*}{Mice.PMM}&0.10&0.05&0.05&0.05&0.05&0.05&0.05&0.00\\

\hline  &0.01&0.00&0.00&0.00&0.00&0.00&0.00&0.00\\
  &0.05&0.00&0.00&0.00&0.00&0.00&0.00&0.00\\
\multirow{-5}{*}{Mice.RF}&0.10&0.00&0.00&0.00&0.00&0.00&0.00&0.00\\

\hline Naive &0.01&0.01&0.01&0.01&0.01&0.01&0.01&0.00\\
  Imputation&0.05&0.04&0.05&0.05&0.05&0.05&0.05&0.00\\
 &0.10&0.09&0.09&0.09&0.09&0.09&0.10&0.00\\

\hline  &0.01&0.00&0.00&0.00&0.00&0.01&0.01&0.00\\
  &0.05&0.02&0.02&0.02&0.02&0.03&0.03&0.00\\
\multirow{-5}{*}{MissRanger}&0.10&0.05&0.05&0.05&0.05&0.05&0.07&0.00\\

\hline 
\end{tabular} 
\end{minipage} 
\end{table}

\newpage 
\begin{table}[H] 
\caption{minimum ($q_0$), first quartile ($q_{0.25}$),
               median ($q_{0.5}$), mean ($\bar{x}$), third quartile ($q_{0.75}$),
               maximum ($q_{1}$) and standard deviation (sd) of the observed values of the realization of the Kolmogorow-Smirnow-statistic for the variable \texttt{ef21} for the different
               imputation methods and missing rates (r)
               creating missing values using the MAR-mechanism for the employee data (VSE, Kolmogorow-Smirnow, \texttt{ef21}, MAR)}\label{Tab:vse.vals.ksmirn.MAR.ef21 }\begin{minipage}[t]{\textwidth} 

\centering 
\begin{tabular}[t]{|l|c|c|c|c|c|c|c|c|}\hline
 method&r&$q_{0}$& $q_{0.25}$&$q_{0.5}$&$\bar{x}$&$q_{0.75}$& $q_{1}$&sd\\ \hline 
\hline  &0.01&0.00&0.00&0.00&0.00&0.00&0.00&0.00\\
  &0.05&0.00&0.00&0.00&0.00&0.00&0.00&0.00\\
\multirow{-5}{*}{Amelia}&0.10&0.00&0.00&0.00&0.00&0.00&0.00&0.00\\

\hline  &0.01&0.00&0.00&0.00&0.00&0.00&0.00&0.00\\
  &0.05&0.00&0.00&0.00&0.00&0.00&0.00&0.00\\
\multirow{-5}{*}{Mice.Norm}&0.10&0.00&0.00&0.00&0.00&0.00&0.00&0.00\\

\hline  &0.01&0.00&0.00&0.00&0.00&0.00&0.00&0.00\\
  &0.05&0.00&0.00&0.00&0.00&0.00&0.00&0.00\\
\multirow{-5}{*}{Mice.PMM}&0.10&0.00&0.00&0.00&0.00&0.00&0.00&0.00\\

\hline  &0.01&0.00&0.00&0.00&0.00&0.00&0.00&0.00\\
  &0.05&0.00&0.00&0.00&0.00&0.00&0.00&0.00\\
\multirow{-5}{*}{Mice.RF}&0.10&0.00&0.00&0.00&0.00&0.00&0.01&0.00\\

\hline Naive &0.01&0.00&0.01&0.01&0.01&0.01&0.01&0.00\\
  Imputation&0.05&0.03&0.03&0.03&0.03&0.03&0.03&0.00\\
 &0.10&0.05&0.06&0.06&0.06&0.06&0.06&0.00\\

\hline  &0.01&0.00&0.00&0.00&0.00&0.00&0.00&0.00\\
  &0.05&0.00&0.00&0.00&0.00&0.00&0.00&0.00\\
\multirow{-5}{*}{MissRanger}&0.10&0.00&0.00&0.00&0.00&0.00&0.01&0.00\\

\hline 
\end{tabular} 
\end{minipage} 
\end{table}

\newpage 
\begin{table}[H] 
\caption{minimum ($q_0$), first quartile ($q_{0.25}$),
               median ($q_{0.5}$), mean ($\bar{x}$), third quartile ($q_{0.75}$),
               maximum ($q_{1}$) and standard deviation (sd) of the observed values of the realization of the Kolmogorow-Smirnow-statistic for the variable \texttt{ef21} for the different
               imputation methods and missing rates (r)
               creating missing values using the MCAR-mechanism for the employee data (VSE, Kolmogorow-Smirnow, \texttt{ef21}, MCAR)}\label{Tab:vse.vals.ksmirn.MCAR.ef21 }\begin{minipage}[t]{\textwidth} 

\centering 
\begin{tabular}[t]{|l|c|c|c|c|c|c|c|c|}\hline
 method&r&$q_{0}$& $q_{0.25}$&$q_{0.5}$&$\bar{x}$&$q_{0.75}$& $q_{1}$&sd\\ \hline 
\hline  &0.01&0.00&0.00&0.00&0.00&0.00&0.00&0.00\\
  &0.05&0.00&0.00&0.00&0.00&0.00&0.00&0.00\\
\multirow{-5}{*}{Amelia}&0.10&0.00&0.00&0.00&0.00&0.00&0.00&0.00\\

\hline  &0.01&0.00&0.00&0.00&0.00&0.00&0.00&0.00\\
  &0.05&0.00&0.00&0.00&0.00&0.00&0.00&0.00\\
\multirow{-5}{*}{Mice.Norm}&0.10&0.00&0.00&0.00&0.00&0.00&0.00&0.00\\

\hline  &0.01&0.00&0.00&0.00&0.00&0.00&0.00&0.00\\
  &0.05&0.00&0.00&0.00&0.00&0.00&0.00&0.00\\
\multirow{-5}{*}{Mice.PMM}&0.10&0.00&0.00&0.00&0.00&0.00&0.00&0.00\\

\hline  &0.01&0.00&0.00&0.00&0.00&0.00&0.00&0.00\\
  &0.05&0.00&0.00&0.00&0.00&0.00&0.00&0.00\\
\multirow{-5}{*}{Mice.RF}&0.10&0.00&0.00&0.00&0.00&0.01&0.01&0.00\\

\hline Naive &0.01&0.00&0.01&0.01&0.01&0.01&0.01&0.00\\
  Imputation&0.05&0.03&0.03&0.03&0.03&0.03&0.03&0.00\\
 &0.10&0.05&0.05&0.06&0.06&0.06&0.06&0.00\\

\hline  &0.01&0.00&0.00&0.00&0.00&0.00&0.00&0.00\\
  &0.05&0.00&0.00&0.00&0.00&0.00&0.00&0.00\\
\multirow{-5}{*}{MissRanger}&0.10&0.00&0.00&0.00&0.00&0.00&0.01&0.00\\

\hline 
\end{tabular} 
\end{minipage} 
\end{table}

\newpage 
\begin{table}[H] 
\caption{minimum ($q_0$), first quartile ($q_{0.25}$),
               median ($q_{0.5}$), mean ($\bar{x}$), third quartile ($q_{0.75}$),
               maximum ($q_{1}$) and standard deviation (sd) of the observed values of the realization of the Kolmogorow-Smirnow-statistic for the variable \texttt{ef22} for the different
               imputation methods and missing rates (r)
               creating missing values using the MAR-mechanism for the employee data (VSE, Kolmogorow-Smirnow, \texttt{ef22}, MAR)}\label{Tab:vse.vals.ksmirn.MAR.ef22 }\begin{minipage}[t]{\textwidth} 

\centering 
\begin{tabular}[t]{|l|c|c|c|c|c|c|c|c|}\hline
 method&r&$q_{0}$& $q_{0.25}$&$q_{0.5}$&$\bar{x}$&$q_{0.75}$& $q_{1}$&sd\\ \hline 
\hline  &0.01&0.00&0.00&0.00&0.00&0.01&0.01&0.00\\
  &0.05&0.02&0.02&0.02&0.02&0.02&0.03&0.00\\
\multirow{-5}{*}{Amelia}&0.10&0.04&0.05&0.05&0.05&0.05&0.05&0.00\\

\hline  &0.01&0.00&0.00&0.00&0.00&0.00&0.00&0.00\\
  &0.05&0.00&0.00&0.00&0.00&0.00&0.00&0.00\\
\multirow{-5}{*}{Mice.Norm}&0.10&0.00&0.00&0.00&0.00&0.00&0.00&0.00\\

\hline  &0.01&0.00&0.00&0.01&0.01&0.01&0.01&0.00\\
  &0.05&0.02&0.02&0.02&0.02&0.03&0.03&0.00\\
\multirow{-5}{*}{Mice.PMM}&0.10&0.05&0.05&0.05&0.05&0.05&0.05&0.00\\

\hline  &0.01&0.00&0.00&0.00&0.00&0.00&0.00&0.00\\
  &0.05&0.00&0.00&0.00&0.00&0.00&0.00&0.00\\
\multirow{-5}{*}{Mice.RF}&0.10&0.00&0.00&0.00&0.00&0.00&0.00&0.00\\

\hline Naive &0.01&0.01&0.01&0.01&0.01&0.01&0.01&0.00\\
  Imputation&0.05&0.04&0.05&0.05&0.05&0.05&0.05&0.00\\
 &0.10&0.09&0.09&0.09&0.09&0.09&0.10&0.00\\

\hline  &0.01&0.00&0.01&0.01&0.01&0.01&0.01&0.00\\
  &0.05&0.02&0.03&0.03&0.03&0.03&0.04&0.00\\
\multirow{-5}{*}{MissRanger}&0.10&0.05&0.05&0.06&0.06&0.06&0.08&0.01\\

\hline 
\end{tabular} 
\end{minipage} 
\end{table}

\newpage 
\begin{table}[H] 
\caption{minimum ($q_0$), first quartile ($q_{0.25}$),
               median ($q_{0.5}$), mean ($\bar{x}$), third quartile ($q_{0.75}$),
               maximum ($q_{1}$) and standard deviation (sd) of the observed values of the realization of the Kolmogorow-Smirnow-statistic for the variable \texttt{ef22} for the different
               imputation methods and missing rates (r)
               creating missing values using the MCAR-mechanism for the employee data (VSE, Kolmogorow-Smirnow, \texttt{ef22}, MCAR)}\label{Tab:vse.vals.ksmirn.MCAR.ef22 }\begin{minipage}[t]{\textwidth} 

\centering 
\begin{tabular}[t]{|l|c|c|c|c|c|c|c|c|}\hline
 method&r&$q_{0}$& $q_{0.25}$&$q_{0.5}$&$\bar{x}$&$q_{0.75}$& $q_{1}$&sd\\ \hline 
\hline  &0.01&0.00&0.00&0.00&0.00&0.01&0.01&0.00\\
  &0.05&0.02&0.02&0.02&0.02&0.02&0.02&0.00\\
\multirow{-5}{*}{Amelia}&0.10&0.04&0.05&0.05&0.05&0.05&0.05&0.00\\

\hline  &0.01&0.00&0.00&0.00&0.00&0.00&0.00&0.00\\
  &0.05&0.00&0.00&0.00&0.00&0.00&0.00&0.00\\
\multirow{-5}{*}{Mice.Norm}&0.10&0.00&0.00&0.00&0.00&0.00&0.00&0.00\\

\hline  &0.01&0.00&0.00&0.00&0.01&0.01&0.01&0.00\\
  &0.05&0.02&0.02&0.02&0.02&0.03&0.03&0.00\\
\multirow{-5}{*}{Mice.PMM}&0.10&0.05&0.05&0.05&0.05&0.05&0.05&0.00\\

\hline  &0.01&0.00&0.00&0.00&0.00&0.00&0.00&0.00\\
  &0.05&0.00&0.00&0.00&0.00&0.00&0.00&0.00\\
\multirow{-5}{*}{Mice.RF}&0.10&0.00&0.00&0.00&0.00&0.00&0.00&0.00\\

\hline Naive &0.01&0.01&0.01&0.01&0.01&0.01&0.01&0.00\\
  Imputation&0.05&0.04&0.05&0.05&0.05&0.05&0.05&0.00\\
 &0.10&0.09&0.09&0.09&0.09&0.09&0.10&0.00\\

\hline  &0.01&0.00&0.01&0.01&0.01&0.01&0.01&0.00\\
  &0.05&0.02&0.03&0.03&0.03&0.03&0.04&0.00\\
\multirow{-5}{*}{MissRanger}&0.10&0.05&0.05&0.06&0.06&0.06&0.09&0.01\\

\hline 
\end{tabular} 
\end{minipage} 
\end{table}

\newpage 
\begin{table}[H] 
\caption{minimum ($q_0$), first quartile ($q_{0.25}$),
               median ($q_{0.5}$), mean ($\bar{x}$), third quartile ($q_{0.75}$),
               maximum ($q_{1}$) and standard deviation (sd) of the observed values of the realization of the Kolmogorow-Smirnow-statistic for the variable \texttt{ef23} for the different
               imputation methods and missing rates (r)
               creating missing values using the MAR-mechanism for the employee data (VSE, Kolmogorow-Smirnow, \texttt{ef23}, MAR)}\label{Tab:vse.vals.ksmirn.MAR.ef23 }\begin{minipage}[t]{\textwidth} 

\centering 
\begin{tabular}[t]{|l|c|c|c|c|c|c|c|c|}\hline
 method&r&$q_{0}$& $q_{0.25}$&$q_{0.5}$&$\bar{x}$&$q_{0.75}$& $q_{1}$&sd\\ \hline 
\hline  &0.01&0.00&0.00&0.00&0.00&0.00&0.01&0.00\\
  &0.05&0.02&0.02&0.02&0.02&0.02&0.02&0.00\\
\multirow{-5}{*}{Amelia}&0.10&0.04&0.04&0.05&0.05&0.05&0.05&0.00\\

\hline  &0.01&0.00&0.00&0.00&0.00&0.00&0.00&0.00\\
  &0.05&0.00&0.00&0.00&0.00&0.00&0.00&0.00\\
\multirow{-5}{*}{Mice.Norm}&0.10&0.00&0.00&0.00&0.00&0.00&0.00&0.00\\

\hline  &0.01&0.00&0.00&0.00&0.00&0.00&0.01&0.00\\
  &0.05&0.02&0.02&0.02&0.02&0.02&0.03&0.00\\
\multirow{-5}{*}{Mice.PMM}&0.10&0.04&0.04&0.04&0.04&0.05&0.05&0.00\\

\hline  &0.01&0.00&0.00&0.00&0.00&0.00&0.00&0.00\\
  &0.05&0.00&0.00&0.00&0.00&0.00&0.00&0.00\\
\multirow{-5}{*}{Mice.RF}&0.10&0.00&0.00&0.00&0.00&0.00&0.00&0.00\\

\hline Naive &0.01&0.01&0.01&0.01&0.01&0.01&0.01&0.00\\
  Imputation&0.05&0.04&0.04&0.05&0.05&0.05&0.05&0.00\\
 &0.10&0.08&0.09&0.09&0.09&0.09&0.10&0.00\\

\hline  &0.01&0.01&0.01&0.01&0.01&0.01&0.01&0.00\\
  &0.05&0.03&0.03&0.03&0.03&0.04&0.04&0.00\\
\multirow{-5}{*}{MissRanger}&0.10&0.07&0.07&0.07&0.07&0.07&0.07&0.00\\

\hline 
\end{tabular} 
\end{minipage} 
\end{table}

\newpage 
\begin{table}[H] 
\caption{minimum ($q_0$), first quartile ($q_{0.25}$),
               median ($q_{0.5}$), mean ($\bar{x}$), third quartile ($q_{0.75}$),
               maximum ($q_{1}$) and standard deviation (sd) of the observed values of the realization of the Kolmogorow-Smirnow-statistic for the variable \texttt{ef23} for the different
               imputation methods and missing rates (r)
               creating missing values using the MCAR-mechanism for the employee data (VSE, Kolmogorow-Smirnow, \texttt{ef23}, MCAR)}\label{Tab:vse.vals.ksmirn.MCAR.ef23 }\begin{minipage}[t]{\textwidth} 

\centering 
\begin{tabular}[t]{|l|c|c|c|c|c|c|c|c|}\hline
 method&r&$q_{0}$& $q_{0.25}$&$q_{0.5}$&$\bar{x}$&$q_{0.75}$& $q_{1}$&sd\\ \hline 
\hline  &0.01&0.00&0.00&0.00&0.00&0.00&0.01&0.00\\
  &0.05&0.02&0.02&0.02&0.02&0.02&0.02&0.00\\
\multirow{-5}{*}{Amelia}&0.10&0.04&0.04&0.04&0.04&0.04&0.04&0.00\\

\hline  &0.01&0.00&0.00&0.00&0.00&0.00&0.00&0.00\\
  &0.05&0.00&0.00&0.00&0.00&0.00&0.00&0.00\\
\multirow{-5}{*}{Mice.Norm}&0.10&0.00&0.00&0.00&0.00&0.00&0.00&0.00\\

\hline  &0.01&0.00&0.00&0.00&0.00&0.00&0.01&0.00\\
  &0.05&0.02&0.02&0.02&0.02&0.02&0.02&0.00\\
\multirow{-5}{*}{Mice.PMM}&0.10&0.04&0.04&0.04&0.04&0.04&0.05&0.00\\

\hline  &0.01&0.00&0.00&0.00&0.00&0.00&0.00&0.00\\
  &0.05&0.00&0.00&0.00&0.00&0.00&0.00&0.00\\
\multirow{-5}{*}{Mice.RF}&0.10&0.00&0.01&0.01&0.01&0.01&0.01&0.00\\

\hline Naive &0.01&0.01&0.01&0.01&0.01&0.01&0.01&0.00\\
  Imputation&0.05&0.04&0.04&0.04&0.04&0.04&0.05&0.00\\
 &0.10&0.08&0.08&0.09&0.09&0.09&0.09&0.00\\

\hline  &0.01&0.01&0.01&0.01&0.01&0.01&0.01&0.00\\
  &0.05&0.03&0.03&0.03&0.04&0.04&0.04&0.00\\
\multirow{-5}{*}{MissRanger}&0.10&0.07&0.07&0.07&0.07&0.07&0.08&0.00\\

\hline 
\end{tabular} 
\end{minipage} 
\end{table}

\newpage 
\begin{table}[H] 
\caption{minimum ($q_0$), first quartile ($q_{0.25}$),
               median ($q_{0.5}$), mean ($\bar{x}$), third quartile ($q_{0.75}$),
               maximum ($q_{1}$) and standard deviation (sd) of the observed values of the realization of the Kolmogorow-Smirnow-statistic for the variable \texttt{ef24} for the different
               imputation methods and missing rates (r)
               creating missing values using the MAR-mechanism for the employee data (VSE, Kolmogorow-Smirnow, \texttt{ef24}, MAR)}\label{Tab:vse.vals.ksmirn.MAR.ef24 }\begin{minipage}[t]{\textwidth} 

\centering 
\begin{tabular}[t]{|l|c|c|c|c|c|c|c|c|}\hline
 method&r&$q_{0}$& $q_{0.25}$&$q_{0.5}$&$\bar{x}$&$q_{0.75}$& $q_{1}$&sd\\ \hline 
\hline  &0.01&0.00&0.00&0.00&0.00&0.00&0.00&0.00\\
  &0.05&0.01&0.01&0.01&0.01&0.01&0.01&0.00\\
\multirow{-5}{*}{Amelia}&0.10&0.02&0.02&0.02&0.02&0.02&0.02&0.00\\

\hline  &0.01&0.00&0.00&0.00&0.00&0.00&0.00&0.00\\
  &0.05&0.00&0.00&0.00&0.00&0.00&0.00&0.00\\
\multirow{-5}{*}{Mice.Norm}&0.10&0.00&0.00&0.00&0.00&0.00&0.00&0.00\\

\hline  &0.01&0.00&0.00&0.00&0.00&0.00&0.00&0.00\\
  &0.05&0.01&0.01&0.01&0.01&0.01&0.01&0.00\\
\multirow{-5}{*}{Mice.PMM}&0.10&0.02&0.02&0.02&0.02&0.02&0.02&0.00\\

\hline  &0.01&0.00&0.00&0.00&0.00&0.00&0.00&0.00\\
  &0.05&0.00&0.00&0.00&0.00&0.00&0.00&0.00\\
\multirow{-5}{*}{Mice.RF}&0.10&0.00&0.01&0.01&0.01&0.01&0.01&0.00\\

\hline Naive &0.01&0.00&0.01&0.01&0.01&0.01&0.01&0.00\\
  Imputation&0.05&0.03&0.03&0.03&0.03&0.03&0.03&0.00\\
 &0.10&0.06&0.06&0.06&0.06&0.06&0.07&0.00\\

\hline  &0.01&0.00&0.00&0.00&0.00&0.00&0.00&0.00\\
  &0.05&0.00&0.00&0.01&0.01&0.01&0.01&0.00\\
\multirow{-5}{*}{MissRanger}&0.10&0.01&0.01&0.01&0.01&0.01&0.03&0.00\\

\hline 
\end{tabular} 
\end{minipage} 
\end{table}

\newpage 
\begin{table}[H] 
\caption{minimum ($q_0$), first quartile ($q_{0.25}$),
               median ($q_{0.5}$), mean ($\bar{x}$), third quartile ($q_{0.75}$),
               maximum ($q_{1}$) and standard deviation (sd) of the observed values of the realization of the Kolmogorow-Smirnow-statistic for the variable \texttt{ef24} for the different
               imputation methods and missing rates (r)
               creating missing values using the MCAR-mechanism for the employee data (VSE, Kolmogorow-Smirnow, \texttt{ef24}, MCAR)}\label{Tab:vse.vals.ksmirn.MCAR.ef24 }\begin{minipage}[t]{\textwidth} 

\centering 
\begin{tabular}[t]{|l|c|c|c|c|c|c|c|c|}\hline
 method&r&$q_{0}$& $q_{0.25}$&$q_{0.5}$&$\bar{x}$&$q_{0.75}$& $q_{1}$&sd\\ \hline 
\hline  &0.01&0.00&0.00&0.00&0.00&0.00&0.00&0.00\\
  &0.05&0.01&0.01&0.01&0.01&0.01&0.01&0.00\\
\multirow{-5}{*}{Amelia}&0.10&0.02&0.02&0.02&0.02&0.02&0.02&0.00\\

\hline  &0.01&0.00&0.00&0.00&0.00&0.00&0.00&0.00\\
  &0.05&0.00&0.00&0.00&0.00&0.00&0.00&0.00\\
\multirow{-5}{*}{Mice.Norm}&0.10&0.00&0.00&0.00&0.00&0.00&0.00&0.00\\

\hline  &0.01&0.00&0.00&0.00&0.00&0.00&0.00&0.00\\
  &0.05&0.01&0.01&0.01&0.01&0.01&0.01&0.00\\
\multirow{-5}{*}{Mice.PMM}&0.10&0.02&0.02&0.02&0.02&0.02&0.02&0.00\\

\hline  &0.01&0.00&0.00&0.00&0.00&0.00&0.00&0.00\\
  &0.05&0.00&0.00&0.00&0.00&0.00&0.00&0.00\\
\multirow{-5}{*}{Mice.RF}&0.10&0.01&0.01&0.01&0.01&0.01&0.01&0.00\\

\hline Naive &0.01&0.00&0.01&0.01&0.01&0.01&0.01&0.00\\
  Imputation&0.05&0.03&0.03&0.03&0.03&0.03&0.03&0.00\\
 &0.10&0.06&0.06&0.06&0.06&0.06&0.07&0.00\\

\hline  &0.01&0.00&0.00&0.00&0.00&0.00&0.00&0.00\\
  &0.05&0.00&0.00&0.01&0.01&0.01&0.01&0.00\\
\multirow{-5}{*}{MissRanger}&0.10&0.01&0.01&0.01&0.01&0.01&0.03&0.00\\

\hline 
\end{tabular} 
\end{minipage} 
\end{table}

\newpage 
\begin{table}[H] 
\caption{minimum ($q_0$), first quartile ($q_{0.25}$),
               median ($q_{0.5}$), mean ($\bar{x}$), third quartile ($q_{0.75}$),
               maximum ($q_{1}$) and standard deviation (sd) of the observed values of the realization of the Kolmogorow-Smirnow-statistic for the variable \texttt{ef25} for the different
               imputation methods and missing rates (r)
               creating missing values using the MAR-mechanism for the employee data (VSE, Kolmogorow-Smirnow, \texttt{ef25}, MAR)}\label{Tab:vse.vals.ksmirn.MAR.ef25 }\begin{minipage}[t]{\textwidth} 

\centering 
\begin{tabular}[t]{|l|c|c|c|c|c|c|c|c|}\hline
 method&r&$q_{0}$& $q_{0.25}$&$q_{0.5}$&$\bar{x}$&$q_{0.75}$& $q_{1}$&sd\\ \hline 
\hline  &0.01&0.00&0.00&0.00&0.00&0.00&0.00&0.00\\
  &0.05&0.00&0.00&0.00&0.00&0.00&0.00&0.00\\
\multirow{-5}{*}{Amelia}&0.10&0.01&0.01&0.01&0.01&0.01&0.01&0.00\\

\hline  &0.01&0.00&0.00&0.00&0.00&0.00&0.00&0.00\\
  &0.05&0.00&0.00&0.00&0.00&0.00&0.00&0.00\\
\multirow{-5}{*}{Mice.Norm}&0.10&0.00&0.00&0.00&0.00&0.00&0.00&0.00\\

\hline  &0.01&0.00&0.00&0.00&0.00&0.00&0.00&0.00\\
  &0.05&0.00&0.00&0.00&0.00&0.00&0.00&0.00\\
\multirow{-5}{*}{Mice.PMM}&0.10&0.01&0.01&0.01&0.01&0.01&0.01&0.00\\

\hline  &0.01&0.00&0.00&0.00&0.00&0.00&0.00&0.00\\
  &0.05&0.00&0.00&0.00&0.00&0.00&0.00&0.00\\
\multirow{-5}{*}{Mice.RF}&0.10&0.00&0.00&0.00&0.00&0.00&0.01&0.00\\

\hline Naive &0.01&0.00&0.01&0.01&0.01&0.01&0.01&0.00\\
  Imputation&0.05&0.02&0.03&0.03&0.03&0.03&0.03&0.00\\
 &0.10&0.05&0.05&0.05&0.05&0.05&0.06&0.00\\

\hline  &0.01&0.00&0.00&0.00&0.00&0.00&0.00&0.00\\
  &0.05&0.00&0.00&0.00&0.00&0.00&0.01&0.00\\
\multirow{-5}{*}{MissRanger}&0.10&0.01&0.01&0.01&0.01&0.01&0.02&0.00\\

\hline 
\end{tabular} 
\end{minipage} 
\end{table}

\newpage 
\begin{table}[H] 
\caption{minimum ($q_0$), first quartile ($q_{0.25}$),
               median ($q_{0.5}$), mean ($\bar{x}$), third quartile ($q_{0.75}$),
               maximum ($q_{1}$) and standard deviation (sd) of the observed values of the realization of the Kolmogorow-Smirnow-statistic for the variable \texttt{ef25} for the different
               imputation methods and missing rates (r)
               creating missing values using the MCAR-mechanism for the employee data (VSE, Kolmogorow-Smirnow, \texttt{ef25}, MCAR)}\label{Tab:vse.vals.ksmirn.MCAR.ef25 }\begin{minipage}[t]{\textwidth} 

\centering 
\begin{tabular}[t]{|l|c|c|c|c|c|c|c|c|}\hline
 method&r&$q_{0}$& $q_{0.25}$&$q_{0.5}$&$\bar{x}$&$q_{0.75}$& $q_{1}$&sd\\ \hline 
\hline  &0.01&0.00&0.00&0.00&0.00&0.00&0.00&0.00\\
  &0.05&0.00&0.00&0.00&0.00&0.00&0.00&0.00\\
\multirow{-5}{*}{Amelia}&0.10&0.01&0.01&0.01&0.01&0.01&0.01&0.00\\

\hline  &0.01&0.00&0.00&0.00&0.00&0.00&0.00&0.00\\
  &0.05&0.00&0.00&0.00&0.00&0.00&0.00&0.00\\
\multirow{-5}{*}{Mice.Norm}&0.10&0.00&0.00&0.00&0.00&0.00&0.00&0.00\\

\hline  &0.01&0.00&0.00&0.00&0.00&0.00&0.00&0.00\\
  &0.05&0.00&0.00&0.00&0.00&0.00&0.00&0.00\\
\multirow{-5}{*}{Mice.PMM}&0.10&0.01&0.01&0.01&0.01&0.01&0.01&0.00\\

\hline  &0.01&0.00&0.00&0.00&0.00&0.00&0.00&0.00\\
  &0.05&0.00&0.00&0.00&0.00&0.00&0.00&0.00\\
\multirow{-5}{*}{Mice.RF}&0.10&0.00&0.00&0.00&0.00&0.00&0.01&0.00\\

\hline Naive &0.01&0.00&0.01&0.01&0.01&0.01&0.01&0.00\\
  Imputation&0.05&0.02&0.03&0.03&0.03&0.03&0.03&0.00\\
 &0.10&0.05&0.05&0.05&0.05&0.05&0.05&0.00\\

\hline  &0.01&0.00&0.00&0.00&0.00&0.00&0.00&0.00\\
  &0.05&0.00&0.00&0.00&0.00&0.00&0.01&0.00\\
\multirow{-5}{*}{MissRanger}&0.10&0.01&0.01&0.01&0.01&0.01&0.02&0.00\\

\hline 
\end{tabular} 
\end{minipage} 
\end{table}

\newpage 
\begin{table}[H] 
\caption{minimum ($q_0$), first quartile ($q_{0.25}$),
               median ($q_{0.5}$), mean ($\bar{x}$), third quartile ($q_{0.75}$),
               maximum ($q_{1}$) and standard deviation (sd) of the observed values of the realization of the Kolmogorow-Smirnow-statistic for the variable \texttt{ef26} for the different
               imputation methods and missing rates (r)
               creating missing values using the MAR-mechanism for the employee data (VSE, Kolmogorow-Smirnow, \texttt{ef26}, MAR)}\label{Tab:vse.vals.ksmirn.MAR.ef26 }\begin{minipage}[t]{\textwidth} 

\centering 
\begin{tabular}[t]{|l|c|c|c|c|c|c|c|c|}\hline
 method&r&$q_{0}$& $q_{0.25}$&$q_{0.5}$&$\bar{x}$&$q_{0.75}$& $q_{1}$&sd\\ \hline 
\hline  &0.01&0.00&0.00&0.00&0.00&0.00&0.01&0.00\\
  &0.05&0.02&0.02&0.02&0.02&0.02&0.02&0.00\\
\multirow{-5}{*}{Amelia}&0.10&0.04&0.04&0.04&0.04&0.05&0.05&0.00\\

\hline  &0.01&0.00&0.00&0.00&0.00&0.00&0.00&0.00\\
  &0.05&0.00&0.00&0.00&0.00&0.00&0.00&0.00\\
\multirow{-5}{*}{Mice.Norm}&0.10&0.00&0.00&0.00&0.00&0.00&0.00&0.00\\

\hline  &0.01&0.00&0.00&0.00&0.00&0.00&0.01&0.00\\
  &0.05&0.02&0.02&0.02&0.02&0.02&0.02&0.00\\
\multirow{-5}{*}{Mice.PMM}&0.10&0.04&0.04&0.04&0.04&0.05&0.05&0.00\\

\hline  &0.01&0.00&0.00&0.00&0.00&0.00&0.00&0.00\\
  &0.05&0.00&0.00&0.00&0.00&0.00&0.00&0.00\\
\multirow{-5}{*}{Mice.RF}&0.10&0.01&0.01&0.01&0.01&0.01&0.01&0.00\\

\hline Naive &0.01&0.01&0.01&0.01&0.01&0.01&0.01&0.00\\
  Imputation&0.05&0.04&0.04&0.04&0.04&0.04&0.04&0.00\\
 &0.10&0.08&0.08&0.08&0.08&0.09&0.09&0.00\\

\hline  &0.01&0.01&0.01&0.01&0.01&0.01&0.01&0.00\\
  &0.05&0.03&0.04&0.04&0.04&0.04&0.04&0.00\\
\multirow{-5}{*}{MissRanger}&0.10&0.07&0.07&0.07&0.07&0.08&0.08&0.00\\

\hline 
\end{tabular} 
\end{minipage} 
\end{table}

\newpage 
\begin{table}[H] 
\caption{minimum ($q_0$), first quartile ($q_{0.25}$),
               median ($q_{0.5}$), mean ($\bar{x}$), third quartile ($q_{0.75}$),
               maximum ($q_{1}$) and standard deviation (sd) of the observed values of the realization of the Kolmogorow-Smirnow-statistic for the variable \texttt{ef26} for the different
               imputation methods and missing rates (r)
               creating missing values using the MCAR-mechanism for the employee data (VSE, Kolmogorow-Smirnow, \texttt{ef26}, MCAR)}\label{Tab:vse.vals.ksmirn.MCAR.ef26 }\begin{minipage}[t]{\textwidth} 

\centering 
\begin{tabular}[t]{|l|c|c|c|c|c|c|c|c|}\hline
 method&r&$q_{0}$& $q_{0.25}$&$q_{0.5}$&$\bar{x}$&$q_{0.75}$& $q_{1}$&sd\\ \hline 
\hline  &0.01&0.00&0.00&0.00&0.00&0.00&0.01&0.00\\
  &0.05&0.02&0.02&0.02&0.02&0.02&0.02&0.00\\
\multirow{-5}{*}{Amelia}&0.10&0.04&0.04&0.04&0.04&0.05&0.05&0.00\\

\hline  &0.01&0.00&0.00&0.00&0.00&0.00&0.00&0.00\\
  &0.05&0.00&0.00&0.00&0.00&0.00&0.00&0.00\\
\multirow{-5}{*}{Mice.Norm}&0.10&0.00&0.00&0.00&0.00&0.00&0.00&0.00\\

\hline  &0.01&0.00&0.00&0.00&0.00&0.00&0.01&0.00\\
  &0.05&0.02&0.02&0.02&0.02&0.02&0.02&0.00\\
\multirow{-5}{*}{Mice.PMM}&0.10&0.04&0.04&0.04&0.04&0.05&0.05&0.00\\

\hline  &0.01&0.00&0.00&0.00&0.00&0.00&0.00&0.00\\
  &0.05&0.00&0.00&0.00&0.00&0.00&0.01&0.00\\
\multirow{-5}{*}{Mice.RF}&0.10&0.00&0.01&0.01&0.01&0.01&0.01&0.00\\

\hline Naive &0.01&0.01&0.01&0.01&0.01&0.01&0.01&0.00\\
  Imputation&0.05&0.04&0.04&0.04&0.04&0.04&0.05&0.00\\
 &0.10&0.08&0.08&0.08&0.08&0.08&0.09&0.00\\

\hline  &0.01&0.01&0.01&0.01&0.01&0.01&0.01&0.00\\
  &0.05&0.03&0.04&0.04&0.04&0.04&0.04&0.00\\
\multirow{-5}{*}{MissRanger}&0.10&0.07&0.07&0.07&0.07&0.07&0.08&0.00\\

\hline 
\end{tabular} 
\end{minipage} 
\end{table}

\newpage 
\begin{table}[H] 
\caption{minimum ($q_0$), first quartile ($q_{0.25}$),
               median ($q_{0.5}$), mean ($\bar{x}$), third quartile ($q_{0.75}$),
               maximum ($q_{1}$) and standard deviation (sd) of the observed values of the realization of the Kolmogorow-Smirnow-statistic for the variable \texttt{ef29} for the different
               imputation methods and missing rates (r)
               creating missing values using the MAR-mechanism for the employee data (VSE, Kolmogorow-Smirnow, \texttt{ef29}, MAR)}\label{Tab:vse.vals.ksmirn.MAR.ef29 }\begin{minipage}[t]{\textwidth} 

\centering 
\begin{tabular}[t]{|l|c|c|c|c|c|c|c|c|}\hline
 method&r&$q_{0}$& $q_{0.25}$&$q_{0.5}$&$\bar{x}$&$q_{0.75}$& $q_{1}$&sd\\ \hline 
\hline  &0.01&0.00&0.00&0.00&0.00&0.00&0.00&0.00\\
  &0.05&0.01&0.01&0.01&0.01&0.01&0.01&0.00\\
\multirow{-5}{*}{Amelia}&0.10&0.02&0.02&0.02&0.02&0.02&0.02&0.00\\

\hline  &0.01&0.00&0.00&0.00&0.00&0.00&0.00&0.00\\
  &0.05&0.00&0.00&0.00&0.00&0.00&0.00&0.00\\
\multirow{-5}{*}{Mice.Norm}&0.10&0.00&0.00&0.00&0.00&0.00&0.00&0.00\\

\hline  &0.01&0.00&0.00&0.00&0.00&0.00&0.00&0.00\\
  &0.05&0.01&0.01&0.01&0.01&0.01&0.01&0.00\\
\multirow{-5}{*}{Mice.PMM}&0.10&0.02&0.02&0.02&0.02&0.02&0.02&0.00\\

\hline  &0.01&0.00&0.00&0.00&0.00&0.00&0.00&0.00\\
  &0.05&0.00&0.00&0.00&0.01&0.01&0.01&0.00\\
\multirow{-5}{*}{Mice.RF}&0.10&0.01&0.01&0.01&0.01&0.01&0.01&0.00\\

\hline Naive &0.01&0.01&0.01&0.01&0.01&0.01&0.01&0.00\\
  Imputation&0.05&0.03&0.03&0.03&0.03&0.03&0.04&0.00\\
 &0.10&0.06&0.06&0.06&0.07&0.07&0.07&0.00\\

\hline  &0.01&0.00&0.00&0.00&0.00&0.00&0.00&0.00\\
  &0.05&0.01&0.01&0.01&0.02&0.02&0.02&0.00\\
\multirow{-5}{*}{MissRanger}&0.10&0.03&0.03&0.03&0.03&0.03&0.05&0.00\\

\hline 
\end{tabular} 
\end{minipage} 
\end{table}

\newpage 
\begin{table}[H] 
\caption{minimum ($q_0$), first quartile ($q_{0.25}$),
               median ($q_{0.5}$), mean ($\bar{x}$), third quartile ($q_{0.75}$),
               maximum ($q_{1}$) and standard deviation (sd) of the observed values of the realization of the Kolmogorow-Smirnow-statistic for the variable \texttt{ef29} for the different
               imputation methods and missing rates (r)
               creating missing values using the MCAR-mechanism for the employee data (VSE, Kolmogorow-Smirnow, \texttt{ef29}, MCAR)}\label{Tab:vse.vals.ksmirn.MCAR.ef29 }\begin{minipage}[t]{\textwidth} 

\centering 
\begin{tabular}[t]{|l|c|c|c|c|c|c|c|c|}\hline
 method&r&$q_{0}$& $q_{0.25}$&$q_{0.5}$&$\bar{x}$&$q_{0.75}$& $q_{1}$&sd\\ \hline 
\hline  &0.01&0.00&0.00&0.00&0.00&0.00&0.00&0.00\\
  &0.05&0.01&0.01&0.01&0.01&0.01&0.01&0.00\\
\multirow{-5}{*}{Amelia}&0.10&0.02&0.02&0.02&0.02&0.02&0.02&0.00\\

\hline  &0.01&0.00&0.00&0.00&0.00&0.00&0.00&0.00\\
  &0.05&0.00&0.00&0.00&0.00&0.00&0.00&0.00\\
\multirow{-5}{*}{Mice.Norm}&0.10&0.00&0.00&0.00&0.00&0.00&0.00&0.00\\

\hline  &0.01&0.00&0.00&0.00&0.00&0.00&0.00&0.00\\
  &0.05&0.01&0.01&0.01&0.01&0.01&0.01&0.00\\
\multirow{-5}{*}{Mice.PMM}&0.10&0.02&0.02&0.02&0.02&0.02&0.02&0.00\\

\hline  &0.01&0.00&0.00&0.00&0.00&0.00&0.00&0.00\\
  &0.05&0.00&0.00&0.00&0.01&0.01&0.01&0.00\\
\multirow{-5}{*}{Mice.RF}&0.10&0.01&0.01&0.01&0.01&0.01&0.01&0.00\\

\hline Naive &0.01&0.01&0.01&0.01&0.01&0.01&0.01&0.00\\
  Imputation&0.05&0.03&0.03&0.03&0.03&0.03&0.03&0.00\\
 &0.10&0.06&0.06&0.07&0.07&0.07&0.07&0.00\\

\hline  &0.01&0.00&0.00&0.00&0.00&0.00&0.00&0.00\\
  &0.05&0.01&0.01&0.01&0.02&0.02&0.02&0.00\\
\multirow{-5}{*}{MissRanger}&0.10&0.03&0.03&0.03&0.03&0.03&0.04&0.00\\

\hline 
\end{tabular} 
\end{minipage} 
\end{table}

\newpage 
\begin{table}[H] 
\caption{minimum ($q_0$), first quartile ($q_{0.25}$),
               median ($q_{0.5}$), mean ($\bar{x}$), third quartile ($q_{0.75}$),
               maximum ($q_{1}$) and standard deviation (sd) of the observed values of the realization of the Kolmogorow-Smirnow-statistic for the variable \texttt{ef40} for the different
               imputation methods and missing rates (r)
               creating missing values using the MAR-mechanism for the employee data (VSE, Kolmogorow-Smirnow, \texttt{ef40}, MAR)}\label{Tab:vse.vals.ksmirn.MAR.ef40 }\begin{minipage}[t]{\textwidth} 

\centering 
\begin{tabular}[t]{|l|c|c|c|c|c|c|c|c|}\hline
 method&r&$q_{0}$& $q_{0.25}$&$q_{0.5}$&$\bar{x}$&$q_{0.75}$& $q_{1}$&sd\\ \hline 
\hline  &0.01&0.00&0.00&0.00&0.00&0.00&0.00&0.00\\
  &0.05&0.01&0.01&0.01&0.01&0.01&0.01&0.00\\
\multirow{-5}{*}{Amelia}&0.10&0.02&0.02&0.02&0.02&0.02&0.02&0.00\\

\hline  &0.01&0.00&0.00&0.00&0.00&0.00&0.00&0.00\\
  &0.05&0.00&0.00&0.00&0.00&0.00&0.00&0.00\\
\multirow{-5}{*}{Mice.Norm}&0.10&0.00&0.00&0.00&0.00&0.00&0.00&0.00\\

\hline  &0.01&0.00&0.00&0.00&0.00&0.00&0.00&0.00\\
  &0.05&0.01&0.01&0.01&0.01&0.01&0.01&0.00\\
\multirow{-5}{*}{Mice.PMM}&0.10&0.02&0.02&0.02&0.02&0.02&0.02&0.00\\

\hline  &0.01&0.00&0.00&0.00&0.00&0.00&0.00&0.00\\
  &0.05&0.00&0.00&0.00&0.00&0.00&0.00&0.00\\
\multirow{-5}{*}{Mice.RF}&0.10&0.00&0.00&0.00&0.00&0.01&0.01&0.00\\

\hline Naive &0.01&0.01&0.01&0.01&0.01&0.01&0.01&0.00\\
  Imputation&0.05&0.03&0.03&0.03&0.03&0.03&0.03&0.00\\
 &0.10&0.06&0.06&0.06&0.06&0.06&0.07&0.00\\

\hline  &0.01&0.00&0.00&0.00&0.00&0.00&0.00&0.00\\
  &0.05&0.01&0.01&0.01&0.01&0.01&0.02&0.00\\
\multirow{-5}{*}{MissRanger}&0.10&0.02&0.02&0.02&0.02&0.02&0.04&0.00\\

\hline 
\end{tabular} 
\end{minipage} 
\end{table}

\newpage 
\begin{table}[H] 
\caption{minimum ($q_0$), first quartile ($q_{0.25}$),
               median ($q_{0.5}$), mean ($\bar{x}$), third quartile ($q_{0.75}$),
               maximum ($q_{1}$) and standard deviation (sd) of the observed values of the realization of the Kolmogorow-Smirnow-statistic for the variable \texttt{ef40} for the different
               imputation methods and missing rates (r)
               creating missing values using the MCAR-mechanism for the employee data (VSE, Kolmogorow-Smirnow, \texttt{ef40}, MCAR)}\label{Tab:vse.vals.ksmirn.MCAR.ef40 }\begin{minipage}[t]{\textwidth} 

\centering 
\begin{tabular}[t]{|l|c|c|c|c|c|c|c|c|}\hline
 method&r&$q_{0}$& $q_{0.25}$&$q_{0.5}$&$\bar{x}$&$q_{0.75}$& $q_{1}$&sd\\ \hline 
\hline  &0.01&0.00&0.00&0.00&0.00&0.00&0.00&0.00\\
  &0.05&0.01&0.01&0.01&0.01&0.01&0.01&0.00\\
\multirow{-5}{*}{Amelia}&0.10&0.02&0.02&0.02&0.02&0.02&0.02&0.00\\

\hline  &0.01&0.00&0.00&0.00&0.00&0.00&0.00&0.00\\
  &0.05&0.00&0.00&0.00&0.00&0.00&0.00&0.00\\
\multirow{-5}{*}{Mice.Norm}&0.10&0.00&0.00&0.00&0.00&0.00&0.00&0.00\\

\hline  &0.01&0.00&0.00&0.00&0.00&0.00&0.00&0.00\\
  &0.05&0.01&0.01&0.01&0.01&0.01&0.01&0.00\\
\multirow{-5}{*}{Mice.PMM}&0.10&0.02&0.02&0.02&0.02&0.02&0.02&0.00\\

\hline  &0.01&0.00&0.00&0.00&0.00&0.00&0.00&0.00\\
  &0.05&0.00&0.00&0.00&0.00&0.00&0.00&0.00\\
\multirow{-5}{*}{Mice.RF}&0.10&0.00&0.00&0.00&0.00&0.01&0.01&0.00\\

\hline Naive &0.01&0.01&0.01&0.01&0.01&0.01&0.01&0.00\\
  Imputation&0.05&0.03&0.03&0.03&0.03&0.03&0.03&0.00\\
 &0.10&0.06&0.06&0.06&0.06&0.06&0.07&0.00\\

\hline  &0.01&0.00&0.00&0.00&0.00&0.00&0.00&0.00\\
  &0.05&0.01&0.01&0.01&0.01&0.01&0.02&0.00\\
\multirow{-5}{*}{MissRanger}&0.10&0.02&0.02&0.02&0.02&0.02&0.04&0.00\\

\hline 
\end{tabular} 
\end{minipage} 
\end{table}

\newpage 
\begin{table}[H] 
\caption{minimum ($q_0$), first quartile ($q_{0.25}$),
               median ($q_{0.5}$), mean ($\bar{x}$), third quartile ($q_{0.75}$),
               maximum ($q_{1}$) and standard deviation (sd) of the observed values of the realization of the Kolmogorow-Smirnow-statistic for the variable \texttt{ef41} for the different
               imputation methods and missing rates (r)
               creating missing values using the MAR-mechanism for the employee data (VSE, Kolmogorow-Smirnow, \texttt{ef41}, MAR)}\label{Tab:vse.vals.ksmirn.MAR.ef41 }\begin{minipage}[t]{\textwidth} 

\centering 
\begin{tabular}[t]{|l|c|c|c|c|c|c|c|c|}\hline
 method&r&$q_{0}$& $q_{0.25}$&$q_{0.5}$&$\bar{x}$&$q_{0.75}$& $q_{1}$&sd\\ \hline 
\hline  &0.01&0.00&0.00&0.00&0.00&0.00&0.00&0.00\\
  &0.05&0.00&0.00&0.00&0.00&0.00&0.01&0.00\\
\multirow{-5}{*}{Amelia}&0.10&0.01&0.01&0.01&0.01&0.01&0.01&0.00\\

\hline  &0.01&0.00&0.00&0.00&0.00&0.00&0.00&0.00\\
  &0.05&0.00&0.00&0.00&0.00&0.00&0.00&0.00\\
\multirow{-5}{*}{Mice.Norm}&0.10&0.00&0.00&0.00&0.00&0.00&0.00&0.00\\

\hline  &0.01&0.00&0.00&0.00&0.00&0.00&0.00&0.00\\
  &0.05&0.00&0.00&0.00&0.00&0.00&0.01&0.00\\
\multirow{-5}{*}{Mice.PMM}&0.10&0.01&0.01&0.01&0.01&0.01&0.01&0.00\\

\hline  &0.01&0.00&0.00&0.00&0.00&0.00&0.00&0.00\\
  &0.05&0.00&0.00&0.00&0.00&0.00&0.00&0.00\\
\multirow{-5}{*}{Mice.RF}&0.10&0.00&0.00&0.00&0.00&0.00&0.00&0.00\\

\hline Naive &0.01&0.00&0.01&0.01&0.01&0.01&0.01&0.00\\
  Imputation&0.05&0.02&0.03&0.03&0.03&0.03&0.03&0.00\\
 &0.10&0.05&0.05&0.05&0.05&0.05&0.06&0.00\\

\hline  &0.01&0.00&0.00&0.00&0.00&0.00&0.00&0.00\\
  &0.05&0.01&0.01&0.01&0.01&0.01&0.01&0.00\\
\multirow{-5}{*}{MissRanger}&0.10&0.01&0.02&0.02&0.02&0.02&0.03&0.00\\

\hline 
\end{tabular} 
\end{minipage} 
\end{table}

\newpage 
\begin{table}[H] 
\caption{minimum ($q_0$), first quartile ($q_{0.25}$),
               median ($q_{0.5}$), mean ($\bar{x}$), third quartile ($q_{0.75}$),
               maximum ($q_{1}$) and standard deviation (sd) of the observed values of the realization of the Kolmogorow-Smirnow-statistic for the variable \texttt{ef41} for the different
               imputation methods and missing rates (r)
               creating missing values using the MCAR-mechanism for the employee data (VSE, Kolmogorow-Smirnow, \texttt{ef41}, MCAR)}\label{Tab:vse.vals.ksmirn.MCAR.ef41 }\begin{minipage}[t]{\textwidth} 

\centering 
\begin{tabular}[t]{|l|c|c|c|c|c|c|c|c|}\hline
 method&r&$q_{0}$& $q_{0.25}$&$q_{0.5}$&$\bar{x}$&$q_{0.75}$& $q_{1}$&sd\\ \hline 
\hline  &0.01&0.00&0.00&0.00&0.00&0.00&0.00&0.00\\
  &0.05&0.00&0.00&0.00&0.00&0.00&0.00&0.00\\
\multirow{-5}{*}{Amelia}&0.10&0.01&0.01&0.01&0.01&0.01&0.01&0.00\\

\hline  &0.01&0.00&0.00&0.00&0.00&0.00&0.00&0.00\\
  &0.05&0.00&0.00&0.00&0.00&0.00&0.00&0.00\\
\multirow{-5}{*}{Mice.Norm}&0.10&0.00&0.00&0.00&0.00&0.00&0.00&0.00\\

\hline  &0.01&0.00&0.00&0.00&0.00&0.00&0.00&0.00\\
  &0.05&0.00&0.00&0.00&0.00&0.00&0.01&0.00\\
\multirow{-5}{*}{Mice.PMM}&0.10&0.01&0.01&0.01&0.01&0.01&0.01&0.00\\

\hline  &0.01&0.00&0.00&0.00&0.00&0.00&0.00&0.00\\
  &0.05&0.00&0.00&0.00&0.00&0.00&0.00&0.00\\
\multirow{-5}{*}{Mice.RF}&0.10&0.00&0.00&0.00&0.00&0.00&0.00&0.00\\

\hline Naive &0.01&0.00&0.01&0.01&0.01&0.01&0.01&0.00\\
  Imputation&0.05&0.02&0.03&0.03&0.03&0.03&0.03&0.00\\
 &0.10&0.05&0.05&0.05&0.05&0.05&0.06&0.00\\

\hline  &0.01&0.00&0.00&0.00&0.00&0.00&0.00&0.00\\
  &0.05&0.01&0.01&0.01&0.01&0.01&0.01&0.00\\
\multirow{-5}{*}{MissRanger}&0.10&0.01&0.02&0.02&0.02&0.02&0.03&0.00\\

\hline 
\end{tabular} 
\end{minipage} 
\end{table}

\newpage 
\begin{table}[H] 
\caption{minimum ($q_0$), first quartile ($q_{0.25}$),
               median ($q_{0.5}$), mean ($\bar{x}$), third quartile ($q_{0.75}$),
               maximum ($q_{1}$) and standard deviation (sd) of the observed values of the realization of the Kolmogorow-Smirnow-statistic for the variable \texttt{ef44} for the different
               imputation methods and missing rates (r)
               creating missing values using the MAR-mechanism for the employee data (VSE, Kolmogorow-Smirnow, \texttt{ef44}, MAR)}\label{Tab:vse.vals.ksmirn.MAR.ef44 }\begin{minipage}[t]{\textwidth} 

\centering 
\begin{tabular}[t]{|l|c|c|c|c|c|c|c|c|}\hline
 method&r&$q_{0}$& $q_{0.25}$&$q_{0.5}$&$\bar{x}$&$q_{0.75}$& $q_{1}$&sd\\ \hline 
\hline  &0.01&0.00&0.00&0.00&0.00&0.00&0.00&0.00\\
  &0.05&0.00&0.00&0.00&0.00&0.00&0.00&0.00\\
\multirow{-5}{*}{Amelia}&0.10&0.00&0.00&0.00&0.00&0.00&0.00&0.00\\

\hline  &0.01&0.00&0.00&0.00&0.00&0.00&0.00&0.00\\
  &0.05&0.00&0.00&0.00&0.00&0.00&0.00&0.00\\
\multirow{-5}{*}{Mice.Norm}&0.10&0.00&0.00&0.00&0.00&0.00&0.00&0.00\\

\hline  &0.01&0.00&0.00&0.00&0.00&0.00&0.00&0.00\\
  &0.05&0.00&0.00&0.00&0.00&0.00&0.00&0.00\\
\multirow{-5}{*}{Mice.PMM}&0.10&0.00&0.00&0.00&0.00&0.00&0.00&0.00\\

\hline  &0.01&0.00&0.00&0.00&0.00&0.00&0.00&0.00\\
  &0.05&0.00&0.00&0.00&0.00&0.00&0.00&0.00\\
\multirow{-5}{*}{Mice.RF}&0.10&0.00&0.00&0.00&0.00&0.00&0.00&0.00\\

\hline Naive &0.01&0.00&0.01&0.01&0.01&0.01&0.01&0.00\\
  Imputation&0.05&0.02&0.03&0.03&0.03&0.03&0.03&0.00\\
 &0.10&0.05&0.05&0.05&0.05&0.05&0.06&0.00\\

\hline  &0.01&0.00&0.00&0.00&0.00&0.00&0.00&0.00\\
  &0.05&0.00&0.00&0.00&0.00&0.00&0.00&0.00\\
\multirow{-5}{*}{MissRanger}&0.10&0.00&0.00&0.00&0.00&0.00&0.01&0.00\\

\hline 
\end{tabular} 
\end{minipage} 
\end{table}

\newpage 
\begin{table}[H] 
\caption{minimum ($q_0$), first quartile ($q_{0.25}$),
               median ($q_{0.5}$), mean ($\bar{x}$), third quartile ($q_{0.75}$),
               maximum ($q_{1}$) and standard deviation (sd) of the observed values of the realization of the Kolmogorow-Smirnow-statistic for the variable \texttt{ef44} for the different
               imputation methods and missing rates (r)
               creating missing values using the MCAR-mechanism for the employee data (VSE, Kolmogorow-Smirnow, \texttt{ef44}, MCAR)}\label{Tab:vse.vals.ksmirn.MCAR.ef44 }\begin{minipage}[t]{\textwidth} 

\centering 
\begin{tabular}[t]{|l|c|c|c|c|c|c|c|c|}\hline
 method&r&$q_{0}$& $q_{0.25}$&$q_{0.5}$&$\bar{x}$&$q_{0.75}$& $q_{1}$&sd\\ \hline 
\hline  &0.01&0.00&0.00&0.00&0.00&0.00&0.00&0.00\\
  &0.05&0.00&0.00&0.00&0.00&0.00&0.00&0.00\\
\multirow{-5}{*}{Amelia}&0.10&0.00&0.00&0.00&0.00&0.00&0.00&0.00\\

\hline  &0.01&0.00&0.00&0.00&0.00&0.00&0.00&0.00\\
  &0.05&0.00&0.00&0.00&0.00&0.00&0.00&0.00\\
\multirow{-5}{*}{Mice.Norm}&0.10&0.00&0.00&0.00&0.00&0.00&0.00&0.00\\

\hline  &0.01&0.00&0.00&0.00&0.00&0.00&0.00&0.00\\
  &0.05&0.00&0.00&0.00&0.00&0.00&0.00&0.00\\
\multirow{-5}{*}{Mice.PMM}&0.10&0.00&0.00&0.00&0.00&0.00&0.00&0.00\\

\hline  &0.01&0.00&0.00&0.00&0.00&0.00&0.00&0.00\\
  &0.05&0.00&0.00&0.00&0.00&0.00&0.00&0.00\\
\multirow{-5}{*}{Mice.RF}&0.10&0.00&0.00&0.00&0.00&0.00&0.00&0.00\\

\hline Naive &0.01&0.00&0.01&0.01&0.01&0.01&0.01&0.00\\
  Imputation&0.05&0.02&0.03&0.03&0.03&0.03&0.03&0.00\\
 &0.10&0.05&0.05&0.05&0.05&0.05&0.06&0.00\\

\hline  &0.01&0.00&0.00&0.00&0.00&0.00&0.00&0.00\\
  &0.05&0.00&0.00&0.00&0.00&0.00&0.01&0.00\\
\multirow{-5}{*}{MissRanger}&0.10&0.00&0.00&0.00&0.00&0.00&0.01&0.00\\

\hline 
\end{tabular} 
\end{minipage} 
\end{table}

\newpage 
\begin{table}[H] 
\caption{minimum ($q_0$), first quartile ($q_{0.25}$),
               median ($q_{0.5}$), mean ($\bar{x}$), third quartile ($q_{0.75}$),
               maximum ($q_{1}$) and standard deviation (sd) of the observed values of the realization of the Kolmogorow-Smirnow-statistic for the variable \texttt{ef45} for the different
               imputation methods and missing rates (r)
               creating missing values using the MAR-mechanism for the employee data (VSE, Kolmogorow-Smirnow, \texttt{ef45}, MAR)}\label{Tab:vse.vals.ksmirn.MAR.ef45 }\begin{minipage}[t]{\textwidth} 

\centering 
\begin{tabular}[t]{|l|c|c|c|c|c|c|c|c|}\hline
 method&r&$q_{0}$& $q_{0.25}$&$q_{0.5}$&$\bar{x}$&$q_{0.75}$& $q_{1}$&sd\\ \hline 
\hline  &0.01&0.00&0.00&0.00&0.00&0.00&0.00&0.00\\
  &0.05&0.00&0.00&0.00&0.00&0.00&0.00&0.00\\
\multirow{-5}{*}{Amelia}&0.10&0.00&0.00&0.00&0.00&0.00&0.01&0.00\\

\hline  &0.01&0.00&0.00&0.00&0.00&0.00&0.00&0.00\\
  &0.05&0.00&0.00&0.00&0.00&0.00&0.00&0.00\\
\multirow{-5}{*}{Mice.Norm}&0.10&0.00&0.00&0.00&0.00&0.00&0.00&0.00\\

\hline  &0.01&0.00&0.00&0.00&0.00&0.00&0.00&0.00\\
  &0.05&0.00&0.00&0.00&0.00&0.00&0.00&0.00\\
\multirow{-5}{*}{Mice.PMM}&0.10&0.00&0.00&0.00&0.00&0.00&0.00&0.00\\

\hline  &0.01&0.00&0.00&0.00&0.00&0.00&0.00&0.00\\
  &0.05&0.00&0.00&0.00&0.00&0.00&0.01&0.00\\
\multirow{-5}{*}{Mice.RF}&0.10&0.00&0.01&0.01&0.01&0.01&0.01&0.00\\

\hline Naive &0.01&0.00&0.01&0.01&0.01&0.01&0.01&0.00\\
  Imputation&0.05&0.03&0.03&0.03&0.03&0.03&0.03&0.00\\
 &0.10&0.06&0.06&0.06&0.06&0.06&0.07&0.00\\

\hline  &0.01&0.00&0.00&0.00&0.00&0.00&0.00&0.00\\
  &0.05&0.00&0.00&0.00&0.00&0.00&0.01&0.00\\
\multirow{-5}{*}{MissRanger}&0.10&0.00&0.00&0.00&0.00&0.00&0.01&0.00\\

\hline 
\end{tabular} 
\end{minipage} 
\end{table}

\newpage 
\begin{table}[H] 
\caption{minimum ($q_0$), first quartile ($q_{0.25}$),
               median ($q_{0.5}$), mean ($\bar{x}$), third quartile ($q_{0.75}$),
               maximum ($q_{1}$) and standard deviation (sd) of the observed values of the realization of the Kolmogorow-Smirnow-statistic for the variable \texttt{ef45} for the different
               imputation methods and missing rates (r)
               creating missing values using the MCAR-mechanism for the employee data (VSE, Kolmogorow-Smirnow, \texttt{ef45}, MCAR)}\label{Tab:vse.vals.ksmirn.MCAR.ef45 }\begin{minipage}[t]{\textwidth} 

\centering 
\begin{tabular}[t]{|l|c|c|c|c|c|c|c|c|}\hline
 method&r&$q_{0}$& $q_{0.25}$&$q_{0.5}$&$\bar{x}$&$q_{0.75}$& $q_{1}$&sd\\ \hline 
\hline  &0.01&0.00&0.00&0.00&0.00&0.00&0.00&0.00\\
  &0.05&0.00&0.00&0.00&0.00&0.00&0.00&0.00\\
\multirow{-5}{*}{Amelia}&0.10&0.00&0.00&0.00&0.00&0.00&0.00&0.00\\

\hline  &0.01&0.00&0.00&0.00&0.00&0.00&0.00&0.00\\
  &0.05&0.00&0.00&0.00&0.00&0.00&0.00&0.00\\
\multirow{-5}{*}{Mice.Norm}&0.10&0.00&0.00&0.00&0.00&0.00&0.00&0.00\\

\hline  &0.01&0.00&0.00&0.00&0.00&0.00&0.00&0.00\\
  &0.05&0.00&0.00&0.00&0.00&0.00&0.00&0.00\\
\multirow{-5}{*}{Mice.PMM}&0.10&0.00&0.00&0.00&0.00&0.00&0.00&0.00\\

\hline  &0.01&0.00&0.00&0.00&0.00&0.00&0.00&0.00\\
  &0.05&0.00&0.00&0.00&0.00&0.00&0.00&0.00\\
\multirow{-5}{*}{Mice.RF}&0.10&0.00&0.00&0.00&0.00&0.00&0.01&0.00\\

\hline Naive &0.01&0.00&0.01&0.01&0.01&0.01&0.01&0.00\\
  Imputation&0.05&0.03&0.03&0.03&0.03&0.03&0.03&0.00\\
 &0.10&0.05&0.06&0.06&0.06&0.06&0.06&0.00\\

\hline  &0.01&0.00&0.00&0.00&0.00&0.00&0.00&0.00\\
  &0.05&0.00&0.00&0.00&0.00&0.00&0.01&0.00\\
\multirow{-5}{*}{MissRanger}&0.10&0.00&0.00&0.00&0.00&0.00&0.01&0.00\\

\hline 
\end{tabular} 
\end{minipage} 
\end{table}

\newpage 
\begin{table}[H] 
\caption{minimum ($q_0$), first quartile ($q_{0.25}$),
               median ($q_{0.5}$), mean ($\bar{x}$), third quartile ($q_{0.75}$),
               maximum ($q_{1}$) and standard deviation (sd) of the observed values of the realization of the Kolmogorow-Smirnow-statistic for the variable \texttt{ef47} for the different
               imputation methods and missing rates (r)
               creating missing values using the MAR-mechanism for the employee data (VSE, Kolmogorow-Smirnow, \texttt{ef47}, MAR)}\label{Tab:vse.vals.ksmirn.MAR.ef47 }\begin{minipage}[t]{\textwidth} 

\centering 
\begin{tabular}[t]{|l|c|c|c|c|c|c|c|c|}\hline
 method&r&$q_{0}$& $q_{0.25}$&$q_{0.5}$&$\bar{x}$&$q_{0.75}$& $q_{1}$&sd\\ \hline 
\hline  &0.01&0.00&0.00&0.00&0.00&0.00&0.00&0.00\\
  &0.05&0.01&0.01&0.01&0.01&0.01&0.01&0.00\\
\multirow{-5}{*}{Amelia}&0.10&0.02&0.02&0.02&0.02&0.02&0.03&0.00\\

\hline  &0.01&0.00&0.00&0.00&0.00&0.00&0.00&0.00\\
  &0.05&0.00&0.00&0.00&0.00&0.00&0.00&0.00\\
\multirow{-5}{*}{Mice.Norm}&0.10&0.00&0.00&0.00&0.00&0.00&0.00&0.00\\

\hline  &0.01&0.00&0.00&0.00&0.00&0.00&0.00&0.00\\
  &0.05&0.01&0.01&0.01&0.01&0.01&0.01&0.00\\
\multirow{-5}{*}{Mice.PMM}&0.10&0.02&0.02&0.02&0.02&0.02&0.03&0.00\\

\hline  &0.01&0.00&0.00&0.00&0.00&0.00&0.00&0.00\\
  &0.05&0.00&0.00&0.00&0.00&0.00&0.01&0.00\\
\multirow{-5}{*}{Mice.RF}&0.10&0.00&0.01&0.01&0.01&0.01&0.01&0.00\\

\hline Naive &0.01&0.00&0.01&0.01&0.01&0.01&0.01&0.00\\
  Imputation&0.05&0.03&0.03&0.03&0.03&0.03&0.03&0.00\\
 &0.10&0.06&0.06&0.06&0.06&0.06&0.07&0.00\\

\hline  &0.01&0.00&0.00&0.00&0.00&0.00&0.00&0.00\\
  &0.05&0.01&0.01&0.01&0.01&0.01&0.02&0.00\\
\multirow{-5}{*}{MissRanger}&0.10&0.02&0.03&0.03&0.03&0.03&0.03&0.00\\

\hline 
\end{tabular} 
\end{minipage} 
\end{table}

\newpage 
\begin{table}[H] 
\caption{minimum ($q_0$), first quartile ($q_{0.25}$),
               median ($q_{0.5}$), mean ($\bar{x}$), third quartile ($q_{0.75}$),
               maximum ($q_{1}$) and standard deviation (sd) of the observed values of the realization of the Kolmogorow-Smirnow-statistic for the variable \texttt{ef47} for the different
               imputation methods and missing rates (r)
               creating missing values using the MCAR-mechanism for the employee data (VSE, Kolmogorow-Smirnow, \texttt{ef47}, MCAR)}\label{Tab:vse.vals.ksmirn.MCAR.ef47 }\begin{minipage}[t]{\textwidth} 

\centering 
\begin{tabular}[t]{|l|c|c|c|c|c|c|c|c|}\hline
 method&r&$q_{0}$& $q_{0.25}$&$q_{0.5}$&$\bar{x}$&$q_{0.75}$& $q_{1}$&sd\\ \hline 
\hline  &0.01&0.00&0.00&0.00&0.00&0.00&0.00&0.00\\
  &0.05&0.01&0.01&0.01&0.01&0.01&0.01&0.00\\
\multirow{-5}{*}{Amelia}&0.10&0.02&0.02&0.02&0.02&0.02&0.03&0.00\\

\hline  &0.01&0.00&0.00&0.00&0.00&0.00&0.00&0.00\\
  &0.05&0.00&0.00&0.00&0.00&0.00&0.00&0.00\\
\multirow{-5}{*}{Mice.Norm}&0.10&0.00&0.00&0.00&0.00&0.00&0.00&0.00\\

\hline  &0.01&0.00&0.00&0.00&0.00&0.00&0.00&0.00\\
  &0.05&0.01&0.01&0.01&0.01&0.01&0.01&0.00\\
\multirow{-5}{*}{Mice.PMM}&0.10&0.02&0.02&0.02&0.02&0.02&0.03&0.00\\

\hline  &0.01&0.00&0.00&0.00&0.00&0.00&0.00&0.00\\
  &0.05&0.00&0.00&0.00&0.00&0.00&0.01&0.00\\
\multirow{-5}{*}{Mice.RF}&0.10&0.00&0.00&0.01&0.01&0.01&0.01&0.00\\

\hline Naive &0.01&0.01&0.01&0.01&0.01&0.01&0.01&0.00\\
  Imputation&0.05&0.03&0.03&0.03&0.03&0.03&0.03&0.00\\
 &0.10&0.06&0.06&0.06&0.06&0.07&0.07&0.00\\

\hline  &0.01&0.00&0.00&0.00&0.00&0.00&0.00&0.00\\
  &0.05&0.01&0.01&0.01&0.01&0.01&0.02&0.00\\
\multirow{-5}{*}{MissRanger}&0.10&0.03&0.03&0.03&0.03&0.03&0.03&0.00\\

\hline 
\end{tabular} 
\end{minipage} 
\end{table}

\newpage 
\begin{table}[H] 
\caption{minimum ($q_0$), first quartile ($q_{0.25}$),
               median ($q_{0.5}$), mean ($\bar{x}$), third quartile ($q_{0.75}$),
               maximum ($q_{1}$) and standard deviation (sd) of the observed values of the realization of the Kolmogorow-Smirnow-statistic for the variable \texttt{ef53} for the different
               imputation methods and missing rates (r)
               creating missing values using the MAR-mechanism for the employee data (VSE, Kolmogorow-Smirnow, \texttt{ef53}, MAR)}\label{Tab:vse.vals.ksmirn.MAR.ef53 }\begin{minipage}[t]{\textwidth} 

\centering 
\begin{tabular}[t]{|l|c|c|c|c|c|c|c|c|}\hline
 method&r&$q_{0}$& $q_{0.25}$&$q_{0.5}$&$\bar{x}$&$q_{0.75}$& $q_{1}$&sd\\ \hline 
\hline  &0.01&0.00&0.00&0.00&0.00&0.00&0.00&0.00\\
  &0.05&0.01&0.01&0.01&0.01&0.01&0.01&0.00\\
\multirow{-5}{*}{Amelia}&0.10&0.02&0.02&0.02&0.02&0.02&0.02&0.00\\

\hline  &0.01&0.00&0.00&0.00&0.00&0.00&0.00&0.00\\
  &0.05&0.00&0.00&0.00&0.00&0.00&0.00&0.00\\
\multirow{-5}{*}{Mice.Norm}&0.10&0.00&0.00&0.00&0.00&0.00&0.00&0.00\\

\hline  &0.01&0.00&0.00&0.00&0.00&0.00&0.00&0.00\\
  &0.05&0.01&0.01&0.01&0.01&0.01&0.01&0.00\\
\multirow{-5}{*}{Mice.PMM}&0.10&0.02&0.02&0.02&0.02&0.02&0.02&0.00\\

\hline  &0.01&0.00&0.00&0.00&0.00&0.00&0.00&0.00\\
  &0.05&0.00&0.00&0.00&0.00&0.00&0.00&0.00\\
\multirow{-5}{*}{Mice.RF}&0.10&0.00&0.01&0.01&0.01&0.01&0.01&0.00\\

\hline Naive &0.01&0.01&0.01&0.01&0.01&0.01&0.01&0.00\\
  Imputation&0.05&0.03&0.03&0.03&0.03&0.04&0.04&0.00\\
 &0.10&0.07&0.07&0.07&0.07&0.07&0.07&0.00\\

\hline  &0.01&0.00&0.00&0.00&0.00&0.00&0.00&0.00\\
  &0.05&0.01&0.01&0.01&0.01&0.01&0.02&0.00\\
\multirow{-5}{*}{MissRanger}&0.10&0.02&0.02&0.02&0.02&0.03&0.04&0.01\\

\hline 
\end{tabular} 
\end{minipage} 
\end{table}

\newpage 
\begin{table}[H] 
\caption{minimum ($q_0$), first quartile ($q_{0.25}$),
               median ($q_{0.5}$), mean ($\bar{x}$), third quartile ($q_{0.75}$),
               maximum ($q_{1}$) and standard deviation (sd) of the observed values of the realization of the Kolmogorow-Smirnow-statistic for the variable \texttt{ef53} for the different
               imputation methods and missing rates (r)
               creating missing values using the MCAR-mechanism for the employee data (VSE, Kolmogorow-Smirnow, \texttt{ef53}, MCAR)}\label{Tab:vse.vals.ksmirn.MCAR.ef53 }\begin{minipage}[t]{\textwidth} 

\centering 
\begin{tabular}[t]{|l|c|c|c|c|c|c|c|c|}\hline
 method&r&$q_{0}$& $q_{0.25}$&$q_{0.5}$&$\bar{x}$&$q_{0.75}$& $q_{1}$&sd\\ \hline 
\hline  &0.01&0.00&0.00&0.00&0.00&0.00&0.00&0.00\\
  &0.05&0.01&0.01&0.01&0.01&0.01&0.01&0.00\\
\multirow{-5}{*}{Amelia}&0.10&0.02&0.02&0.02&0.02&0.02&0.02&0.00\\

\hline  &0.01&0.00&0.00&0.00&0.00&0.00&0.00&0.00\\
  &0.05&0.00&0.00&0.00&0.00&0.00&0.00&0.00\\
\multirow{-5}{*}{Mice.Norm}&0.10&0.00&0.00&0.00&0.00&0.00&0.00&0.00\\

\hline  &0.01&0.00&0.00&0.00&0.00&0.00&0.00&0.00\\
  &0.05&0.01&0.01&0.01&0.01&0.01&0.01&0.00\\
\multirow{-5}{*}{Mice.PMM}&0.10&0.02&0.02&0.02&0.02&0.02&0.02&0.00\\

\hline  &0.01&0.00&0.00&0.00&0.00&0.00&0.00&0.00\\
  &0.05&0.00&0.00&0.00&0.00&0.00&0.00&0.00\\
\multirow{-5}{*}{Mice.RF}&0.10&0.00&0.00&0.01&0.01&0.01&0.01&0.00\\

\hline Naive &0.01&0.01&0.01&0.01&0.01&0.01&0.01&0.00\\
  Imputation&0.05&0.03&0.03&0.03&0.03&0.04&0.04&0.00\\
 &0.10&0.07&0.07&0.07&0.07&0.07&0.07&0.00\\

\hline  &0.01&0.00&0.00&0.00&0.00&0.00&0.00&0.00\\
  &0.05&0.01&0.01&0.01&0.01&0.01&0.02&0.00\\
\multirow{-5}{*}{MissRanger}&0.10&0.02&0.02&0.02&0.02&0.03&0.05&0.00\\

\hline 
\end{tabular} 
\end{minipage} 
\end{table}

\newpage 

%% file: table_fivepoint_CramerVonMises_vse.txt
\begin{table}[H] 
\caption{minimum ($q_0$), first quartile ($q_{0.25}$),
               median ($q_{0.5}$), mean ($\bar{x}$), third quartile ($q_{0.75}$),
               maximum ($q_{1}$) and standard deviation (sd) of the observed values of the realization of the Cramer-von-Mises-statistic for the variable \texttt{ef19} for the different
               imputation methods and missing rates (r)
               creating missing values using the MAR-mechanism for the employee data (VSE, Cramer-von-Mises, \texttt{ef19}, MAR)}\label{Tab:vse.vals.cramervonmises.MAR.ef19 }\begin{minipage}[t]{\textwidth} 

\centering 
\begin{tabular}[t]{|l|c|c|c|c|c|c|c|c|}\hline
 method&r&$q_{0}$& $q_{0.25}$&$q_{0.5}$&$\bar{x}$&$q_{0.75}$& $q_{1}$&sd\\ \hline 
\hline  &0.01&0.00&0.01&0.01&0.01&0.01&0.01&0.00\\
  &0.05&0.16&0.20&0.21&0.21&0.22&0.28&0.02\\
\multirow{-5}{*}{Amelia}&0.10&0.72&0.82&0.87&0.87&0.91&1.02&0.06\\

\hline  &0.01&0.00&0.00&0.00&0.00&0.00&0.00&0.00\\
  &0.05&0.00&0.00&0.00&0.00&0.00&0.01&0.00\\
\multirow{-5}{*}{Mice.Norm}&0.10&0.00&0.01&0.01&0.01&0.01&0.03&0.00\\

\hline  &0.01&0.01&0.01&0.01&0.01&0.02&0.02&0.00\\
  &0.05&0.27&0.32&0.34&0.34&0.36&0.42&0.03\\
\multirow{-5}{*}{Mice.PMM}&0.10&1.17&1.36&1.43&1.43&1.49&1.77&0.10\\

\hline  &0.01&0.00&0.00&0.00&0.00&0.00&0.00&0.00\\
  &0.05&0.01&0.02&0.03&0.03&0.04&0.06&0.01\\
\multirow{-5}{*}{Mice.RF}&0.10&0.03&0.07&0.09&0.10&0.12&0.21&0.04\\

\hline Naive &0.01&0.10&0.13&0.14&0.14&0.16&0.19&0.02\\
  Imputation&0.05&3.26&3.68&3.84&3.83&3.98&4.44&0.25\\
 &0.10&15.07&16.06&16.49&16.50&16.90&18.79&0.72\\

\hline  &0.01&0.00&0.00&0.00&0.01&0.01&0.06&0.02\\
  &0.05&0.01&0.01&0.02&0.17&0.15&1.72&0.32\\
\multirow{-5}{*}{MissRanger}&0.10&0.04&0.06&0.09&0.87&0.92&6.24&1.51\\

\hline 
\end{tabular} 
\end{minipage} 
\end{table}

\newpage 
\begin{table}[H] 
\caption{minimum ($q_0$), first quartile ($q_{0.25}$),
               median ($q_{0.5}$), mean ($\bar{x}$), third quartile ($q_{0.75}$),
               maximum ($q_{1}$) and standard deviation (sd) of the observed values of the realization of the Cramer-von-Mises-statistic for the variable \texttt{ef19} for the different
               imputation methods and missing rates (r)
               creating missing values using the MCAR-mechanism for the employee data (VSE, Cramer-von-Mises, \texttt{ef19}, MCAR)}\label{Tab:vse.vals.cramervonmises.MCAR.ef19 }\begin{minipage}[t]{\textwidth} 

\centering 
\begin{tabular}[t]{|l|c|c|c|c|c|c|c|c|}\hline
 method&r&$q_{0}$& $q_{0.25}$&$q_{0.5}$&$\bar{x}$&$q_{0.75}$& $q_{1}$&sd\\ \hline 
\hline  &0.01&0.00&0.01&0.01&0.01&0.01&0.01&0.00\\
  &0.05&0.17&0.20&0.21&0.21&0.23&0.26&0.02\\
\multirow{-5}{*}{Amelia}&0.10&0.70&0.84&0.87&0.88&0.93&1.09&0.06\\

\hline  &0.01&0.00&0.00&0.00&0.00&0.00&0.00&0.00\\
  &0.05&0.00&0.00&0.00&0.00&0.00&0.01&0.00\\
\multirow{-5}{*}{Mice.Norm}&0.10&0.00&0.01&0.01&0.01&0.01&0.02&0.00\\

\hline  &0.01&0.01&0.01&0.01&0.01&0.02&0.02&0.00\\
  &0.05&0.26&0.32&0.33&0.33&0.36&0.40&0.03\\
\multirow{-5}{*}{Mice.PMM}&0.10&1.20&1.37&1.44&1.43&1.49&1.73&0.09\\

\hline  &0.01&0.00&0.00&0.00&0.00&0.00&0.01&0.00\\
  &0.05&0.01&0.02&0.03&0.03&0.03&0.08&0.01\\
\multirow{-5}{*}{Mice.RF}&0.10&0.05&0.09&0.10&0.11&0.13&0.23&0.04\\

\hline Naive &0.01&0.09&0.13&0.15&0.15&0.16&0.20&0.02\\
  Imputation&0.05&3.18&3.64&3.79&3.79&3.96&4.47&0.24\\
 &0.10&14.05&16.05&16.57&16.56&17.08&18.41&0.80\\

\hline  &0.01&0.00&0.00&0.00&0.01&0.00&0.06&0.01\\
  &0.05&0.01&0.01&0.03&0.19&0.18&1.42&0.33\\
\multirow{-5}{*}{MissRanger}&0.10&0.05&0.06&0.12&0.59&0.55&5.80&1.10\\

\hline 
\end{tabular} 
\end{minipage} 
\end{table}

\newpage 
\begin{table}[H] 
\caption{minimum ($q_0$), first quartile ($q_{0.25}$),
               median ($q_{0.5}$), mean ($\bar{x}$), third quartile ($q_{0.75}$),
               maximum ($q_{1}$) and standard deviation (sd) of the observed values of the realization of the Cramer-von-Mises-statistic for the variable \texttt{ef20} for the different
               imputation methods and missing rates (r)
               creating missing values using the MAR-mechanism for the employee data (VSE, Cramer-von-Mises, \texttt{ef20}, MAR)}\label{Tab:vse.vals.cramervonmises.MAR.ef20 }\begin{minipage}[t]{\textwidth} 

\centering 
\begin{tabular}[t]{|l|c|c|c|c|c|c|c|c|}\hline
 method&r&$q_{0}$& $q_{0.25}$&$q_{0.5}$&$\bar{x}$&$q_{0.75}$& $q_{1}$&sd\\ \hline 
\hline  &0.01&0.18&0.24&0.27&0.27&0.29&0.39&0.04\\
  &0.05&5.64&6.43&6.69&6.70&6.96&7.83&0.44\\
\multirow{-5}{*}{Amelia}&0.10&24.05&25.60&26.22&26.38&27.21&29.10&1.12\\

\hline  &0.01&0.00&0.00&0.00&0.00&0.00&0.00&0.00\\
  &0.05&0.00&0.00&0.00&0.00&0.00&0.02&0.00\\
\multirow{-5}{*}{Mice.Norm}&0.10&0.00&0.00&0.01&0.01&0.01&0.02&0.00\\

\hline  &0.01&0.18&0.28&0.31&0.31&0.33&0.43&0.04\\
  &0.05&6.43&7.18&7.56&7.53&7.89&8.78&0.51\\
\multirow{-5}{*}{Mice.PMM}&0.10&26.81&28.98&29.82&29.81&30.70&32.60&1.25\\

\hline  &0.01&0.00&0.00&0.00&0.00&0.00&0.01&0.00\\
  &0.05&0.01&0.03&0.03&0.04&0.05&0.10&0.02\\
\multirow{-5}{*}{Mice.RF}&0.10&0.06&0.10&0.12&0.13&0.16&0.24&0.05\\

\hline Naive &0.01&0.67&0.93&1.00&1.00&1.06&1.44&0.12\\
  Imputation&0.05&21.68&23.59&24.60&24.62&25.56&28.58&1.49\\
 &0.10&86.91&93.90&95.85&96.10&98.88&105.89&3.68\\

\hline  &0.01&0.17&0.25&0.28&0.29&0.33&0.48&0.06\\
  &0.05&5.69&6.69&7.09&7.49&7.89&12.71&1.38\\
\multirow{-5}{*}{MissRanger}&0.10&24.59&27.71&29.33&29.72&30.69&50.00&3.48\\

\hline 
\end{tabular} 
\end{minipage} 
\end{table}

\newpage 
\begin{table}[H] 
\caption{minimum ($q_0$), first quartile ($q_{0.25}$),
               median ($q_{0.5}$), mean ($\bar{x}$), third quartile ($q_{0.75}$),
               maximum ($q_{1}$) and standard deviation (sd) of the observed values of the realization of the Cramer-von-Mises-statistic for the variable \texttt{ef20} for the different
               imputation methods and missing rates (r)
               creating missing values using the MCAR-mechanism for the employee data (VSE, Cramer-von-Mises, \texttt{ef20}, MCAR)}\label{Tab:vse.vals.cramervonmises.MCAR.ef20 }\begin{minipage}[t]{\textwidth} 

\centering 
\begin{tabular}[t]{|l|c|c|c|c|c|c|c|c|}\hline
 method&r&$q_{0}$& $q_{0.25}$&$q_{0.5}$&$\bar{x}$&$q_{0.75}$& $q_{1}$&sd\\ \hline 
\hline  &0.01&0.20&0.25&0.28&0.27&0.29&0.35&0.03\\
  &0.05&6.07&6.63&6.82&6.84&7.10&7.75&0.37\\
\multirow{-5}{*}{Amelia}&0.10&23.77&26.13&26.78&26.78&27.59&29.20&1.18\\

\hline  &0.01&0.00&0.00&0.00&0.00&0.00&0.00&0.00\\
  &0.05&0.00&0.00&0.00&0.00&0.00&0.02&0.00\\
\multirow{-5}{*}{Mice.Norm}&0.10&0.00&0.00&0.00&0.01&0.01&0.03&0.01\\

\hline  &0.01&0.22&0.28&0.32&0.31&0.34&0.40&0.04\\
  &0.05&6.82&7.35&7.70&7.74&8.07&8.88&0.47\\
\multirow{-5}{*}{Mice.PMM}&0.10&26.76&29.19&30.20&30.19&31.43&33.78&1.50\\

\hline  &0.01&0.00&0.00&0.00&0.00&0.00&0.01&0.00\\
  &0.05&0.01&0.02&0.03&0.04&0.04&0.08&0.02\\
\multirow{-5}{*}{Mice.RF}&0.10&0.04&0.09&0.11&0.12&0.15&0.29&0.05\\

\hline Naive &0.01&0.80&0.92&1.02&1.01&1.08&1.24&0.11\\
  Imputation&0.05&22.37&24.23&25.12&25.14&25.97&28.02&1.26\\
 &0.10&87.03&94.26&97.35&97.37&100.87&106.47&4.14\\

\hline  &0.01&0.21&0.26&0.29&0.30&0.32&0.54&0.06\\
  &0.05&5.53&6.62&7.04&7.16&7.57&12.47&0.94\\
\multirow{-5}{*}{MissRanger}&0.10&25.17&27.85&29.35&29.84&30.48&52.23&3.70\\

\hline 
\end{tabular} 
\end{minipage} 
\end{table}

\newpage 
\begin{table}[H] 
\caption{minimum ($q_0$), first quartile ($q_{0.25}$),
               median ($q_{0.5}$), mean ($\bar{x}$), third quartile ($q_{0.75}$),
               maximum ($q_{1}$) and standard deviation (sd) of the observed values of the realization of the Cramer-von-Mises-statistic for the variable \texttt{ef21} for the different
               imputation methods and missing rates (r)
               creating missing values using the MAR-mechanism for the employee data (VSE, Cramer-von-Mises, \texttt{ef21}, MAR)}\label{Tab:vse.vals.cramervonmises.MAR.ef21 }\begin{minipage}[t]{\textwidth} 

\centering 
\begin{tabular}[t]{|l|c|c|c|c|c|c|c|c|}\hline
 method&r&$q_{0}$& $q_{0.25}$&$q_{0.5}$&$\bar{x}$&$q_{0.75}$& $q_{1}$&sd\\ \hline 
\hline  &0.01&0.00&0.00&0.00&0.00&0.00&0.00&0.00\\
  &0.05&0.00&0.00&0.00&0.00&0.00&0.00&0.00\\
\multirow{-5}{*}{Amelia}&0.10&0.01&0.01&0.01&0.01&0.01&0.01&0.00\\

\hline  &0.01&0.00&0.00&0.00&0.00&0.00&0.00&0.00\\
  &0.05&0.00&0.00&0.00&0.00&0.00&0.00&0.00\\
\multirow{-5}{*}{Mice.Norm}&0.10&0.00&0.00&0.00&0.00&0.00&0.00&0.00\\

\hline  &0.01&0.00&0.00&0.00&0.00&0.00&0.00&0.00\\
  &0.05&0.00&0.00&0.00&0.00&0.00&0.00&0.00\\
\multirow{-5}{*}{Mice.PMM}&0.10&0.00&0.00&0.00&0.00&0.00&0.00&0.00\\

\hline  &0.01&0.00&0.00&0.00&0.00&0.00&0.00&0.00\\
  &0.05&0.01&0.02&0.02&0.02&0.03&0.04&0.01\\
\multirow{-5}{*}{Mice.RF}&0.10&0.02&0.06&0.08&0.08&0.10&0.15&0.03\\

\hline Naive &0.01&0.08&0.10&0.11&0.11&0.13&0.15&0.02\\
  Imputation&0.05&2.54&2.82&2.93&2.93&3.04&3.46&0.17\\
 &0.10&10.51&11.63&12.05&12.04&12.39&13.27&0.56\\

\hline  &0.01&0.00&0.00&0.00&0.00&0.00&0.00&0.00\\
  &0.05&0.00&0.00&0.00&0.00&0.00&0.11&0.01\\
\multirow{-5}{*}{MissRanger}&0.10&0.00&0.00&0.00&0.01&0.00&0.51&0.05\\

\hline 
\end{tabular} 
\end{minipage} 
\end{table}

\newpage 
\begin{table}[H] 
\caption{minimum ($q_0$), first quartile ($q_{0.25}$),
               median ($q_{0.5}$), mean ($\bar{x}$), third quartile ($q_{0.75}$),
               maximum ($q_{1}$) and standard deviation (sd) of the observed values of the realization of the Cramer-von-Mises-statistic for the variable \texttt{ef21} for the different
               imputation methods and missing rates (r)
               creating missing values using the MCAR-mechanism for the employee data (VSE, Cramer-von-Mises, \texttt{ef21}, MCAR)}\label{Tab:vse.vals.cramervonmises.MCAR.ef21 }\begin{minipage}[t]{\textwidth} 

\centering 
\begin{tabular}[t]{|l|c|c|c|c|c|c|c|c|}\hline
 method&r&$q_{0}$& $q_{0.25}$&$q_{0.5}$&$\bar{x}$&$q_{0.75}$& $q_{1}$&sd\\ \hline 
\hline  &0.01&0.00&0.00&0.00&0.00&0.00&0.00&0.00\\
  &0.05&0.00&0.00&0.00&0.00&0.00&0.00&0.00\\
\multirow{-5}{*}{Amelia}&0.10&0.01&0.01&0.01&0.01&0.01&0.01&0.00\\

\hline  &0.01&0.00&0.00&0.00&0.00&0.00&0.00&0.00\\
  &0.05&0.00&0.00&0.00&0.00&0.00&0.00&0.00\\
\multirow{-5}{*}{Mice.Norm}&0.10&0.00&0.00&0.00&0.00&0.00&0.00&0.00\\

\hline  &0.01&0.00&0.00&0.00&0.00&0.00&0.00&0.00\\
  &0.05&0.00&0.00&0.00&0.00&0.00&0.00&0.00\\
\multirow{-5}{*}{Mice.PMM}&0.10&0.00&0.00&0.00&0.00&0.00&0.00&0.00\\

\hline  &0.01&0.00&0.00&0.00&0.00&0.00&0.00&0.00\\
  &0.05&0.01&0.02&0.02&0.02&0.03&0.05&0.01\\
\multirow{-5}{*}{Mice.RF}&0.10&0.04&0.07&0.09&0.09&0.11&0.18&0.03\\

\hline Naive &0.01&0.08&0.10&0.11&0.11&0.12&0.16&0.02\\
  Imputation&0.05&2.45&2.81&2.95&2.93&3.05&3.35&0.19\\
 &0.10&10.81&11.57&11.90&11.94&12.28&13.53&0.52\\

\hline  &0.01&0.00&0.00&0.00&0.00&0.00&0.00&0.00\\
  &0.05&0.00&0.00&0.00&0.00&0.00&0.03&0.01\\
\multirow{-5}{*}{MissRanger}&0.10&0.00&0.00&0.00&0.01&0.01&0.13&0.02\\

\hline 
\end{tabular} 
\end{minipage} 
\end{table}

\newpage 
\begin{table}[H] 
\caption{minimum ($q_0$), first quartile ($q_{0.25}$),
               median ($q_{0.5}$), mean ($\bar{x}$), third quartile ($q_{0.75}$),
               maximum ($q_{1}$) and standard deviation (sd) of the observed values of the realization of the Cramer-von-Mises-statistic for the variable \texttt{ef22} for the different
               imputation methods and missing rates (r)
               creating missing values using the MAR-mechanism for the employee data (VSE, Cramer-von-Mises, \texttt{ef22}, MAR)}\label{Tab:vse.vals.cramervonmises.MAR.ef22 }\begin{minipage}[t]{\textwidth} 

\centering 
\begin{tabular}[t]{|l|c|c|c|c|c|c|c|c|}\hline
 method&r&$q_{0}$& $q_{0.25}$&$q_{0.5}$&$\bar{x}$&$q_{0.75}$& $q_{1}$&sd\\ \hline 
\hline  &0.01&0.19&0.23&0.26&0.26&0.28&0.36&0.04\\
  &0.05&4.90&5.96&6.25&6.26&6.56&7.33&0.41\\
\multirow{-5}{*}{Amelia}&0.10&22.17&23.99&24.61&24.67&25.37&27.53&1.08\\

\hline  &0.01&0.00&0.00&0.00&0.00&0.00&0.00&0.00\\
  &0.05&0.00&0.00&0.00&0.00&0.00&0.01&0.00\\
\multirow{-5}{*}{Mice.Norm}&0.10&0.00&0.00&0.00&0.01&0.01&0.04&0.01\\

\hline  &0.01&0.22&0.28&0.30&0.30&0.33&0.40&0.04\\
  &0.05&6.16&7.07&7.32&7.35&7.72&8.39&0.49\\
\multirow{-5}{*}{Mice.PMM}&0.10&25.58&28.08&29.09&28.87&29.70&32.61&1.34\\

\hline  &0.01&0.00&0.00&0.00&0.00&0.00&0.01&0.00\\
  &0.05&0.00&0.02&0.03&0.03&0.04&0.08&0.02\\
\multirow{-5}{*}{Mice.RF}&0.10&0.02&0.10&0.12&0.12&0.15&0.24&0.04\\

\hline Naive &0.01&0.80&0.94&1.02&1.03&1.11&1.32&0.12\\
  Imputation&0.05&20.51&23.95&24.79&24.86&25.75&27.67&1.41\\
 &0.10&86.73&94.71&96.65&96.79&98.72&106.91&3.70\\

\hline  &0.01&0.25&0.35&0.41&0.44&0.47&0.90&0.13\\
  &0.05&7.17&8.43&9.86&10.53&11.22&22.82&3.00\\
\multirow{-5}{*}{MissRanger}&0.10&29.21&33.63&35.82&41.10&44.37&82.91&12.12\\

\hline 
\end{tabular} 
\end{minipage} 
\end{table}

\newpage 
\begin{table}[H] 
\caption{minimum ($q_0$), first quartile ($q_{0.25}$),
               median ($q_{0.5}$), mean ($\bar{x}$), third quartile ($q_{0.75}$),
               maximum ($q_{1}$) and standard deviation (sd) of the observed values of the realization of the Cramer-von-Mises-statistic for the variable \texttt{ef22} for the different
               imputation methods and missing rates (r)
               creating missing values using the MCAR-mechanism for the employee data (VSE, Cramer-von-Mises, \texttt{ef22}, MCAR)}\label{Tab:vse.vals.cramervonmises.MCAR.ef22 }\begin{minipage}[t]{\textwidth} 

\centering 
\begin{tabular}[t]{|l|c|c|c|c|c|c|c|c|}\hline
 method&r&$q_{0}$& $q_{0.25}$&$q_{0.5}$&$\bar{x}$&$q_{0.75}$& $q_{1}$&sd\\ \hline 
\hline  &0.01&0.17&0.23&0.26&0.26&0.28&0.38&0.04\\
  &0.05&5.30&5.97&6.23&6.22&6.55&7.04&0.38\\
\multirow{-5}{*}{Amelia}&0.10&22.31&23.79&24.76&24.63&25.34&27.39&1.02\\

\hline  &0.01&0.00&0.00&0.00&0.00&0.00&0.00&0.00\\
  &0.05&0.00&0.00&0.00&0.00&0.00&0.01&0.00\\
\multirow{-5}{*}{Mice.Norm}&0.10&0.00&0.00&0.01&0.01&0.01&0.03&0.00\\

\hline  &0.01&0.20&0.27&0.30&0.30&0.33&0.44&0.05\\
  &0.05&6.32&7.00&7.32&7.32&7.70&8.38&0.46\\
\multirow{-5}{*}{Mice.PMM}&0.10&26.11&27.87&28.75&28.71&29.53&31.75&1.22\\

\hline  &0.01&0.00&0.00&0.00&0.00&0.00&0.01&0.00\\
  &0.05&0.00&0.02&0.03&0.03&0.04&0.08&0.02\\
\multirow{-5}{*}{Mice.RF}&0.10&0.04&0.09&0.11&0.13&0.16&0.27&0.05\\

\hline Naive &0.01&0.72&0.94&1.00&1.02&1.09&1.50&0.13\\
  Imputation&0.05&21.60&23.80&24.71&24.71&25.47&27.50&1.25\\
 &0.10&88.54&94.09&96.53&96.37&98.50&105.51&3.29\\

\hline  &0.01&0.24&0.35&0.40&0.44&0.50&0.84&0.13\\
  &0.05&6.73&8.41&9.10&10.40&11.32&20.39&3.05\\
\multirow{-5}{*}{MissRanger}&0.10&30.30&32.97&35.48&40.10&42.06&84.85&11.64\\

\hline 
\end{tabular} 
\end{minipage} 
\end{table}

\newpage 
\begin{table}[H] 
\caption{minimum ($q_0$), first quartile ($q_{0.25}$),
               median ($q_{0.5}$), mean ($\bar{x}$), third quartile ($q_{0.75}$),
               maximum ($q_{1}$) and standard deviation (sd) of the observed values of the realization of the Cramer-von-Mises-statistic for the variable \texttt{ef23} for the different
               imputation methods and missing rates (r)
               creating missing values using the MAR-mechanism for the employee data (VSE, Cramer-von-Mises, \texttt{ef23}, MAR)}\label{Tab:vse.vals.cramervonmises.MAR.ef23 }\begin{minipage}[t]{\textwidth} 

\centering 
\begin{tabular}[t]{|l|c|c|c|c|c|c|c|c|}\hline
 method&r&$q_{0}$& $q_{0.25}$&$q_{0.5}$&$\bar{x}$&$q_{0.75}$& $q_{1}$&sd\\ \hline 
\hline  &0.01&0.14&0.19&0.20&0.21&0.23&0.28&0.03\\
  &0.05&4.18&4.88&5.05&5.07&5.29&5.83&0.35\\
\multirow{-5}{*}{Amelia}&0.10&17.28&19.25&19.94&19.84&20.52&22.40&1.03\\

\hline  &0.01&0.00&0.00&0.00&0.00&0.00&0.00&0.00\\
  &0.05&0.00&0.00&0.01&0.01&0.01&0.05&0.01\\
\multirow{-5}{*}{Mice.Norm}&0.10&0.00&0.02&0.03&0.04&0.05&0.13&0.03\\

\hline  &0.01&0.16&0.20&0.22&0.22&0.24&0.28&0.03\\
  &0.05&4.44&5.21&5.43&5.40&5.60&6.27&0.36\\
\multirow{-5}{*}{Mice.PMM}&0.10&17.70&20.33&21.18&21.04&21.73&23.54&1.08\\

\hline  &0.01&0.00&0.00&0.00&0.00&0.00&0.01&0.00\\
  &0.05&0.00&0.00&0.01&0.01&0.01&0.03&0.01\\
\multirow{-5}{*}{Mice.RF}&0.10&0.00&0.01&0.02&0.02&0.03&0.11&0.02\\

\hline Naive &0.01&0.67&0.78&0.86&0.87&0.94&1.13&0.10\\
  Imputation&0.05&18.52&20.43&21.47&21.34&22.12&24.68&1.20\\
 &0.10&71.12&80.38&82.72&82.47&85.10&90.98&3.69\\

\hline  &0.01&0.38&0.48&0.51&0.51&0.54&0.66&0.06\\
  &0.05&11.07&12.16&12.57&12.75&13.28&19.20&1.02\\
\multirow{-5}{*}{MissRanger}&0.10&44.98&49.51&50.84&50.85&52.45&55.69&2.14\\

\hline 
\end{tabular} 
\end{minipage} 
\end{table}

\newpage 
\begin{table}[H] 
\caption{minimum ($q_0$), first quartile ($q_{0.25}$),
               median ($q_{0.5}$), mean ($\bar{x}$), third quartile ($q_{0.75}$),
               maximum ($q_{1}$) and standard deviation (sd) of the observed values of the realization of the Cramer-von-Mises-statistic for the variable \texttt{ef23} for the different
               imputation methods and missing rates (r)
               creating missing values using the MCAR-mechanism for the employee data (VSE, Cramer-von-Mises, \texttt{ef23}, MCAR)}\label{Tab:vse.vals.cramervonmises.MCAR.ef23 }\begin{minipage}[t]{\textwidth} 

\centering 
\begin{tabular}[t]{|l|c|c|c|c|c|c|c|c|}\hline
 method&r&$q_{0}$& $q_{0.25}$&$q_{0.5}$&$\bar{x}$&$q_{0.75}$& $q_{1}$&sd\\ \hline 
\hline  &0.01&0.11&0.17&0.19&0.19&0.21&0.26&0.03\\
  &0.05&3.71&4.31&4.56&4.55&4.77&5.67&0.36\\
\multirow{-5}{*}{Amelia}&0.10&15.89&17.60&18.22&18.19&18.74&20.25&0.96\\

\hline  &0.01&0.00&0.00&0.00&0.00&0.00&0.01&0.00\\
  &0.05&0.00&0.00&0.00&0.01&0.01&0.02&0.00\\
\multirow{-5}{*}{Mice.Norm}&0.10&0.00&0.01&0.01&0.01&0.01&0.04&0.01\\

\hline  &0.01&0.12&0.18&0.20&0.20&0.22&0.28&0.03\\
  &0.05&4.03&4.58&4.90&4.87&5.16&6.10&0.38\\
\multirow{-5}{*}{Mice.PMM}&0.10&16.64&18.42&19.32&19.21&19.97&23.24&1.12\\

\hline  &0.01&0.00&0.00&0.01&0.01&0.01&0.02&0.00\\
  &0.05&0.04&0.10&0.13&0.13&0.16&0.22&0.04\\
\multirow{-5}{*}{Mice.RF}&0.10&0.25&0.43&0.52&0.52&0.59&0.88&0.13\\

\hline Naive &0.01&0.51&0.67&0.74&0.74&0.81&1.01&0.10\\
  Imputation&0.05&15.38&17.37&18.07&18.10&18.85&21.53&1.20\\
 &0.10&64.56&69.04&71.74&71.40&73.25&80.13&3.13\\

\hline  &0.01&0.39&0.50&0.54&0.54&0.58&0.67&0.06\\
  &0.05&10.70&12.14&12.71&12.91&13.45&16.69&1.15\\
\multirow{-5}{*}{MissRanger}&0.10&43.81&49.41&50.66&51.47&52.82&62.64&3.55\\

\hline 
\end{tabular} 
\end{minipage} 
\end{table}

\newpage 
\begin{table}[H] 
\caption{minimum ($q_0$), first quartile ($q_{0.25}$),
               median ($q_{0.5}$), mean ($\bar{x}$), third quartile ($q_{0.75}$),
               maximum ($q_{1}$) and standard deviation (sd) of the observed values of the realization of the Cramer-von-Mises-statistic for the variable \texttt{ef24} for the different
               imputation methods and missing rates (r)
               creating missing values using the MAR-mechanism for the employee data (VSE, Cramer-von-Mises, \texttt{ef24}, MAR)}\label{Tab:vse.vals.cramervonmises.MAR.ef24 }\begin{minipage}[t]{\textwidth} 

\centering 
\begin{tabular}[t]{|l|c|c|c|c|c|c|c|c|}\hline
 method&r&$q_{0}$& $q_{0.25}$&$q_{0.5}$&$\bar{x}$&$q_{0.75}$& $q_{1}$&sd\\ \hline 
\hline  &0.01&0.00&0.00&0.00&0.00&0.00&0.01&0.00\\
  &0.05&0.04&0.06&0.06&0.06&0.07&0.09&0.01\\
\multirow{-5}{*}{Amelia}&0.10&0.18&0.25&0.26&0.26&0.27&0.32&0.03\\

\hline  &0.01&0.00&0.00&0.00&0.00&0.00&0.00&0.00\\
  &0.05&0.00&0.00&0.00&0.00&0.00&0.01&0.00\\
\multirow{-5}{*}{Mice.Norm}&0.10&0.00&0.01&0.01&0.01&0.01&0.02&0.00\\

\hline  &0.01&0.00&0.00&0.00&0.00&0.00&0.01&0.00\\
  &0.05&0.05&0.06&0.07&0.07&0.07&0.10&0.01\\
\multirow{-5}{*}{Mice.PMM}&0.10&0.20&0.25&0.27&0.28&0.30&0.36&0.03\\

\hline  &0.01&0.00&0.00&0.00&0.00&0.00&0.00&0.00\\
  &0.05&0.01&0.04&0.05&0.05&0.06&0.10&0.02\\
\multirow{-5}{*}{Mice.RF}&0.10&0.09&0.16&0.22&0.22&0.27&0.44&0.07\\

\hline Naive &0.01&0.08&0.12&0.14&0.13&0.14&0.18&0.02\\
  Imputation&0.05&2.89&3.27&3.37&3.38&3.47&3.81&0.19\\
 &0.10&12.49&13.21&13.51&13.53&13.86&15.48&0.53\\

\hline  &0.01&0.00&0.00&0.00&0.00&0.00&0.01&0.00\\
  &0.05&0.04&0.06&0.06&0.10&0.08&0.58&0.09\\
\multirow{-5}{*}{MissRanger}&0.10&0.18&0.24&0.26&0.45&0.35&3.14&0.57\\

\hline 
\end{tabular} 
\end{minipage} 
\end{table}

\newpage 
\begin{table}[H] 
\caption{minimum ($q_0$), first quartile ($q_{0.25}$),
               median ($q_{0.5}$), mean ($\bar{x}$), third quartile ($q_{0.75}$),
               maximum ($q_{1}$) and standard deviation (sd) of the observed values of the realization of the Cramer-von-Mises-statistic for the variable \texttt{ef24} for the different
               imputation methods and missing rates (r)
               creating missing values using the MCAR-mechanism for the employee data (VSE, Cramer-von-Mises, \texttt{ef24}, MCAR)}\label{Tab:vse.vals.cramervonmises.MCAR.ef24 }\begin{minipage}[t]{\textwidth} 

\centering 
\begin{tabular}[t]{|l|c|c|c|c|c|c|c|c|}\hline
 method&r&$q_{0}$& $q_{0.25}$&$q_{0.5}$&$\bar{x}$&$q_{0.75}$& $q_{1}$&sd\\ \hline 
\hline  &0.01&0.00&0.00&0.00&0.00&0.00&0.01&0.00\\
  &0.05&0.04&0.06&0.06&0.06&0.07&0.08&0.01\\
\multirow{-5}{*}{Amelia}&0.10&0.18&0.24&0.25&0.25&0.27&0.33&0.03\\

\hline  &0.01&0.00&0.00&0.00&0.00&0.00&0.00&0.00\\
  &0.05&0.00&0.00&0.00&0.00&0.00&0.01&0.00\\
\multirow{-5}{*}{Mice.Norm}&0.10&0.00&0.01&0.01&0.01&0.01&0.02&0.00\\

\hline  &0.01&0.00&0.00&0.00&0.00&0.00&0.01&0.00\\
  &0.05&0.04&0.06&0.07&0.07&0.07&0.10&0.01\\
\multirow{-5}{*}{Mice.PMM}&0.10&0.19&0.25&0.27&0.27&0.29&0.36&0.03\\

\hline  &0.01&0.00&0.00&0.00&0.00&0.00&0.01&0.00\\
  &0.05&0.02&0.04&0.05&0.05&0.06&0.09&0.02\\
\multirow{-5}{*}{Mice.RF}&0.10&0.09&0.19&0.23&0.24&0.28&0.44&0.07\\

\hline Naive &0.01&0.07&0.12&0.13&0.13&0.14&0.21&0.02\\
  Imputation&0.05&2.87&3.23&3.35&3.36&3.45&3.95&0.21\\
 &0.10&12.24&12.97&13.34&13.39&13.78&15.10&0.60\\

\hline  &0.01&0.00&0.00&0.00&0.00&0.00&0.02&0.00\\
  &0.05&0.04&0.06&0.06&0.11&0.09&0.77&0.13\\
\multirow{-5}{*}{MissRanger}&0.10&0.17&0.23&0.26&0.47&0.42&2.97&0.51\\

\hline 
\end{tabular} 
\end{minipage} 
\end{table}

\newpage 
\begin{table}[H] 
\caption{minimum ($q_0$), first quartile ($q_{0.25}$),
               median ($q_{0.5}$), mean ($\bar{x}$), third quartile ($q_{0.75}$),
               maximum ($q_{1}$) and standard deviation (sd) of the observed values of the realization of the Cramer-von-Mises-statistic for the variable \texttt{ef25} for the different
               imputation methods and missing rates (r)
               creating missing values using the MAR-mechanism for the employee data (VSE, Cramer-von-Mises, \texttt{ef25}, MAR)}\label{Tab:vse.vals.cramervonmises.MAR.ef25 }\begin{minipage}[t]{\textwidth} 

\centering 
\begin{tabular}[t]{|l|c|c|c|c|c|c|c|c|}\hline
 method&r&$q_{0}$& $q_{0.25}$&$q_{0.5}$&$\bar{x}$&$q_{0.75}$& $q_{1}$&sd\\ \hline 
\hline  &0.01&0.00&0.00&0.00&0.00&0.00&0.00&0.00\\
  &0.05&0.01&0.01&0.02&0.02&0.02&0.03&0.00\\
\multirow{-5}{*}{Amelia}&0.10&0.04&0.05&0.06&0.06&0.06&0.09&0.01\\

\hline  &0.01&0.00&0.00&0.00&0.00&0.00&0.00&0.00\\
  &0.05&0.00&0.00&0.00&0.00&0.00&0.00&0.00\\
\multirow{-5}{*}{Mice.Norm}&0.10&0.00&0.00&0.00&0.00&0.00&0.01&0.00\\

\hline  &0.01&0.00&0.00&0.00&0.00&0.00&0.00&0.00\\
  &0.05&0.01&0.02&0.02&0.02&0.02&0.02&0.00\\
\multirow{-5}{*}{Mice.PMM}&0.10&0.05&0.06&0.07&0.07&0.07&0.09&0.01\\

\hline  &0.01&0.00&0.00&0.00&0.00&0.00&0.00&0.00\\
  &0.05&0.00&0.01&0.01&0.01&0.02&0.03&0.01\\
\multirow{-5}{*}{Mice.RF}&0.10&0.02&0.03&0.05&0.06&0.08&0.16&0.03\\

\hline Naive &0.01&0.08&0.11&0.12&0.12&0.12&0.16&0.01\\
  Imputation&0.05&2.43&2.76&2.87&2.87&2.99&3.39&0.17\\
 &0.10&10.60&11.50&11.84&11.87&12.20&13.90&0.56\\

\hline  &0.01&0.00&0.00&0.00&0.00&0.00&0.01&0.00\\
  &0.05&0.01&0.02&0.02&0.04&0.03&0.32&0.05\\
\multirow{-5}{*}{MissRanger}&0.10&0.06&0.08&0.09&0.17&0.20&1.32&0.17\\

\hline 
\end{tabular} 
\end{minipage} 
\end{table}

\newpage 
\begin{table}[H] 
\caption{minimum ($q_0$), first quartile ($q_{0.25}$),
               median ($q_{0.5}$), mean ($\bar{x}$), third quartile ($q_{0.75}$),
               maximum ($q_{1}$) and standard deviation (sd) of the observed values of the realization of the Cramer-von-Mises-statistic for the variable \texttt{ef25} for the different
               imputation methods and missing rates (r)
               creating missing values using the MCAR-mechanism for the employee data (VSE, Cramer-von-Mises, \texttt{ef25}, MCAR)}\label{Tab:vse.vals.cramervonmises.MCAR.ef25 }\begin{minipage}[t]{\textwidth} 

\centering 
\begin{tabular}[t]{|l|c|c|c|c|c|c|c|c|}\hline
 method&r&$q_{0}$& $q_{0.25}$&$q_{0.5}$&$\bar{x}$&$q_{0.75}$& $q_{1}$&sd\\ \hline 
\hline  &0.01&0.00&0.00&0.00&0.00&0.00&0.00&0.00\\
  &0.05&0.01&0.01&0.02&0.02&0.02&0.03&0.00\\
\multirow{-5}{*}{Amelia}&0.10&0.04&0.06&0.06&0.06&0.07&0.08&0.01\\

\hline  &0.01&0.00&0.00&0.00&0.00&0.00&0.00&0.00\\
  &0.05&0.00&0.00&0.00&0.00&0.00&0.00&0.00\\
\multirow{-5}{*}{Mice.Norm}&0.10&0.00&0.00&0.00&0.00&0.00&0.01&0.00\\

\hline  &0.01&0.00&0.00&0.00&0.00&0.00&0.00&0.00\\
  &0.05&0.01&0.02&0.02&0.02&0.02&0.03&0.00\\
\multirow{-5}{*}{Mice.PMM}&0.10&0.05&0.06&0.07&0.07&0.07&0.09&0.01\\

\hline  &0.01&0.00&0.00&0.00&0.00&0.00&0.00&0.00\\
  &0.05&0.00&0.01&0.01&0.01&0.02&0.04&0.01\\
\multirow{-5}{*}{Mice.RF}&0.10&0.02&0.04&0.05&0.06&0.07&0.18&0.03\\

\hline Naive &0.01&0.08&0.10&0.11&0.11&0.12&0.15&0.01\\
  Imputation&0.05&2.49&2.75&2.85&2.86&2.97&3.20&0.16\\
 &0.10&10.49&11.57&11.87&11.91&12.22&13.42&0.51\\

\hline  &0.01&0.00&0.00&0.00&0.00&0.00&0.01&0.00\\
  &0.05&0.01&0.02&0.02&0.04&0.04&0.33&0.05\\
\multirow{-5}{*}{MissRanger}&0.10&0.06&0.08&0.09&0.17&0.12&1.33&0.21\\

\hline 
\end{tabular} 
\end{minipage} 
\end{table}

\newpage 
\begin{table}[H] 
\caption{minimum ($q_0$), first quartile ($q_{0.25}$),
               median ($q_{0.5}$), mean ($\bar{x}$), third quartile ($q_{0.75}$),
               maximum ($q_{1}$) and standard deviation (sd) of the observed values of the realization of the Cramer-von-Mises-statistic for the variable \texttt{ef26} for the different
               imputation methods and missing rates (r)
               creating missing values using the MAR-mechanism for the employee data (VSE, Cramer-von-Mises, \texttt{ef26}, MAR)}\label{Tab:vse.vals.cramervonmises.MAR.ef26 }\begin{minipage}[t]{\textwidth} 

\centering 
\begin{tabular}[t]{|l|c|c|c|c|c|c|c|c|}\hline
 method&r&$q_{0}$& $q_{0.25}$&$q_{0.5}$&$\bar{x}$&$q_{0.75}$& $q_{1}$&sd\\ \hline 
\hline  &0.01&0.11&0.14&0.15&0.15&0.17&0.22&0.02\\
  &0.05&3.13&3.50&3.69&3.68&3.86&4.19&0.23\\
\multirow{-5}{*}{Amelia}&0.10&13.04&14.02&14.56&14.55&14.97&17.05&0.72\\

\hline  &0.01&0.00&0.00&0.00&0.00&0.00&0.00&0.00\\
  &0.05&0.00&0.00&0.00&0.00&0.00&0.00&0.00\\
\multirow{-5}{*}{Mice.Norm}&0.10&0.00&0.00&0.00&0.00&0.00&0.00&0.00\\

\hline  &0.01&0.10&0.13&0.15&0.15&0.16&0.20&0.02\\
  &0.05&3.06&3.48&3.64&3.63&3.82&4.14&0.24\\
\multirow{-5}{*}{Mice.PMM}&0.10&12.97&13.76&14.25&14.24&14.72&15.86&0.65\\

\hline  &0.01&0.00&0.00&0.00&0.00&0.00&0.00&0.00\\
  &0.05&0.00&0.01&0.01&0.01&0.01&0.02&0.00\\
\multirow{-5}{*}{Mice.RF}&0.10&0.03&0.04&0.05&0.05&0.06&0.08&0.01\\

\hline Naive &0.01&0.02&0.03&0.03&0.03&0.04&0.04&0.00\\
  Imputation&0.05&1.00&1.19&1.23&1.24&1.30&1.45&0.09\\
 &0.10&6.31&6.92&7.18&7.18&7.44&8.20&0.35\\

\hline  &0.01&0.00&0.00&0.00&0.00&0.01&0.01&0.00\\
  &0.05&0.11&0.14&0.16&0.19&0.21&0.49&0.09\\
\multirow{-5}{*}{MissRanger}&0.10&0.87&0.97&1.10&1.23&1.39&2.48&0.34\\

\hline 
\end{tabular} 
\end{minipage} 
\end{table}

\newpage 
\begin{table}[H] 
\caption{minimum ($q_0$), first quartile ($q_{0.25}$),
               median ($q_{0.5}$), mean ($\bar{x}$), third quartile ($q_{0.75}$),
               maximum ($q_{1}$) and standard deviation (sd) of the observed values of the realization of the Cramer-von-Mises-statistic for the variable \texttt{ef26} for the different
               imputation methods and missing rates (r)
               creating missing values using the MCAR-mechanism for the employee data (VSE, Cramer-von-Mises, \texttt{ef26}, MCAR)}\label{Tab:vse.vals.cramervonmises.MCAR.ef26 }\begin{minipage}[t]{\textwidth} 

\centering 
\begin{tabular}[t]{|l|c|c|c|c|c|c|c|c|}\hline
 method&r&$q_{0}$& $q_{0.25}$&$q_{0.5}$&$\bar{x}$&$q_{0.75}$& $q_{1}$&sd\\ \hline 
\hline  &0.01&0.10&0.13&0.15&0.15&0.16&0.20&0.02\\
  &0.05&3.12&3.51&3.67&3.69&3.82&4.39&0.26\\
\multirow{-5}{*}{Amelia}&0.10&13.03&13.99&14.39&14.41&14.85&15.68&0.60\\

\hline  &0.01&0.00&0.00&0.00&0.00&0.00&0.00&0.00\\
  &0.05&0.00&0.00&0.00&0.00&0.00&0.00&0.00\\
\multirow{-5}{*}{Mice.Norm}&0.10&0.00&0.00&0.00&0.00&0.00&0.01&0.00\\

\hline  &0.01&0.11&0.13&0.14&0.14&0.16&0.21&0.02\\
  &0.05&3.20&3.50&3.67&3.65&3.78&4.41&0.22\\
\multirow{-5}{*}{Mice.PMM}&0.10&13.12&13.80&14.08&14.18&14.52&15.90&0.56\\

\hline  &0.01&0.00&0.00&0.00&0.00&0.00&0.00&0.00\\
  &0.05&0.01&0.01&0.01&0.01&0.02&0.03&0.00\\
\multirow{-5}{*}{Mice.RF}&0.10&0.02&0.04&0.05&0.05&0.06&0.08&0.01\\

\hline Naive &0.01&0.02&0.03&0.03&0.03&0.03&0.04&0.00\\
  Imputation&0.05&1.08&1.19&1.25&1.25&1.31&1.47&0.09\\
 &0.10&6.40&6.96&7.15&7.15&7.37&8.07&0.32\\

\hline  &0.01&0.00&0.00&0.00&0.00&0.01&0.01&0.00\\
  &0.05&0.10&0.13&0.15&0.18&0.21&0.48&0.07\\
\multirow{-5}{*}{MissRanger}&0.10&0.87&0.96&1.12&1.21&1.36&2.48&0.33\\

\hline 
\end{tabular} 
\end{minipage} 
\end{table}

\newpage 
\begin{table}[H] 
\caption{minimum ($q_0$), first quartile ($q_{0.25}$),
               median ($q_{0.5}$), mean ($\bar{x}$), third quartile ($q_{0.75}$),
               maximum ($q_{1}$) and standard deviation (sd) of the observed values of the realization of the Cramer-von-Mises-statistic for the variable \texttt{ef29} for the different
               imputation methods and missing rates (r)
               creating missing values using the MAR-mechanism for the employee data (VSE, Cramer-von-Mises, \texttt{ef29}, MAR)}\label{Tab:vse.vals.cramervonmises.MAR.ef29 }\begin{minipage}[t]{\textwidth} 

\centering 
\begin{tabular}[t]{|l|c|c|c|c|c|c|c|c|}\hline
 method&r&$q_{0}$& $q_{0.25}$&$q_{0.5}$&$\bar{x}$&$q_{0.75}$& $q_{1}$&sd\\ \hline 
\hline  &0.01&0.01&0.02&0.03&0.03&0.03&0.04&0.01\\
  &0.05&0.46&0.62&0.66&0.66&0.70&0.86&0.07\\
\multirow{-5}{*}{Amelia}&0.10&2.27&2.47&2.57&2.58&2.68&3.04&0.15\\

\hline  &0.01&0.00&0.00&0.00&0.00&0.00&0.00&0.00\\
  &0.05&0.00&0.00&0.00&0.00&0.00&0.02&0.00\\
\multirow{-5}{*}{Mice.Norm}&0.10&0.00&0.00&0.01&0.01&0.01&0.02&0.00\\

\hline  &0.01&0.02&0.02&0.03&0.03&0.03&0.05&0.01\\
  &0.05&0.41&0.64&0.68&0.67&0.70&0.81&0.06\\
\multirow{-5}{*}{Mice.PMM}&0.10&2.23&2.53&2.63&2.63&2.74&3.05&0.15\\

\hline  &0.01&0.00&0.00&0.00&0.00&0.01&0.01&0.00\\
  &0.05&0.05&0.08&0.09&0.10&0.11&0.18&0.02\\
\multirow{-5}{*}{Mice.RF}&0.10&0.21&0.30&0.35&0.36&0.40&0.56&0.07\\

\hline Naive &0.01&0.06&0.08&0.09&0.09&0.10&0.13&0.01\\
  Imputation&0.05&2.04&2.45&2.55&2.56&2.66&3.03&0.18\\
 &0.10&10.32&11.15&11.35&11.42&11.72&12.80&0.45\\

\hline  &0.01&0.00&0.01&0.01&0.01&0.01&0.03&0.00\\
  &0.05&0.10&0.18&0.21&0.23&0.24&0.74&0.11\\
\multirow{-5}{*}{MissRanger}&0.10&0.39&0.72&0.81&0.85&0.88&2.52&0.32\\

\hline 
\end{tabular} 
\end{minipage} 
\end{table}

\newpage 
\begin{table}[H] 
\caption{minimum ($q_0$), first quartile ($q_{0.25}$),
               median ($q_{0.5}$), mean ($\bar{x}$), third quartile ($q_{0.75}$),
               maximum ($q_{1}$) and standard deviation (sd) of the observed values of the realization of the Cramer-von-Mises-statistic for the variable \texttt{ef29} for the different
               imputation methods and missing rates (r)
               creating missing values using the MCAR-mechanism for the employee data (VSE, Cramer-von-Mises, \texttt{ef29}, MCAR)}\label{Tab:vse.vals.cramervonmises.MCAR.ef29 }\begin{minipage}[t]{\textwidth} 

\centering 
\begin{tabular}[t]{|l|c|c|c|c|c|c|c|c|}\hline
 method&r&$q_{0}$& $q_{0.25}$&$q_{0.5}$&$\bar{x}$&$q_{0.75}$& $q_{1}$&sd\\ \hline 
\hline  &0.01&0.02&0.02&0.03&0.03&0.03&0.04&0.01\\
  &0.05&0.54&0.61&0.65&0.65&0.69&0.80&0.05\\
\multirow{-5}{*}{Amelia}&0.10&2.11&2.46&2.58&2.58&2.70&3.04&0.18\\

\hline  &0.01&0.00&0.00&0.00&0.00&0.00&0.00&0.00\\
  &0.05&0.00&0.00&0.00&0.00&0.00&0.01&0.00\\
\multirow{-5}{*}{Mice.Norm}&0.10&0.00&0.01&0.01&0.01&0.01&0.02&0.00\\

\hline  &0.01&0.02&0.02&0.03&0.03&0.03&0.04&0.01\\
  &0.05&0.56&0.63&0.66&0.66&0.69&0.83&0.05\\
\multirow{-5}{*}{Mice.PMM}&0.10&2.27&2.52&2.66&2.65&2.78&3.10&0.17\\

\hline  &0.01&0.00&0.00&0.00&0.00&0.01&0.01&0.00\\
  &0.05&0.05&0.08&0.09&0.09&0.11&0.15&0.02\\
\multirow{-5}{*}{Mice.RF}&0.10&0.20&0.31&0.35&0.36&0.41&0.58&0.07\\

\hline Naive &0.01&0.06&0.08&0.09&0.09&0.10&0.13&0.01\\
  Imputation&0.05&2.13&2.46&2.54&2.55&2.62&3.03&0.15\\
 &0.10&10.07&11.06&11.42&11.42&11.79&12.56&0.54\\

\hline  &0.01&0.00&0.01&0.01&0.01&0.01&0.03&0.00\\
  &0.05&0.09&0.17&0.19&0.21&0.23&0.77&0.09\\
\multirow{-5}{*}{MissRanger}&0.10&0.33&0.69&0.82&0.84&0.90&2.42&0.28\\

\hline 
\end{tabular} 
\end{minipage} 
\end{table}

\newpage 
\begin{table}[H] 
\caption{minimum ($q_0$), first quartile ($q_{0.25}$),
               median ($q_{0.5}$), mean ($\bar{x}$), third quartile ($q_{0.75}$),
               maximum ($q_{1}$) and standard deviation (sd) of the observed values of the realization of the Cramer-von-Mises-statistic for the variable \texttt{ef40} for the different
               imputation methods and missing rates (r)
               creating missing values using the MAR-mechanism for the employee data (VSE, Cramer-von-Mises, \texttt{ef40}, MAR)}\label{Tab:vse.vals.cramervonmises.MAR.ef40 }\begin{minipage}[t]{\textwidth} 

\centering 
\begin{tabular}[t]{|l|c|c|c|c|c|c|c|c|}\hline
 method&r&$q_{0}$& $q_{0.25}$&$q_{0.5}$&$\bar{x}$&$q_{0.75}$& $q_{1}$&sd\\ \hline 
\hline  &0.01&0.01&0.01&0.01&0.01&0.01&0.02&0.00\\
  &0.05&0.19&0.23&0.25&0.25&0.27&0.38&0.03\\
\multirow{-5}{*}{Amelia}&0.10&0.76&0.97&1.03&1.03&1.11&1.22&0.10\\

\hline  &0.01&0.00&0.00&0.00&0.00&0.00&0.00&0.00\\
  &0.05&0.00&0.00&0.01&0.01&0.01&0.02&0.00\\
\multirow{-5}{*}{Mice.Norm}&0.10&0.00&0.01&0.01&0.01&0.01&0.04&0.01\\

\hline  &0.01&0.01&0.01&0.01&0.01&0.02&0.03&0.00\\
  &0.05&0.23&0.28&0.31&0.31&0.34&0.49&0.04\\
\multirow{-5}{*}{Mice.PMM}&0.10&0.86&1.17&1.26&1.25&1.33&1.56&0.12\\

\hline  &0.01&0.00&0.00&0.00&0.00&0.00&0.01&0.00\\
  &0.05&0.01&0.02&0.03&0.03&0.04&0.09&0.02\\
\multirow{-5}{*}{Mice.RF}&0.10&0.03&0.08&0.12&0.14&0.17&0.47&0.07\\

\hline Naive &0.01&0.10&0.13&0.15&0.14&0.16&0.20&0.02\\
  Imputation&0.05&2.94&3.45&3.54&3.56&3.72&4.08&0.24\\
 &0.10&13.02&14.24&14.61&14.59&15.01&16.03&0.58\\

\hline  &0.01&0.01&0.01&0.02&0.02&0.02&0.05&0.01\\
  &0.05&0.31&0.38&0.44&0.51&0.55&1.38&0.20\\
\multirow{-5}{*}{MissRanger}&0.10&1.26&1.53&1.70&1.96&1.93&6.33&0.81\\

\hline 
\end{tabular} 
\end{minipage} 
\end{table}

\newpage 
\begin{table}[H] 
\caption{minimum ($q_0$), first quartile ($q_{0.25}$),
               median ($q_{0.5}$), mean ($\bar{x}$), third quartile ($q_{0.75}$),
               maximum ($q_{1}$) and standard deviation (sd) of the observed values of the realization of the Cramer-von-Mises-statistic for the variable \texttt{ef40} for the different
               imputation methods and missing rates (r)
               creating missing values using the MCAR-mechanism for the employee data (VSE, Cramer-von-Mises, \texttt{ef40}, MCAR)}\label{Tab:vse.vals.cramervonmises.MCAR.ef40 }\begin{minipage}[t]{\textwidth} 

\centering 
\begin{tabular}[t]{|l|c|c|c|c|c|c|c|c|}\hline
 method&r&$q_{0}$& $q_{0.25}$&$q_{0.5}$&$\bar{x}$&$q_{0.75}$& $q_{1}$&sd\\ \hline 
\hline  &0.01&0.00&0.01&0.01&0.01&0.01&0.03&0.00\\
  &0.05&0.18&0.23&0.26&0.26&0.28&0.40&0.04\\
\multirow{-5}{*}{Amelia}&0.10&0.73&0.94&1.01&1.02&1.10&1.40&0.11\\

\hline  &0.01&0.00&0.00&0.00&0.00&0.00&0.00&0.00\\
  &0.05&0.00&0.00&0.00&0.01&0.01&0.01&0.00\\
\multirow{-5}{*}{Mice.Norm}&0.10&0.00&0.01&0.01&0.01&0.02&0.04&0.01\\

\hline  &0.01&0.01&0.01&0.01&0.01&0.02&0.03&0.00\\
  &0.05&0.20&0.28&0.31&0.31&0.34&0.45&0.04\\
\multirow{-5}{*}{Mice.PMM}&0.10&0.93&1.17&1.27&1.26&1.36&1.62&0.13\\

\hline  &0.01&0.00&0.00&0.00&0.00&0.00&0.01&0.00\\
  &0.05&0.01&0.03&0.04&0.04&0.05&0.09&0.02\\
\multirow{-5}{*}{Mice.RF}&0.10&0.02&0.10&0.13&0.14&0.17&0.37&0.06\\

\hline Naive &0.01&0.09&0.13&0.14&0.15&0.16&0.22&0.02\\
  Imputation&0.05&3.13&3.47&3.59&3.60&3.74&4.18&0.20\\
 &0.10&12.22&14.26&14.67&14.58&15.00&16.24&0.70\\

\hline  &0.01&0.01&0.02&0.02&0.02&0.02&0.06&0.01\\
  &0.05&0.30&0.37&0.42&0.46&0.48&1.33&0.16\\
\multirow{-5}{*}{MissRanger}&0.10&1.29&1.51&1.67&1.98&2.03&5.50&0.81\\

\hline 
\end{tabular} 
\end{minipage} 
\end{table}

\newpage 
\begin{table}[H] 
\caption{minimum ($q_0$), first quartile ($q_{0.25}$),
               median ($q_{0.5}$), mean ($\bar{x}$), third quartile ($q_{0.75}$),
               maximum ($q_{1}$) and standard deviation (sd) of the observed values of the realization of the Cramer-von-Mises-statistic for the variable \texttt{ef41} for the different
               imputation methods and missing rates (r)
               creating missing values using the MAR-mechanism for the employee data (VSE, Cramer-von-Mises, \texttt{ef41}, MAR)}\label{Tab:vse.vals.cramervonmises.MAR.ef41 }\begin{minipage}[t]{\textwidth} 

\centering 
\begin{tabular}[t]{|l|c|c|c|c|c|c|c|c|}\hline
 method&r&$q_{0}$& $q_{0.25}$&$q_{0.5}$&$\bar{x}$&$q_{0.75}$& $q_{1}$&sd\\ \hline 
\hline  &0.01&0.00&0.00&0.00&0.00&0.00&0.01&0.00\\
  &0.05&0.02&0.03&0.04&0.04&0.04&0.06&0.01\\
\multirow{-5}{*}{Amelia}&0.10&0.11&0.14&0.15&0.15&0.16&0.20&0.02\\

\hline  &0.01&0.00&0.00&0.00&0.00&0.00&0.00&0.00\\
  &0.05&0.00&0.00&0.01&0.01&0.01&0.03&0.00\\
\multirow{-5}{*}{Mice.Norm}&0.10&0.00&0.01&0.01&0.01&0.01&0.03&0.01\\

\hline  &0.01&0.00&0.00&0.00&0.00&0.00&0.01&0.00\\
  &0.05&0.02&0.04&0.04&0.04&0.05&0.06&0.01\\
\multirow{-5}{*}{Mice.PMM}&0.10&0.11&0.14&0.16&0.16&0.17&0.25&0.02\\

\hline  &0.01&0.00&0.00&0.00&0.00&0.00&0.00&0.00\\
  &0.05&0.00&0.01&0.01&0.01&0.01&0.02&0.00\\
\multirow{-5}{*}{Mice.RF}&0.10&0.01&0.02&0.02&0.03&0.03&0.09&0.02\\

\hline Naive &0.01&0.08&0.10&0.11&0.11&0.12&0.15&0.01\\
  Imputation&0.05&2.51&2.75&2.85&2.87&2.98&3.36&0.17\\
 &0.10&10.93&11.75&12.05&12.06&12.31&13.63&0.48\\

\hline  &0.01&0.00&0.01&0.01&0.01&0.01&0.04&0.01\\
  &0.05&0.16&0.20&0.25&0.32&0.36&0.94&0.18\\
\multirow{-5}{*}{MissRanger}&0.10&0.69&0.82&1.02&1.37&1.62&4.50&0.78\\

\hline 
\end{tabular} 
\end{minipage} 
\end{table}

\newpage 
\begin{table}[H] 
\caption{minimum ($q_0$), first quartile ($q_{0.25}$),
               median ($q_{0.5}$), mean ($\bar{x}$), third quartile ($q_{0.75}$),
               maximum ($q_{1}$) and standard deviation (sd) of the observed values of the realization of the Cramer-von-Mises-statistic for the variable \texttt{ef41} for the different
               imputation methods and missing rates (r)
               creating missing values using the MCAR-mechanism for the employee data (VSE, Cramer-von-Mises, \texttt{ef41}, MCAR)}\label{Tab:vse.vals.cramervonmises.MCAR.ef41 }\begin{minipage}[t]{\textwidth} 

\centering 
\begin{tabular}[t]{|l|c|c|c|c|c|c|c|c|}\hline
 method&r&$q_{0}$& $q_{0.25}$&$q_{0.5}$&$\bar{x}$&$q_{0.75}$& $q_{1}$&sd\\ \hline 
\hline  &0.01&0.00&0.00&0.00&0.00&0.00&0.01&0.00\\
  &0.05&0.02&0.03&0.04&0.04&0.05&0.06&0.01\\
\multirow{-5}{*}{Amelia}&0.10&0.10&0.14&0.15&0.15&0.17&0.21&0.02\\

\hline  &0.01&0.00&0.00&0.00&0.00&0.00&0.00&0.00\\
  &0.05&0.00&0.00&0.00&0.01&0.01&0.03&0.00\\
\multirow{-5}{*}{Mice.Norm}&0.10&0.00&0.01&0.01&0.01&0.01&0.04&0.01\\

\hline  &0.01&0.00&0.00&0.00&0.00&0.00&0.00&0.00\\
  &0.05&0.03&0.04&0.04&0.04&0.05&0.07&0.01\\
\multirow{-5}{*}{Mice.PMM}&0.10&0.11&0.15&0.16&0.16&0.18&0.23&0.02\\

\hline  &0.01&0.00&0.00&0.00&0.00&0.00&0.00&0.00\\
  &0.05&0.00&0.01&0.01&0.01&0.01&0.03&0.01\\
\multirow{-5}{*}{Mice.RF}&0.10&0.01&0.02&0.03&0.03&0.04&0.07&0.01\\

\hline Naive &0.01&0.07&0.10&0.11&0.11&0.12&0.15&0.02\\
  Imputation&0.05&2.41&2.79&2.89&2.88&3.00&3.29&0.16\\
 &0.10&11.12&11.77&12.06&12.09&12.36&13.05&0.43\\

\hline  &0.01&0.00&0.01&0.01&0.01&0.01&0.03&0.01\\
  &0.05&0.14&0.20&0.26&0.30&0.37&0.97&0.14\\
\multirow{-5}{*}{MissRanger}&0.10&0.67&0.89&1.22&1.31&1.46&3.73&0.58\\

\hline 
\end{tabular} 
\end{minipage} 
\end{table}

\newpage 
\begin{table}[H] 
\caption{minimum ($q_0$), first quartile ($q_{0.25}$),
               median ($q_{0.5}$), mean ($\bar{x}$), third quartile ($q_{0.75}$),
               maximum ($q_{1}$) and standard deviation (sd) of the observed values of the realization of the Cramer-von-Mises-statistic for the variable \texttt{ef44} for the different
               imputation methods and missing rates (r)
               creating missing values using the MAR-mechanism for the employee data (VSE, Cramer-von-Mises, \texttt{ef44}, MAR)}\label{Tab:vse.vals.cramervonmises.MAR.ef44 }\begin{minipage}[t]{\textwidth} 

\centering 
\begin{tabular}[t]{|l|c|c|c|c|c|c|c|c|}\hline
 method&r&$q_{0}$& $q_{0.25}$&$q_{0.5}$&$\bar{x}$&$q_{0.75}$& $q_{1}$&sd\\ \hline 
\hline  &0.01&0.00&0.00&0.00&0.00&0.00&0.00&0.00\\
  &0.05&0.00&0.00&0.00&0.00&0.00&0.01&0.00\\
\multirow{-5}{*}{Amelia}&0.10&0.01&0.01&0.01&0.01&0.01&0.02&0.00\\

\hline  &0.01&0.00&0.00&0.00&0.00&0.00&0.00&0.00\\
  &0.05&0.00&0.00&0.00&0.00&0.00&0.00&0.00\\
\multirow{-5}{*}{Mice.Norm}&0.10&0.00&0.00&0.00&0.00&0.00&0.00&0.00\\

\hline  &0.01&0.00&0.00&0.00&0.00&0.00&0.00&0.00\\
  &0.05&0.00&0.00&0.00&0.00&0.00&0.00&0.00\\
\multirow{-5}{*}{Mice.PMM}&0.10&0.00&0.01&0.01&0.01&0.01&0.01&0.00\\

\hline  &0.01&0.00&0.00&0.00&0.00&0.00&0.00&0.00\\
  &0.05&0.00&0.01&0.01&0.01&0.02&0.04&0.01\\
\multirow{-5}{*}{Mice.RF}&0.10&0.01&0.02&0.03&0.04&0.05&0.09&0.02\\

\hline Naive &0.01&0.08&0.10&0.11&0.11&0.12&0.16&0.02\\
  Imputation&0.05&2.55&2.76&2.85&2.85&2.95&3.23&0.14\\
 &0.10&10.46&11.49&11.87&11.82&12.14&12.97&0.48\\

\hline  &0.01&0.00&0.00&0.00&0.00&0.00&0.00&0.00\\
  &0.05&0.00&0.00&0.00&0.01&0.00&0.14&0.02\\
\multirow{-5}{*}{MissRanger}&0.10&0.00&0.00&0.00&0.03&0.01&0.71&0.11\\

\hline 
\end{tabular} 
\end{minipage} 
\end{table}

\newpage 
\begin{table}[H] 
\caption{minimum ($q_0$), first quartile ($q_{0.25}$),
               median ($q_{0.5}$), mean ($\bar{x}$), third quartile ($q_{0.75}$),
               maximum ($q_{1}$) and standard deviation (sd) of the observed values of the realization of the Cramer-von-Mises-statistic for the variable \texttt{ef44} for the different
               imputation methods and missing rates (r)
               creating missing values using the MCAR-mechanism for the employee data (VSE, Cramer-von-Mises, \texttt{ef44}, MCAR)}\label{Tab:vse.vals.cramervonmises.MCAR.ef44 }\begin{minipage}[t]{\textwidth} 

\centering 
\begin{tabular}[t]{|l|c|c|c|c|c|c|c|c|}\hline
 method&r&$q_{0}$& $q_{0.25}$&$q_{0.5}$&$\bar{x}$&$q_{0.75}$& $q_{1}$&sd\\ \hline 
\hline  &0.01&0.00&0.00&0.00&0.00&0.00&0.00&0.00\\
  &0.05&0.00&0.00&0.00&0.00&0.00&0.00&0.00\\
\multirow{-5}{*}{Amelia}&0.10&0.01&0.01&0.01&0.01&0.01&0.02&0.00\\

\hline  &0.01&0.00&0.00&0.00&0.00&0.00&0.00&0.00\\
  &0.05&0.00&0.00&0.00&0.00&0.00&0.00&0.00\\
\multirow{-5}{*}{Mice.Norm}&0.10&0.00&0.00&0.00&0.00&0.00&0.00&0.00\\

\hline  &0.01&0.00&0.00&0.00&0.00&0.00&0.00&0.00\\
  &0.05&0.00&0.00&0.00&0.00&0.00&0.00&0.00\\
\multirow{-5}{*}{Mice.PMM}&0.10&0.00&0.01&0.01&0.01&0.01&0.01&0.00\\

\hline  &0.01&0.00&0.00&0.00&0.00&0.00&0.00&0.00\\
  &0.05&0.00&0.01&0.01&0.01&0.01&0.04&0.01\\
\multirow{-5}{*}{Mice.RF}&0.10&0.01&0.02&0.03&0.04&0.05&0.11&0.02\\

\hline Naive &0.01&0.07&0.10&0.11&0.11&0.12&0.16&0.02\\
  Imputation&0.05&2.45&2.76&2.86&2.86&2.97&3.26&0.17\\
 &0.10&10.72&11.57&11.76&11.83&12.11&12.74&0.43\\

\hline  &0.01&0.00&0.00&0.00&0.00&0.00&0.01&0.00\\
  &0.05&0.00&0.00&0.00&0.01&0.00&0.16&0.02\\
\multirow{-5}{*}{MissRanger}&0.10&0.00&0.00&0.00&0.02&0.01&0.18&0.03\\

\hline 
\end{tabular} 
\end{minipage} 
\end{table}

\newpage 
\begin{table}[H] 
\caption{minimum ($q_0$), first quartile ($q_{0.25}$),
               median ($q_{0.5}$), mean ($\bar{x}$), third quartile ($q_{0.75}$),
               maximum ($q_{1}$) and standard deviation (sd) of the observed values of the realization of the Cramer-von-Mises-statistic for the variable \texttt{ef45} for the different
               imputation methods and missing rates (r)
               creating missing values using the MAR-mechanism for the employee data (VSE, Cramer-von-Mises, \texttt{ef45}, MAR)}\label{Tab:vse.vals.cramervonmises.MAR.ef45 }\begin{minipage}[t]{\textwidth} 

\centering 
\begin{tabular}[t]{|l|c|c|c|c|c|c|c|c|}\hline
 method&r&$q_{0}$& $q_{0.25}$&$q_{0.5}$&$\bar{x}$&$q_{0.75}$& $q_{1}$&sd\\ \hline 
\hline  &0.01&0.00&0.00&0.00&0.00&0.00&0.00&0.00\\
  &0.05&0.00&0.00&0.00&0.00&0.00&0.01&0.00\\
\multirow{-5}{*}{Amelia}&0.10&0.01&0.01&0.01&0.01&0.02&0.02&0.00\\

\hline  &0.01&0.00&0.00&0.00&0.00&0.00&0.00&0.00\\
  &0.05&0.00&0.00&0.00&0.00&0.00&0.00&0.00\\
\multirow{-5}{*}{Mice.Norm}&0.10&0.00&0.00&0.00&0.00&0.00&0.00&0.00\\

\hline  &0.01&0.00&0.00&0.00&0.00&0.00&0.00&0.00\\
  &0.05&0.00&0.00&0.00&0.00&0.00&0.00&0.00\\
\multirow{-5}{*}{Mice.PMM}&0.10&0.01&0.01&0.01&0.01&0.01&0.01&0.00\\

\hline  &0.01&0.00&0.00&0.00&0.00&0.00&0.01&0.00\\
  &0.05&0.01&0.03&0.04&0.04&0.05&0.09&0.01\\
\multirow{-5}{*}{Mice.RF}&0.10&0.06&0.11&0.14&0.15&0.18&0.37&0.06\\

\hline Naive &0.01&0.09&0.12&0.13&0.13&0.14&0.19&0.02\\
  Imputation&0.05&2.78&3.08&3.21&3.24&3.38&3.70&0.20\\
 &0.10&11.85&12.96&13.32&13.33&13.72&14.50&0.58\\

\hline  &0.01&0.00&0.00&0.00&0.00&0.00&0.01&0.00\\
  &0.05&0.00&0.00&0.00&0.01&0.00&0.27&0.04\\
\multirow{-5}{*}{MissRanger}&0.10&0.00&0.00&0.00&0.02&0.01&0.41&0.05\\

\hline 
\end{tabular} 
\end{minipage} 
\end{table}

\newpage 
\begin{table}[H] 
\caption{minimum ($q_0$), first quartile ($q_{0.25}$),
               median ($q_{0.5}$), mean ($\bar{x}$), third quartile ($q_{0.75}$),
               maximum ($q_{1}$) and standard deviation (sd) of the observed values of the realization of the Cramer-von-Mises-statistic for the variable \texttt{ef45} for the different
               imputation methods and missing rates (r)
               creating missing values using the MCAR-mechanism for the employee data (VSE, Cramer-von-Mises, \texttt{ef45}, MCAR)}\label{Tab:vse.vals.cramervonmises.MCAR.ef45 }\begin{minipage}[t]{\textwidth} 

\centering 
\begin{tabular}[t]{|l|c|c|c|c|c|c|c|c|}\hline
 method&r&$q_{0}$& $q_{0.25}$&$q_{0.5}$&$\bar{x}$&$q_{0.75}$& $q_{1}$&sd\\ \hline 
\hline  &0.01&0.00&0.00&0.00&0.00&0.00&0.00&0.00\\
  &0.05&0.00&0.00&0.00&0.00&0.00&0.00&0.00\\
\multirow{-5}{*}{Amelia}&0.10&0.01&0.01&0.01&0.01&0.01&0.02&0.00\\

\hline  &0.01&0.00&0.00&0.00&0.00&0.00&0.00&0.00\\
  &0.05&0.00&0.00&0.00&0.00&0.00&0.00&0.00\\
\multirow{-5}{*}{Mice.Norm}&0.10&0.00&0.00&0.00&0.00&0.00&0.00&0.00\\

\hline  &0.01&0.00&0.00&0.00&0.00&0.00&0.00&0.00\\
  &0.05&0.00&0.00&0.00&0.00&0.00&0.00&0.00\\
\multirow{-5}{*}{Mice.PMM}&0.10&0.01&0.01&0.01&0.01&0.01&0.01&0.00\\

\hline  &0.01&0.00&0.00&0.00&0.00&0.00&0.00&0.00\\
  &0.05&0.01&0.02&0.02&0.02&0.02&0.03&0.01\\
\multirow{-5}{*}{Mice.RF}&0.10&0.03&0.05&0.07&0.07&0.09&0.15&0.03\\

\hline Naive &0.01&0.09&0.10&0.11&0.11&0.12&0.16&0.01\\
  Imputation&0.05&2.56&2.81&2.93&2.93&3.03&3.37&0.17\\
 &0.10&10.99&11.76&12.02&12.04&12.32&13.26&0.47\\

\hline  &0.01&0.00&0.00&0.00&0.00&0.00&0.01&0.00\\
  &0.05&0.00&0.00&0.00&0.01&0.00&0.24&0.04\\
\multirow{-5}{*}{MissRanger}&0.10&0.00&0.00&0.00&0.02&0.01&0.39&0.06\\

\hline 
\end{tabular} 
\end{minipage} 
\end{table}

\newpage 
\begin{table}[H] 
\caption{minimum ($q_0$), first quartile ($q_{0.25}$),
               median ($q_{0.5}$), mean ($\bar{x}$), third quartile ($q_{0.75}$),
               maximum ($q_{1}$) and standard deviation (sd) of the observed values of the realization of the Cramer-von-Mises-statistic for the variable \texttt{ef47} for the different
               imputation methods and missing rates (r)
               creating missing values using the MAR-mechanism for the employee data (VSE, Cramer-von-Mises, \texttt{ef47}, MAR)}\label{Tab:vse.vals.cramervonmises.MAR.ef47 }\begin{minipage}[t]{\textwidth} 

\centering 
\begin{tabular}[t]{|l|c|c|c|c|c|c|c|c|}\hline
 method&r&$q_{0}$& $q_{0.25}$&$q_{0.5}$&$\bar{x}$&$q_{0.75}$& $q_{1}$&sd\\ \hline 
\hline  &0.01&0.00&0.01&0.01&0.01&0.01&0.02&0.00\\
  &0.05&0.13&0.18&0.21&0.21&0.24&0.29&0.04\\
\multirow{-5}{*}{Amelia}&0.10&0.61&0.75&0.80&0.82&0.90&1.00&0.09\\

\hline  &0.01&0.00&0.00&0.00&0.00&0.00&0.00&0.00\\
  &0.05&0.00&0.00&0.01&0.01&0.01&0.02&0.00\\
\multirow{-5}{*}{Mice.Norm}&0.10&0.00&0.01&0.01&0.01&0.01&0.07&0.01\\

\hline  &0.01&0.00&0.01&0.01&0.01&0.01&0.02&0.00\\
  &0.05&0.14&0.20&0.23&0.24&0.27&0.37&0.05\\
\multirow{-5}{*}{Mice.PMM}&0.10&0.68&0.86&0.92&0.94&1.01&1.26&0.12\\

\hline  &0.01&0.00&0.00&0.00&0.00&0.00&0.02&0.00\\
  &0.05&0.01&0.04&0.06&0.07&0.08&0.21&0.04\\
\multirow{-5}{*}{Mice.RF}&0.10&0.03&0.21&0.33&0.38&0.50&1.40&0.24\\

\hline Naive &0.01&0.10&0.14&0.15&0.16&0.17&0.23&0.03\\
  Imputation&0.05&3.27&3.78&3.99&3.98&4.14&4.71&0.27\\
 &0.10&14.26&15.41&15.77&15.80&16.28&17.54&0.68\\

\hline  &0.01&0.02&0.03&0.04&0.04&0.04&0.06&0.01\\
  &0.05&0.73&0.84&0.91&0.97&1.03&1.81&0.20\\
\multirow{-5}{*}{MissRanger}&0.10&2.85&3.37&3.54&3.70&3.75&6.31&0.66\\

\hline 
\end{tabular} 
\end{minipage} 
\end{table}

\newpage 
\begin{table}[H] 
\caption{minimum ($q_0$), first quartile ($q_{0.25}$),
               median ($q_{0.5}$), mean ($\bar{x}$), third quartile ($q_{0.75}$),
               maximum ($q_{1}$) and standard deviation (sd) of the observed values of the realization of the Cramer-von-Mises-statistic for the variable \texttt{ef47} for the different
               imputation methods and missing rates (r)
               creating missing values using the MCAR-mechanism for the employee data (VSE, Cramer-von-Mises, \texttt{ef47}, MCAR)}\label{Tab:vse.vals.cramervonmises.MCAR.ef47 }\begin{minipage}[t]{\textwidth} 

\centering 
\begin{tabular}[t]{|l|c|c|c|c|c|c|c|c|}\hline
 method&r&$q_{0}$& $q_{0.25}$&$q_{0.5}$&$\bar{x}$&$q_{0.75}$& $q_{1}$&sd\\ \hline 
\hline  &0.01&0.00&0.01&0.01&0.01&0.01&0.02&0.00\\
  &0.05&0.12&0.18&0.20&0.20&0.22&0.27&0.03\\
\multirow{-5}{*}{Amelia}&0.10&0.64&0.77&0.82&0.83&0.88&1.09&0.09\\

\hline  &0.01&0.00&0.00&0.00&0.00&0.00&0.01&0.00\\
  &0.05&0.00&0.00&0.01&0.01&0.01&0.02&0.00\\
\multirow{-5}{*}{Mice.Norm}&0.10&0.00&0.01&0.01&0.01&0.01&0.05&0.01\\

\hline  &0.01&0.00&0.01&0.01&0.01&0.01&0.03&0.00\\
  &0.05&0.14&0.21&0.23&0.23&0.26&0.33&0.04\\
\multirow{-5}{*}{Mice.PMM}&0.10&0.72&0.87&0.95&0.95&1.03&1.23&0.12\\

\hline  &0.01&0.00&0.00&0.00&0.00&0.00&0.01&0.00\\
  &0.05&0.01&0.05&0.06&0.07&0.09&0.26&0.04\\
\multirow{-5}{*}{Mice.RF}&0.10&0.05&0.19&0.32&0.35&0.48&0.94&0.20\\

\hline Naive &0.01&0.11&0.14&0.16&0.16&0.17&0.23&0.03\\
  Imputation&0.05&3.26&3.73&3.91&3.90&4.11&4.53&0.27\\
 &0.10&14.17&15.35&15.97&15.91&16.49&17.50&0.69\\

\hline  &0.01&0.02&0.03&0.04&0.04&0.05&0.08&0.01\\
  &0.05&0.67&0.86&0.92&0.96&0.99&1.67&0.18\\
\multirow{-5}{*}{MissRanger}&0.10&3.05&3.44&3.58&3.77&3.79&7.16&0.69\\

\hline 
\end{tabular} 
\end{minipage} 
\end{table}

\newpage 
\begin{table}[H] 
\caption{minimum ($q_0$), first quartile ($q_{0.25}$),
               median ($q_{0.5}$), mean ($\bar{x}$), third quartile ($q_{0.75}$),
               maximum ($q_{1}$) and standard deviation (sd) of the observed values of the realization of the Cramer-von-Mises-statistic for the variable \texttt{ef53} for the different
               imputation methods and missing rates (r)
               creating missing values using the MAR-mechanism for the employee data (VSE, Cramer-von-Mises, \texttt{ef53}, MAR)}\label{Tab:vse.vals.cramervonmises.MAR.ef53 }\begin{minipage}[t]{\textwidth} 

\centering 
\begin{tabular}[t]{|l|c|c|c|c|c|c|c|c|}\hline
 method&r&$q_{0}$& $q_{0.25}$&$q_{0.5}$&$\bar{x}$&$q_{0.75}$& $q_{1}$&sd\\ \hline 
\hline  &0.01&0.01&0.02&0.02&0.02&0.02&0.03&0.00\\
  &0.05&0.35&0.43&0.46&0.46&0.49&0.56&0.05\\
\multirow{-5}{*}{Amelia}&0.10&1.48&1.76&1.83&1.83&1.89&2.21&0.11\\

\hline  &0.01&0.00&0.00&0.00&0.00&0.00&0.00&0.00\\
  &0.05&0.00&0.00&0.00&0.00&0.00&0.01&0.00\\
\multirow{-5}{*}{Mice.Norm}&0.10&0.00&0.00&0.01&0.01&0.01&0.02&0.00\\

\hline  &0.01&0.01&0.02&0.02&0.02&0.03&0.04&0.00\\
  &0.05&0.41&0.52&0.57&0.56&0.60&0.72&0.06\\
\multirow{-5}{*}{Mice.PMM}&0.10&1.92&2.17&2.26&2.26&2.33&2.62&0.13\\

\hline  &0.01&0.00&0.00&0.00&0.00&0.00&0.01&0.00\\
  &0.05&0.01&0.02&0.03&0.03&0.04&0.07&0.01\\
\multirow{-5}{*}{Mice.RF}&0.10&0.05&0.10&0.12&0.13&0.15&0.29&0.05\\

\hline Naive &0.01&0.09&0.11&0.12&0.12&0.14&0.17&0.02\\
  Imputation&0.05&2.75&3.14&3.29&3.31&3.47&3.94&0.25\\
 &0.10&13.04&14.13&14.54&14.53&14.92&16.08&0.66\\

\hline  &0.01&0.00&0.01&0.01&0.01&0.01&0.03&0.01\\
  &0.05&0.02&0.10&0.15&0.18&0.19&1.00&0.18\\
\multirow{-5}{*}{MissRanger}&0.10&0.14&0.31&0.57&0.58&0.68&2.74&0.43\\

\hline 
\end{tabular} 
\end{minipage} 
\end{table}

\newpage 
\begin{table}[H] 
\caption{minimum ($q_0$), first quartile ($q_{0.25}$),
               median ($q_{0.5}$), mean ($\bar{x}$), third quartile ($q_{0.75}$),
               maximum ($q_{1}$) and standard deviation (sd) of the observed values of the realization of the Cramer-von-Mises-statistic for the variable \texttt{ef53} for the different
               imputation methods and missing rates (r)
               creating missing values using the MCAR-mechanism for the employee data (VSE, Cramer-von-Mises, \texttt{ef53}, MCAR)}\label{Tab:vse.vals.cramervonmises.MCAR.ef53 }\begin{minipage}[t]{\textwidth} 

\centering 
\begin{tabular}[t]{|l|c|c|c|c|c|c|c|c|}\hline
 method&r&$q_{0}$& $q_{0.25}$&$q_{0.5}$&$\bar{x}$&$q_{0.75}$& $q_{1}$&sd\\ \hline 
\hline  &0.01&0.01&0.02&0.02&0.02&0.02&0.03&0.00\\
  &0.05&0.36&0.44&0.46&0.46&0.49&0.59&0.04\\
\multirow{-5}{*}{Amelia}&0.10&1.60&1.76&1.84&1.85&1.91&2.16&0.12\\

\hline  &0.01&0.00&0.00&0.00&0.00&0.00&0.00&0.00\\
  &0.05&0.00&0.00&0.00&0.00&0.00&0.01&0.00\\
\multirow{-5}{*}{Mice.Norm}&0.10&0.00&0.01&0.01&0.01&0.01&0.02&0.00\\

\hline  &0.01&0.01&0.02&0.02&0.02&0.03&0.04&0.00\\
  &0.05&0.46&0.52&0.56&0.56&0.59&0.68&0.05\\
\multirow{-5}{*}{Mice.PMM}&0.10&1.96&2.19&2.27&2.27&2.34&2.62&0.12\\

\hline  &0.01&0.00&0.00&0.00&0.00&0.00&0.01&0.00\\
  &0.05&0.01&0.02&0.03&0.03&0.04&0.06&0.01\\
\multirow{-5}{*}{Mice.RF}&0.10&0.04&0.10&0.12&0.12&0.15&0.24&0.04\\

\hline Naive &0.01&0.09&0.11&0.12&0.12&0.13&0.17&0.02\\
  Imputation&0.05&2.85&3.18&3.30&3.33&3.46&4.02&0.21\\
 &0.10&12.59&14.17&14.58&14.54&14.91&15.80&0.62\\

\hline  &0.01&0.00&0.00&0.01&0.01&0.01&0.05&0.01\\
  &0.05&0.03&0.11&0.16&0.16&0.18&0.70&0.11\\
\multirow{-5}{*}{MissRanger}&0.10&0.11&0.33&0.57&0.58&0.68&4.67&0.51\\

\hline 
\end{tabular} 
\end{minipage} 
\end{table}

\newpage 

%% file: table_fivepoint_KullbackLeibler_vse.txt
\begin{table}[H] 
\caption{minimum ($q_0$), first quartile ($q_{0.25}$),
               median ($q_{0.5}$), mean ($\bar{x}$), third quartile ($q_{0.75}$),
               maximum ($q_{1}$) and standard deviation (sd) of the observed values of the Kullback-Leibler divergence for the variable \texttt{ef19} for the different
               imputation methods and missing rates (r)
               creating missing values using the MAR-mechanism for the employee data (VSE, Kullback-Leibler, \texttt{ef19}, MAR)}\label{Tab:vse.vals.kullbackleibler.MAR.ef19 }\begin{minipage}[t]{\textwidth} 

\centering 
\begin{tabular}[t]{|l|c|c|c|c|c|c|c|c|}\hline
 method&r&$q_{0}$& $q_{0.25}$&$q_{0.5}$&$\bar{x}$&$q_{0.75}$& $q_{1}$&sd\\ \hline 
\hline  &0.01&0.00&0.00&0.00&0.00&0.00&0.00&0.00\\
  &0.05&0.00&0.00&0.00&0.00&0.00&0.00&0.00\\
\multirow{-5}{*}{Amelia}&0.10&0.01&0.01&0.01&0.01&0.01&0.01&0.00\\

\hline  &0.01&0.00&0.00&0.00&0.00&0.00&0.00&0.00\\
  &0.05&0.00&0.00&0.00&0.00&0.00&0.00&0.00\\
\multirow{-5}{*}{Mice.Norm}&0.10&0.00&0.00&0.00&0.00&0.00&0.00&0.00\\

\hline  &0.01&0.00&0.00&0.00&0.00&0.00&0.00&0.00\\
  &0.05&0.01&0.01&0.01&0.01&0.01&0.01&0.00\\
\multirow{-5}{*}{Mice.PMM}&0.10&0.02&0.02&0.02&0.02&0.02&0.02&0.00\\

\hline  &0.01&0.00&0.00&0.00&0.00&0.00&0.00&0.00\\
  &0.05&0.00&0.00&0.00&0.00&0.00&0.00&0.00\\
\multirow{-5}{*}{Mice.RF}&0.10&0.00&0.00&0.00&0.00&0.00&0.00&0.00\\

\hline Naive &0.01&0.00&0.00&0.00&0.00&0.00&0.00&0.00\\
  Imputation&0.05&0.04&0.05&0.05&0.05&0.05&0.05&0.00\\
 &0.10&0.13&0.14&0.14&0.14&0.14&0.15&0.00\\

\hline  &0.01&0.00&0.00&0.00&0.00&0.00&0.00&0.00\\
  &0.05&0.00&0.00&0.00&0.00&0.00&0.01&0.00\\
\multirow{-5}{*}{MissRanger}&0.10&0.00&0.00&0.00&0.00&0.00&0.02&0.00\\

\hline 
\end{tabular} 
\end{minipage} 
\end{table}

\newpage 
\begin{table}[H] 
\caption{minimum ($q_0$), first quartile ($q_{0.25}$),
               median ($q_{0.5}$), mean ($\bar{x}$), third quartile ($q_{0.75}$),
               maximum ($q_{1}$) and standard deviation (sd) of the observed values of the Kullback-Leibler divergence for the variable \texttt{ef19} for the different
               imputation methods and missing rates (r)
               creating missing values using the MCAR-mechanism for the employee data (VSE, Kullback-Leibler, \texttt{ef19}, MCAR)}\label{Tab:vse.vals.kullbackleibler.MCAR.ef19 }\begin{minipage}[t]{\textwidth} 

\centering 
\begin{tabular}[t]{|l|c|c|c|c|c|c|c|c|}\hline
 method&r&$q_{0}$& $q_{0.25}$&$q_{0.5}$&$\bar{x}$&$q_{0.75}$& $q_{1}$&sd\\ \hline 
\hline  &0.01&0.00&0.00&0.00&0.00&0.00&0.00&0.00\\
  &0.05&0.00&0.00&0.00&0.00&0.00&0.00&0.00\\
\multirow{-5}{*}{Amelia}&0.10&0.01&0.01&0.01&0.01&0.01&0.01&0.00\\

\hline  &0.01&0.00&0.00&0.00&0.00&0.00&0.00&0.00\\
  &0.05&0.00&0.00&0.00&0.00&0.00&0.00&0.00\\
\multirow{-5}{*}{Mice.Norm}&0.10&0.00&0.00&0.00&0.00&0.00&0.00&0.00\\

\hline  &0.01&0.00&0.00&0.00&0.00&0.00&0.00&0.00\\
  &0.05&0.01&0.01&0.01&0.01&0.01&0.01&0.00\\
\multirow{-5}{*}{Mice.PMM}&0.10&0.02&0.02&0.02&0.02&0.02&0.02&0.00\\

\hline  &0.01&0.00&0.00&0.00&0.00&0.00&0.00&0.00\\
  &0.05&0.00&0.00&0.00&0.00&0.00&0.00&0.00\\
\multirow{-5}{*}{Mice.RF}&0.10&0.00&0.00&0.00&0.00&0.00&0.00&0.00\\

\hline Naive &0.01&0.00&0.00&0.00&0.00&0.00&0.00&0.00\\
  Imputation&0.05&0.04&0.04&0.05&0.05&0.05&0.05&0.00\\
 &0.10&0.13&0.14&0.14&0.14&0.15&0.15&0.00\\

\hline  &0.01&0.00&0.00&0.00&0.00&0.00&0.00&0.00\\
  &0.05&0.00&0.00&0.00&0.00&0.00&0.00&0.00\\
\multirow{-5}{*}{MissRanger}&0.10&0.00&0.00&0.00&0.00&0.00&0.02&0.00\\

\hline 
\end{tabular} 
\end{minipage} 
\end{table}

\newpage 
\begin{table}[H] 
\caption{minimum ($q_0$), first quartile ($q_{0.25}$),
               median ($q_{0.5}$), mean ($\bar{x}$), third quartile ($q_{0.75}$),
               maximum ($q_{1}$) and standard deviation (sd) of the observed values of the Kullback-Leibler divergence for the variable \texttt{ef20} for the different
               imputation methods and missing rates (r)
               creating missing values using the MAR-mechanism for the employee data (VSE, Kullback-Leibler, \texttt{ef20}, MAR)}\label{Tab:vse.vals.kullbackleibler.MAR.ef20 }\begin{minipage}[t]{\textwidth} 

\centering 
\begin{tabular}[t]{|l|c|c|c|c|c|c|c|c|}\hline
 method&r&$q_{0}$& $q_{0.25}$&$q_{0.5}$&$\bar{x}$&$q_{0.75}$& $q_{1}$&sd\\ \hline 
\hline  &0.01&0.01&0.02&0.02&0.02&0.02&0.02&0.00\\
  &0.05&0.10&0.11&0.12&0.12&0.12&0.13&0.01\\
\multirow{-5}{*}{Amelia}&0.10&0.26&0.28&0.28&0.28&0.29&0.31&0.01\\

\hline  &0.01&0.00&0.00&0.00&0.00&0.00&0.00&0.00\\
  &0.05&0.00&0.00&0.00&0.00&0.00&0.00&0.00\\
\multirow{-5}{*}{Mice.Norm}&0.10&0.00&0.00&0.00&0.00&0.00&0.00&0.00\\

\hline  &0.01&0.02&0.03&0.03&0.03&0.03&0.04&0.00\\
  &0.05&0.17&0.18&0.18&0.18&0.19&0.20&0.01\\
\multirow{-5}{*}{Mice.PMM}&0.10&0.40&0.42&0.42&0.42&0.43&0.45&0.01\\

\hline  &0.01&0.00&0.00&0.00&0.00&0.00&0.00&0.00\\
  &0.05&0.00&0.00&0.00&0.00&0.00&0.01&0.00\\
\multirow{-5}{*}{Mice.RF}&0.10&0.00&0.00&0.00&0.00&0.01&0.01&0.00\\

\hline Naive &0.01&0.00&0.00&0.00&0.00&0.00&0.00&0.00\\
  Imputation&0.05&0.05&0.06&0.06&0.06&0.07&0.07&0.00\\
 &0.10&0.18&0.20&0.20&0.20&0.21&0.22&0.01\\

\hline  &0.01&0.00&0.00&0.00&0.00&0.00&0.00&0.00\\
  &0.05&0.00&0.00&0.00&0.00&0.00&0.00&0.00\\
\multirow{-5}{*}{MissRanger}&0.10&0.00&0.00&0.00&0.00&0.00&0.02&0.00\\

\hline 
\end{tabular} 
\end{minipage} 
\end{table}

\newpage 
\begin{table}[H] 
\caption{minimum ($q_0$), first quartile ($q_{0.25}$),
               median ($q_{0.5}$), mean ($\bar{x}$), third quartile ($q_{0.75}$),
               maximum ($q_{1}$) and standard deviation (sd) of the observed values of the Kullback-Leibler divergence for the variable \texttt{ef20} for the different
               imputation methods and missing rates (r)
               creating missing values using the MCAR-mechanism for the employee data (VSE, Kullback-Leibler, \texttt{ef20}, MCAR)}\label{Tab:vse.vals.kullbackleibler.MCAR.ef20 }\begin{minipage}[t]{\textwidth} 

\centering 
\begin{tabular}[t]{|l|c|c|c|c|c|c|c|c|}\hline
 method&r&$q_{0}$& $q_{0.25}$&$q_{0.5}$&$\bar{x}$&$q_{0.75}$& $q_{1}$&sd\\ \hline 
\hline  &0.01&0.02&0.02&0.02&0.02&0.02&0.02&0.00\\
  &0.05&0.11&0.12&0.12&0.12&0.12&0.13&0.00\\
\multirow{-5}{*}{Amelia}&0.10&0.26&0.28&0.28&0.28&0.29&0.31&0.01\\

\hline  &0.01&0.00&0.00&0.00&0.00&0.00&0.00&0.00\\
  &0.05&0.00&0.00&0.00&0.00&0.00&0.00&0.00\\
\multirow{-5}{*}{Mice.Norm}&0.10&0.00&0.00&0.00&0.00&0.00&0.00&0.00\\

\hline  &0.01&0.02&0.03&0.03&0.03&0.03&0.03&0.00\\
  &0.05&0.17&0.18&0.18&0.18&0.19&0.20&0.01\\
\multirow{-5}{*}{Mice.PMM}&0.10&0.40&0.42&0.43&0.43&0.44&0.46&0.01\\

\hline  &0.01&0.00&0.00&0.00&0.00&0.00&0.00&0.00\\
  &0.05&0.00&0.00&0.00&0.00&0.00&0.01&0.00\\
\multirow{-5}{*}{Mice.RF}&0.10&0.00&0.00&0.00&0.00&0.01&0.01&0.00\\

\hline Naive &0.01&0.00&0.00&0.00&0.00&0.00&0.00&0.00\\
  Imputation&0.05&0.06&0.06&0.06&0.06&0.07&0.07&0.00\\
 &0.10&0.19&0.20&0.21&0.21&0.21&0.23&0.01\\

\hline  &0.01&0.00&0.00&0.00&0.00&0.00&0.00&0.00\\
  &0.05&0.00&0.00&0.00&0.00&0.00&0.00&0.00\\
\multirow{-5}{*}{MissRanger}&0.10&0.00&0.00&0.00&0.00&0.00&0.03&0.00\\

\hline 
\end{tabular} 
\end{minipage} 
\end{table}

\newpage 
\begin{table}[H] 
\caption{minimum ($q_0$), first quartile ($q_{0.25}$),
               median ($q_{0.5}$), mean ($\bar{x}$), third quartile ($q_{0.75}$),
               maximum ($q_{1}$) and standard deviation (sd) of the observed values of the Kullback-Leibler divergence for the variable \texttt{ef21} for the different
               imputation methods and missing rates (r)
               creating missing values using the MAR-mechanism for the employee data (VSE, Kullback-Leibler, \texttt{ef21}, MAR)}\label{Tab:vse.vals.kullbackleibler.MAR.ef21 }\begin{minipage}[t]{\textwidth} 

\centering 
\begin{tabular}[t]{|l|c|c|c|c|c|c|c|c|}\hline
 method&r&$q_{0}$& $q_{0.25}$&$q_{0.5}$&$\bar{x}$&$q_{0.75}$& $q_{1}$&sd\\ \hline 
\hline  &0.01&0.00&0.00&0.00&0.00&0.00&0.00&0.00\\
  &0.05&0.00&0.00&0.00&0.00&0.00&0.00&0.00\\
\multirow{-5}{*}{Amelia}&0.10&0.00&0.00&0.00&0.00&0.00&0.00&0.00\\

\hline  &0.01&0.00&0.00&0.00&0.00&0.00&0.00&0.00\\
  &0.05&0.00&0.00&0.00&0.00&0.00&0.00&0.00\\
\multirow{-5}{*}{Mice.Norm}&0.10&0.00&0.00&0.00&0.00&0.00&0.00&0.00\\

\hline  &0.01&0.00&0.00&0.00&0.00&0.00&0.00&0.00\\
  &0.05&0.00&0.00&0.00&0.00&0.00&0.00&0.00\\
\multirow{-5}{*}{Mice.PMM}&0.10&0.00&0.00&0.00&0.00&0.00&0.00&0.00\\

\hline  &0.01&0.00&0.00&0.00&0.00&0.00&0.00&0.00\\
  &0.05&0.00&0.00&0.00&0.00&0.00&0.00&0.00\\
\multirow{-5}{*}{Mice.RF}&0.10&0.00&0.00&0.00&0.00&0.00&0.00&0.00\\

\hline Naive &0.01&0.00&0.00&0.00&0.00&0.00&0.00&0.00\\
  Imputation&0.05&0.00&0.00&0.00&0.00&0.00&0.00&0.00\\
 &0.10&0.00&0.00&0.00&0.00&0.00&0.00&0.00\\

\hline  &0.01&0.00&0.00&0.00&0.00&0.00&0.00&0.00\\
  &0.05&0.00&0.00&0.00&0.00&0.00&0.00&0.00\\
\multirow{-5}{*}{MissRanger}&0.10&0.00&0.00&0.00&0.00&0.00&0.00&0.00\\

\hline 
\end{tabular} 
\end{minipage} 
\end{table}

\newpage 
\begin{table}[H] 
\caption{minimum ($q_0$), first quartile ($q_{0.25}$),
               median ($q_{0.5}$), mean ($\bar{x}$), third quartile ($q_{0.75}$),
               maximum ($q_{1}$) and standard deviation (sd) of the observed values of the Kullback-Leibler divergence for the variable \texttt{ef21} for the different
               imputation methods and missing rates (r)
               creating missing values using the MCAR-mechanism for the employee data (VSE, Kullback-Leibler, \texttt{ef21}, MCAR)}\label{Tab:vse.vals.kullbackleibler.MCAR.ef21 }\begin{minipage}[t]{\textwidth} 

\centering 
\begin{tabular}[t]{|l|c|c|c|c|c|c|c|c|}\hline
 method&r&$q_{0}$& $q_{0.25}$&$q_{0.5}$&$\bar{x}$&$q_{0.75}$& $q_{1}$&sd\\ \hline 
\hline  &0.01&0.00&0.00&0.00&0.00&0.00&0.00&0.00\\
  &0.05&0.00&0.00&0.00&0.00&0.00&0.00&0.00\\
\multirow{-5}{*}{Amelia}&0.10&0.00&0.00&0.00&0.00&0.00&0.00&0.00\\

\hline  &0.01&0.00&0.00&0.00&0.00&0.00&0.00&0.00\\
  &0.05&0.00&0.00&0.00&0.00&0.00&0.00&0.00\\
\multirow{-5}{*}{Mice.Norm}&0.10&0.00&0.00&0.00&0.00&0.00&0.00&0.00\\

\hline  &0.01&0.00&0.00&0.00&0.00&0.00&0.00&0.00\\
  &0.05&0.00&0.00&0.00&0.00&0.00&0.00&0.00\\
\multirow{-5}{*}{Mice.PMM}&0.10&0.00&0.00&0.00&0.00&0.00&0.00&0.00\\

\hline  &0.01&0.00&0.00&0.00&0.00&0.00&0.00&0.00\\
  &0.05&0.00&0.00&0.00&0.00&0.00&0.00&0.00\\
\multirow{-5}{*}{Mice.RF}&0.10&0.00&0.00&0.00&0.00&0.00&0.00&0.00\\

\hline Naive &0.01&0.00&0.00&0.00&0.00&0.00&0.00&0.00\\
  Imputation&0.05&0.00&0.00&0.00&0.00&0.00&0.00&0.00\\
 &0.10&0.00&0.00&0.00&0.00&0.00&0.00&0.00\\

\hline  &0.01&0.00&0.00&0.00&0.00&0.00&0.00&0.00\\
  &0.05&0.00&0.00&0.00&0.00&0.00&0.00&0.00\\
\multirow{-5}{*}{MissRanger}&0.10&0.00&0.00&0.00&0.00&0.00&0.00&0.00\\

\hline 
\end{tabular} 
\end{minipage} 
\end{table}

\newpage 
\begin{table}[H] 
\caption{minimum ($q_0$), first quartile ($q_{0.25}$),
               median ($q_{0.5}$), mean ($\bar{x}$), third quartile ($q_{0.75}$),
               maximum ($q_{1}$) and standard deviation (sd) of the observed values of the Kullback-Leibler divergence for the variable \texttt{ef22} for the different
               imputation methods and missing rates (r)
               creating missing values using the MAR-mechanism for the employee data (VSE, Kullback-Leibler, \texttt{ef22}, MAR)}\label{Tab:vse.vals.kullbackleibler.MAR.ef22 }\begin{minipage}[t]{\textwidth} 

\centering 
\begin{tabular}[t]{|l|c|c|c|c|c|c|c|c|}\hline
 method&r&$q_{0}$& $q_{0.25}$&$q_{0.5}$&$\bar{x}$&$q_{0.75}$& $q_{1}$&sd\\ \hline 
\hline  &0.01&$\infty$&$\infty$&$\infty$&$\infty$&$\infty$&$\infty$&NaN\\
  &0.05&$\infty$&$\infty$&$\infty$&$\infty$&$\infty$&$\infty$&NaN\\
\multirow{-5}{*}{Amelia}&0.10&$\infty$&$\infty$&$\infty$&$\infty$&$\infty$&$\infty$&NaN\\

\hline  &0.01&$\infty$&$\infty$&$\infty$&$\infty$&$\infty$&$\infty$&NaN\\
  &0.05&$\infty$&$\infty$&$\infty$&$\infty$&$\infty$&$\infty$&NaN\\
\multirow{-5}{*}{Mice.Norm}&0.10&$\infty$&$\infty$&$\infty$&$\infty$&$\infty$&$\infty$&NaN\\

\hline  &0.01&$\infty$&$\infty$&$\infty$&$\infty$&$\infty$&$\infty$&NaN\\
  &0.05&$\infty$&$\infty$&$\infty$&$\infty$&$\infty$&$\infty$&NaN\\
\multirow{-5}{*}{Mice.PMM}&0.10&$\infty$&$\infty$&$\infty$&$\infty$&$\infty$&$\infty$&NaN\\

\hline  &0.01&$\infty$&$\infty$&$\infty$&$\infty$&$\infty$&$\infty$&NaN\\
  &0.05&$\infty$&$\infty$&$\infty$&$\infty$&$\infty$&$\infty$&NaN\\
\multirow{-5}{*}{Mice.RF}&0.10&$\infty$&$\infty$&$\infty$&$\infty$&$\infty$&$\infty$&NaN\\

\hline Naive &0.01&0.00&$\infty$&$\infty$&$\infty$&$\infty$&$\infty$&NaN\\
  Imputation&0.05&0.00&$\infty$&$\infty$&$\infty$&$\infty$&$\infty$&NaN\\
 &0.10&0.00&$\infty$&$\infty$&$\infty$&$\infty$&$\infty$&NaN\\

\hline  &0.01&0.00&$\infty$&$\infty$&$\infty$&$\infty$&$\infty$&NaN\\
  &0.05&0.00&$\infty$&$\infty$&$\infty$&$\infty$&$\infty$&NaN\\
\multirow{-5}{*}{MissRanger}&0.10&0.00&$\infty$&$\infty$&$\infty$&$\infty$&$\infty$&NaN\\

\hline 
\end{tabular} 
\end{minipage} 
\end{table}

\newpage 
\begin{table}[H] 
\caption{minimum ($q_0$), first quartile ($q_{0.25}$),
               median ($q_{0.5}$), mean ($\bar{x}$), third quartile ($q_{0.75}$),
               maximum ($q_{1}$) and standard deviation (sd) of the observed values of the Kullback-Leibler divergence for the variable \texttt{ef22} for the different
               imputation methods and missing rates (r)
               creating missing values using the MCAR-mechanism for the employee data (VSE, Kullback-Leibler, \texttt{ef22}, MCAR)}\label{Tab:vse.vals.kullbackleibler.MCAR.ef22 }\begin{minipage}[t]{\textwidth} 

\centering 
\begin{tabular}[t]{|l|c|c|c|c|c|c|c|c|}\hline
 method&r&$q_{0}$& $q_{0.25}$&$q_{0.5}$&$\bar{x}$&$q_{0.75}$& $q_{1}$&sd\\ \hline 
\hline  &0.01&$\infty$&$\infty$&$\infty$&$\infty$&$\infty$&$\infty$&NaN\\
  &0.05&$\infty$&$\infty$&$\infty$&$\infty$&$\infty$&$\infty$&NaN\\
\multirow{-5}{*}{Amelia}&0.10&$\infty$&$\infty$&$\infty$&$\infty$&$\infty$&$\infty$&NaN\\

\hline  &0.01&$\infty$&$\infty$&$\infty$&$\infty$&$\infty$&$\infty$&NaN\\
  &0.05&$\infty$&$\infty$&$\infty$&$\infty$&$\infty$&$\infty$&NaN\\
\multirow{-5}{*}{Mice.Norm}&0.10&$\infty$&$\infty$&$\infty$&$\infty$&$\infty$&$\infty$&NaN\\

\hline  &0.01&$\infty$&$\infty$&$\infty$&$\infty$&$\infty$&$\infty$&NaN\\
  &0.05&$\infty$&$\infty$&$\infty$&$\infty$&$\infty$&$\infty$&NaN\\
\multirow{-5}{*}{Mice.PMM}&0.10&$\infty$&$\infty$&$\infty$&$\infty$&$\infty$&$\infty$&NaN\\

\hline  &0.01&$\infty$&$\infty$&$\infty$&$\infty$&$\infty$&$\infty$&NaN\\
  &0.05&$\infty$&$\infty$&$\infty$&$\infty$&$\infty$&$\infty$&NaN\\
\multirow{-5}{*}{Mice.RF}&0.10&$\infty$&$\infty$&$\infty$&$\infty$&$\infty$&$\infty$&NaN\\

\hline Naive &0.01&0.00&$\infty$&$\infty$&$\infty$&$\infty$&$\infty$&NaN\\
  Imputation&0.05&0.00&$\infty$&$\infty$&$\infty$&$\infty$&$\infty$&NaN\\
 &0.10&0.00&$\infty$&$\infty$&$\infty$&$\infty$&$\infty$&NaN\\

\hline  &0.01&0.00&$\infty$&$\infty$&$\infty$&$\infty$&$\infty$&NaN\\
  &0.05&0.00&$\infty$&$\infty$&$\infty$&$\infty$&$\infty$&NaN\\
\multirow{-5}{*}{MissRanger}&0.10&0.00&$\infty$&$\infty$&$\infty$&$\infty$&$\infty$&NaN\\

\hline 
\end{tabular} 
\end{minipage} 
\end{table}

\newpage 
\begin{table}[H] 
\caption{minimum ($q_0$), first quartile ($q_{0.25}$),
               median ($q_{0.5}$), mean ($\bar{x}$), third quartile ($q_{0.75}$),
               maximum ($q_{1}$) and standard deviation (sd) of the observed values of the Kullback-Leibler divergence for the variable \texttt{ef23} for the different
               imputation methods and missing rates (r)
               creating missing values using the MAR-mechanism for the employee data (VSE, Kullback-Leibler, \texttt{ef23}, MAR)}\label{Tab:vse.vals.kullbackleibler.MAR.ef23 }\begin{minipage}[t]{\textwidth} 

\centering 
\begin{tabular}[t]{|l|c|c|c|c|c|c|c|c|}\hline
 method&r&$q_{0}$& $q_{0.25}$&$q_{0.5}$&$\bar{x}$&$q_{0.75}$& $q_{1}$&sd\\ \hline 
\hline  &0.01&$\infty$&$\infty$&$\infty$&$\infty$&$\infty$&$\infty$&NaN\\
  &0.05&$\infty$&$\infty$&$\infty$&$\infty$&$\infty$&$\infty$&NaN\\
\multirow{-5}{*}{Amelia}&0.10&$\infty$&$\infty$&$\infty$&$\infty$&$\infty$&$\infty$&NaN\\

\hline  &0.01&$\infty$&$\infty$&$\infty$&$\infty$&$\infty$&$\infty$&NaN\\
  &0.05&$\infty$&$\infty$&$\infty$&$\infty$&$\infty$&$\infty$&NaN\\
\multirow{-5}{*}{Mice.Norm}&0.10&$\infty$&$\infty$&$\infty$&$\infty$&$\infty$&$\infty$&NaN\\

\hline  &0.01&$\infty$&$\infty$&$\infty$&$\infty$&$\infty$&$\infty$&NaN\\
  &0.05&$\infty$&$\infty$&$\infty$&$\infty$&$\infty$&$\infty$&NaN\\
\multirow{-5}{*}{Mice.PMM}&0.10&$\infty$&$\infty$&$\infty$&$\infty$&$\infty$&$\infty$&NaN\\

\hline  &0.01&$\infty$&$\infty$&$\infty$&$\infty$&$\infty$&$\infty$&NaN\\
  &0.05&$\infty$&$\infty$&$\infty$&$\infty$&$\infty$&$\infty$&NaN\\
\multirow{-5}{*}{Mice.RF}&0.10&$\infty$&$\infty$&$\infty$&$\infty$&$\infty$&$\infty$&NaN\\

\hline Naive &0.01&0.00&$\infty$&$\infty$&$\infty$&$\infty$&$\infty$&NaN\\
  Imputation&0.05&$\infty$&$\infty$&$\infty$&$\infty$&$\infty$&$\infty$&NaN\\
 &0.10&$\infty$&$\infty$&$\infty$&$\infty$&$\infty$&$\infty$&NaN\\

\hline  &0.01&0.00&$\infty$&$\infty$&$\infty$&$\infty$&$\infty$&NaN\\
  &0.05&0.00&$\infty$&$\infty$&$\infty$&$\infty$&$\infty$&NaN\\
\multirow{-5}{*}{MissRanger}&0.10&$\infty$&$\infty$&$\infty$&$\infty$&$\infty$&$\infty$&NaN\\

\hline 
\end{tabular} 
\end{minipage} 
\end{table}

\newpage 
\begin{table}[H] 
\caption{minimum ($q_0$), first quartile ($q_{0.25}$),
               median ($q_{0.5}$), mean ($\bar{x}$), third quartile ($q_{0.75}$),
               maximum ($q_{1}$) and standard deviation (sd) of the observed values of the Kullback-Leibler divergence for the variable \texttt{ef23} for the different
               imputation methods and missing rates (r)
               creating missing values using the MCAR-mechanism for the employee data (VSE, Kullback-Leibler, \texttt{ef23}, MCAR)}\label{Tab:vse.vals.kullbackleibler.MCAR.ef23 }\begin{minipage}[t]{\textwidth} 

\centering 
\begin{tabular}[t]{|l|c|c|c|c|c|c|c|c|}\hline
 method&r&$q_{0}$& $q_{0.25}$&$q_{0.5}$&$\bar{x}$&$q_{0.75}$& $q_{1}$&sd\\ \hline 
\hline  &0.01&$\infty$&$\infty$&$\infty$&$\infty$&$\infty$&$\infty$&NaN\\
  &0.05&$\infty$&$\infty$&$\infty$&$\infty$&$\infty$&$\infty$&NaN\\
\multirow{-5}{*}{Amelia}&0.10&$\infty$&$\infty$&$\infty$&$\infty$&$\infty$&$\infty$&NaN\\

\hline  &0.01&$\infty$&$\infty$&$\infty$&$\infty$&$\infty$&$\infty$&NaN\\
  &0.05&$\infty$&$\infty$&$\infty$&$\infty$&$\infty$&$\infty$&NaN\\
\multirow{-5}{*}{Mice.Norm}&0.10&$\infty$&$\infty$&$\infty$&$\infty$&$\infty$&$\infty$&NaN\\

\hline  &0.01&$\infty$&$\infty$&$\infty$&$\infty$&$\infty$&$\infty$&NaN\\
  &0.05&$\infty$&$\infty$&$\infty$&$\infty$&$\infty$&$\infty$&NaN\\
\multirow{-5}{*}{Mice.PMM}&0.10&$\infty$&$\infty$&$\infty$&$\infty$&$\infty$&$\infty$&NaN\\

\hline  &0.01&$\infty$&$\infty$&$\infty$&$\infty$&$\infty$&$\infty$&NaN\\
  &0.05&$\infty$&$\infty$&$\infty$&$\infty$&$\infty$&$\infty$&NaN\\
\multirow{-5}{*}{Mice.RF}&0.10&$\infty$&$\infty$&$\infty$&$\infty$&$\infty$&$\infty$&NaN\\

\hline Naive &0.01&$\infty$&$\infty$&$\infty$&$\infty$&$\infty$&$\infty$&NaN\\
  Imputation&0.05&0.00&$\infty$&$\infty$&$\infty$&$\infty$&$\infty$&NaN\\
 &0.10&$\infty$&$\infty$&$\infty$&$\infty$&$\infty$&$\infty$&NaN\\

\hline  &0.01&$\infty$&$\infty$&$\infty$&$\infty$&$\infty$&$\infty$&NaN\\
  &0.05&$\infty$&$\infty$&$\infty$&$\infty$&$\infty$&$\infty$&NaN\\
\multirow{-5}{*}{MissRanger}&0.10&$\infty$&$\infty$&$\infty$&$\infty$&$\infty$&$\infty$&NaN\\

\hline 
\end{tabular} 
\end{minipage} 
\end{table}

\newpage 
\begin{table}[H] 
\caption{minimum ($q_0$), first quartile ($q_{0.25}$),
               median ($q_{0.5}$), mean ($\bar{x}$), third quartile ($q_{0.75}$),
               maximum ($q_{1}$) and standard deviation (sd) of the observed values of the Kullback-Leibler divergence for the variable \texttt{ef24} for the different
               imputation methods and missing rates (r)
               creating missing values using the MAR-mechanism for the employee data (VSE, Kullback-Leibler, \texttt{ef24}, MAR)}\label{Tab:vse.vals.kullbackleibler.MAR.ef24 }\begin{minipage}[t]{\textwidth} 

\centering 
\begin{tabular}[t]{|l|c|c|c|c|c|c|c|c|}\hline
 method&r&$q_{0}$& $q_{0.25}$&$q_{0.5}$&$\bar{x}$&$q_{0.75}$& $q_{1}$&sd\\ \hline 
\hline  &0.01&0.00&0.00&0.00&0.00&0.00&0.00&0.00\\
  &0.05&0.00&0.00&0.00&0.00&0.00&0.00&0.00\\
\multirow{-5}{*}{Amelia}&0.10&0.00&0.00&0.00&$\infty$&0.00&$\infty$&NaN\\

\hline  &0.01&0.00&0.00&0.00&0.00&0.00&0.00&0.00\\
  &0.05&0.00&0.00&0.00&0.00&0.00&0.00&0.00\\
\multirow{-5}{*}{Mice.Norm}&0.10&0.00&0.00&0.00&$\infty$&0.00&$\infty$&NaN\\

\hline  &0.01&0.00&0.00&0.00&0.00&0.00&0.00&0.00\\
  &0.05&0.00&0.00&0.00&0.00&0.00&0.00&0.00\\
\multirow{-5}{*}{Mice.PMM}&0.10&0.00&0.00&0.00&$\infty$&0.00&$\infty$&NaN\\

\hline  &0.01&0.00&0.00&0.00&0.00&0.00&0.00&0.00\\
  &0.05&0.00&0.00&0.00&0.00&0.00&0.00&0.00\\
\multirow{-5}{*}{Mice.RF}&0.10&0.00&0.00&0.00&$\infty$&0.00&$\infty$&NaN\\

\hline Naive &0.01&0.00&0.00&0.00&0.00&0.00&0.00&0.00\\
  Imputation&0.05&0.00&0.00&0.00&0.00&0.00&0.00&0.00\\
 &0.10&0.00&0.00&0.00&$\infty$&0.00&$\infty$&NaN\\

\hline  &0.01&0.00&0.00&0.00&0.00&0.00&0.00&0.00\\
  &0.05&0.00&0.00&0.00&0.00&0.00&0.00&0.00\\
\multirow{-5}{*}{MissRanger}&0.10&0.00&0.00&0.00&$\infty$&0.00&$\infty$&NaN\\

\hline 
\end{tabular} 
\end{minipage} 
\end{table}

\newpage 
\begin{table}[H] 
\caption{minimum ($q_0$), first quartile ($q_{0.25}$),
               median ($q_{0.5}$), mean ($\bar{x}$), third quartile ($q_{0.75}$),
               maximum ($q_{1}$) and standard deviation (sd) of the observed values of the Kullback-Leibler divergence for the variable \texttt{ef24} for the different
               imputation methods and missing rates (r)
               creating missing values using the MCAR-mechanism for the employee data (VSE, Kullback-Leibler, \texttt{ef24}, MCAR)}\label{Tab:vse.vals.kullbackleibler.MCAR.ef24 }\begin{minipage}[t]{\textwidth} 

\centering 
\begin{tabular}[t]{|l|c|c|c|c|c|c|c|c|}\hline
 method&r&$q_{0}$& $q_{0.25}$&$q_{0.5}$&$\bar{x}$&$q_{0.75}$& $q_{1}$&sd\\ \hline 
\hline  &0.01&0.00&0.00&0.00&0.00&0.00&0.00&0.00\\
  &0.05&0.00&0.00&0.00&0.00&0.00&0.00&0.00\\
\multirow{-5}{*}{Amelia}&0.10&0.00&0.00&0.00&$\infty$&0.00&$\infty$&NaN\\

\hline  &0.01&0.00&0.00&0.00&0.00&0.00&0.00&0.00\\
  &0.05&0.00&0.00&0.00&0.00&0.00&0.00&0.00\\
\multirow{-5}{*}{Mice.Norm}&0.10&0.00&0.00&0.00&$\infty$&0.00&$\infty$&NaN\\

\hline  &0.01&0.00&0.00&0.00&0.00&0.00&0.00&0.00\\
  &0.05&0.00&0.00&0.00&0.00&0.00&0.00&0.00\\
\multirow{-5}{*}{Mice.PMM}&0.10&0.00&0.00&0.00&$\infty$&0.00&$\infty$&NaN\\

\hline  &0.01&0.00&0.00&0.00&0.00&0.00&0.00&0.00\\
  &0.05&0.00&0.00&0.00&0.00&0.00&0.00&0.00\\
\multirow{-5}{*}{Mice.RF}&0.10&0.00&0.00&0.00&$\infty$&0.00&$\infty$&NaN\\

\hline Naive &0.01&0.00&0.00&0.00&0.00&0.00&0.00&0.00\\
  Imputation&0.05&0.00&0.00&0.00&0.00&0.00&0.00&0.00\\
 &0.10&0.00&0.00&0.00&$\infty$&0.00&$\infty$&NaN\\

\hline  &0.01&0.00&0.00&0.00&0.00&0.00&0.00&0.00\\
  &0.05&0.00&0.00&0.00&0.00&0.00&0.00&0.00\\
\multirow{-5}{*}{MissRanger}&0.10&0.00&0.00&0.00&$\infty$&0.00&$\infty$&NaN\\

\hline 
\end{tabular} 
\end{minipage} 
\end{table}

\newpage 
\begin{table}[H] 
\caption{minimum ($q_0$), first quartile ($q_{0.25}$),
               median ($q_{0.5}$), mean ($\bar{x}$), third quartile ($q_{0.75}$),
               maximum ($q_{1}$) and standard deviation (sd) of the observed values of the Kullback-Leibler divergence for the variable \texttt{ef25} for the different
               imputation methods and missing rates (r)
               creating missing values using the MAR-mechanism for the employee data (VSE, Kullback-Leibler, \texttt{ef25}, MAR)}\label{Tab:vse.vals.kullbackleibler.MAR.ef25 }\begin{minipage}[t]{\textwidth} 

\centering 
\begin{tabular}[t]{|l|c|c|c|c|c|c|c|c|}\hline
 method&r&$q_{0}$& $q_{0.25}$&$q_{0.5}$&$\bar{x}$&$q_{0.75}$& $q_{1}$&sd\\ \hline 
\hline  &0.01&0.00&0.00&0.00&0.00&0.00&0.00&0.00\\
  &0.05&0.00&0.00&0.00&0.00&0.00&0.00&0.00\\
\multirow{-5}{*}{Amelia}&0.10&0.00&0.00&0.00&0.00&0.00&0.00&0.00\\

\hline  &0.01&0.00&0.00&0.00&0.00&0.00&0.00&0.00\\
  &0.05&0.00&0.00&0.00&0.00&0.00&0.00&0.00\\
\multirow{-5}{*}{Mice.Norm}&0.10&0.00&0.00&0.00&0.00&0.00&0.00&0.00\\

\hline  &0.01&0.00&0.00&0.00&0.00&0.00&0.00&0.00\\
  &0.05&0.00&0.00&0.00&0.00&0.00&0.00&0.00\\
\multirow{-5}{*}{Mice.PMM}&0.10&0.00&0.00&0.00&0.00&0.00&0.00&0.00\\

\hline  &0.01&0.00&0.00&0.00&0.00&0.00&0.00&0.00\\
  &0.05&0.00&0.00&0.00&0.00&0.00&0.00&0.00\\
\multirow{-5}{*}{Mice.RF}&0.10&0.00&0.00&0.00&0.00&0.00&0.00&0.00\\

\hline Naive &0.01&0.00&0.00&0.00&0.00&0.00&0.00&0.00\\
  Imputation&0.05&0.00&0.00&0.00&0.00&0.00&0.00&0.00\\
 &0.10&0.01&0.01&0.01&0.01&0.01&0.01&0.00\\

\hline  &0.01&0.00&0.00&0.00&0.00&0.00&0.00&0.00\\
  &0.05&0.00&0.00&0.00&0.00&0.00&0.00&0.00\\
\multirow{-5}{*}{MissRanger}&0.10&0.00&0.00&0.00&0.00&0.00&0.00&0.00\\

\hline 
\end{tabular} 
\end{minipage} 
\end{table}

\newpage 
\begin{table}[H] 
\caption{minimum ($q_0$), first quartile ($q_{0.25}$),
               median ($q_{0.5}$), mean ($\bar{x}$), third quartile ($q_{0.75}$),
               maximum ($q_{1}$) and standard deviation (sd) of the observed values of the Kullback-Leibler divergence for the variable \texttt{ef25} for the different
               imputation methods and missing rates (r)
               creating missing values using the MCAR-mechanism for the employee data (VSE, Kullback-Leibler, \texttt{ef25}, MCAR)}\label{Tab:vse.vals.kullbackleibler.MCAR.ef25 }\begin{minipage}[t]{\textwidth} 

\centering 
\begin{tabular}[t]{|l|c|c|c|c|c|c|c|c|}\hline
 method&r&$q_{0}$& $q_{0.25}$&$q_{0.5}$&$\bar{x}$&$q_{0.75}$& $q_{1}$&sd\\ \hline 
\hline  &0.01&0.00&0.00&0.00&0.00&0.00&0.00&0.00\\
  &0.05&0.00&0.00&0.00&0.00&0.00&0.00&0.00\\
\multirow{-5}{*}{Amelia}&0.10&0.00&0.00&0.00&0.00&0.00&0.00&0.00\\

\hline  &0.01&0.00&0.00&0.00&0.00&0.00&0.00&0.00\\
  &0.05&0.00&0.00&0.00&0.00&0.00&0.00&0.00\\
\multirow{-5}{*}{Mice.Norm}&0.10&0.00&0.00&0.00&0.00&0.00&0.00&0.00\\

\hline  &0.01&0.00&0.00&0.00&0.00&0.00&0.00&0.00\\
  &0.05&0.00&0.00&0.00&0.00&0.00&0.00&0.00\\
\multirow{-5}{*}{Mice.PMM}&0.10&0.00&0.00&0.00&0.00&0.00&0.00&0.00\\

\hline  &0.01&0.00&0.00&0.00&0.00&0.00&0.00&0.00\\
  &0.05&0.00&0.00&0.00&0.00&0.00&0.00&0.00\\
\multirow{-5}{*}{Mice.RF}&0.10&0.00&0.00&0.00&0.00&0.00&0.00&0.00\\

\hline Naive &0.01&0.00&0.00&0.00&0.00&0.00&0.00&0.00\\
  Imputation&0.05&0.00&0.00&0.00&0.00&0.00&0.00&0.00\\
 &0.10&0.01&0.01&0.01&0.01&0.01&0.01&0.00\\

\hline  &0.01&0.00&0.00&0.00&0.00&0.00&0.00&0.00\\
  &0.05&0.00&0.00&0.00&0.00&0.00&0.00&0.00\\
\multirow{-5}{*}{MissRanger}&0.10&0.00&0.00&0.00&0.00&0.00&0.00&0.00\\

\hline 
\end{tabular} 
\end{minipage} 
\end{table}

\newpage 
\begin{table}[H] 
\caption{minimum ($q_0$), first quartile ($q_{0.25}$),
               median ($q_{0.5}$), mean ($\bar{x}$), third quartile ($q_{0.75}$),
               maximum ($q_{1}$) and standard deviation (sd) of the observed values of the Kullback-Leibler divergence for the variable \texttt{ef26} for the different
               imputation methods and missing rates (r)
               creating missing values using the MAR-mechanism for the employee data (VSE, Kullback-Leibler, \texttt{ef26}, MAR)}\label{Tab:vse.vals.kullbackleibler.MAR.ef26 }\begin{minipage}[t]{\textwidth} 

\centering 
\begin{tabular}[t]{|l|c|c|c|c|c|c|c|c|}\hline
 method&r&$q_{0}$& $q_{0.25}$&$q_{0.5}$&$\bar{x}$&$q_{0.75}$& $q_{1}$&sd\\ \hline 
\hline  &0.01&0.00&0.00&0.00&0.00&0.00&0.00&0.00\\
  &0.05&0.02&0.02&0.02&0.02&0.02&0.02&0.00\\
\multirow{-5}{*}{Amelia}&0.10&0.04&0.04&0.04&0.04&0.04&0.05&0.00\\

\hline  &0.01&0.00&0.00&0.00&0.00&0.00&0.00&0.00\\
  &0.05&0.00&0.00&0.00&0.00&0.00&0.00&0.00\\
\multirow{-5}{*}{Mice.Norm}&0.10&0.00&0.00&0.00&0.00&0.00&0.00&0.00\\

\hline  &0.01&0.00&0.00&0.00&0.00&0.00&0.00&0.00\\
  &0.05&0.02&0.02&0.02&0.02&0.02&0.02&0.00\\
\multirow{-5}{*}{Mice.PMM}&0.10&0.04&0.04&0.04&0.04&0.04&0.05&0.00\\

\hline  &0.01&0.00&0.00&0.00&0.00&0.00&0.00&0.00\\
  &0.05&0.00&0.00&0.00&0.00&0.00&0.00&0.00\\
\multirow{-5}{*}{Mice.RF}&0.10&0.00&0.00&0.00&0.00&0.00&0.00&0.00\\

\hline Naive &0.01&0.00&0.00&0.00&0.00&0.00&0.00&0.00\\
  Imputation&0.05&0.02&0.02&0.02&0.02&0.02&0.02&0.00\\
 &0.10&$\infty$&$\infty$&$\infty$&$\infty$&$\infty$&$\infty$&NaN\\

\hline  &0.01&0.00&0.00&0.00&0.00&0.00&0.00&0.00\\
  &0.05&0.00&0.00&0.00&0.00&0.00&0.01&0.00\\
\multirow{-5}{*}{MissRanger}&0.10&$\infty$&$\infty$&$\infty$&$\infty$&$\infty$&$\infty$&NaN\\

\hline 
\end{tabular} 
\end{minipage} 
\end{table}

\newpage 
\begin{table}[H] 
\caption{minimum ($q_0$), first quartile ($q_{0.25}$),
               median ($q_{0.5}$), mean ($\bar{x}$), third quartile ($q_{0.75}$),
               maximum ($q_{1}$) and standard deviation (sd) of the observed values of the Kullback-Leibler divergence for the variable \texttt{ef26} for the different
               imputation methods and missing rates (r)
               creating missing values using the MCAR-mechanism for the employee data (VSE, Kullback-Leibler, \texttt{ef26}, MCAR)}\label{Tab:vse.vals.kullbackleibler.MCAR.ef26 }\begin{minipage}[t]{\textwidth} 

\centering 
\begin{tabular}[t]{|l|c|c|c|c|c|c|c|c|}\hline
 method&r&$q_{0}$& $q_{0.25}$&$q_{0.5}$&$\bar{x}$&$q_{0.75}$& $q_{1}$&sd\\ \hline 
\hline  &0.01&0.00&0.00&0.00&0.00&0.00&0.00&0.00\\
  &0.05&0.02&0.02&0.02&0.02&0.02&0.02&0.00\\
\multirow{-5}{*}{Amelia}&0.10&0.04&0.04&0.04&0.04&0.04&0.04&0.00\\

\hline  &0.01&0.00&0.00&0.00&0.00&0.00&0.00&0.00\\
  &0.05&0.00&0.00&0.00&0.00&0.00&0.00&0.00\\
\multirow{-5}{*}{Mice.Norm}&0.10&0.00&0.00&0.00&0.00&0.00&0.00&0.00\\

\hline  &0.01&0.00&0.00&0.00&0.00&0.00&0.00&0.00\\
  &0.05&0.02&0.02&0.02&0.02&0.02&0.02&0.00\\
\multirow{-5}{*}{Mice.PMM}&0.10&0.04&0.04&0.04&0.04&0.04&0.05&0.00\\

\hline  &0.01&0.00&0.00&0.00&0.00&0.00&0.00&0.00\\
  &0.05&0.00&0.00&0.00&0.00&0.00&0.00&0.00\\
\multirow{-5}{*}{Mice.RF}&0.10&0.00&0.00&0.00&0.00&0.00&0.00&0.00\\

\hline Naive &0.01&0.00&0.00&0.00&0.00&0.00&0.00&0.00\\
  Imputation&0.05&0.02&0.02&0.02&0.02&0.02&0.02&0.00\\
 &0.10&$\infty$&$\infty$&$\infty$&$\infty$&$\infty$&$\infty$&NaN\\

\hline  &0.01&0.00&0.00&0.00&0.00&0.00&0.00&0.00\\
  &0.05&0.00&0.00&0.00&0.00&0.00&0.01&0.00\\
\multirow{-5}{*}{MissRanger}&0.10&$\infty$&$\infty$&$\infty$&$\infty$&$\infty$&$\infty$&NaN\\

\hline 
\end{tabular} 
\end{minipage} 
\end{table}

\newpage 
\begin{table}[H] 
\caption{minimum ($q_0$), first quartile ($q_{0.25}$),
               median ($q_{0.5}$), mean ($\bar{x}$), third quartile ($q_{0.75}$),
               maximum ($q_{1}$) and standard deviation (sd) of the observed values of the Kullback-Leibler divergence for the variable \texttt{ef29} for the different
               imputation methods and missing rates (r)
               creating missing values using the MAR-mechanism for the employee data (VSE, Kullback-Leibler, \texttt{ef29}, MAR)}\label{Tab:vse.vals.kullbackleibler.MAR.ef29 }\begin{minipage}[t]{\textwidth} 

\centering 
\begin{tabular}[t]{|l|c|c|c|c|c|c|c|c|}\hline
 method&r&$q_{0}$& $q_{0.25}$&$q_{0.5}$&$\bar{x}$&$q_{0.75}$& $q_{1}$&sd\\ \hline 
\hline  &0.01&0.00&0.00&0.00&0.00&0.00&0.00&0.00\\
  &0.05&0.01&0.01&0.02&0.02&0.02&0.02&0.00\\
\multirow{-5}{*}{Amelia}&0.10&0.04&0.05&0.05&0.05&0.05&0.06&0.00\\

\hline  &0.01&0.00&0.00&0.00&0.00&0.00&0.00&0.00\\
  &0.05&0.00&0.00&0.00&0.00&0.00&0.00&0.00\\
\multirow{-5}{*}{Mice.Norm}&0.10&0.00&0.00&0.00&0.00&0.00&0.00&0.00\\

\hline  &0.01&0.00&0.00&0.00&0.00&0.00&0.00&0.00\\
  &0.05&0.01&0.02&0.02&0.02&0.02&0.02&0.00\\
\multirow{-5}{*}{Mice.PMM}&0.10&0.05&0.05&0.05&0.05&0.05&0.06&0.00\\

\hline  &0.01&0.00&0.00&0.00&0.00&0.00&0.00&0.00\\
  &0.05&0.00&0.00&0.00&0.00&0.00&0.00&0.00\\
\multirow{-5}{*}{Mice.RF}&0.10&0.00&0.00&0.00&0.00&0.00&0.00&0.00\\

\hline Naive &0.01&0.00&0.00&0.00&0.00&0.00&0.01&0.00\\
  Imputation&0.05&0.08&0.09&0.10&0.10&0.10&0.11&0.01\\
 &0.10&0.31&0.37&0.38&0.38&0.39&0.41&0.01\\

\hline  &0.01&0.00&0.00&0.00&0.00&0.00&0.00&0.00\\
  &0.05&0.00&0.00&0.00&0.00&0.00&0.03&0.01\\
\multirow{-5}{*}{MissRanger}&0.10&0.00&0.01&0.01&0.01&0.01&0.09&0.02\\

\hline 
\end{tabular} 
\end{minipage} 
\end{table}

\newpage 
\begin{table}[H] 
\caption{minimum ($q_0$), first quartile ($q_{0.25}$),
               median ($q_{0.5}$), mean ($\bar{x}$), third quartile ($q_{0.75}$),
               maximum ($q_{1}$) and standard deviation (sd) of the observed values of the Kullback-Leibler divergence for the variable \texttt{ef29} for the different
               imputation methods and missing rates (r)
               creating missing values using the MCAR-mechanism for the employee data (VSE, Kullback-Leibler, \texttt{ef29}, MCAR)}\label{Tab:vse.vals.kullbackleibler.MCAR.ef29 }\begin{minipage}[t]{\textwidth} 

\centering 
\begin{tabular}[t]{|l|c|c|c|c|c|c|c|c|}\hline
 method&r&$q_{0}$& $q_{0.25}$&$q_{0.5}$&$\bar{x}$&$q_{0.75}$& $q_{1}$&sd\\ \hline 
\hline  &0.01&0.00&0.00&0.00&0.00&0.00&0.00&0.00\\
  &0.05&0.01&0.01&0.02&0.02&0.02&0.02&0.00\\
\multirow{-5}{*}{Amelia}&0.10&0.04&0.05&0.05&0.05&0.05&0.06&0.00\\

\hline  &0.01&0.00&0.00&0.00&0.00&0.00&0.00&0.00\\
  &0.05&0.00&0.00&0.00&0.00&0.00&0.00&0.00\\
\multirow{-5}{*}{Mice.Norm}&0.10&0.00&0.00&0.00&0.00&0.00&0.00&0.00\\

\hline  &0.01&0.00&0.00&0.00&0.00&0.00&0.00&0.00\\
  &0.05&0.01&0.02&0.02&0.02&0.02&0.02&0.00\\
\multirow{-5}{*}{Mice.PMM}&0.10&0.04&0.05&0.05&0.05&0.05&0.06&0.00\\

\hline  &0.01&0.00&0.00&0.00&0.00&0.00&0.00&0.00\\
  &0.05&0.00&0.00&0.00&0.00&0.00&0.00&0.00\\
\multirow{-5}{*}{Mice.RF}&0.10&0.00&0.00&0.00&0.00&0.00&0.00&0.00\\

\hline Naive &0.01&0.00&0.00&0.00&0.00&0.00&0.01&0.00\\
  Imputation&0.05&0.08&0.09&0.10&0.10&0.10&0.11&0.00\\
 &0.10&0.29&0.37&0.38&0.38&0.39&0.41&0.02\\

\hline  &0.01&0.00&0.00&0.00&0.00&0.00&0.00&0.00\\
  &0.05&0.00&0.00&0.00&0.00&0.00&0.03&0.00\\
\multirow{-5}{*}{MissRanger}&0.10&0.00&0.01&0.01&0.01&0.01&0.08&0.02\\

\hline 
\end{tabular} 
\end{minipage} 
\end{table}

\newpage 
\begin{table}[H] 
\caption{minimum ($q_0$), first quartile ($q_{0.25}$),
               median ($q_{0.5}$), mean ($\bar{x}$), third quartile ($q_{0.75}$),
               maximum ($q_{1}$) and standard deviation (sd) of the observed values of the Kullback-Leibler divergence for the variable \texttt{ef40} for the different
               imputation methods and missing rates (r)
               creating missing values using the MAR-mechanism for the employee data (VSE, Kullback-Leibler, \texttt{ef40}, MAR)}\label{Tab:vse.vals.kullbackleibler.MAR.ef40 }\begin{minipage}[t]{\textwidth} 

\centering 
\begin{tabular}[t]{|l|c|c|c|c|c|c|c|c|}\hline
 method&r&$q_{0}$& $q_{0.25}$&$q_{0.5}$&$\bar{x}$&$q_{0.75}$& $q_{1}$&sd\\ \hline 
\hline  &0.01&0.01&0.01&0.01&0.01&0.01&0.01&0.00\\
  &0.05&0.04&0.05&0.05&0.05&0.05&0.06&0.00\\
\multirow{-5}{*}{Amelia}&0.10&0.09&0.10&0.11&0.11&0.11&0.11&0.00\\

\hline  &0.01&0.00&0.00&0.00&0.00&0.00&0.00&0.00\\
  &0.05&0.00&0.00&0.00&0.00&0.00&0.00&0.00\\
\multirow{-5}{*}{Mice.Norm}&0.10&0.00&0.00&0.00&0.00&0.00&0.00&0.00\\

\hline  &0.01&0.01&0.01&0.01&0.01&0.01&0.01&0.00\\
  &0.05&0.04&0.05&0.05&0.05&0.05&0.06&0.00\\
\multirow{-5}{*}{Mice.PMM}&0.10&0.10&0.11&0.11&0.11&0.12&0.12&0.00\\

\hline  &0.01&0.00&0.00&0.00&0.00&0.00&0.00&0.00\\
  &0.05&0.00&0.00&0.00&0.00&0.00&0.00&0.00\\
\multirow{-5}{*}{Mice.RF}&0.10&0.00&0.00&0.00&0.00&0.00&0.00&0.00\\

\hline Naive &0.01&0.00&0.00&0.00&0.00&0.00&0.00&0.00\\
  Imputation&0.05&0.07&0.07&0.08&0.08&0.08&0.09&0.01\\
 &0.10&0.25&0.27&0.27&0.28&0.29&0.32&0.02\\

\hline  &0.01&0.00&0.00&0.00&0.00&0.00&0.00&0.00\\
  &0.05&0.00&0.00&0.00&0.00&0.00&0.01&0.00\\
\multirow{-5}{*}{MissRanger}&0.10&0.01&0.01&0.01&0.01&0.01&0.06&0.01\\

\hline 
\end{tabular} 
\end{minipage} 
\end{table}

\newpage 
\begin{table}[H] 
\caption{minimum ($q_0$), first quartile ($q_{0.25}$),
               median ($q_{0.5}$), mean ($\bar{x}$), third quartile ($q_{0.75}$),
               maximum ($q_{1}$) and standard deviation (sd) of the observed values of the Kullback-Leibler divergence for the variable \texttt{ef40} for the different
               imputation methods and missing rates (r)
               creating missing values using the MCAR-mechanism for the employee data (VSE, Kullback-Leibler, \texttt{ef40}, MCAR)}\label{Tab:vse.vals.kullbackleibler.MCAR.ef40 }\begin{minipage}[t]{\textwidth} 

\centering 
\begin{tabular}[t]{|l|c|c|c|c|c|c|c|c|}\hline
 method&r&$q_{0}$& $q_{0.25}$&$q_{0.5}$&$\bar{x}$&$q_{0.75}$& $q_{1}$&sd\\ \hline 
\hline  &0.01&0.01&0.01&0.01&0.01&0.01&0.01&0.00\\
  &0.05&0.04&0.05&0.05&0.05&0.05&0.06&0.00\\
\multirow{-5}{*}{Amelia}&0.10&0.09&0.10&0.11&0.11&0.11&0.11&0.00\\

\hline  &0.01&0.00&0.00&0.00&0.00&0.00&0.00&0.00\\
  &0.05&0.00&0.00&0.00&0.00&0.00&0.00&0.00\\
\multirow{-5}{*}{Mice.Norm}&0.10&0.00&0.00&0.00&0.00&0.00&0.00&0.00\\

\hline  &0.01&0.01&0.01&0.01&0.01&0.01&0.01&0.00\\
  &0.05&0.04&0.05&0.05&0.05&0.05&0.06&0.00\\
\multirow{-5}{*}{Mice.PMM}&0.10&0.10&0.11&0.11&0.11&0.12&0.13&0.00\\

\hline  &0.01&0.00&0.00&0.00&0.00&0.00&0.00&0.00\\
  &0.05&0.00&0.00&0.00&0.00&0.00&0.00&0.00\\
\multirow{-5}{*}{Mice.RF}&0.10&0.00&0.00&0.00&0.00&0.00&0.00&0.00\\

\hline Naive &0.01&0.00&0.00&0.00&0.00&0.00&0.00&0.00\\
  Imputation&0.05&0.07&0.07&0.08&0.08&0.08&0.09&0.01\\
 &0.10&0.24&0.27&0.27&0.28&0.28&0.32&0.02\\

\hline  &0.01&0.00&0.00&0.00&0.00&0.00&0.00&0.00\\
  &0.05&0.00&0.00&0.00&0.00&0.00&0.01&0.00\\
\multirow{-5}{*}{MissRanger}&0.10&0.01&0.01&0.01&0.01&0.01&0.06&0.01\\

\hline 
\end{tabular} 
\end{minipage} 
\end{table}

\newpage 
\begin{table}[H] 
\caption{minimum ($q_0$), first quartile ($q_{0.25}$),
               median ($q_{0.5}$), mean ($\bar{x}$), third quartile ($q_{0.75}$),
               maximum ($q_{1}$) and standard deviation (sd) of the observed values of the Kullback-Leibler divergence for the variable \texttt{ef41} for the different
               imputation methods and missing rates (r)
               creating missing values using the MAR-mechanism for the employee data (VSE, Kullback-Leibler, \texttt{ef41}, MAR)}\label{Tab:vse.vals.kullbackleibler.MAR.ef41 }\begin{minipage}[t]{\textwidth} 

\centering 
\begin{tabular}[t]{|l|c|c|c|c|c|c|c|c|}\hline
 method&r&$q_{0}$& $q_{0.25}$&$q_{0.5}$&$\bar{x}$&$q_{0.75}$& $q_{1}$&sd\\ \hline 
\hline  &0.01&0.00&0.00&0.00&0.00&0.00&0.00&0.00\\
  &0.05&0.01&0.01&0.01&0.01&0.01&0.01&0.00\\
\multirow{-5}{*}{Amelia}&0.10&0.02&0.02&0.02&0.02&0.02&0.03&0.00\\

\hline  &0.01&0.00&0.00&0.00&0.00&0.00&0.00&0.00\\
  &0.05&0.00&0.00&0.00&0.00&0.00&0.00&0.00\\
\multirow{-5}{*}{Mice.Norm}&0.10&0.00&0.00&0.00&0.00&0.00&0.00&0.00\\

\hline  &0.01&0.00&0.00&0.00&0.00&0.00&0.00&0.00\\
  &0.05&0.01&0.01&0.01&0.01&0.01&0.01&0.00\\
\multirow{-5}{*}{Mice.PMM}&0.10&0.02&0.03&0.03&0.03&0.03&0.03&0.00\\

\hline  &0.01&0.00&0.00&0.00&0.00&0.00&0.00&0.00\\
  &0.05&0.00&0.00&0.00&0.00&0.00&0.00&0.00\\
\multirow{-5}{*}{Mice.RF}&0.10&0.00&0.00&0.00&0.00&0.00&0.00&0.00\\

\hline Naive &0.01&0.00&0.00&0.00&0.00&0.00&0.00&0.00\\
  Imputation&0.05&0.05&0.06&0.06&0.06&0.06&0.07&0.00\\
 &0.10&0.20&0.20&0.21&0.21&0.21&0.23&0.01\\

\hline  &0.01&0.00&0.00&0.00&0.00&0.00&0.00&0.00\\
  &0.05&0.00&0.00&0.00&0.01&0.01&0.02&0.00\\
\multirow{-5}{*}{MissRanger}&0.10&0.01&0.02&0.02&0.02&0.02&0.08&0.01\\

\hline 
\end{tabular} 
\end{minipage} 
\end{table}

\newpage 
\begin{table}[H] 
\caption{minimum ($q_0$), first quartile ($q_{0.25}$),
               median ($q_{0.5}$), mean ($\bar{x}$), third quartile ($q_{0.75}$),
               maximum ($q_{1}$) and standard deviation (sd) of the observed values of the Kullback-Leibler divergence for the variable \texttt{ef41} for the different
               imputation methods and missing rates (r)
               creating missing values using the MCAR-mechanism for the employee data (VSE, Kullback-Leibler, \texttt{ef41}, MCAR)}\label{Tab:vse.vals.kullbackleibler.MCAR.ef41 }\begin{minipage}[t]{\textwidth} 

\centering 
\begin{tabular}[t]{|l|c|c|c|c|c|c|c|c|}\hline
 method&r&$q_{0}$& $q_{0.25}$&$q_{0.5}$&$\bar{x}$&$q_{0.75}$& $q_{1}$&sd\\ \hline 
\hline  &0.01&0.00&0.00&0.00&0.00&0.00&0.00&0.00\\
  &0.05&0.01&0.01&0.01&0.01&0.01&0.01&0.00\\
\multirow{-5}{*}{Amelia}&0.10&0.02&0.02&0.02&0.02&0.03&0.03&0.00\\

\hline  &0.01&0.00&0.00&0.00&0.00&0.00&0.00&0.00\\
  &0.05&0.00&0.00&0.00&0.00&0.00&0.00&0.00\\
\multirow{-5}{*}{Mice.Norm}&0.10&0.00&0.00&0.00&0.00&0.00&0.00&0.00\\

\hline  &0.01&0.00&0.00&0.00&0.00&0.00&0.00&0.00\\
  &0.05&0.01&0.01&0.01&0.01&0.01&0.02&0.00\\
\multirow{-5}{*}{Mice.PMM}&0.10&0.02&0.03&0.03&0.03&0.03&0.03&0.00\\

\hline  &0.01&0.00&0.00&0.00&0.00&0.00&0.00&0.00\\
  &0.05&0.00&0.00&0.00&0.00&0.00&0.00&0.00\\
\multirow{-5}{*}{Mice.RF}&0.10&0.00&0.00&0.00&0.00&0.00&0.00&0.00\\

\hline Naive &0.01&0.00&0.00&0.00&0.00&0.00&0.00&0.00\\
  Imputation&0.05&0.05&0.06&0.06&0.06&0.06&0.07&0.00\\
 &0.10&0.19&0.20&0.21&0.21&0.21&0.22&0.01\\

\hline  &0.01&0.00&0.00&0.00&0.00&0.00&0.00&0.00\\
  &0.05&0.00&0.00&0.00&0.01&0.01&0.02&0.00\\
\multirow{-5}{*}{MissRanger}&0.10&0.01&0.02&0.02&0.02&0.02&0.06&0.01\\

\hline 
\end{tabular} 
\end{minipage} 
\end{table}

\newpage 
\begin{table}[H] 
\caption{minimum ($q_0$), first quartile ($q_{0.25}$),
               median ($q_{0.5}$), mean ($\bar{x}$), third quartile ($q_{0.75}$),
               maximum ($q_{1}$) and standard deviation (sd) of the observed values of the Kullback-Leibler divergence for the variable \texttt{ef44} for the different
               imputation methods and missing rates (r)
               creating missing values using the MAR-mechanism for the employee data (VSE, Kullback-Leibler, \texttt{ef44}, MAR)}\label{Tab:vse.vals.kullbackleibler.MAR.ef44 }\begin{minipage}[t]{\textwidth} 

\centering 
\begin{tabular}[t]{|l|c|c|c|c|c|c|c|c|}\hline
 method&r&$q_{0}$& $q_{0.25}$&$q_{0.5}$&$\bar{x}$&$q_{0.75}$& $q_{1}$&sd\\ \hline 
\hline  &0.01&0.00&0.00&0.00&0.00&0.00&0.00&0.00\\
  &0.05&0.00&0.00&0.00&0.00&0.00&0.00&0.00\\
\multirow{-5}{*}{Amelia}&0.10&0.00&0.00&0.00&0.00&0.00&0.00&0.00\\

\hline  &0.01&0.00&0.00&0.00&0.00&0.00&0.00&0.00\\
  &0.05&0.00&0.00&0.00&0.00&0.00&0.00&0.00\\
\multirow{-5}{*}{Mice.Norm}&0.10&0.00&0.00&0.00&0.00&0.00&0.00&0.00\\

\hline  &0.01&0.00&0.00&0.00&0.00&0.00&0.00&0.00\\
  &0.05&0.00&0.00&0.00&0.00&0.00&0.00&0.00\\
\multirow{-5}{*}{Mice.PMM}&0.10&0.00&0.00&0.00&0.00&0.00&0.00&0.00\\

\hline  &0.01&0.00&0.00&0.00&0.00&0.00&0.00&0.00\\
  &0.05&0.00&0.00&0.00&0.00&0.00&0.00&0.00\\
\multirow{-5}{*}{Mice.RF}&0.10&0.00&0.00&0.00&0.00&0.00&0.00&0.00\\

\hline Naive &0.01&0.00&0.00&0.00&0.00&0.00&0.00&0.00\\
  Imputation&0.05&0.00&0.00&0.00&0.00&0.00&0.00&0.00\\
 &0.10&0.00&0.00&0.00&0.00&0.00&0.00&0.00\\

\hline  &0.01&0.00&0.00&0.00&0.00&0.00&0.00&0.00\\
  &0.05&0.00&0.00&0.00&0.00&0.00&0.00&0.00\\
\multirow{-5}{*}{MissRanger}&0.10&0.00&0.00&0.00&0.00&0.00&0.00&0.00\\

\hline 
\end{tabular} 
\end{minipage} 
\end{table}

\newpage 
\begin{table}[H] 
\caption{minimum ($q_0$), first quartile ($q_{0.25}$),
               median ($q_{0.5}$), mean ($\bar{x}$), third quartile ($q_{0.75}$),
               maximum ($q_{1}$) and standard deviation (sd) of the observed values of the Kullback-Leibler divergence for the variable \texttt{ef44} for the different
               imputation methods and missing rates (r)
               creating missing values using the MCAR-mechanism for the employee data (VSE, Kullback-Leibler, \texttt{ef44}, MCAR)}\label{Tab:vse.vals.kullbackleibler.MCAR.ef44 }\begin{minipage}[t]{\textwidth} 

\centering 
\begin{tabular}[t]{|l|c|c|c|c|c|c|c|c|}\hline
 method&r&$q_{0}$& $q_{0.25}$&$q_{0.5}$&$\bar{x}$&$q_{0.75}$& $q_{1}$&sd\\ \hline 
\hline  &0.01&0.00&0.00&0.00&0.00&0.00&0.00&0.00\\
  &0.05&0.00&0.00&0.00&0.00&0.00&0.00&0.00\\
\multirow{-5}{*}{Amelia}&0.10&0.00&0.00&0.00&0.00&0.00&0.00&0.00\\

\hline  &0.01&0.00&0.00&0.00&0.00&0.00&0.00&0.00\\
  &0.05&0.00&0.00&0.00&0.00&0.00&0.00&0.00\\
\multirow{-5}{*}{Mice.Norm}&0.10&0.00&0.00&0.00&0.00&0.00&0.00&0.00\\

\hline  &0.01&0.00&0.00&0.00&0.00&0.00&0.00&0.00\\
  &0.05&0.00&0.00&0.00&0.00&0.00&0.00&0.00\\
\multirow{-5}{*}{Mice.PMM}&0.10&0.00&0.00&0.00&0.00&0.00&0.00&0.00\\

\hline  &0.01&0.00&0.00&0.00&0.00&0.00&0.00&0.00\\
  &0.05&0.00&0.00&0.00&0.00&0.00&0.00&0.00\\
\multirow{-5}{*}{Mice.RF}&0.10&0.00&0.00&0.00&0.00&0.00&0.00&0.00\\

\hline Naive &0.01&0.00&0.00&0.00&0.00&0.00&0.00&0.00\\
  Imputation&0.05&0.00&0.00&0.00&0.00&0.00&0.00&0.00\\
 &0.10&0.00&0.00&0.00&0.00&0.00&0.00&0.00\\

\hline  &0.01&0.00&0.00&0.00&0.00&0.00&0.00&0.00\\
  &0.05&0.00&0.00&0.00&0.00&0.00&0.00&0.00\\
\multirow{-5}{*}{MissRanger}&0.10&0.00&0.00&0.00&0.00&0.00&0.00&0.00\\

\hline 
\end{tabular} 
\end{minipage} 
\end{table}

\newpage 
\begin{table}[H] 
\caption{minimum ($q_0$), first quartile ($q_{0.25}$),
               median ($q_{0.5}$), mean ($\bar{x}$), third quartile ($q_{0.75}$),
               maximum ($q_{1}$) and standard deviation (sd) of the observed values of the Kullback-Leibler divergence for the variable \texttt{ef45} for the different
               imputation methods and missing rates (r)
               creating missing values using the MAR-mechanism for the employee data (VSE, Kullback-Leibler, \texttt{ef45}, MAR)}\label{Tab:vse.vals.kullbackleibler.MAR.ef45 }\begin{minipage}[t]{\textwidth} 

\centering 
\begin{tabular}[t]{|l|c|c|c|c|c|c|c|c|}\hline
 method&r&$q_{0}$& $q_{0.25}$&$q_{0.5}$&$\bar{x}$&$q_{0.75}$& $q_{1}$&sd\\ \hline 
\hline  &0.01&0.00&0.00&0.00&0.00&0.00&0.00&0.00\\
  &0.05&0.00&0.00&0.00&0.00&0.00&0.00&0.00\\
\multirow{-5}{*}{Amelia}&0.10&0.00&0.00&0.00&0.00&0.00&0.00&0.00\\

\hline  &0.01&0.00&0.00&0.00&0.00&0.00&0.00&0.00\\
  &0.05&0.00&0.00&0.00&0.00&0.00&0.00&0.00\\
\multirow{-5}{*}{Mice.Norm}&0.10&0.00&0.00&0.00&0.00&0.00&0.00&0.00\\

\hline  &0.01&0.00&0.00&0.00&0.00&0.00&0.00&0.00\\
  &0.05&0.00&0.00&0.00&0.00&0.00&0.00&0.00\\
\multirow{-5}{*}{Mice.PMM}&0.10&0.00&0.00&0.00&0.00&0.00&0.00&0.00\\

\hline  &0.01&0.00&0.00&0.00&0.00&0.00&0.00&0.00\\
  &0.05&0.00&0.00&0.00&0.00&0.00&0.00&0.00\\
\multirow{-5}{*}{Mice.RF}&0.10&0.00&0.00&0.00&0.00&0.00&0.00&0.00\\

\hline Naive &0.01&0.00&0.00&0.00&0.00&0.00&0.00&0.00\\
  Imputation&0.05&0.00&0.00&0.00&0.00&0.00&0.00&0.00\\
 &0.10&0.00&0.00&0.00&0.00&0.00&0.00&0.00\\

\hline  &0.01&0.00&0.00&0.00&0.00&0.00&0.00&0.00\\
  &0.05&0.00&0.00&0.00&0.00&0.00&0.00&0.00\\
\multirow{-5}{*}{MissRanger}&0.10&0.00&0.00&0.00&0.00&0.00&0.00&0.00\\

\hline 
\end{tabular} 
\end{minipage} 
\end{table}

\newpage 
\begin{table}[H] 
\caption{minimum ($q_0$), first quartile ($q_{0.25}$),
               median ($q_{0.5}$), mean ($\bar{x}$), third quartile ($q_{0.75}$),
               maximum ($q_{1}$) and standard deviation (sd) of the observed values of the Kullback-Leibler divergence for the variable \texttt{ef45} for the different
               imputation methods and missing rates (r)
               creating missing values using the MCAR-mechanism for the employee data (VSE, Kullback-Leibler, \texttt{ef45}, MCAR)}\label{Tab:vse.vals.kullbackleibler.MCAR.ef45 }\begin{minipage}[t]{\textwidth} 

\centering 
\begin{tabular}[t]{|l|c|c|c|c|c|c|c|c|}\hline
 method&r&$q_{0}$& $q_{0.25}$&$q_{0.5}$&$\bar{x}$&$q_{0.75}$& $q_{1}$&sd\\ \hline 
\hline  &0.01&0.00&0.00&0.00&0.00&0.00&0.00&0.00\\
  &0.05&0.00&0.00&0.00&0.00&0.00&0.00&0.00\\
\multirow{-5}{*}{Amelia}&0.10&0.00&0.00&0.00&0.00&0.00&0.00&0.00\\

\hline  &0.01&0.00&0.00&0.00&0.00&0.00&0.00&0.00\\
  &0.05&0.00&0.00&0.00&0.00&0.00&0.00&0.00\\
\multirow{-5}{*}{Mice.Norm}&0.10&0.00&0.00&0.00&0.00&0.00&0.00&0.00\\

\hline  &0.01&0.00&0.00&0.00&0.00&0.00&0.00&0.00\\
  &0.05&0.00&0.00&0.00&0.00&0.00&0.00&0.00\\
\multirow{-5}{*}{Mice.PMM}&0.10&0.00&0.00&0.00&0.00&0.00&0.00&0.00\\

\hline  &0.01&0.00&0.00&0.00&0.00&0.00&0.00&0.00\\
  &0.05&0.00&0.00&0.00&0.00&0.00&0.00&0.00\\
\multirow{-5}{*}{Mice.RF}&0.10&0.00&0.00&0.00&0.00&0.00&0.00&0.00\\

\hline Naive &0.01&0.00&0.00&0.00&0.00&0.00&0.00&0.00\\
  Imputation&0.05&0.00&0.00&0.00&0.00&0.00&0.00&0.00\\
 &0.10&0.00&0.00&0.00&0.00&0.00&0.00&0.00\\

\hline  &0.01&0.00&0.00&0.00&0.00&0.00&0.00&0.00\\
  &0.05&0.00&0.00&0.00&0.00&0.00&0.00&0.00\\
\multirow{-5}{*}{MissRanger}&0.10&0.00&0.00&0.00&0.00&0.00&0.00&0.00\\

\hline 
\end{tabular} 
\end{minipage} 
\end{table}

\newpage 
\begin{table}[H] 
\caption{minimum ($q_0$), first quartile ($q_{0.25}$),
               median ($q_{0.5}$), mean ($\bar{x}$), third quartile ($q_{0.75}$),
               maximum ($q_{1}$) and standard deviation (sd) of the observed values of the Kullback-Leibler divergence for the variable \texttt{ef47} for the different
               imputation methods and missing rates (r)
               creating missing values using the MAR-mechanism for the employee data (VSE, Kullback-Leibler, \texttt{ef47}, MAR)}\label{Tab:vse.vals.kullbackleibler.MAR.ef47 }\begin{minipage}[t]{\textwidth} 

\centering 
\begin{tabular}[t]{|l|c|c|c|c|c|c|c|c|}\hline
 method&r&$q_{0}$& $q_{0.25}$&$q_{0.5}$&$\bar{x}$&$q_{0.75}$& $q_{1}$&sd\\ \hline 
\hline  &0.01&$\infty$&$\infty$&$\infty$&$\infty$&$\infty$&$\infty$&NaN\\
  &0.05&$\infty$&$\infty$&$\infty$&$\infty$&$\infty$&$\infty$&NaN\\
\multirow{-5}{*}{Amelia}&0.10&$\infty$&$\infty$&$\infty$&$\infty$&$\infty$&$\infty$&NaN\\

\hline  &0.01&$\infty$&$\infty$&$\infty$&$\infty$&$\infty$&$\infty$&NaN\\
  &0.05&$\infty$&$\infty$&$\infty$&$\infty$&$\infty$&$\infty$&NaN\\
\multirow{-5}{*}{Mice.Norm}&0.10&$\infty$&$\infty$&$\infty$&$\infty$&$\infty$&$\infty$&NaN\\

\hline  &0.01&$\infty$&$\infty$&$\infty$&$\infty$&$\infty$&$\infty$&NaN\\
  &0.05&$\infty$&$\infty$&$\infty$&$\infty$&$\infty$&$\infty$&NaN\\
\multirow{-5}{*}{Mice.PMM}&0.10&$\infty$&$\infty$&$\infty$&$\infty$&$\infty$&$\infty$&NaN\\

\hline  &0.01&$\infty$&$\infty$&$\infty$&$\infty$&$\infty$&$\infty$&NaN\\
  &0.05&$\infty$&$\infty$&$\infty$&$\infty$&$\infty$&$\infty$&NaN\\
\multirow{-5}{*}{Mice.RF}&0.10&$\infty$&$\infty$&$\infty$&$\infty$&$\infty$&$\infty$&NaN\\

\hline Naive &0.01&0.00&$\infty$&$\infty$&$\infty$&$\infty$&$\infty$&NaN\\
  Imputation&0.05&0.00&$\infty$&$\infty$&$\infty$&$\infty$&$\infty$&NaN\\
 &0.10&$\infty$&$\infty$&$\infty$&$\infty$&$\infty$&$\infty$&NaN\\

\hline  &0.01&0.00&$\infty$&$\infty$&$\infty$&$\infty$&$\infty$&NaN\\
  &0.05&0.00&$\infty$&$\infty$&$\infty$&$\infty$&$\infty$&NaN\\
\multirow{-5}{*}{MissRanger}&0.10&0.00&$\infty$&$\infty$&$\infty$&$\infty$&$\infty$&NaN\\

\hline 
\end{tabular} 
\end{minipage} 
\end{table}

\newpage 
\begin{table}[H] 
\caption{minimum ($q_0$), first quartile ($q_{0.25}$),
               median ($q_{0.5}$), mean ($\bar{x}$), third quartile ($q_{0.75}$),
               maximum ($q_{1}$) and standard deviation (sd) of the observed values of the Kullback-Leibler divergence for the variable \texttt{ef47} for the different
               imputation methods and missing rates (r)
               creating missing values using the MCAR-mechanism for the employee data (VSE, Kullback-Leibler, \texttt{ef47}, MCAR)}\label{Tab:vse.vals.kullbackleibler.MCAR.ef47 }\begin{minipage}[t]{\textwidth} 

\centering 
\begin{tabular}[t]{|l|c|c|c|c|c|c|c|c|}\hline
 method&r&$q_{0}$& $q_{0.25}$&$q_{0.5}$&$\bar{x}$&$q_{0.75}$& $q_{1}$&sd\\ \hline 
\hline  &0.01&$\infty$&$\infty$&$\infty$&$\infty$&$\infty$&$\infty$&NaN\\
  &0.05&$\infty$&$\infty$&$\infty$&$\infty$&$\infty$&$\infty$&NaN\\
\multirow{-5}{*}{Amelia}&0.10&$\infty$&$\infty$&$\infty$&$\infty$&$\infty$&$\infty$&NaN\\

\hline  &0.01&$\infty$&$\infty$&$\infty$&$\infty$&$\infty$&$\infty$&NaN\\
  &0.05&$\infty$&$\infty$&$\infty$&$\infty$&$\infty$&$\infty$&NaN\\
\multirow{-5}{*}{Mice.Norm}&0.10&$\infty$&$\infty$&$\infty$&$\infty$&$\infty$&$\infty$&NaN\\

\hline  &0.01&$\infty$&$\infty$&$\infty$&$\infty$&$\infty$&$\infty$&NaN\\
  &0.05&$\infty$&$\infty$&$\infty$&$\infty$&$\infty$&$\infty$&NaN\\
\multirow{-5}{*}{Mice.PMM}&0.10&$\infty$&$\infty$&$\infty$&$\infty$&$\infty$&$\infty$&NaN\\

\hline  &0.01&$\infty$&$\infty$&$\infty$&$\infty$&$\infty$&$\infty$&NaN\\
  &0.05&$\infty$&$\infty$&$\infty$&$\infty$&$\infty$&$\infty$&NaN\\
\multirow{-5}{*}{Mice.RF}&0.10&$\infty$&$\infty$&$\infty$&$\infty$&$\infty$&$\infty$&NaN\\

\hline Naive &0.01&0.00&$\infty$&$\infty$&$\infty$&$\infty$&$\infty$&NaN\\
  Imputation&0.05&0.00&$\infty$&$\infty$&$\infty$&$\infty$&$\infty$&NaN\\
 &0.10&0.00&$\infty$&$\infty$&$\infty$&$\infty$&$\infty$&NaN\\

\hline  &0.01&0.00&$\infty$&$\infty$&$\infty$&$\infty$&$\infty$&NaN\\
  &0.05&0.00&$\infty$&$\infty$&$\infty$&$\infty$&$\infty$&NaN\\
\multirow{-5}{*}{MissRanger}&0.10&0.00&$\infty$&$\infty$&$\infty$&$\infty$&$\infty$&NaN\\

\hline 
\end{tabular} 
\end{minipage} 
\end{table}

\newpage 
\begin{table}[H] 
\caption{minimum ($q_0$), first quartile ($q_{0.25}$),
               median ($q_{0.5}$), mean ($\bar{x}$), third quartile ($q_{0.75}$),
               maximum ($q_{1}$) and standard deviation (sd) of the observed values of the Kullback-Leibler divergence for the variable \texttt{ef53} for the different
               imputation methods and missing rates (r)
               creating missing values using the MAR-mechanism for the employee data (VSE, Kullback-Leibler, \texttt{ef53}, MAR)}\label{Tab:vse.vals.kullbackleibler.MAR.ef53 }\begin{minipage}[t]{\textwidth} 

\centering 
\begin{tabular}[t]{|l|c|c|c|c|c|c|c|c|}\hline
 method&r&$q_{0}$& $q_{0.25}$&$q_{0.5}$&$\bar{x}$&$q_{0.75}$& $q_{1}$&sd\\ \hline 
\hline  &0.01&0.00&0.00&0.00&0.00&0.00&0.00&0.00\\
  &0.05&0.00&0.00&0.00&0.00&0.00&0.01&0.00\\
\multirow{-5}{*}{Amelia}&0.10&0.02&0.02&0.03&0.03&0.03&0.04&0.01\\

\hline  &0.01&0.00&0.00&0.00&0.00&0.00&0.00&0.00\\
  &0.05&0.00&0.00&0.00&0.00&0.00&0.00&0.00\\
\multirow{-5}{*}{Mice.Norm}&0.10&0.00&0.00&0.00&0.00&0.00&0.01&0.00\\

\hline  &0.01&0.00&0.00&0.00&0.00&0.00&0.00&0.00\\
  &0.05&0.01&0.02&0.02&0.02&0.03&0.04&0.01\\
\multirow{-5}{*}{Mice.PMM}&0.10&0.05&0.07&0.07&0.07&0.07&0.08&0.01\\

\hline  &0.01&0.00&0.00&0.00&0.00&0.00&0.00&0.00\\
  &0.05&0.00&0.00&0.00&0.00&0.00&0.00&0.00\\
\multirow{-5}{*}{Mice.RF}&0.10&0.00&0.00&0.01&0.01&0.01&0.02&0.00\\

\hline Naive &0.01&0.01&0.01&0.01&0.01&0.01&0.02&0.00\\
  Imputation&0.05&0.22&0.24&0.25&0.25&0.25&0.27&0.01\\
 &0.10&0.57&0.59&0.60&0.60&0.61&0.65&0.01\\

\hline  &0.01&0.00&0.00&0.00&0.00&0.00&0.00&0.00\\
  &0.05&0.00&0.00&0.00&0.00&0.00&0.02&0.00\\
\multirow{-5}{*}{MissRanger}&0.10&0.00&0.00&0.00&0.01&0.01&0.05&0.01\\

\hline 
\end{tabular} 
\end{minipage} 
\end{table}

\newpage 
\begin{table}[H] 
\caption{minimum ($q_0$), first quartile ($q_{0.25}$),
               median ($q_{0.5}$), mean ($\bar{x}$), third quartile ($q_{0.75}$),
               maximum ($q_{1}$) and standard deviation (sd) of the observed values of the Kullback-Leibler divergence for the variable \texttt{ef53} for the different
               imputation methods and missing rates (r)
               creating missing values using the MCAR-mechanism for the employee data (VSE, Kullback-Leibler, \texttt{ef53}, MCAR)}\label{Tab:vse.vals.kullbackleibler.MCAR.ef53 }\begin{minipage}[t]{\textwidth} 

\centering 
\begin{tabular}[t]{|l|c|c|c|c|c|c|c|c|}\hline
 method&r&$q_{0}$& $q_{0.25}$&$q_{0.5}$&$\bar{x}$&$q_{0.75}$& $q_{1}$&sd\\ \hline 
\hline  &0.01&0.00&0.00&0.00&0.00&0.00&0.00&0.00\\
  &0.05&0.00&0.00&0.00&0.00&0.00&0.01&0.00\\
\multirow{-5}{*}{Amelia}&0.10&0.02&0.02&0.03&0.03&0.03&0.04&0.00\\

\hline  &0.01&0.00&0.00&0.00&0.00&0.00&0.00&0.00\\
  &0.05&0.00&0.00&0.00&0.00&0.00&0.00&0.00\\
\multirow{-5}{*}{Mice.Norm}&0.10&0.00&0.00&0.00&0.00&0.00&0.01&0.00\\

\hline  &0.01&0.00&0.00&0.00&0.00&0.00&0.00&0.00\\
  &0.05&0.01&0.02&0.02&0.02&0.03&0.04&0.01\\
\multirow{-5}{*}{Mice.PMM}&0.10&0.06&0.07&0.07&0.07&0.07&0.08&0.00\\

\hline  &0.01&0.00&0.00&0.00&0.00&0.00&0.00&0.00\\
  &0.05&0.00&0.00&0.00&0.00&0.00&0.00&0.00\\
\multirow{-5}{*}{Mice.RF}&0.10&0.00&0.00&0.01&0.01&0.01&0.02&0.00\\

\hline Naive &0.01&0.01&0.01&0.01&0.01&0.01&0.02&0.00\\
  Imputation&0.05&0.23&0.24&0.25&0.25&0.25&0.27&0.01\\
 &0.10&0.57&0.59&0.60&0.60&0.61&0.64&0.01\\

\hline  &0.01&0.00&0.00&0.00&0.00&0.00&0.00&0.00\\
  &0.05&0.00&0.00&0.00&0.00&0.00&0.01&0.00\\
\multirow{-5}{*}{MissRanger}&0.10&0.00&0.00&0.00&0.00&0.00&0.07&0.01\\

\hline 
\end{tabular} 
\end{minipage} 
\end{table}

\newpage 

%% file: table_fivepoint_Mallows_vse.txt
\begin{table}[H] 
\caption{minimum ($q_0$), first quartile ($q_{0.25}$),
               median ($q_{0.5}$), mean ($\bar{x}$), third quartile ($q_{0.75}$),
               maximum ($q_{1}$) and standard deviation (sd) of the observed values of Mallows $L^2$-distance for the variable \texttt{ef19} for the different
               imputation methods and missing rates (r)
               creating missing values using the MAR-mechanism for the employee data (VSE, Mallows, \texttt{ef19}, MAR)}\label{Tab:vse.vals.mallows.MAR.ef19 }\begin{minipage}[t]{\textwidth} 

\centering 
\begin{tabular}[t]{|l|c|c|c|c|c|c|c|c|}\hline
 method&r&$q_{0}$& $q_{0.25}$&$q_{0.5}$&$\bar{x}$&$q_{0.75}$& $q_{1}$&sd\\ \hline 
\hline  &0.01&0.05&0.07&0.08&0.08&0.09&0.13&0.02\\
  &0.05&0.33&0.42&0.44&0.45&0.48&0.59&0.05\\
\multirow{-5}{*}{Amelia}&0.10&0.99&1.12&1.17&1.17&1.22&1.46&0.08\\

\hline  &0.01&0.02&0.02&0.02&0.02&0.02&0.03&0.00\\
  &0.05&0.04&0.05&0.05&0.05&0.06&0.10&0.01\\
\multirow{-5}{*}{Mice.Norm}&0.10&0.05&0.07&0.07&0.08&0.09&0.14&0.02\\

\hline  &0.01&0.14&0.20&0.22&0.22&0.24&0.31&0.03\\
  &0.05&1.19&1.31&1.38&1.38&1.44&1.62&0.10\\
\multirow{-5}{*}{Mice.PMM}&0.10&3.17&3.49&3.63&3.66&3.86&4.28&0.25\\

\hline  &0.01&0.02&0.03&0.04&0.04&0.05&0.08&0.01\\
  &0.05&0.10&0.18&0.23&0.24&0.29&0.49&0.08\\
\multirow{-5}{*}{Mice.RF}&0.10&0.33&0.56&0.75&0.77&0.93&1.79&0.28\\

\hline Naive &0.01&0.67&0.80&0.87&0.89&0.97&1.13&0.10\\
  Imputation&0.05&11.20&12.17&12.71&12.68&13.13&14.20&0.63\\
 &0.10&35.05&37.50&38.30&38.52&39.44&43.39&1.55\\

\hline  &0.01&0.01&0.01&0.02&0.04&0.04&0.21&0.05\\
  &0.05&0.02&0.03&0.06&0.18&0.18&2.28&0.30\\
\multirow{-5}{*}{MissRanger}&0.10&0.05&0.08&0.11&0.63&0.62&7.64&1.28\\

\hline 
\end{tabular} 
\end{minipage} 
\end{table}

\newpage 
\begin{table}[H] 
\caption{minimum ($q_0$), first quartile ($q_{0.25}$),
               median ($q_{0.5}$), mean ($\bar{x}$), third quartile ($q_{0.75}$),
               maximum ($q_{1}$) and standard deviation (sd) of the observed values of Mallows $L^2$-distance for the variable \texttt{ef19} for the different
               imputation methods and missing rates (r)
               creating missing values using the MCAR-mechanism for the employee data (VSE, Mallows, \texttt{ef19}, MCAR)}\label{Tab:vse.vals.mallows.MCAR.ef19 }\begin{minipage}[t]{\textwidth} 

\centering 
\begin{tabular}[t]{|l|c|c|c|c|c|c|c|c|}\hline
 method&r&$q_{0}$& $q_{0.25}$&$q_{0.5}$&$\bar{x}$&$q_{0.75}$& $q_{1}$&sd\\ \hline 
\hline  &0.01&0.05&0.07&0.08&0.08&0.09&0.15&0.02\\
  &0.05&0.36&0.42&0.44&0.44&0.47&0.54&0.04\\
\multirow{-5}{*}{Amelia}&0.10&0.96&1.11&1.17&1.17&1.23&1.41&0.09\\

\hline  &0.01&0.01&0.02&0.02&0.02&0.02&0.03&0.00\\
  &0.05&0.03&0.04&0.05&0.05&0.05&0.08&0.01\\
\multirow{-5}{*}{Mice.Norm}&0.10&0.05&0.07&0.08&0.08&0.09&0.13&0.02\\

\hline  &0.01&0.16&0.20&0.22&0.22&0.24&0.30&0.03\\
  &0.05&1.16&1.30&1.37&1.37&1.43&1.69&0.10\\
\multirow{-5}{*}{Mice.PMM}&0.10&3.12&3.50&3.66&3.69&3.87&4.42&0.26\\

\hline  &0.01&0.02&0.03&0.04&0.04&0.05&0.08&0.01\\
  &0.05&0.12&0.19&0.24&0.26&0.31&0.73&0.10\\
\multirow{-5}{*}{Mice.RF}&0.10&0.39&0.63&0.77&0.84&1.00&1.85&0.29\\

\hline Naive &0.01&0.68&0.83&0.89&0.89&0.95&1.16&0.10\\
  Imputation&0.05&11.01&12.20&12.75&12.69&13.10&14.25&0.66\\
 &0.10&35.47&37.80&38.64&38.71&39.74&43.01&1.51\\

\hline  &0.01&0.01&0.01&0.01&0.03&0.03&0.19&0.03\\
  &0.05&0.02&0.03&0.06&0.18&0.19&1.67&0.29\\
\multirow{-5}{*}{MissRanger}&0.10&0.06&0.08&0.13&0.38&0.39&6.10&0.75\\

\hline 
\end{tabular} 
\end{minipage} 
\end{table}

\newpage 
\begin{table}[H] 
\caption{minimum ($q_0$), first quartile ($q_{0.25}$),
               median ($q_{0.5}$), mean ($\bar{x}$), third quartile ($q_{0.75}$),
               maximum ($q_{1}$) and standard deviation (sd) of the observed values of Mallows $L^2$-distance for the variable \texttt{ef20} for the different
               imputation methods and missing rates (r)
               creating missing values using the MAR-mechanism for the employee data (VSE, Mallows, \texttt{ef20}, MAR)}\label{Tab:vse.vals.mallows.MAR.ef20 }\begin{minipage}[t]{\textwidth} 

\centering 
\begin{tabular}[t]{|l|c|c|c|c|c|c|c|c|}\hline
 method&r&$q_{0}$& $q_{0.25}$&$q_{0.5}$&$\bar{x}$&$q_{0.75}$& $q_{1}$&sd\\ \hline 
\hline  &0.01&0.04&0.04&0.05&0.06&0.05&0.24&0.03\\
  &0.05&0.26&0.30&0.31&0.33&0.36&0.51&0.05\\
\multirow{-5}{*}{Amelia}&0.10&0.69&0.75&0.79&0.81&0.86&1.05&0.08\\

\hline  &0.01&0.00&0.00&0.00&0.01&0.01&0.01&0.00\\
  &0.05&0.01&0.01&0.01&0.01&0.01&0.04&0.00\\
\multirow{-5}{*}{Mice.Norm}&0.10&0.01&0.01&0.02&0.02&0.02&0.04&0.01\\

\hline  &0.01&0.15&0.17&0.18&0.18&0.19&0.23&0.02\\
  &0.05&1.01&1.10&1.13&1.13&1.16&1.27&0.05\\
\multirow{-5}{*}{Mice.PMM}&0.10&2.42&2.54&2.58&2.59&2.65&2.77&0.07\\

\hline  &0.01&0.00&0.00&0.01&0.01&0.01&0.02&0.00\\
  &0.05&0.01&0.03&0.04&0.04&0.04&0.09&0.01\\
\multirow{-5}{*}{Mice.RF}&0.10&0.04&0.06&0.08&0.08&0.10&0.22&0.03\\

\hline Naive &0.01&0.02&0.03&0.03&0.03&0.03&0.05&0.00\\
  Imputation&0.05&0.14&0.18&0.19&0.19&0.20&0.28&0.02\\
 &0.10&0.44&0.48&0.51&0.52&0.56&0.67&0.05\\

\hline  &0.01&0.00&0.00&0.00&0.00&0.01&0.01&0.00\\
  &0.05&0.00&0.01&0.02&0.02&0.03&0.08&0.02\\
\multirow{-5}{*}{MissRanger}&0.10&0.01&0.03&0.04&0.05&0.06&0.23&0.04\\

\hline 
\end{tabular} 
\end{minipage} 
\end{table}

\newpage 
\begin{table}[H] 
\caption{minimum ($q_0$), first quartile ($q_{0.25}$),
               median ($q_{0.5}$), mean ($\bar{x}$), third quartile ($q_{0.75}$),
               maximum ($q_{1}$) and standard deviation (sd) of the observed values of Mallows $L^2$-distance for the variable \texttt{ef20} for the different
               imputation methods and missing rates (r)
               creating missing values using the MCAR-mechanism for the employee data (VSE, Mallows, \texttt{ef20}, MCAR)}\label{Tab:vse.vals.mallows.MCAR.ef20 }\begin{minipage}[t]{\textwidth} 

\centering 
\begin{tabular}[t]{|l|c|c|c|c|c|c|c|c|}\hline
 method&r&$q_{0}$& $q_{0.25}$&$q_{0.5}$&$\bar{x}$&$q_{0.75}$& $q_{1}$&sd\\ \hline 
\hline  &0.01&0.04&0.04&0.05&0.06&0.05&0.21&0.03\\
  &0.05&0.27&0.30&0.32&0.34&0.35&0.51&0.05\\
\multirow{-5}{*}{Amelia}&0.10&0.67&0.74&0.79&0.81&0.84&1.16&0.09\\

\hline  &0.01&0.00&0.00&0.00&0.01&0.01&0.01&0.00\\
  &0.05&0.01&0.01&0.01&0.01&0.02&0.04&0.00\\
\multirow{-5}{*}{Mice.Norm}&0.10&0.01&0.01&0.02&0.02&0.02&0.04&0.01\\

\hline  &0.01&0.15&0.17&0.18&0.18&0.19&0.22&0.01\\
  &0.05&1.01&1.11&1.14&1.14&1.17&1.26&0.05\\
\multirow{-5}{*}{Mice.PMM}&0.10&2.41&2.54&2.61&2.61&2.66&2.78&0.08\\

\hline  &0.01&0.00&0.00&0.01&0.01&0.01&0.02&0.00\\
  &0.05&0.02&0.03&0.04&0.04&0.04&0.09&0.01\\
\multirow{-5}{*}{Mice.RF}&0.10&0.03&0.06&0.08&0.08&0.10&0.15&0.03\\

\hline Naive &0.01&0.02&0.03&0.03&0.03&0.03&0.04&0.00\\
  Imputation&0.05&0.16&0.18&0.19&0.19&0.21&0.31&0.02\\
 &0.10&0.41&0.49&0.52&0.52&0.55&0.65&0.05\\

\hline  &0.01&0.00&0.00&0.00&0.00&0.01&0.01&0.00\\
  &0.05&0.00&0.01&0.01&0.02&0.02&0.06&0.01\\
\multirow{-5}{*}{MissRanger}&0.10&0.01&0.03&0.03&0.05&0.06&0.19&0.03\\

\hline 
\end{tabular} 
\end{minipage} 
\end{table}

\newpage 
\begin{table}[H] 
\caption{minimum ($q_0$), first quartile ($q_{0.25}$),
               median ($q_{0.5}$), mean ($\bar{x}$), third quartile ($q_{0.75}$),
               maximum ($q_{1}$) and standard deviation (sd) of the observed values of Mallows $L^2$-distance $\left(\text{with factor } \frac{1}{10}\right)$ for the variable \texttt{ef21} for the different
               imputation methods and missing rates (r)
               creating missing values using the MAR-mechanism for the employee data (VSE, Mallows, \texttt{ef21}, MAR)}\label{Tab:vse.vals.mallows.MAR.ef21 }\begin{minipage}[t]{\textwidth} 

\centering 
\begin{tabular}[t]{|l|c|c|c|c|c|c|c|c|}\hline
 method&r&$q_{0}$& $q_{0.25}$&$q_{0.5}$&$\bar{x}$&$q_{0.75}$& $q_{1}$&sd\\ \hline 
\hline  &0.01&0.18&0.55&0.74&0.79&0.99&1.91&0.32\\
  &0.05&3.94&4.84&5.38&5.71&6.58&8.46&1.15\\
\multirow{-5}{*}{Amelia}&0.10&11.82&15.41&16.48&16.58&17.80&24.65&2.08\\

\hline  &0.01&0.01&0.01&0.01&0.01&0.01&0.03&<0.01\\
  &0.05&0.10&0.12&0.14&0.14&0.15&0.27&0.02\\
\multirow{-5}{*}{Mice.Norm}&0.10&0.34&0.44&0.50&0.51&0.56&0.95&0.10\\

\hline  &0.01&0.02&0.03&0.04&0.05&0.05&0.24&0.04\\
  &0.05&0.59&0.84&1.06&1.98&2.01&19.69&2.70\\
\multirow{-5}{*}{Mice.PMM}&0.10&3.26&5.17&6.83&8.22&9.48&25.29&4.65\\

\hline  &0.01&0.25&0.52&0.73&0.78&0.98&2.28&0.37\\
  &0.05&3.26&7.60&9.63&10.34&12.76&22.65&3.88\\
\multirow{-5}{*}{Mice.RF}&0.10&8.95&26.00&33.74&35.79&42.90&72.21&13.09\\

\hline Naive &0.01&14.67&18.19&20.13&20.31&22.32&26.20&2.88\\
  Imputation&0.05&426.26&473.03&499.26&497.21&518.67&570.52&30.61\\
 &0.10&1734.52&1924.91&1990.07&1996.39&2069.56&2227.06&105.19\\

\hline  &0.01&0.02&0.04&0.06&0.26&0.16&2.20&0.48\\
  &0.05&0.16&0.32&0.50&2.48&1.49&37.01&5.56\\
\multirow{-5}{*}{MissRanger}&0.10&0.75&1.36&1.76&5.65&3.08&194.72&20.09\\

\hline 
\end{tabular} 
\end{minipage} 
\end{table}

\newpage 
\begin{table}[H] 
\caption{minimum ($q_0$), first quartile ($q_{0.25}$),
               median ($q_{0.5}$), mean ($\bar{x}$), third quartile ($q_{0.75}$),
               maximum ($q_{1}$) and standard deviation (sd) of the observed values of Mallows $L^2$-distance $\left(\text{with factor } \frac{1}{10}\right)$ for the variable \texttt{ef21} for the different
               imputation methods and missing rates (r)
               creating missing values using the MCAR-mechanism for the employee data (VSE, Mallows, \texttt{ef21}, MCAR)}\label{Tab:vse.vals.mallows.MCAR.ef21 }\begin{minipage}[t]{\textwidth} 

\centering 
\begin{tabular}[t]{|l|c|c|c|c|c|c|c|c|}\hline
 method&r&$q_{0}$& $q_{0.25}$&$q_{0.5}$&$\bar{x}$&$q_{0.75}$& $q_{1}$&sd\\ \hline 
\hline  &0.01&0.32&0.59&0.74&0.81&0.99&1.71&0.29\\
  &0.05&3.45&4.86&5.50&5.61&6.29&8.90&0.99\\
\multirow{-5}{*}{Amelia}&0.10&11.28&14.35&15.84&16.06&17.46&24.82&2.60\\

\hline  &0.01&0.01&0.01&0.01&0.01&0.01&0.03&<0.01\\
  &0.05&0.10&0.13&0.15&0.15&0.17&0.27&0.03\\
\multirow{-5}{*}{Mice.Norm}&0.10&0.30&0.43&0.48&0.50&0.56&0.92&0.11\\

\hline  &0.01&0.02&0.03&0.04&0.08&0.05&1.29&0.16\\
  &0.05&0.55&0.97&1.24&2.15&2.36&14.26&2.36\\
\multirow{-5}{*}{Mice.PMM}&0.10&2.81&5.14&7.07&9.68&11.01&50.69&7.77\\

\hline  &0.01&0.30&0.54&0.75&0.83&1.08&2.04&0.37\\
  &0.05&3.13&7.47&9.74&10.66&13.56&23.74&4.44\\
\multirow{-5}{*}{Mice.RF}&0.10&14.87&29.44&38.66&39.06&45.62&82.54&12.46\\

\hline Naive &0.01&14.49&18.05&20.23&20.57&22.43&28.98&3.36\\
  Imputation&0.05&399.88&472.62&497.60&498.39&522.44&581.66&38.31\\
 &0.10&1763.39&1913.09&1970.26&1976.40&2035.99&2216.01&90.29\\

\hline  &0.01&0.03&0.04&0.06&0.15&0.11&1.39&0.24\\
  &0.05&0.14&0.33&0.52&1.54&1.07&14.25&2.87\\
\multirow{-5}{*}{MissRanger}&0.10&0.44&1.15&1.71&4.78&5.76&63.25&8.06\\

\hline 
\end{tabular} 
\end{minipage} 
\end{table}

\newpage 
\begin{table}[H] 
\caption{minimum ($q_0$), first quartile ($q_{0.25}$),
               median ($q_{0.5}$), mean ($\bar{x}$), third quartile ($q_{0.75}$),
               maximum ($q_{1}$) and standard deviation (sd) of the observed values of Mallows $L^2$-distance $\left(\text{with factor } \frac{1}{10}\right)$ for the variable \texttt{ef22} for the different
               imputation methods and missing rates (r)
               creating missing values using the MAR-mechanism for the employee data (VSE, Mallows, \texttt{ef22}, MAR)}\label{Tab:vse.vals.mallows.MAR.ef22 }\begin{minipage}[t]{\textwidth} 

\centering 
\begin{tabular}[t]{|l|c|c|c|c|c|c|c|c|}\hline
 method&r&$q_{0}$& $q_{0.25}$&$q_{0.5}$&$\bar{x}$&$q_{0.75}$& $q_{1}$&sd\\ \hline 
\hline  &0.01&1.24&1.55&1.72&1.75&1.94&2.49&0.27\\
  &0.05&9.62&10.65&11.20&11.24&11.78&13.76&0.77\\
\multirow{-5}{*}{Amelia}&0.10&23.16&26.15&27.29&27.12&28.09&29.81&1.37\\

\hline  &0.01&0.01&0.04&0.07&0.13&0.14&0.89&0.17\\
  &0.05&0.08&0.29&0.40&0.48&0.60&1.45&0.29\\
\multirow{-5}{*}{Mice.Norm}&0.10&0.26&0.69&0.87&0.94&1.09&2.14&0.37\\

\hline  &0.01&4.51&5.40&5.83&5.82&6.21&7.48&0.58\\
  &0.05&30.87&35.91&36.81&36.79&37.66&40.64&1.58\\
\multirow{-5}{*}{Mice.PMM}&0.10&78.85&82.72&84.22&84.46&86.12&91.03&2.59\\

\hline  &0.01&0.01&0.04&0.07&0.13&0.13&0.93&0.17\\
  &0.05&0.14&0.49&0.68&0.80&0.97&2.58&0.48\\
\multirow{-5}{*}{Mice.RF}&0.10&0.36&1.65&2.18&2.25&2.91&5.16&0.90\\

\hline Naive &0.01&0.40&0.50&0.56&0.61&0.63&1.44&0.19\\
  Imputation&0.05&3.69&4.29&4.73&4.84&5.26&6.88&0.69\\
 &0.10&9.92&12.34&13.33&13.33&14.19&16.38&1.32\\

\hline  &0.01&<0.01&0.01&0.04&0.09&0.08&0.86&0.16\\
  &0.05&0.04&0.16&0.36&0.49&0.67&1.86&0.42\\
\multirow{-5}{*}{MissRanger}&0.10&0.14&0.66&1.17&1.48&1.87&4.75&1.11\\

\hline 
\end{tabular} 
\end{minipage} 
\end{table}

\newpage 
\begin{table}[H] 
\caption{minimum ($q_0$), first quartile ($q_{0.25}$),
               median ($q_{0.5}$), mean ($\bar{x}$), third quartile ($q_{0.75}$),
               maximum ($q_{1}$) and standard deviation (sd) of the observed values of Mallows $L^2$-distance $\left(\text{with factor } \frac{1}{10}\right)$ for the variable \texttt{ef22} for the different
               imputation methods and missing rates (r)
               creating missing values using the MCAR-mechanism for the employee data (VSE, Mallows, \texttt{ef22}, MCAR)}\label{Tab:vse.vals.mallows.MCAR.ef22 }\begin{minipage}[t]{\textwidth} 

\centering 
\begin{tabular}[t]{|l|c|c|c|c|c|c|c|c|}\hline
 method&r&$q_{0}$& $q_{0.25}$&$q_{0.5}$&$\bar{x}$&$q_{0.75}$& $q_{1}$&sd\\ \hline 
\hline  &0.01&1.33&1.54&1.62&1.68&1.74&2.93&0.25\\
  &0.05&9.39&10.55&10.99&11.02&11.51&12.93&0.68\\
\multirow{-5}{*}{Amelia}&0.10&24.06&25.68&27.11&26.93&27.94&30.49&1.45\\

\hline  &0.01&0.01&0.04&0.07&0.12&0.11&1.02&0.19\\
  &0.05&0.14&0.32&0.39&0.47&0.58&1.68&0.26\\
\multirow{-5}{*}{Mice.Norm}&0.10&0.42&0.61&0.84&0.95&1.12&3.39&0.51\\

\hline  &0.01&4.78&5.46&5.86&5.88&6.17&7.81&0.58\\
  &0.05&32.66&35.84&36.82&36.67&37.74&39.61&1.31\\
\multirow{-5}{*}{Mice.PMM}&0.10&77.26&82.21&84.63&84.15&85.72&89.86&2.55\\

\hline  &0.01&0.01&0.04&0.07&0.12&0.10&1.02&0.18\\
  &0.05&0.19&0.48&0.72&0.79&0.95&3.98&0.51\\
\multirow{-5}{*}{Mice.RF}&0.10&0.82&1.52&2.03&2.24&2.71&5.40&1.02\\

\hline Naive &0.01&0.43&0.52&0.55&0.61&0.60&1.68&0.21\\
  Imputation&0.05&3.66&4.26&4.68&4.76&4.96&8.65&0.72\\
 &0.10&10.86&12.06&13.15&13.34&14.18&18.03&1.65\\

\hline  &0.01&<0.01&0.01&0.04&0.09&0.07&1.01&0.19\\
  &0.05&0.04&0.16&0.41&0.53&0.81&2.29&0.44\\
\multirow{-5}{*}{MissRanger}&0.10&0.20&0.67&1.04&1.40&1.72&5.74&1.18\\

\hline 
\end{tabular} 
\end{minipage} 
\end{table}

\newpage 
\begin{table}[H] 
\caption{minimum ($q_0$), first quartile ($q_{0.25}$),
               median ($q_{0.5}$), mean ($\bar{x}$), third quartile ($q_{0.75}$),
               maximum ($q_{1}$) and standard deviation (sd) of the observed values of Mallows $L^2$-distance $\left(\text{with factor } \frac{1}{10}\right)$ for the variable \texttt{ef23} for the different
               imputation methods and missing rates (r)
               creating missing values using the MAR-mechanism for the employee data (VSE, Mallows, \texttt{ef23}, MAR)}\label{Tab:vse.vals.mallows.MAR.ef23 }\begin{minipage}[t]{\textwidth} 

\centering 
\begin{tabular}[t]{|l|c|c|c|c|c|c|c|c|}\hline
 method&r&$q_{0}$& $q_{0.25}$&$q_{0.5}$&$\bar{x}$&$q_{0.75}$& $q_{1}$&sd\\ \hline 
\hline  &0.01&2.51&3.05&3.23&3.27&3.48&4.74&0.37\\
  &0.05&15.13&17.41&18.12&18.12&18.86&21.33&1.12\\
\multirow{-5}{*}{Amelia}&0.10&35.27&39.11&40.35&40.27&41.58&46.69&1.72\\

\hline  &0.01&<0.01&0.01&0.01&0.08&0.02&1.92&0.25\\
  &0.05&0.02&0.06&0.14&0.34&0.48&1.89&0.41\\
\multirow{-5}{*}{Mice.Norm}&0.10&0.07&0.25&0.48&0.76&0.93&5.75&0.89\\

\hline  &0.01&3.44&4.10&4.29&4.38&4.64&5.89&0.45\\
  &0.05&21.83&24.09&24.82&24.82&25.53&27.81&1.21\\
\multirow{-5}{*}{Mice.PMM}&0.10&50.16&54.52&55.67&55.58&56.84&60.93&1.96\\

\hline  &0.01&0.01&0.01&0.02&0.12&0.04&1.91&0.28\\
  &0.05&0.03&0.23&0.44&0.55&0.82&1.93&0.41\\
\multirow{-5}{*}{Mice.RF}&0.10&0.09&0.66&1.00&1.14&1.49&4.41&0.69\\

\hline Naive &0.01&0.15&0.18&0.20&0.26&0.22&2.08&0.26\\
  Imputation&0.05&3.05&3.26&3.41&3.57&3.61&5.34&0.46\\
 &0.10&8.32&8.89&9.16&9.43&9.77&14.04&0.92\\

\hline  &0.01&<0.01&0.01&0.01&0.06&0.01&1.90&0.23\\
  &0.05&0.04&0.08&0.11&0.21&0.19&2.16&0.30\\
\multirow{-5}{*}{MissRanger}&0.10&0.12&0.27&0.43&0.66&0.66&4.66&0.70\\

\hline 
\end{tabular} 
\end{minipage} 
\end{table}

\newpage 
\begin{table}[H] 
\caption{minimum ($q_0$), first quartile ($q_{0.25}$),
               median ($q_{0.5}$), mean ($\bar{x}$), third quartile ($q_{0.75}$),
               maximum ($q_{1}$) and standard deviation (sd) of the observed values of Mallows $L^2$-distance $\left(\text{with factor } \frac{1}{10}\right)$ for the variable \texttt{ef23} for the different
               imputation methods and missing rates (r)
               creating missing values using the MCAR-mechanism for the employee data (VSE, Mallows, \texttt{ef23}, MCAR)}\label{Tab:vse.vals.mallows.MCAR.ef23 }\begin{minipage}[t]{\textwidth} 

\centering 
\begin{tabular}[t]{|l|c|c|c|c|c|c|c|c|}\hline
 method&r&$q_{0}$& $q_{0.25}$&$q_{0.5}$&$\bar{x}$&$q_{0.75}$& $q_{1}$&sd\\ \hline 
\hline  &0.01&2.07&2.67&2.96&3.08&3.28&6.68&0.69\\
  &0.05&14.32&15.64&16.24&16.42&17.10&20.11&1.17\\
\multirow{-5}{*}{Amelia}&0.10&31.42&35.32&36.46&36.59&37.85&43.79&1.98\\

\hline  &0.01&0.01&0.02&0.03&0.28&0.35&3.37&0.50\\
  &0.05&0.07&0.41&0.64&0.80&0.96&3.42&0.66\\
\multirow{-5}{*}{Mice.Norm}&0.10&0.20&0.76&1.17&1.62&2.23&6.92&1.25\\

\hline  &0.01&3.07&3.74&3.96&4.13&4.36&7.89&0.78\\
  &0.05&18.85&21.34&21.99&22.13&22.89&26.18&1.25\\
\multirow{-5}{*}{Mice.PMM}&0.10&46.86&49.30&50.47&50.48&51.54&55.17&1.68\\

\hline  &0.01&0.01&0.02&0.06&0.36&0.40&4.28&0.65\\
  &0.05&0.13&0.59&0.98&1.08&1.32&4.13&0.73\\
\multirow{-5}{*}{Mice.RF}&0.10&0.79&1.57&2.04&2.56&3.21&7.75&1.41\\

\hline Naive &0.01&0.15&0.20&0.23&0.52&0.33&4.46&0.70\\
  Imputation&0.05&3.09&3.71&4.20&4.40&4.76&8.13&0.96\\
 &0.10&10.01&11.53&12.60&13.10&14.11&20.22&2.08\\

\hline  &0.01&0.01&0.02&0.03&0.20&0.09&2.27&0.45\\
  &0.05&0.16&0.29&0.41&0.68&0.88&4.32&0.65\\
\multirow{-5}{*}{MissRanger}&0.10&0.44&1.15&1.59&2.05&2.77&6.90&1.31\\

\hline 
\end{tabular} 
\end{minipage} 
\end{table}

\newpage 
\begin{table}[H] 
\caption{minimum ($q_0$), first quartile ($q_{0.25}$),
               median ($q_{0.5}$), mean ($\bar{x}$), third quartile ($q_{0.75}$),
               maximum ($q_{1}$) and standard deviation (sd) of the observed values of Mallows $L^2$-distance $\left(\text{with factor } \frac{1}{10}\right)$ for the variable \texttt{ef24} for the different
               imputation methods and missing rates (r)
               creating missing values using the MAR-mechanism for the employee data (VSE, Mallows, \texttt{ef24}, MAR)}\label{Tab:vse.vals.mallows.MAR.ef24 }\begin{minipage}[t]{\textwidth} 

\centering 
\begin{tabular}[t]{|l|c|c|c|c|c|c|c|c|}\hline
 method&r&$q_{0}$& $q_{0.25}$&$q_{0.5}$&$\bar{x}$&$q_{0.75}$& $q_{1}$&sd\\ \hline 
\hline  &0.01&1.70&2.32&2.62&2.64&2.91&3.87&0.45\\
  &0.05&12.65&14.02&14.59&14.72&15.35&17.61&1.05\\
\multirow{-5}{*}{Amelia}&0.10&29.14&31.61&32.93&32.95&34.16&38.21&1.97\\

\hline  &0.01&0.02&0.03&0.04&0.05&0.05&0.18&0.03\\
  &0.05&0.09&0.14&0.20&0.24&0.26&0.82&0.16\\
\multirow{-5}{*}{Mice.Norm}&0.10&0.17&0.34&0.47&0.52&0.61&1.74&0.27\\

\hline  &0.01&2.04&2.85&3.30&3.29&3.74&4.37&0.56\\
  &0.05&14.24&16.54&17.20&17.24&17.98&21.20&1.11\\
\multirow{-5}{*}{Mice.PMM}&0.10&32.46&35.82&37.05&36.99&38.02&43.18&1.83\\

\hline  &0.01&0.03&0.05&0.06&0.07&0.08&0.29&0.04\\
  &0.05&0.16&0.28&0.37&0.41&0.51&1.05&0.19\\
\multirow{-5}{*}{Mice.RF}&0.10&0.56&0.89&1.11&1.17&1.33&2.52&0.39\\

\hline Naive &0.01&0.88&1.07&1.20&1.23&1.34&1.82&0.20\\
  Imputation&0.05&24.67&27.80&29.27&29.10&30.50&35.16&1.93\\
 &0.10&103.97&112.43&115.43&115.45&117.87&129.60&5.04\\

\hline  &0.01&0.01&0.03&0.03&0.05&0.06&0.17&0.03\\
  &0.05&0.10&0.17&0.29&0.63&0.60&6.42&1.00\\
\multirow{-5}{*}{MissRanger}&0.10&0.44&0.74&1.12&2.95&1.97&29.95&5.41\\

\hline 
\end{tabular} 
\end{minipage} 
\end{table}

\newpage 
\begin{table}[H] 
\caption{minimum ($q_0$), first quartile ($q_{0.25}$),
               median ($q_{0.5}$), mean ($\bar{x}$), third quartile ($q_{0.75}$),
               maximum ($q_{1}$) and standard deviation (sd) of the observed values of Mallows $L^2$-distance $\left(\text{with factor } \frac{1}{10}\right)$ for the variable \texttt{ef24} for the different
               imputation methods and missing rates (r)
               creating missing values using the MCAR-mechanism for the employee data (VSE, Mallows, \texttt{ef24}, MCAR)}\label{Tab:vse.vals.mallows.MCAR.ef24 }\begin{minipage}[t]{\textwidth} 

\centering 
\begin{tabular}[t]{|l|c|c|c|c|c|c|c|c|}\hline
 method&r&$q_{0}$& $q_{0.25}$&$q_{0.5}$&$\bar{x}$&$q_{0.75}$& $q_{1}$&sd\\ \hline 
\hline  &0.01&1.64&2.34&2.60&2.63&2.91&4.64&0.51\\
  &0.05&11.64&13.88&14.45&14.51&15.20&18.12&1.12\\
\multirow{-5}{*}{Amelia}&0.10&29.10&31.18&32.42&32.54&33.67&38.54&1.86\\

\hline  &0.01&0.02&0.03&0.04&0.05&0.05&0.62&0.07\\
  &0.05&0.08&0.14&0.21&0.24&0.27&0.80&0.13\\
\multirow{-5}{*}{Mice.Norm}&0.10&0.18&0.35&0.46&0.55&0.72&2.42&0.34\\

\hline  &0.01&2.20&2.74&3.19&3.26&3.63&4.86&0.58\\
  &0.05&14.54&16.39&17.29&17.24&18.16&20.10&1.25\\
\multirow{-5}{*}{Mice.PMM}&0.10&31.85&35.08&36.17&36.45&37.70&40.90&1.84\\

\hline  &0.01&0.03&0.05&0.06&0.08&0.08&0.61&0.07\\
  &0.05&0.16&0.29&0.38&0.39&0.48&0.75&0.14\\
\multirow{-5}{*}{Mice.RF}&0.10&0.65&1.02&1.22&1.31&1.55&3.07&0.47\\

\hline Naive &0.01&0.61&1.09&1.22&1.24&1.35&2.07&0.23\\
  Imputation&0.05&24.60&27.80&28.81&29.06&30.24&35.36&2.04\\
 &0.10&102.44&110.99&114.57&114.50&117.74&127.45&5.29\\

\hline  &0.01&0.01&0.03&0.04&0.06&0.07&0.60&0.07\\
  &0.05&0.09&0.17&0.28&0.69&0.63&7.55&1.26\\
\multirow{-5}{*}{MissRanger}&0.10&0.30&0.70&1.12&2.99&2.08&33.11&4.92\\

\hline 
\end{tabular} 
\end{minipage} 
\end{table}

\newpage 
\begin{table}[H] 
\caption{minimum ($q_0$), first quartile ($q_{0.25}$),
               median ($q_{0.5}$), mean ($\bar{x}$), third quartile ($q_{0.75}$),
               maximum ($q_{1}$) and standard deviation (sd) of the observed values of Mallows $L^2$-distance $\left(\text{with factor } \frac{1}{10}\right)$ for the variable \texttt{ef25} for the different
               imputation methods and missing rates (r)
               creating missing values using the MAR-mechanism for the employee data (VSE, Mallows, \texttt{ef25}, MAR)}\label{Tab:vse.vals.mallows.MAR.ef25 }\begin{minipage}[t]{\textwidth} 

\centering 
\begin{tabular}[t]{|l|c|c|c|c|c|c|c|c|}\hline
 method&r&$q_{0}$& $q_{0.25}$&$q_{0.5}$&$\bar{x}$&$q_{0.75}$& $q_{1}$&sd\\ \hline 
\hline  &0.01&0.18&0.36&0.51&0.77&1.15&2.54&0.57\\
  &0.05&1.64&3.13&4.03&4.20&5.04&7.55&1.35\\
\multirow{-5}{*}{Amelia}&0.10&5.47&8.20&9.39&9.55&10.55&18.16&2.41\\

\hline  &0.01&0.01&0.02&0.02&0.02&0.03&0.07&0.01\\
  &0.05&0.03&0.06&0.06&0.07&0.09&0.20&0.03\\
\multirow{-5}{*}{Mice.Norm}&0.10&0.08&0.09&0.12&0.13&0.15&0.29&0.05\\

\hline  &0.01&0.17&0.36&0.47&0.75&1.12&2.48&0.53\\
  &0.05&1.70&3.15&3.99&4.05&4.77&6.61&1.18\\
\multirow{-5}{*}{Mice.PMM}&0.10&5.37&7.75&8.67&9.00&10.09&15.54&1.89\\

\hline  &0.01&0.02&0.03&0.03&0.03&0.04&0.07&0.01\\
  &0.05&0.07&0.13&0.18&0.19&0.24&0.45&0.08\\
\multirow{-5}{*}{Mice.RF}&0.10&0.22&0.41&0.59&0.70&0.91&1.81&0.35\\

\hline Naive &0.01&0.51&0.69&0.75&0.75&0.80&0.99&0.09\\
  Imputation&0.05&13.85&15.66&16.32&16.35&17.00&19.64&1.04\\
 &0.10&56.54&61.24&63.36&63.24&65.10&73.33&2.97\\

\hline  &0.01&<0.01&0.01&0.02&0.02&0.03&0.16&0.03\\
  &0.05&0.03&0.07&0.11&0.35&0.20&3.16&0.60\\
\multirow{-5}{*}{MissRanger}&0.10&0.09&0.25&0.39&1.26&1.24&12.17&1.88\\

\hline 
\end{tabular} 
\end{minipage} 
\end{table}

\newpage 
\begin{table}[H] 
\caption{minimum ($q_0$), first quartile ($q_{0.25}$),
               median ($q_{0.5}$), mean ($\bar{x}$), third quartile ($q_{0.75}$),
               maximum ($q_{1}$) and standard deviation (sd) of the observed values of Mallows $L^2$-distance $\left(\text{with factor } \frac{1}{10}\right)$ for the variable \texttt{ef25} for the different
               imputation methods and missing rates (r)
               creating missing values using the MCAR-mechanism for the employee data (VSE, Mallows, \texttt{ef25}, MCAR)}\label{Tab:vse.vals.mallows.MCAR.ef25 }\begin{minipage}[t]{\textwidth} 

\centering 
\begin{tabular}[t]{|l|c|c|c|c|c|c|c|c|}\hline
 method&r&$q_{0}$& $q_{0.25}$&$q_{0.5}$&$\bar{x}$&$q_{0.75}$& $q_{1}$&sd\\ \hline 
\hline  &0.01&0.19&0.35&0.46&0.84&1.30&2.22&0.60\\
  &0.05&1.67&3.42&4.12&4.37&5.18&9.02&1.41\\
\multirow{-5}{*}{Amelia}&0.10&4.87&7.52&8.79&8.93&10.11&14.77&2.04\\

\hline  &0.01&0.01&0.02&0.02&0.02&0.03&0.06&0.01\\
  &0.05&0.03&0.05&0.06&0.07&0.08&0.17&0.03\\
\multirow{-5}{*}{Mice.Norm}&0.10&0.06&0.09&0.11&0.12&0.13&0.30&0.04\\

\hline  &0.01&0.20&0.34&0.47&0.79&1.20&2.19&0.55\\
  &0.05&1.94&3.38&3.92&4.16&4.90&7.89&1.18\\
\multirow{-5}{*}{Mice.PMM}&0.10&5.29&7.26&8.41&8.46&9.41&12.27&1.54\\

\hline  &0.01&0.01&0.03&0.03&0.03&0.04&0.07&0.01\\
  &0.05&0.06&0.13&0.18&0.20&0.23&0.47&0.09\\
\multirow{-5}{*}{Mice.RF}&0.10&0.25&0.46&0.61&0.68&0.76&2.08&0.33\\

\hline Naive &0.01&0.48&0.66&0.71&0.71&0.77&1.00&0.10\\
  Imputation&0.05&13.98&15.70&16.17&16.31&16.99&18.38&0.96\\
 &0.10&55.41&61.58&63.18&63.35&65.25&71.06&2.78\\

\hline  &0.01&<0.01&0.01&0.01&0.02&0.03&0.15&0.03\\
  &0.05&0.03&0.07&0.11&0.34&0.24&3.19&0.62\\
\multirow{-5}{*}{MissRanger}&0.10&0.08&0.26&0.33&1.19&0.72&12.53&2.20\\

\hline 
\end{tabular} 
\end{minipage} 
\end{table}

\newpage 
\begin{table}[H] 
\caption{minimum ($q_0$), first quartile ($q_{0.25}$),
               median ($q_{0.5}$), mean ($\bar{x}$), third quartile ($q_{0.75}$),
               maximum ($q_{1}$) and standard deviation (sd) of the observed values of Mallows $L^2$-distance $\left(\text{with factor } \frac{1}{10}\right)$ for the variable \texttt{ef26} for the different
               imputation methods and missing rates (r)
               creating missing values using the MAR-mechanism for the employee data (VSE, Mallows, \texttt{ef26}, MAR)}\label{Tab:vse.vals.mallows.MAR.ef26 }\begin{minipage}[t]{\textwidth} 

\centering 
\begin{tabular}[t]{|l|c|c|c|c|c|c|c|c|}\hline
 method&r&$q_{0}$& $q_{0.25}$&$q_{0.5}$&$\bar{x}$&$q_{0.75}$& $q_{1}$&sd\\ \hline 
\hline  &0.01&0.70&0.89&0.95&0.96&1.02&1.27&0.11\\
  &0.05&5.29&5.70&5.84&5.87&6.04&6.43&0.25\\
\multirow{-5}{*}{Amelia}&0.10&12.61&13.31&13.57&13.59&13.90&14.57&0.41\\

\hline  &0.01&<0.01&<0.01&0.01&0.01&0.01&0.03&<0.01\\
  &0.05&0.01&0.02&0.03&0.03&0.04&0.11&0.02\\
\multirow{-5}{*}{Mice.Norm}&0.10&0.02&0.04&0.06&0.06&0.07&0.18&0.03\\

\hline  &0.01&0.88&1.00&1.09&1.09&1.16&1.41&0.11\\
  &0.05&5.74&6.53&6.70&6.71&6.88&7.26&0.29\\
\multirow{-5}{*}{Mice.PMM}&0.10&14.36&15.21&15.50&15.50&15.83&16.46&0.43\\

\hline  &0.01&0.01&0.02&0.03&0.03&0.03&0.06&0.01\\
  &0.05&0.17&0.32&0.38&0.40&0.45&0.76&0.11\\
\multirow{-5}{*}{Mice.RF}&0.10&0.94&1.22&1.43&1.44&1.64&2.22&0.27\\

\hline Naive &0.01&0.18&0.23&0.26&0.26&0.28&0.40&0.04\\
  Imputation&0.05&3.13&3.57&3.70&3.70&3.84&4.36&0.23\\
 &0.10&8.69&9.12&9.37&9.34&9.56&10.48&0.33\\

\hline  &0.01&<0.01&0.01&0.02&0.03&0.06&0.11&0.03\\
  &0.05&0.05&0.11&0.22&0.44&0.59&1.87&0.46\\
\multirow{-5}{*}{MissRanger}&0.10&0.19&0.42&0.78&1.44&2.59&5.04&1.27\\

\hline 
\end{tabular} 
\end{minipage} 
\end{table}

\newpage 
\begin{table}[H] 
\caption{minimum ($q_0$), first quartile ($q_{0.25}$),
               median ($q_{0.5}$), mean ($\bar{x}$), third quartile ($q_{0.75}$),
               maximum ($q_{1}$) and standard deviation (sd) of the observed values of Mallows $L^2$-distance $\left(\text{with factor } \frac{1}{10}\right)$ for the variable \texttt{ef26} for the different
               imputation methods and missing rates (r)
               creating missing values using the MCAR-mechanism for the employee data (VSE, Mallows, \texttt{ef26}, MCAR)}\label{Tab:vse.vals.mallows.MCAR.ef26 }\begin{minipage}[t]{\textwidth} 

\centering 
\begin{tabular}[t]{|l|c|c|c|c|c|c|c|c|}\hline
 method&r&$q_{0}$& $q_{0.25}$&$q_{0.5}$&$\bar{x}$&$q_{0.75}$& $q_{1}$&sd\\ \hline 
\hline  &0.01&0.74&0.85&0.93&0.93&0.99&1.15&0.10\\
  &0.05&5.19&5.68&5.86&5.85&6.00&6.56&0.27\\
\multirow{-5}{*}{Amelia}&0.10&12.42&13.25&13.46&13.47&13.75&14.38&0.37\\

\hline  &0.01&<0.01&0.01&0.01&0.01&0.01&0.02&<0.01\\
  &0.05&0.01&0.02&0.03&0.03&0.04&0.08&0.01\\
\multirow{-5}{*}{Mice.Norm}&0.10&0.02&0.03&0.05&0.06&0.07&0.22&0.03\\

\hline  &0.01&0.82&0.97&1.04&1.05&1.12&1.37&0.11\\
  &0.05&6.07&6.57&6.74&6.74&6.96&7.60&0.32\\
\multirow{-5}{*}{Mice.PMM}&0.10&14.36&15.20&15.49&15.44&15.73&16.48&0.41\\

\hline  &0.01&<0.01&0.02&0.02&0.03&0.03&0.06&0.01\\
  &0.05&0.21&0.35&0.43&0.43&0.51&0.84&0.11\\
\multirow{-5}{*}{Mice.RF}&0.10&0.82&1.25&1.42&1.44&1.62&2.12&0.29\\

\hline Naive &0.01&0.17&0.23&0.26&0.26&0.28&0.35&0.04\\
  Imputation&0.05&3.18&3.62&3.75&3.77&3.93&4.35&0.22\\
 &0.10&8.61&9.13&9.27&9.32&9.57&10.08&0.33\\

\hline  &0.01&<0.01&0.01&0.02&0.03&0.05&0.10&0.03\\
  &0.05&0.06&0.11&0.23&0.40&0.51&1.89&0.43\\
\multirow{-5}{*}{MissRanger}&0.10&0.21&0.38&0.91&1.44&2.45&5.08&1.25\\

\hline 
\end{tabular} 
\end{minipage} 
\end{table}

\newpage 
\begin{table}[H] 
\caption{minimum ($q_0$), first quartile ($q_{0.25}$),
               median ($q_{0.5}$), mean ($\bar{x}$), third quartile ($q_{0.75}$),
               maximum ($q_{1}$) and standard deviation (sd) of the observed values of Mallows $L^2$-distance for the variable \texttt{ef29} for the different
               imputation methods and missing rates (r)
               creating missing values using the MAR-mechanism for the employee data (VSE, Mallows, \texttt{ef29}, MAR)}\label{Tab:vse.vals.mallows.MAR.ef29 }\begin{minipage}[t]{\textwidth} 

\centering 
\begin{tabular}[t]{|l|c|c|c|c|c|c|c|c|}\hline
 method&r&$q_{0}$& $q_{0.25}$&$q_{0.5}$&$\bar{x}$&$q_{0.75}$& $q_{1}$&sd\\ \hline 
\hline  &0.01&0.02&0.02&0.02&0.03&0.03&0.04&0.00\\
  &0.05&0.11&0.12&0.12&0.12&0.13&0.15&0.01\\
\multirow{-5}{*}{Amelia}&0.10&0.23&0.25&0.26&0.26&0.27&0.29&0.01\\

\hline  &0.01&0.00&0.01&0.01&0.01&0.01&0.01&0.00\\
  &0.05&0.01&0.01&0.01&0.02&0.02&0.02&0.00\\
\multirow{-5}{*}{Mice.Norm}&0.10&0.01&0.02&0.02&0.02&0.02&0.03&0.00\\

\hline  &0.01&0.02&0.02&0.03&0.03&0.03&0.04&0.00\\
  &0.05&0.11&0.12&0.13&0.13&0.13&0.15&0.01\\
\multirow{-5}{*}{Mice.PMM}&0.10&0.26&0.28&0.28&0.29&0.30&0.31&0.01\\

\hline  &0.01&0.01&0.01&0.02&0.02&0.02&0.03&0.00\\
  &0.05&0.05&0.07&0.08&0.08&0.09&0.11&0.01\\
\multirow{-5}{*}{Mice.RF}&0.10&0.12&0.15&0.16&0.16&0.17&0.22&0.02\\

\hline Naive &0.01&0.05&0.06&0.07&0.07&0.07&0.08&0.01\\
  Imputation&0.05&0.35&0.38&0.39&0.40&0.41&0.47&0.02\\
 &0.10&1.13&1.20&1.23&1.23&1.25&1.34&0.04\\

\hline  &0.01&0.01&0.01&0.01&0.01&0.01&0.04&0.01\\
  &0.05&0.03&0.03&0.04&0.06&0.06&0.20&0.04\\
\multirow{-5}{*}{MissRanger}&0.10&0.06&0.07&0.08&0.11&0.13&0.49&0.08\\

\hline 
\end{tabular} 
\end{minipage} 
\end{table}

\newpage 
\begin{table}[H] 
\caption{minimum ($q_0$), first quartile ($q_{0.25}$),
               median ($q_{0.5}$), mean ($\bar{x}$), third quartile ($q_{0.75}$),
               maximum ($q_{1}$) and standard deviation (sd) of the observed values of Mallows $L^2$-distance for the variable \texttt{ef29} for the different
               imputation methods and missing rates (r)
               creating missing values using the MCAR-mechanism for the employee data (VSE, Mallows, \texttt{ef29}, MCAR)}\label{Tab:vse.vals.mallows.MCAR.ef29 }\begin{minipage}[t]{\textwidth} 

\centering 
\begin{tabular}[t]{|l|c|c|c|c|c|c|c|c|}\hline
 method&r&$q_{0}$& $q_{0.25}$&$q_{0.5}$&$\bar{x}$&$q_{0.75}$& $q_{1}$&sd\\ \hline 
\hline  &0.01&0.02&0.02&0.02&0.02&0.03&0.04&0.00\\
  &0.05&0.10&0.12&0.12&0.12&0.13&0.15&0.01\\
\multirow{-5}{*}{Amelia}&0.10&0.23&0.25&0.26&0.26&0.27&0.30&0.02\\

\hline  &0.01&0.00&0.01&0.01&0.01&0.01&0.01&0.00\\
  &0.05&0.01&0.01&0.01&0.01&0.02&0.02&0.00\\
\multirow{-5}{*}{Mice.Norm}&0.10&0.01&0.02&0.02&0.02&0.02&0.03&0.00\\

\hline  &0.01&0.02&0.02&0.03&0.03&0.03&0.04&0.00\\
  &0.05&0.11&0.12&0.13&0.13&0.14&0.15&0.01\\
\multirow{-5}{*}{Mice.PMM}&0.10&0.25&0.28&0.28&0.28&0.29&0.32&0.01\\

\hline  &0.01&0.01&0.01&0.02&0.02&0.02&0.03&0.00\\
  &0.05&0.05&0.07&0.08&0.08&0.09&0.10&0.01\\
\multirow{-5}{*}{Mice.RF}&0.10&0.12&0.15&0.16&0.16&0.17&0.20&0.02\\

\hline Naive &0.01&0.05&0.06&0.07&0.07&0.07&0.08&0.01\\
  Imputation&0.05&0.35&0.38&0.39&0.39&0.41&0.45&0.02\\
 &0.10&1.10&1.19&1.23&1.23&1.25&1.34&0.05\\

\hline  &0.01&0.00&0.01&0.01&0.01&0.01&0.04&0.01\\
  &0.05&0.02&0.03&0.04&0.06&0.06&0.20&0.04\\
\multirow{-5}{*}{MissRanger}&0.10&0.06&0.07&0.08&0.11&0.13&0.46&0.07\\

\hline 
\end{tabular} 
\end{minipage} 
\end{table}

\newpage 
\begin{table}[H] 
\caption{minimum ($q_0$), first quartile ($q_{0.25}$),
               median ($q_{0.5}$), mean ($\bar{x}$), third quartile ($q_{0.75}$),
               maximum ($q_{1}$) and standard deviation (sd) of the observed values of Mallows $L^2$-distance for the variable \texttt{ef40} for the different
               imputation methods and missing rates (r)
               creating missing values using the MAR-mechanism for the employee data (VSE, Mallows, \texttt{ef40}, MAR)}\label{Tab:vse.vals.mallows.MAR.ef40 }\begin{minipage}[t]{\textwidth} 

\centering 
\begin{tabular}[t]{|l|c|c|c|c|c|c|c|c|}\hline
 method&r&$q_{0}$& $q_{0.25}$&$q_{0.5}$&$\bar{x}$&$q_{0.75}$& $q_{1}$&sd\\ \hline 
\hline  &0.01&0.06&0.08&0.09&0.09&0.10&0.13&0.01\\
  &0.05&0.38&0.44&0.46&0.46&0.48&0.53&0.03\\
\multirow{-5}{*}{Amelia}&0.10&0.82&0.88&0.93&0.93&0.97&1.02&0.05\\

\hline  &0.01&0.01&0.01&0.01&0.01&0.01&0.02&0.00\\
  &0.05&0.01&0.02&0.02&0.02&0.02&0.03&0.00\\
\multirow{-5}{*}{Mice.Norm}&0.10&0.02&0.02&0.03&0.03&0.03&0.06&0.01\\

\hline  &0.01&0.08&0.10&0.12&0.12&0.13&0.16&0.02\\
  &0.05&0.45&0.55&0.58&0.57&0.60&0.67&0.04\\
\multirow{-5}{*}{Mice.PMM}&0.10&1.03&1.15&1.19&1.19&1.23&1.34&0.06\\

\hline  &0.01&0.01&0.01&0.01&0.01&0.01&0.02&0.00\\
  &0.05&0.02&0.03&0.04&0.04&0.05&0.08&0.01\\
\multirow{-5}{*}{Mice.RF}&0.10&0.03&0.07&0.08&0.08&0.09&0.16&0.02\\

\hline Naive &0.01&0.08&0.08&0.09&0.09&0.09&0.11&0.01\\
  Imputation&0.05&0.41&0.44&0.46&0.45&0.47&0.49&0.02\\
 &0.10&1.32&1.40&1.43&1.44&1.47&1.57&0.05\\

\hline  &0.01&0.01&0.02&0.02&0.02&0.03&0.05&0.01\\
  &0.05&0.07&0.09&0.11&0.12&0.14&0.25&0.04\\
\multirow{-5}{*}{MissRanger}&0.10&0.15&0.18&0.21&0.23&0.24&0.46&0.06\\

\hline 
\end{tabular} 
\end{minipage} 
\end{table}

\newpage 
\begin{table}[H] 
\caption{minimum ($q_0$), first quartile ($q_{0.25}$),
               median ($q_{0.5}$), mean ($\bar{x}$), third quartile ($q_{0.75}$),
               maximum ($q_{1}$) and standard deviation (sd) of the observed values of Mallows $L^2$-distance for the variable \texttt{ef40} for the different
               imputation methods and missing rates (r)
               creating missing values using the MCAR-mechanism for the employee data (VSE, Mallows, \texttt{ef40}, MCAR)}\label{Tab:vse.vals.mallows.MCAR.ef40 }\begin{minipage}[t]{\textwidth} 

\centering 
\begin{tabular}[t]{|l|c|c|c|c|c|c|c|c|}\hline
 method&r&$q_{0}$& $q_{0.25}$&$q_{0.5}$&$\bar{x}$&$q_{0.75}$& $q_{1}$&sd\\ \hline 
\hline  &0.01&0.07&0.08&0.09&0.09&0.10&0.13&0.01\\
  &0.05&0.40&0.44&0.47&0.46&0.48&0.55&0.03\\
\multirow{-5}{*}{Amelia}&0.10&0.81&0.90&0.93&0.93&0.96&1.03&0.05\\

\hline  &0.01&0.01&0.01&0.01&0.01&0.01&0.01&0.00\\
  &0.05&0.01&0.02&0.02&0.02&0.02&0.03&0.00\\
\multirow{-5}{*}{Mice.Norm}&0.10&0.02&0.02&0.03&0.03&0.03&0.05&0.01\\

\hline  &0.01&0.08&0.10&0.12&0.12&0.13&0.16&0.02\\
  &0.05&0.52&0.56&0.59&0.58&0.60&0.66&0.03\\
\multirow{-5}{*}{Mice.PMM}&0.10&1.07&1.14&1.18&1.18&1.21&1.33&0.06\\

\hline  &0.01&0.01&0.01&0.01&0.01&0.01&0.02&0.00\\
  &0.05&0.02&0.04&0.05&0.05&0.05&0.08&0.01\\
\multirow{-5}{*}{Mice.RF}&0.10&0.04&0.07&0.08&0.08&0.10&0.15&0.02\\

\hline Naive &0.01&0.07&0.09&0.09&0.09&0.09&0.11&0.01\\
  Imputation&0.05&0.43&0.45&0.46&0.46&0.47&0.50&0.02\\
 &0.10&1.26&1.40&1.43&1.43&1.47&1.56&0.06\\

\hline  &0.01&0.01&0.02&0.02&0.02&0.03&0.05&0.01\\
  &0.05&0.06&0.09&0.10&0.11&0.12&0.22&0.03\\
\multirow{-5}{*}{MissRanger}&0.10&0.15&0.17&0.20&0.23&0.25&0.52&0.08\\

\hline 
\end{tabular} 
\end{minipage} 
\end{table}

\newpage 
\begin{table}[H] 
\caption{minimum ($q_0$), first quartile ($q_{0.25}$),
               median ($q_{0.5}$), mean ($\bar{x}$), third quartile ($q_{0.75}$),
               maximum ($q_{1}$) and standard deviation (sd) of the observed values of Mallows $L^2$-distance for the variable \texttt{ef41} for the different
               imputation methods and missing rates (r)
               creating missing values using the MAR-mechanism for the employee data (VSE, Mallows, \texttt{ef41}, MAR)}\label{Tab:vse.vals.mallows.MAR.ef41 }\begin{minipage}[t]{\textwidth} 

\centering 
\begin{tabular}[t]{|l|c|c|c|c|c|c|c|c|}\hline
 method&r&$q_{0}$& $q_{0.25}$&$q_{0.5}$&$\bar{x}$&$q_{0.75}$& $q_{1}$&sd\\ \hline 
\hline  &0.01&0.02&0.03&0.03&0.03&0.03&0.05&0.01\\
  &0.05&0.11&0.13&0.15&0.14&0.16&0.18&0.02\\
\multirow{-5}{*}{Amelia}&0.10&0.24&0.27&0.28&0.29&0.30&0.33&0.02\\

\hline  &0.01&0.01&0.01&0.01&0.01&0.01&0.02&0.00\\
  &0.05&0.02&0.02&0.02&0.02&0.03&0.05&0.01\\
\multirow{-5}{*}{Mice.Norm}&0.10&0.02&0.03&0.03&0.03&0.04&0.06&0.01\\

\hline  &0.01&0.02&0.03&0.04&0.04&0.04&0.06&0.01\\
  &0.05&0.13&0.15&0.17&0.17&0.18&0.21&0.02\\
\multirow{-5}{*}{Mice.PMM}&0.10&0.27&0.32&0.33&0.33&0.35&0.38&0.02\\

\hline  &0.01&0.01&0.01&0.01&0.01&0.01&0.02&0.00\\
  &0.05&0.02&0.03&0.03&0.03&0.04&0.05&0.01\\
\multirow{-5}{*}{Mice.RF}&0.10&0.03&0.05&0.05&0.05&0.06&0.10&0.01\\

\hline Naive &0.01&0.09&0.10&0.10&0.10&0.11&0.12&0.01\\
  Imputation&0.05&0.48&0.50&0.51&0.51&0.53&0.55&0.02\\
 &0.10&1.43&1.51&1.54&1.54&1.57&1.69&0.05\\

\hline  &0.01&0.02&0.03&0.03&0.04&0.04&0.07&0.01\\
  &0.05&0.12&0.14&0.16&0.18&0.20&0.33&0.05\\
\multirow{-5}{*}{MissRanger}&0.10&0.26&0.30&0.34&0.38&0.42&0.77&0.11\\

\hline 
\end{tabular} 
\end{minipage} 
\end{table}

\newpage 
\begin{table}[H] 
\caption{minimum ($q_0$), first quartile ($q_{0.25}$),
               median ($q_{0.5}$), mean ($\bar{x}$), third quartile ($q_{0.75}$),
               maximum ($q_{1}$) and standard deviation (sd) of the observed values of Mallows $L^2$-distance for the variable \texttt{ef41} for the different
               imputation methods and missing rates (r)
               creating missing values using the MCAR-mechanism for the employee data (VSE, Mallows, \texttt{ef41}, MCAR)}\label{Tab:vse.vals.mallows.MCAR.ef41 }\begin{minipage}[t]{\textwidth} 

\centering 
\begin{tabular}[t]{|l|c|c|c|c|c|c|c|c|}\hline
 method&r&$q_{0}$& $q_{0.25}$&$q_{0.5}$&$\bar{x}$&$q_{0.75}$& $q_{1}$&sd\\ \hline 
\hline  &0.01&0.02&0.02&0.03&0.03&0.03&0.05&0.01\\
  &0.05&0.12&0.13&0.14&0.15&0.16&0.19&0.02\\
\multirow{-5}{*}{Amelia}&0.10&0.24&0.28&0.29&0.29&0.31&0.35&0.02\\

\hline  &0.01&0.01&0.01&0.01&0.01&0.01&0.01&0.00\\
  &0.05&0.02&0.02&0.02&0.02&0.03&0.05&0.01\\
\multirow{-5}{*}{Mice.Norm}&0.10&0.02&0.03&0.03&0.03&0.04&0.07&0.01\\

\hline  &0.01&0.02&0.03&0.03&0.03&0.04&0.05&0.01\\
  &0.05&0.12&0.15&0.17&0.17&0.18&0.22&0.02\\
\multirow{-5}{*}{Mice.PMM}&0.10&0.28&0.31&0.33&0.33&0.35&0.40&0.03\\

\hline  &0.01&0.01&0.01&0.01&0.01&0.01&0.02&0.00\\
  &0.05&0.02&0.03&0.03&0.03&0.04&0.06&0.01\\
\multirow{-5}{*}{Mice.RF}&0.10&0.03&0.05&0.06&0.06&0.06&0.09&0.01\\

\hline Naive &0.01&0.08&0.10&0.10&0.10&0.11&0.12&0.01\\
  Imputation&0.05&0.47&0.51&0.52&0.52&0.53&0.57&0.02\\
 &0.10&1.43&1.52&1.54&1.55&1.57&1.67&0.05\\

\hline  &0.01&0.02&0.03&0.03&0.04&0.04&0.06&0.01\\
  &0.05&0.12&0.14&0.17&0.18&0.20&0.33&0.04\\
\multirow{-5}{*}{MissRanger}&0.10&0.26&0.31&0.36&0.37&0.40&0.72&0.08\\

\hline 
\end{tabular} 
\end{minipage} 
\end{table}

\newpage 
\begin{table}[H] 
\caption{minimum ($q_0$), first quartile ($q_{0.25}$),
               median ($q_{0.5}$), mean ($\bar{x}$), third quartile ($q_{0.75}$),
               maximum ($q_{1}$) and standard deviation (sd) of the observed values of Mallows $L^2$-distance $\left(\text{with factor } \frac{1}{10}\right)$ for the variable \texttt{ef44} for the different
               imputation methods and missing rates (r)
               creating missing values using the MAR-mechanism for the employee data (VSE, Mallows, \texttt{ef44}, MAR)}\label{Tab:vse.vals.mallows.MAR.ef44 }\begin{minipage}[t]{\textwidth} 

\centering 
\begin{tabular}[t]{|l|c|c|c|c|c|c|c|c|}\hline
 method&r&$q_{0}$& $q_{0.25}$&$q_{0.5}$&$\bar{x}$&$q_{0.75}$& $q_{1}$&sd\\ \hline 
\hline  &0.01&0.14&0.31&0.45&0.48&0.58&1.54&0.24\\
  &0.05&2.21&3.04&3.48&3.80&4.37&7.45&1.08\\
\multirow{-5}{*}{Amelia}&0.10&6.97&8.89&9.82&10.15&11.27&20.80&2.03\\

\hline  &0.01&0.04&0.06&0.09&0.15&0.16&1.45&0.17\\
  &0.05&0.17&0.33&0.59&0.66&0.81&1.80&0.42\\
\multirow{-5}{*}{Mice.Norm}&0.10&0.49&0.89&1.24&1.29&1.63&3.26&0.53\\

\hline  &0.01&0.09&0.22&0.28&0.33&0.37&1.63&0.19\\
  &0.05&1.06&1.58&1.81&1.92&2.10&3.60&0.55\\
\multirow{-5}{*}{Mice.PMM}&0.10&2.91&4.10&4.47&4.66&5.23&7.14&0.92\\

\hline  &0.01&0.09&0.18&0.25&0.30&0.36&1.56&0.19\\
  &0.05&0.49&1.32&1.69&1.79&2.13&3.54&0.68\\
\multirow{-5}{*}{Mice.RF}&0.10&1.96&3.40&4.20&4.38&5.13&8.04&1.41\\

\hline Naive &0.01&4.60&6.30&6.97&7.00&7.64&10.67&1.10\\
  Imputation&0.05&145.27&159.45&164.24&165.23&171.17&190.71&9.28\\
 &0.10&572.76&644.00&664.60&663.72&681.16&742.03&28.63\\

\hline  &0.01&0.02&0.03&0.05&0.13&0.13&1.87&0.22\\
  &0.05&0.10&0.30&0.62&1.43&1.34&20.26&2.63\\
\multirow{-5}{*}{MissRanger}&0.10&0.29&1.11&2.30&7.72&6.15&109.05&16.90\\

\hline 
\end{tabular} 
\end{minipage} 
\end{table}

\newpage 
\begin{table}[H] 
\caption{minimum ($q_0$), first quartile ($q_{0.25}$),
               median ($q_{0.5}$), mean ($\bar{x}$), third quartile ($q_{0.75}$),
               maximum ($q_{1}$) and standard deviation (sd) of the observed values of Mallows $L^2$-distance $\left(\text{with factor } \frac{1}{10}\right)$ for the variable \texttt{ef44} for the different
               imputation methods and missing rates (r)
               creating missing values using the MCAR-mechanism for the employee data (VSE, Mallows, \texttt{ef44}, MCAR)}\label{Tab:vse.vals.mallows.MCAR.ef44 }\begin{minipage}[t]{\textwidth} 

\centering 
\begin{tabular}[t]{|l|c|c|c|c|c|c|c|c|}\hline
 method&r&$q_{0}$& $q_{0.25}$&$q_{0.5}$&$\bar{x}$&$q_{0.75}$& $q_{1}$&sd\\ \hline 
\hline  &0.01&0.12&0.32&0.45&0.56&0.65&2.38&0.38\\
  &0.05&2.14&2.89&3.31&3.58&3.99&6.54&1.00\\
\multirow{-5}{*}{Amelia}&0.10&6.88&8.69&9.81&10.23&10.82&22.02&2.52\\

\hline  &0.01&0.04&0.06&0.08&0.17&0.16&1.51&0.26\\
  &0.05&0.19&0.36&0.50&0.63&0.74&2.79&0.40\\
\multirow{-5}{*}{Mice.Norm}&0.10&0.49&0.98&1.24&1.42&1.79&3.04&0.60\\

\hline  &0.01&0.09&0.21&0.29&0.35&0.39&1.92&0.29\\
  &0.05&1.08&1.59&1.80&1.91&2.08&4.94&0.58\\
\multirow{-5}{*}{Mice.PMM}&0.10&2.80&4.02&4.47&4.71&5.38&7.41&0.98\\

\hline  &0.01&0.11&0.20&0.24&0.32&0.32&1.77&0.27\\
  &0.05&0.71&1.17&1.54&1.64&1.93&3.82&0.64\\
\multirow{-5}{*}{Mice.RF}&0.10&1.68&3.17&4.15&4.27&4.98&9.19&1.51\\

\hline Naive &0.01&4.22&6.38&7.03&7.01&7.74&10.09&1.08\\
  Imputation&0.05&139.66&158.72&165.94&165.47&172.91&192.45&10.69\\
 &0.10&592.40&647.14&662.43&665.04&683.14&729.58&25.52\\

\hline  &0.01&0.01&0.03&0.06&0.15&0.14&2.93&0.36\\
  &0.05&0.10&0.29&0.51&1.54&1.40&21.83&2.86\\
\multirow{-5}{*}{MissRanger}&0.10&0.53&1.06&2.07&4.72&3.98&43.27&7.63\\

\hline 
\end{tabular} 
\end{minipage} 
\end{table}

\newpage 
\begin{table}[H] 
\caption{minimum ($q_0$), first quartile ($q_{0.25}$),
               median ($q_{0.5}$), mean ($\bar{x}$), third quartile ($q_{0.75}$),
               maximum ($q_{1}$) and standard deviation (sd) of the observed values of Mallows $L^2$-distance $\left(\text{with factor } \frac{1}{10000}\right)$ for the variable \texttt{ef45} for the different
               imputation methods and missing rates (r)
               creating missing values using the MAR-mechanism for the employee data (VSE, Mallows, \texttt{ef45}, MAR)}\label{Tab:vse.vals.mallows.MAR.ef45 }\begin{minipage}[t]{\textwidth} 

\centering 
\begin{tabular}[t]{|l|c|c|c|c|c|c|c|c|}\hline
 method&r&$q_{0}$& $q_{0.25}$&$q_{0.5}$&$\bar{x}$&$q_{0.75}$& $q_{1}$&sd\\ \hline 
\hline  &0.01&0.14&0.34&0.41&0.48&0.50&3.22&0.36\\
  &0.05&1.47&2.30&2.49&2.70&2.83&5.88&0.75\\
\multirow{-5}{*}{Amelia}&0.10&4.89&6.10&6.48&6.76&7.05&10.91&1.12\\

\hline  &0.01&0.01&0.01&0.01&0.02&0.02&0.03&<0.01\\
  &0.05&0.04&0.05&0.06&0.07&0.07&0.16&0.02\\
\multirow{-5}{*}{Mice.Norm}&0.10&0.08&0.12&0.14&0.15&0.16&0.42&0.05\\

\hline  &0.01&0.09&0.19&0.25&0.31&0.32&4.99&0.49\\
  &0.05&1.10&1.49&1.64&1.96&1.86&6.51&1.16\\
\multirow{-5}{*}{Mice.PMM}&0.10&3.55&4.39&4.69&5.13&5.14&10.01&1.38\\

\hline  &0.01&0.04&0.08&0.10&0.11&0.13&0.25&0.04\\
  &0.05&0.46&0.99&1.25&1.29&1.44&2.70&0.42\\
\multirow{-5}{*}{Mice.RF}&0.10&2.01&3.65&4.53&4.81&5.52&12.52&1.78\\

\hline Naive &0.01&2.40&3.01&3.41&3.43&3.74&4.81&0.54\\
  Imputation&0.05&72.00&80.34&82.90&83.63&86.87&95.38&5.12\\
 &0.10&302.40&329.62&339.84&340.49&351.35&377.81&15.68\\

\hline  &0.01&0.01&0.01&0.01&0.04&0.03&0.44&0.07\\
  &0.05&0.06&0.12&0.16&0.69&0.40&11.23&1.61\\
\multirow{-5}{*}{MissRanger}&0.10&0.21&0.36&0.48&1.34&0.99&18.80&2.60\\

\hline 
\end{tabular} 
\end{minipage} 
\end{table}

\newpage 
\begin{table}[H] 
\caption{minimum ($q_0$), first quartile ($q_{0.25}$),
               median ($q_{0.5}$), mean ($\bar{x}$), third quartile ($q_{0.75}$),
               maximum ($q_{1}$) and standard deviation (sd) of the observed values of Mallows $L^2$-distance $\left(\text{with factor } \frac{1}{10000}\right)$ for the variable \texttt{ef45} for the different
               imputation methods and missing rates (r)
               creating missing values using the MCAR-mechanism for the employee data (VSE, Mallows, \texttt{ef45}, MCAR)}\label{Tab:vse.vals.mallows.MCAR.ef45 }\begin{minipage}[t]{\textwidth} 

\centering 
\begin{tabular}[t]{|l|c|c|c|c|c|c|c|c|}\hline
 method&r&$q_{0}$& $q_{0.25}$&$q_{0.5}$&$\bar{x}$&$q_{0.75}$& $q_{1}$&sd\\ \hline 
\hline  &0.01&0.18&0.28&0.38&0.47&0.48&3.38&0.45\\
  &0.05&1.65&2.28&2.54&2.66&2.85&6.21&0.67\\
\multirow{-5}{*}{Amelia}&0.10&4.69&6.00&6.58&6.65&7.11&10.71&1.06\\

\hline  &0.01&0.01&0.01&0.01&0.02&0.02&0.04&<0.01\\
  &0.05&0.05&0.07&0.08&0.09&0.09&0.24&0.03\\
\multirow{-5}{*}{Mice.Norm}&0.10&0.09&0.13&0.15&0.16&0.18&0.28&0.04\\

\hline  &0.01&0.10&0.17&0.22&0.32&0.27&4.39&0.59\\
  &0.05&1.18&1.52&1.71&1.94&1.98&7.27&1.01\\
\multirow{-5}{*}{Mice.PMM}&0.10&3.30&4.43&4.85&5.22&5.29&10.22&1.43\\

\hline  &0.01&0.04&0.07&0.10&0.11&0.12&0.31&0.05\\
  &0.05&0.36&0.89&1.20&1.27&1.57&3.27&0.54\\
\multirow{-5}{*}{Mice.RF}&0.10&1.20&2.92&3.88&4.30&5.45&9.18&1.82\\

\hline Naive &0.01&2.51&3.12&3.48&3.44&3.76&4.68&0.46\\
  Imputation&0.05&72.44&81.80&85.21&85.61&89.04&102.32&5.85\\
 &0.10&303.07&334.05&343.20&342.75&351.86&381.93&15.44\\

\hline  &0.01&<0.01&0.01&0.02&0.05&0.03&0.64&0.10\\
  &0.05&0.06&0.12&0.20&0.87&0.43&11.94&2.19\\
\multirow{-5}{*}{MissRanger}&0.10&0.23&0.50&0.64&2.08&1.28&26.27&4.09\\

\hline 
\end{tabular} 
\end{minipage} 
\end{table}

\newpage 
\begin{table}[H] 
\caption{minimum ($q_0$), first quartile ($q_{0.25}$),
               median ($q_{0.5}$), mean ($\bar{x}$), third quartile ($q_{0.75}$),
               maximum ($q_{1}$) and standard deviation (sd) of the observed values of Mallows $L^2$-distance $\left(\text{with factor } \frac{1}{10000}\right)$ for the variable \texttt{ef47} for the different
               imputation methods and missing rates (r)
               creating missing values using the MAR-mechanism for the employee data (VSE, Mallows, \texttt{ef47}, MAR)}\label{Tab:vse.vals.mallows.MAR.ef47 }\begin{minipage}[t]{\textwidth} 

\centering 
\begin{tabular}[t]{|l|c|c|c|c|c|c|c|c|}\hline
 method&r&$q_{0}$& $q_{0.25}$&$q_{0.5}$&$\bar{x}$&$q_{0.75}$& $q_{1}$&sd\\ \hline 
\hline  &0.01&0.38&0.50&0.54&0.57&0.61&1.09&0.12\\
  &0.05&2.36&2.86&3.01&3.06&3.21&4.16&0.31\\
\multirow{-5}{*}{Amelia}&0.10&5.51&5.98&6.28&6.30&6.53&8.13&0.47\\

\hline  &0.01&<0.01&<0.01&0.01&0.04&0.01&0.45&0.09\\
  &0.05&0.01&0.03&0.10&0.19&0.30&1.12&0.21\\
\multirow{-5}{*}{Mice.Norm}&0.10&0.03&0.17&0.27&0.34&0.46&1.64&0.27\\

\hline  &0.01&0.41&0.58&0.66&0.70&0.76&1.39&0.17\\
  &0.05&3.03&3.40&3.56&3.67&3.92&5.09&0.39\\
\multirow{-5}{*}{Mice.PMM}&0.10&6.71&7.22&7.59&7.66&8.06&10.07&0.58\\

\hline  &0.01&<0.01&0.01&0.01&0.04&0.02&0.53&0.08\\
  &0.05&0.02&0.11&0.19&0.24&0.31&1.17&0.21\\
\multirow{-5}{*}{Mice.RF}&0.10&0.06&0.31&0.44&0.49&0.55&2.10&0.30\\

\hline Naive &0.01&0.04&0.05&0.06&0.07&0.06&0.58&0.08\\
  Imputation&0.05&1.12&1.24&1.33&1.43&1.51&2.62&0.28\\
 &0.10&4.47&5.00&5.21&5.31&5.50&7.29&0.51\\

\hline  &0.01&<0.01&<0.01&0.01&0.03&0.01&0.54&0.08\\
  &0.05&0.04&0.09&0.15&0.27&0.42&1.28&0.27\\
\multirow{-5}{*}{MissRanger}&0.10&0.18&0.36&0.51&0.70&0.87&2.49&0.49\\

\hline 
\end{tabular} 
\end{minipage} 
\end{table}

\newpage 
\begin{table}[H] 
\caption{minimum ($q_0$), first quartile ($q_{0.25}$),
               median ($q_{0.5}$), mean ($\bar{x}$), third quartile ($q_{0.75}$),
               maximum ($q_{1}$) and standard deviation (sd) of the observed values of Mallows $L^2$-distance $\left(\text{with factor } \frac{1}{10000}\right)$ for the variable \texttt{ef47} for the different
               imputation methods and missing rates (r)
               creating missing values using the MCAR-mechanism for the employee data (VSE, Mallows, \texttt{ef47}, MCAR)}\label{Tab:vse.vals.mallows.MCAR.ef47 }\begin{minipage}[t]{\textwidth} 

\centering 
\begin{tabular}[t]{|l|c|c|c|c|c|c|c|c|}\hline
 method&r&$q_{0}$& $q_{0.25}$&$q_{0.5}$&$\bar{x}$&$q_{0.75}$& $q_{1}$&sd\\ \hline 
\hline  &0.01&0.39&0.50&0.56&0.58&0.63&1.07&0.11\\
  &0.05&2.42&2.76&2.94&2.96&3.12&4.17&0.29\\
\multirow{-5}{*}{Amelia}&0.10&5.47&6.00&6.24&6.35&6.58&8.70&0.52\\

\hline  &0.01&<0.01&<0.01&0.01&0.02&0.01&0.36&0.06\\
  &0.05&0.01&0.03&0.11&0.17&0.24&1.10&0.18\\
\multirow{-5}{*}{Mice.Norm}&0.10&0.02&0.18&0.27&0.35&0.40&3.12&0.37\\

\hline  &0.01&0.41&0.57&0.64&0.67&0.70&1.36&0.15\\
  &0.05&2.78&3.32&3.52&3.52&3.65&4.98&0.34\\
\multirow{-5}{*}{Mice.PMM}&0.10&6.67&7.29&7.60&7.64&7.82&10.23&0.56\\

\hline  &0.01&<0.01&<0.01&0.01&0.05&0.08&0.42&0.08\\
  &0.05&0.01&0.08&0.14&0.19&0.24&1.18&0.19\\
\multirow{-5}{*}{Mice.RF}&0.10&0.07&0.30&0.39&0.48&0.54&3.27&0.40\\

\hline Naive &0.01&0.04&0.05&0.06&0.07&0.06&0.47&0.07\\
  Imputation&0.05&1.01&1.21&1.30&1.35&1.40&2.56&0.24\\
 &0.10&4.78&5.03&5.32&5.38&5.56&8.33&0.54\\

\hline  &0.01&<0.01&<0.01&0.01&0.02&0.01&0.44&0.07\\
  &0.05&0.03&0.08&0.13&0.21&0.23&1.29&0.22\\
\multirow{-5}{*}{MissRanger}&0.10&0.18&0.35&0.57&0.74&0.86&3.80&0.59\\

\hline 
\end{tabular} 
\end{minipage} 
\end{table}

\newpage 
\begin{table}[H] 
\caption{minimum ($q_0$), first quartile ($q_{0.25}$),
               median ($q_{0.5}$), mean ($\bar{x}$), third quartile ($q_{0.75}$),
               maximum ($q_{1}$) and standard deviation (sd) of the observed values of Mallows $L^2$-distance for the variable \texttt{ef53} for the different
               imputation methods and missing rates (r)
               creating missing values using the MAR-mechanism for the employee data (VSE, Mallows, \texttt{ef53}, MAR)}\label{Tab:vse.vals.mallows.MAR.ef53 }\begin{minipage}[t]{\textwidth} 

\centering 
\begin{tabular}[t]{|l|c|c|c|c|c|c|c|c|}\hline
 method&r&$q_{0}$& $q_{0.25}$&$q_{0.5}$&$\bar{x}$&$q_{0.75}$& $q_{1}$&sd\\ \hline 
\hline  &0.01&0.01&0.01&0.01&0.01&0.01&0.01&0.00\\
  &0.05&0.03&0.04&0.04&0.04&0.04&0.04&0.00\\
\multirow{-5}{*}{Amelia}&0.10&0.07&0.08&0.08&0.08&0.08&0.09&0.00\\

\hline  &0.01&0.00&0.00&0.00&0.00&0.00&0.01&0.00\\
  &0.05&0.01&0.01&0.01&0.01&0.01&0.02&0.00\\
\multirow{-5}{*}{Mice.Norm}&0.10&0.01&0.01&0.01&0.01&0.02&0.02&0.00\\

\hline  &0.01&0.01&0.01&0.01&0.01&0.02&0.02&0.00\\
  &0.05&0.07&0.07&0.08&0.08&0.08&0.09&0.00\\
\multirow{-5}{*}{Mice.PMM}&0.10&0.17&0.19&0.19&0.19&0.21&0.23&0.01\\

\hline  &0.01&0.00&0.01&0.01&0.01&0.01&0.01&0.00\\
  &0.05&0.02&0.04&0.04&0.04&0.05&0.07&0.01\\
\multirow{-5}{*}{Mice.RF}&0.10&0.07&0.10&0.11&0.11&0.13&0.20&0.03\\

\hline Naive &0.01&0.08&0.09&0.10&0.10&0.10&0.12&0.01\\
  Imputation&0.05&0.80&0.85&0.88&0.88&0.90&1.01&0.04\\
 &0.10&2.08&2.30&2.34&2.34&2.40&2.57&0.08\\

\hline  &0.01&0.00&0.00&0.00&0.00&0.00&0.02&0.00\\
  &0.05&0.00&0.01&0.01&0.02&0.02&0.15&0.03\\
\multirow{-5}{*}{MissRanger}&0.10&0.01&0.01&0.02&0.04&0.05&0.28&0.05\\

\hline 
\end{tabular} 
\end{minipage} 
\end{table}

\newpage 
\begin{table}[H] 
\caption{minimum ($q_0$), first quartile ($q_{0.25}$),
               median ($q_{0.5}$), mean ($\bar{x}$), third quartile ($q_{0.75}$),
               maximum ($q_{1}$) and standard deviation (sd) of the observed values of Mallows $L^2$-distance for the variable \texttt{ef53} for the different
               imputation methods and missing rates (r)
               creating missing values using the MCAR-mechanism for the employee data (VSE, Mallows, \texttt{ef53}, MCAR)}\label{Tab:vse.vals.mallows.MCAR.ef53 }\begin{minipage}[t]{\textwidth} 

\centering 
\begin{tabular}[t]{|l|c|c|c|c|c|c|c|c|}\hline
 method&r&$q_{0}$& $q_{0.25}$&$q_{0.5}$&$\bar{x}$&$q_{0.75}$& $q_{1}$&sd\\ \hline 
\hline  &0.01&0.01&0.01&0.01&0.01&0.01&0.01&0.00\\
  &0.05&0.03&0.04&0.04&0.04&0.04&0.05&0.00\\
\multirow{-5}{*}{Amelia}&0.10&0.07&0.08&0.08&0.08&0.08&0.09&0.00\\

\hline  &0.01&0.00&0.00&0.00&0.00&0.00&0.01&0.00\\
  &0.05&0.01&0.01&0.01&0.01&0.01&0.01&0.00\\
\multirow{-5}{*}{Mice.Norm}&0.10&0.01&0.01&0.01&0.01&0.02&0.02&0.00\\

\hline  &0.01&0.01&0.01&0.01&0.01&0.02&0.02&0.00\\
  &0.05&0.07&0.07&0.08&0.08&0.08&0.09&0.00\\
\multirow{-5}{*}{Mice.PMM}&0.10&0.17&0.19&0.20&0.20&0.20&0.22&0.01\\

\hline  &0.01&0.00&0.01&0.01&0.01&0.01&0.01&0.00\\
  &0.05&0.02&0.04&0.04&0.04&0.05&0.06&0.01\\
\multirow{-5}{*}{Mice.RF}&0.10&0.06&0.09&0.11&0.11&0.12&0.15&0.02\\

\hline Naive &0.01&0.08&0.09&0.10&0.10&0.10&0.12&0.01\\
  Imputation&0.05&0.78&0.86&0.88&0.88&0.90&1.03&0.04\\
 &0.10&2.15&2.30&2.35&2.34&2.40&2.55&0.09\\

\hline  &0.01&0.00&0.00&0.00&0.00&0.01&0.03&0.01\\
  &0.05&0.00&0.01&0.01&0.02&0.02&0.12&0.02\\
\multirow{-5}{*}{MissRanger}&0.10&0.01&0.01&0.02&0.04&0.04&0.42&0.05\\

\hline 
\end{tabular} 
\end{minipage} 
\end{table}

\newpage

%% file: table_fivepoint_Chisq_drg.txt
\begin{table}[H] 
\caption{minimum ($q_0$), first quartile ($q_{0.25}$),
               median ($q_{0.5}$), mean ($\bar{x}$), third quartile ($q_{0.75}$),
               maximum ($q_{1}$) and standard deviation (sd) of the observed values of the realization of Pearson's chi-squared statistic $\left(\text{with factor} \frac{1}{1000}\right)$ for the variable \texttt{aufn\_anl} for the different
               imputation methods and missing rates (r)
               creating missing values for the hospital data (DRG, $\chi^2$, \texttt{aufn\_anl}, MCAR)}\label{Tab:drg.vals.chisq.aufnanl }\begin{minipage}[t]{\textwidth} 

\centering 
\begin{tabular}[t]{|l|c|c|c|c|c|c|c|c|}\hline
 method&r&$q_{0}$& $q_{0.25}$&$q_{0.5}$&$\bar{x}$&$q_{0.75}$& $q_{1}$&sd\\
\hline 
\hline 
&0.01&29.24&29.44&29.48&29.48&29.52&29.69&0.07\\
  &0.05&27.16&27.38&27.46&27.47&27.56&27.85&0.16\\
  &0.10&24.72&25.02&25.16&25.15&25.28&25.68&0.19\\
  &0.15&22.37&22.92&23.03&23.04&23.20&23.64&0.23\\
\multirow{-5}{*}{Amelia}&0.20&20.35&20.91&21.04&21.03&21.20&21.52&0.24\\

\hline  &0.01&29.46&29.60&29.64&29.64&29.67&29.78&0.06\\
  &0.05&27.81&28.11&28.20&28.19&28.29&28.50&0.13\\
  &0.10&26.04&26.30&26.42&26.42&26.55&26.89&0.18\\
  &0.15&24.14&24.59&24.69&24.69&24.80&25.12&0.19\\
\multirow{-5}{*}{Mice.Norm}&0.20&22.43&22.82&23.00&22.99&23.15&23.49&0.26\\

\hline  &0.01&29.46&29.60&29.64&29.64&29.67&29.79&0.06\\
  &0.05&27.77&28.12&28.21&28.19&28.28&28.46&0.13\\
  &0.10&25.91&26.28&26.41&26.41&26.53&26.82&0.19\\
  &0.15&24.24&24.55&24.69&24.70&24.86&25.28&0.22\\
\multirow{-5}{*}{Mice.PMM}&0.20&22.50&22.84&22.98&23.00&23.20&23.57&0.25\\

\hline  &0.01&29.45&29.60&29.64&29.64&29.67&29.78&0.06\\
  &0.05&27.82&28.13&28.22&28.19&28.26&28.40&0.12\\
  &0.10&25.98&26.29&26.42&26.41&26.52&26.92&0.19\\
  &0.15&24.21&24.57&24.72&24.71&24.85&25.30&0.21\\
\multirow{-5}{*}{Mice.RF}&0.20&22.48&22.80&23.03&23.01&23.18&23.51&0.25\\

\hline Naive &0.01&29.41&29.55&29.61&29.60&29.65&29.79&0.08\\
  Imputation&0.05&27.67&27.97&28.10&28.09&28.22&28.36&0.16\\
  &0.10&25.71&26.12&26.24&26.26&26.43&26.76&0.24\\
  &0.15&23.90&24.38&24.53&24.53&24.66&25.09&0.24\\
 &0.20&22.02&22.63&22.84&22.84&23.04&23.49&0.31\\

\hline  &0.01&29.57&29.75&29.78&29.77&29.81&29.89&0.06\\
  &0.05&28.56&28.80&28.88&28.86&28.95&29.09&0.12\\
  &0.10&27.17&27.58&27.72&27.72&27.83&28.15&0.17\\
  &0.15&26.04&26.49&26.61&26.60&26.76&27.00&0.21\\
\multirow{-5}{*}{MissRanger}&0.20&24.87&25.34&25.50&25.49&25.64&26.03&0.23\\

\hline 
\end{tabular} 
\end{minipage} 
\end{table}

\newpage 
\begin{table}[H] 
\caption{minimum ($q_0$), first quartile ($q_{0.25}$),
               median ($q_{0.5}$), mean ($\bar{x}$), third quartile ($q_{0.75}$),
               maximum ($q_{1}$) and standard deviation (sd) of the observed values of the realization of Pearson's chi-squared statistic $\left(\text{with factor} \frac{1}{1000}\right)$ for the variable \texttt{beatm} for the different
               imputation methods and missing rates (r)
               creating missing values for the hospital data (DRG, $\chi^2$, \texttt{beatm}, MCAR)}\label{Tab:drg.vals.chisq.beatm }\begin{minipage}[t]{\textwidth} 

\centering 
\begin{tabular}[t]{|l|c|c|c|c|c|c|c|c|}\hline
 method&r&$q_{0}$& $q_{0.25}$&$q_{0.5}$&$\bar{x}$&$q_{0.75}$& $q_{1}$&sd\\ \hline 
\hline  &0.01&42.49&46.85&47.36&47.21&47.90&48.77&0.97\\
  &0.05&35.42&38.89&40.04&39.92&40.91&42.90&1.53\\
  &0.10&31.68&33.88&34.85&34.77&35.82&38.78&1.39\\
  &0.15&27.69&30.51&31.24&31.30&32.07&35.05&1.44\\
\multirow{-5}{*}{Amelia}&0.20&25.40&27.68&28.46&28.49&29.29&31.49&1.30\\

\hline  &0.01&44.33&49.21&49.46&49.32&49.65&49.94&0.67\\
  &0.05&42.53&46.04&46.88&46.57&47.48&48.46&1.27\\
  &0.10&38.86&42.38&43.52&43.35&44.52&46.27&1.45\\
  &0.15&35.65&38.70&39.94&39.88&41.38&43.19&1.72\\
\multirow{-5}{*}{Mice.Norm}&0.20&31.23&35.40&36.63&36.65&37.98&40.76&2.09\\

\hline  &0.01&44.47&49.15&49.46&49.29&49.63&49.97&0.68\\
  &0.05&42.47&46.13&46.87&46.56&47.40&48.32&1.27\\
  &0.10&38.81&42.44&43.45&43.33&44.27&45.76&1.37\\
  &0.15&35.82&38.46&39.98&39.95&41.35&43.52&1.80\\
\multirow{-5}{*}{Mice.PMM}&0.20&31.09&35.26&36.92&36.64&37.79&40.48&2.00\\

\hline  &0.01&44.26&49.19&49.49&49.32&49.68&49.91&0.69\\
  &0.05&42.43&46.20&46.88&46.55&47.39&48.43&1.25\\
  &0.10&38.36&42.57&43.50&43.42&44.47&46.02&1.47\\
  &0.15&36.04&38.76&40.13&39.91&41.03&43.64&1.77\\
\multirow{-5}{*}{Mice.RF}&0.20&31.28&35.58&36.90&36.71&38.05&41.49&2.05\\

\hline Naive &0.01&44.53&49.39&49.65&49.49&49.84&50.00&0.67\\
  Imputation&0.05&43.30&47.05&47.82&47.47&48.31&49.15&1.24\\
  &0.10&40.15&44.28&45.18&45.17&46.29&47.57&1.42\\
  &0.15&38.41&41.34&42.71&42.59&43.80&46.33&1.79\\
 &0.20&34.12&38.79&40.18&40.12&41.62&44.10&2.17\\

\hline  &0.01&44.56&49.51&49.71&49.56&49.86&50.00&0.68\\
  &0.05&42.80&47.39&48.09&47.80&48.63&49.24&1.25\\
  &0.10&41.82&44.96&45.95&45.68&46.71&48.20&1.40\\
  &0.15&38.61&41.93&43.55&43.30&44.70&46.90&1.84\\
\multirow{-5}{*}{MissRanger}&0.20&34.81&39.79&41.12&41.04&42.62&45.40&2.19\\

\hline 
\end{tabular} 
\end{minipage} 
\end{table}

\newpage 
\begin{table}[H] 
\caption{minimum ($q_0$), first quartile ($q_{0.25}$),
               median ($q_{0.5}$), mean ($\bar{x}$), third quartile ($q_{0.75}$),
               maximum ($q_{1}$) and standard deviation (sd) of the observed values of the realization of Pearson's chi-squared statistic $\left(\text{with factor} \frac{1}{1000}\right)$ for the variable \texttt{entl\_grd} for the different
               imputation methods and missing rates (r)
               creating missing values for the hospital data (DRG, $\chi^2$, \texttt{entl\_grd}, MCAR)}\label{Tab:drg.vals.chisq.entlgrd }\begin{minipage}[t]{\textwidth} 

\centering 
\begin{tabular}[t]{|l|c|c|c|c|c|c|c|c|}\hline
 method&r&$q_{0}$& $q_{0.25}$&$q_{0.5}$&$\bar{x}$&$q_{0.75}$& $q_{1}$&sd\\ \hline 
\hline  &0.01&47.37&47.92&48.09&48.10&48.26&48.82&0.28\\
  &0.05&39.80&40.88&41.33&41.32&41.67&43.12&0.64\\
  &0.10&32.35&33.84&34.27&34.36&34.94&36.60&0.80\\
  &0.15&27.36&28.61&29.02&29.09&29.52&32.25&0.88\\
\multirow{-5}{*}{Amelia}&0.20&22.80&23.90&24.24&24.49&24.64&27.85&1.06\\

\hline  &0.01&48.45&49.17&49.28&49.25&49.40&49.57&0.20\\
  &0.05&44.89&45.90&46.20&46.20&46.58&47.21&0.47\\
  &0.10&40.92&42.07&42.38&42.43&42.84&43.98&0.60\\
  &0.15&37.82&38.61&39.13&39.03&39.50&41.09&0.65\\
\multirow{-5}{*}{Mice.Norm}&0.20&33.58&34.76&35.30&35.29&35.82&36.71&0.69\\

\hline  &0.01&48.58&49.15&49.27&49.25&49.41&49.66&0.20\\
  &0.05&44.98&45.94&46.20&46.20&46.49&47.25&0.44\\
  &0.10&40.88&42.00&42.39&42.38&42.77&43.85&0.60\\
  &0.15&37.64&38.51&38.99&38.97&39.35&40.53&0.65\\
\multirow{-5}{*}{Mice.PMM}&0.20&33.51&34.81&35.23&35.24&35.68&37.04&0.70\\

\hline  &0.01&48.50&49.13&49.27&49.25&49.37&49.63&0.20\\
  &0.05&45.10&45.88&46.18&46.18&46.49&47.39&0.45\\
  &0.10&40.80&42.04&42.39&42.38&42.73&43.97&0.58\\
  &0.15&37.39&38.50&38.94&38.98&39.38&41.16&0.65\\
\multirow{-5}{*}{Mice.RF}&0.20&33.59&34.87&35.38&35.32&35.85&36.88&0.70\\

\hline Naive &0.01&48.75&49.40&49.53&49.51&49.63&49.83&0.19\\
  Imputation&0.05&46.26&47.19&47.41&47.45&47.79&48.44&0.44\\
  &0.10&43.32&44.43&44.79&44.79&45.13&46.35&0.60\\
  &0.15&40.92&41.96&42.51&42.44&42.90&44.43&0.67\\
 &0.20&37.99&39.22&39.77&39.77&40.26&41.41&0.74\\

\hline  &0.01&48.83&49.41&49.54&49.51&49.65&49.86&0.18\\
  &0.05&46.11&47.17&47.46&47.45&47.77&48.37&0.44\\
  &0.10&43.22&44.43&44.77&44.77&45.09&46.37&0.59\\
  &0.15&40.96&41.95&42.51&42.44&42.87&44.44&0.68\\
\multirow{-5}{*}{MissRanger}&0.20&37.89&39.22&39.82&39.79&40.31&41.27&0.76\\

\hline 
\end{tabular} 
\end{minipage} 
\end{table}

\newpage 
\begin{table}[H] 
\caption{minimum ($q_0$), first quartile ($q_{0.25}$),
               median ($q_{0.5}$), mean ($\bar{x}$), third quartile ($q_{0.75}$),
               maximum ($q_{1}$) and standard deviation (sd) of the observed values of the realization of Pearson's chi-squared statistic $\left(\text{with factor} \frac{1}{1000}\right)$ for the variable \texttt{fab\_max} for the different
               imputation methods and missing rates (r)
               creating missing values for the hospital data (DRG, $\chi^2$, \texttt{fab\_max}, MCAR)}\label{Tab:drg.vals.chisq.fabmax }\begin{minipage}[t]{\textwidth} 

\centering 
\begin{tabular}[t]{|l|c|c|c|c|c|c|c|c|}\hline
 method&r&$q_{0}$& $q_{0.25}$&$q_{0.5}$&$\bar{x}$&$q_{0.75}$& $q_{1}$&sd\\ \hline 
\hline  &0.01&296.67&300.04&300.74&300.75&301.56&303.74&1.30\\
  &0.05&259.97&265.18&267.42&267.29&268.95&274.30&2.76\\
  &0.10&224.06&229.11&233.15&232.84&236.40&241.89&4.50\\
  &0.15&194.44&200.13&202.81&202.97&206.45&212.68&4.33\\
\multirow{-5}{*}{Amelia}&0.20&164.37&174.16&178.80&178.05&181.94&188.25&5.03\\

\hline  &0.01&303.02&304.76&305.38&305.31&305.88&307.08&0.82\\
  &0.05&282.24&285.69&286.86&286.88&288.51&290.89&1.95\\
  &0.10&256.08&262.56&264.67&264.41&266.45&269.65&2.67\\
  &0.15&233.94&240.64&242.61&242.18&243.82&248.95&2.80\\
\multirow{-5}{*}{Mice.Norm}&0.20&212.04&219.07&221.59&221.29&223.64&227.93&3.28\\

\hline  &0.01&302.89&304.72&305.28&305.28&305.90&307.08&0.84\\
  &0.05&282.23&285.65&286.94&286.87&288.27&290.71&1.92\\
  &0.10&256.14&262.51&264.41&264.26&266.32&268.94&2.73\\
  &0.15&234.77&240.34&242.49&242.17&243.99&247.95&2.73\\
\multirow{-5}{*}{Mice.PMM}&0.20&211.76&219.09&221.00&221.12&223.34&227.77&3.30\\

\hline  &0.01&302.96&304.84&305.31&305.31&305.96&306.77&0.80\\
  &0.05&282.36&285.68&286.86&286.78&288.18&290.95&1.89\\
  &0.10&255.68&262.22&264.23&264.27&266.21&269.70&2.74\\
  &0.15&233.04&240.02&242.16&241.90&243.58&249.14&2.86\\
\multirow{-5}{*}{Mice.RF}&0.20&213.13&218.82&221.34&221.22&223.50&227.62&3.24\\

\hline Naive &0.01&304.64&306.09&306.67&306.66&307.22&308.10&0.73\\
  Imputation&0.05&288.89&292.23&293.54&293.47&294.83&297.05&1.77\\
  &0.10&269.45&275.34&277.17&277.29&279.24&282.28&2.57\\
  &0.15&252.39&259.70&261.76&261.30&262.90&267.61&2.71\\
 &0.20&236.25&244.02&246.33&246.16&248.39&252.76&3.32\\

\hline  &0.01&305.36&306.77&307.36&307.36&307.88&308.88&0.75\\
  &0.05&291.65&295.69&296.74&296.80&298.05&300.71&1.80\\
  &0.10&276.50&281.31&283.33&283.28&285.34&288.74&2.53\\
  &0.15&262.14&268.74&270.04&269.78&271.31&276.53&2.58\\
\multirow{-5}{*}{MissRanger}&0.20&248.43&254.79&256.92&256.78&259.07&263.77&3.31\\

\hline 
\end{tabular} 
\end{minipage} 
\end{table}

\newpage 
\begin{table}[H] 
\caption{minimum ($q_0$), first quartile ($q_{0.25}$),
               median ($q_{0.5}$), mean ($\bar{x}$), third quartile ($q_{0.75}$),
               maximum ($q_{1}$) and standard deviation (sd) of the observed values of the realization of Pearson's chi-squared statistic $\left(\text{with factor} \frac{1}{1000}\right)$ for the variable \texttt{sex} for the different
               imputation methods and missing rates (r)
               creating missing values for the hospital data (DRG, $\chi^2$, \texttt{sex}, MCAR)}\label{Tab:drg.vals.chisq.sex }\begin{minipage}[t]{\textwidth} 

\centering 
\begin{tabular}[t]{|l|c|c|c|c|c|c|c|c|}\hline
 method&r&$q_{0}$& $q_{0.25}$&$q_{0.5}$&$\bar{x}$&$q_{0.75}$& $q_{1}$&sd\\ \hline 
\hline  &0.01&9.77&9.80&9.82&9.82&9.83&9.87&0.02\\
  &0.05&9.03&9.11&9.14&9.13&9.16&9.21&0.04\\
  &0.10&8.19&8.29&8.32&8.32&8.35&8.43&0.05\\
  &0.15&7.38&7.47&7.52&7.52&7.56&7.67&0.06\\
\multirow{-5}{*}{Amelia}&0.20&6.63&6.73&6.77&6.77&6.81&6.96&0.07\\

\hline  &0.01&9.75&9.80&9.81&9.81&9.83&9.86&0.02\\
  &0.05&8.95&9.07&9.10&9.09&9.12&9.20&0.04\\
  &0.10&8.10&8.20&8.24&8.23&8.27&8.35&0.05\\
  &0.15&7.26&7.36&7.40&7.40&7.43&7.59&0.06\\
\multirow{-5}{*}{Mice.Norm}&0.20&6.49&6.58&6.62&6.63&6.67&6.77&0.06\\

\hline  &0.01&9.78&9.80&9.81&9.81&9.83&9.85&0.02\\
  &0.05&8.98&9.07&9.10&9.10&9.12&9.20&0.04\\
  &0.10&8.10&8.20&8.24&8.24&8.27&8.35&0.06\\
  &0.15&7.23&7.37&7.40&7.40&7.43&7.54&0.05\\
\multirow{-5}{*}{Mice.PMM}&0.20&6.47&6.57&6.62&6.62&6.66&6.79&0.07\\

\hline  &0.01&9.75&9.80&9.81&9.81&9.83&9.86&0.02\\
  &0.05&8.99&9.07&9.10&9.10&9.13&9.19&0.04\\
  &0.10&8.11&8.20&8.23&8.24&8.28&8.38&0.05\\
  &0.15&7.26&7.36&7.40&7.40&7.44&7.54&0.06\\
\multirow{-5}{*}{Mice.RF}&0.20&6.48&6.57&6.62&6.62&6.67&6.77&0.07\\

\hline Naive &0.01&9.75&9.79&9.81&9.81&9.82&9.87&0.03\\
  Imputation&0.05&8.99&9.06&9.10&9.10&9.13&9.26&0.05\\
  &0.10&8.11&8.24&8.28&8.28&8.33&8.45&0.07\\
  &0.15&7.44&7.52&7.57&7.56&7.60&7.70&0.06\\
 &0.20&6.79&6.90&6.94&6.94&6.98&7.09&0.07\\

\hline  &0.01&9.78&9.83&9.84&9.85&9.86&9.90&0.02\\
  &0.05&9.13&9.23&9.26&9.26&9.29&9.39&0.05\\
  &0.10&8.39&8.50&8.54&8.54&8.58&8.69&0.06\\
  &0.15&7.65&7.78&7.84&7.83&7.88&8.03&0.07\\
\multirow{-5}{*}{MissRanger}&0.20&6.95&7.10&7.17&7.17&7.22&7.53&0.09\\

\hline 
\end{tabular} 
\end{minipage} 
\end{table}

\newpage 
\begin{table}[H] 
\caption{minimum ($q_0$), first quartile ($q_{0.25}$),
               median ($q_{0.5}$), mean ($\bar{x}$), third quartile ($q_{0.75}$),
               maximum ($q_{1}$) and standard deviation (sd) of the observed values of the realization of Pearson's chi-squared statistic $\left(\text{with factor} \frac{1}{1000}\right)$ for the variable \texttt{typ\_alter} for the different
               imputation methods and missing rates (r)
               creating missing values for the hospital data (DRG, $\chi^2$, \texttt{typ\_alter}, MCAR)}\label{Tab:drg.vals.chisq.typalter }\begin{minipage}[t]{\textwidth} 

\centering 
\begin{tabular}[t]{|l|c|c|c|c|c|c|c|c|}\hline
 method&r&$q_{0}$& $q_{0.25}$&$q_{0.5}$&$\bar{x}$&$q_{0.75}$& $q_{1}$&sd\\ \hline 
\hline  &0.01&88.14&88.35&88.47&88.47&88.57&88.81&0.16\\
  &0.05&81.52&82.22&82.42&82.45&82.69&83.10&0.34\\
  &0.10&74.14&74.82&75.20&75.18&75.55&76.13&0.47\\
  &0.15&66.93&67.98&68.25&68.27&68.60&69.63&0.48\\
\multirow{-5}{*}{Amelia}&0.20&60.17&61.51&61.82&61.79&62.19&63.31&0.59\\

\hline  &0.01&88.13&88.51&88.60&88.60&88.69&88.98&0.15\\
  &0.05&82.23&82.90&83.11&83.12&83.35&83.88&0.31\\
  &0.10&75.20&76.12&76.40&76.43&76.74&77.50&0.45\\
  &0.15&68.88&69.65&70.00&69.95&70.19&71.51&0.45\\
\multirow{-5}{*}{Mice.Norm}&0.20&62.57&63.55&63.82&63.83&64.15&65.06&0.53\\

\hline  &0.01&88.20&88.50&88.59&88.60&88.69&88.97&0.15\\
  &0.05&82.33&82.91&83.10&83.12&83.30&83.81&0.31\\
  &0.10&75.42&76.05&76.43&76.41&76.69&77.53&0.45\\
  &0.15&68.73&69.64&69.94&69.94&70.24&71.45&0.48\\
\multirow{-5}{*}{Mice.PMM}&0.20&62.43&63.57&63.85&63.86&64.19&65.15&0.54\\

\hline  &0.01&88.14&88.50&88.61&88.60&88.70&88.93&0.15\\
  &0.05&82.34&82.95&83.12&83.13&83.35&83.70&0.31\\
  &0.10&75.14&76.12&76.40&76.43&76.76&77.56&0.45\\
  &0.15&68.70&69.64&70.00&69.95&70.25&71.49&0.47\\
\multirow{-5}{*}{Mice.RF}&0.20&62.48&63.57&63.92&63.88&64.27&65.02&0.57\\

\hline Naive &0.01&88.37&88.65&88.73&88.73&88.82&89.04&0.14\\
  Imputation&0.05&83.26&83.78&83.97&83.99&84.19&84.57&0.28\\
  &0.10&77.79&78.40&78.67&78.64&78.93&79.36&0.36\\
  &0.15&72.88&73.60&73.86&73.86&74.12&74.95&0.39\\
 &0.20&68.50&69.33&69.59&69.60&69.90&70.60&0.43\\

\hline  &0.01&88.50&88.75&88.83&88.83&88.90&89.09&0.13\\
  &0.05&83.40&84.15&84.26&84.29&84.48&84.96&0.30\\
  &0.10&77.75&78.48&78.71&78.71&78.96&79.92&0.38\\
  &0.15&72.10&72.96&73.35&73.37&73.70&74.55&0.53\\
\multirow{-5}{*}{MissRanger}&0.20&66.83&67.90&68.41&68.33&68.67&69.51&0.56\\

\hline 
\end{tabular} 
\end{minipage} 
\end{table}

\newpage 
\begin{table}[H] 
\caption{minimum ($q_0$), first quartile ($q_{0.25}$),
               median ($q_{0.5}$), mean ($\bar{x}$), third quartile ($q_{0.75}$),
               maximum ($q_{1}$) and standard deviation (sd) of the observed values of the realization of Pearson's chi-squared statistic $\left(\text{with factor} \frac{1}{1000}\right)$ for the variable \texttt{typ\_op} for the different
               imputation methods and missing rates (r)
               creating missing values for the hospital data (DRG, $\chi^2$, \texttt{typ\_op}, MCAR)}\label{Tab:drg.vals.chisq.typop }\begin{minipage}[t]{\textwidth} 

\centering 
\begin{tabular}[t]{|l|c|c|c|c|c|c|c|c|}\hline
 method&r&$q_{0}$& $q_{0.25}$&$q_{0.5}$&$\bar{x}$&$q_{0.75}$& $q_{1}$&sd\\ \hline 
\hline  &0.01&9.96&9.98&9.98&9.98&9.98&9.99&0.01\\
  &0.05&9.88&9.89&9.90&9.90&9.91&9.93&0.01\\
  &0.10&9.75&9.78&9.79&9.79&9.80&9.83&0.02\\
  &0.15&9.61&9.67&9.69&9.69&9.70&9.73&0.02\\
\multirow{-5}{*}{Amelia}&0.20&9.49&9.54&9.57&9.56&9.58&9.62&0.03\\

\hline  &0.01&9.95&9.97&9.97&9.97&9.98&9.99&0.01\\
  &0.05&9.82&9.85&9.86&9.86&9.87&9.90&0.02\\
  &0.10&9.64&9.69&9.71&9.71&9.73&9.80&0.03\\
  &0.15&9.48&9.55&9.57&9.57&9.60&9.69&0.04\\
\multirow{-5}{*}{Mice.Norm}&0.20&9.28&9.37&9.41&9.41&9.44&9.52&0.05\\

\hline  &0.01&9.96&9.97&9.97&9.97&9.98&9.99&0.01\\
  &0.05&9.82&9.85&9.86&9.86&9.87&9.89&0.02\\
  &0.10&9.63&9.70&9.71&9.71&9.74&9.78&0.03\\
  &0.15&9.42&9.54&9.57&9.57&9.59&9.66&0.04\\
\multirow{-5}{*}{Mice.PMM}&0.20&9.28&9.38&9.41&9.42&9.46&9.57&0.06\\

\hline  &0.01&9.96&9.97&9.97&9.97&9.98&9.99&0.01\\
  &0.05&9.82&9.85&9.86&9.86&9.87&9.90&0.02\\
  &0.10&9.66&9.69&9.70&9.71&9.73&9.77&0.03\\
  &0.15&9.45&9.54&9.58&9.57&9.60&9.66&0.04\\
\multirow{-5}{*}{Mice.RF}&0.20&9.30&9.37&9.40&9.41&9.44&9.53&0.05\\

\hline Naive &0.01&9.79&9.82&9.85&9.84&9.86&9.91&0.02\\
  Imputation&0.05&9.10&9.19&9.22&9.22&9.26&9.36&0.05\\
  &0.10&8.35&8.44&8.48&8.48&8.52&8.71&0.07\\
  &0.15&7.58&7.77&7.82&7.81&7.86&7.95&0.08\\
 &0.20&7.02&7.15&7.20&7.21&7.26&7.39&0.08\\

\hline  &0.01&9.99&10.00&10.00&10.00&10.00&10.00&<0.01\\
  &0.05&9.97&9.99&10.00&10.00&10.00&10.00&<0.01\\
  &0.10&9.97&9.98&9.99&9.99&9.99&10.00&0.01\\
  &0.15&9.95&9.98&9.98&9.98&9.99&10.00&0.01\\
\multirow{-5}{*}{MissRanger}&0.20&9.96&9.97&9.98&9.98&9.98&10.00&0.01\\

\hline 
\end{tabular} 
\end{minipage} 
\end{table}

\newpage 
\begin{table}[H] 
\caption{minimum ($q_0$), first quartile ($q_{0.25}$),
               median ($q_{0.5}$), mean ($\bar{x}$), third quartile ($q_{0.75}$),
               maximum ($q_{1}$) and standard deviation (sd) of the observed values of the realization of Pearson's chi-squared statistic $\left(\text{with factor} \frac{1}{1000}\right)$ for the variable \texttt{typ\_vwd} for the different
               imputation methods and missing rates (r)
               creating missing values for the hospital data (DRG, $\chi^2$, \texttt{typ\_vwd}, MCAR)}\label{Tab:drg.vals.chisq.typvwd }\begin{minipage}[t]{\textwidth} 

\centering 
\begin{tabular}[t]{|l|c|c|c|c|c|c|c|c|}\hline
 method&r&$q_{0}$& $q_{0.25}$&$q_{0.5}$&$\bar{x}$&$q_{0.75}$& $q_{1}$&sd\\ \hline 
\hline  &0.01&58.26&58.57&58.68&58.68&58.79&59.03&0.16\\
  &0.05&53.01&53.64&53.84&53.84&54.06&54.64&0.31\\
  &0.10&48.21&48.61&48.80&48.82&49.05&49.45&0.30\\
  &0.15&43.71&44.19&44.55&44.52&44.82&45.59&0.39\\
\multirow{-5}{*}{Amelia}&0.20&39.75&40.45&40.72&40.71&40.91&42.01&0.39\\

\hline  &0.01&59.21&59.41&59.49&59.48&59.56&59.66&0.11\\
  &0.05&56.70&57.24&57.39&57.37&57.52&57.86&0.23\\
  &0.10&53.74&54.50&54.69&54.69&54.90&55.34&0.29\\
  &0.15&51.23&51.76&51.96&52.03&52.29&53.04&0.42\\
\multirow{-5}{*}{Mice.Norm}&0.20&48.19&49.03&49.39&49.35&49.70&50.36&0.47\\

\hline  &0.01&59.20&59.42&59.50&59.48&59.56&59.71&0.11\\
  &0.05&56.64&57.22&57.33&57.35&57.49&57.79&0.22\\
  &0.10&54.02&54.52&54.72&54.71&54.89&55.48&0.27\\
  &0.15&50.95&51.70&52.01&52.02&52.29&53.12&0.43\\
\multirow{-5}{*}{Mice.PMM}&0.20&48.13&49.14&49.40&49.39&49.66&50.65&0.45\\

\hline  &0.01&59.20&59.41&59.49&59.48&59.55&59.72&0.11\\
  &0.05&56.75&57.24&57.35&57.36&57.51&57.85&0.23\\
  &0.10&53.93&54.53&54.74&54.73&54.95&55.52&0.30\\
  &0.15&51.11&51.79&52.03&52.06&52.37&52.99&0.40\\
\multirow{-5}{*}{Mice.RF}&0.20&47.78&49.15&49.44&49.44&49.75&50.93&0.49\\

\hline Naive &0.01&58.96&59.18&59.26&59.25&59.33&59.49&0.12\\
  Imputation&0.05&55.45&56.12&56.28&56.29&56.46&56.88&0.26\\
  &0.10&51.86&52.60&52.77&52.77&52.98&53.46&0.30\\
  &0.15&48.60&49.16&49.38&49.43&49.68&50.53&0.40\\
 &0.20&45.25&46.16&46.37&46.35&46.58&47.70&0.41\\

\hline  &0.01&59.64&59.81&59.85&59.85&59.90&60.00&0.07\\
  &0.05&58.88&59.15&59.28&59.28&59.40&59.60&0.17\\
  &0.10&57.76&58.34&58.51&58.49&58.66&59.00&0.23\\
  &0.15&56.79&57.55&57.79&57.76&58.02&58.45&0.35\\
\multirow{-5}{*}{MissRanger}&0.20&56.05&56.73&56.97&56.98&57.20&57.96&0.35\\

\hline 
\end{tabular} 
\end{minipage} 
\end{table}

\newpage 

%% file: table_fivepoint_CramersV_drg.txt
\begin{table}[H] 
\caption{minimum ($q_0$), first quartile ($q_{0.25}$),
               median ($q_{0.5}$), mean ($\bar{x}$), third quartile ($q_{0.75}$),
               maximum ($q_{1}$) and standard deviation (sd) of the observed values of Cram\'{e}rs $V$ for the variable \texttt{aufn\_anl} for the different
               imputation methods and missing rates (r)
               creating missing values for the hospital data (DRG, Cram\'{e}rs $V$, \texttt{aufn\_anl}, MCAR)}\label{Tab:drg.vals.cramersv.aufnanl }\begin{minipage}[t]{\textwidth} 

\centering 
\begin{tabular}[t]{|l|c|c|c|c|c|c|c|c|}\hline
 method&r&$q_{0}$& $q_{0.25}$&$q_{0.5}$&$\bar{x}$&$q_{0.75}$& $q_{1}$&sd\\ \hline 
\hline  &0.01&0.99&0.99&0.99&0.99&0.99&0.99&0.00\\
  &0.05&0.95&0.96&0.96&0.96&0.96&0.96&0.00\\
  &0.10&0.91&0.91&0.92&0.92&0.92&0.93&0.00\\
  &0.15&0.86&0.87&0.88&0.88&0.88&0.89&0.00\\
\multirow{-5}{*}{Amelia}&0.20&0.82&0.83&0.84&0.84&0.84&0.85&0.00\\

\hline  &0.01&0.99&0.99&0.99&0.99&0.99&1.00&0.00\\
  &0.05&0.96&0.97&0.97&0.97&0.97&0.97&0.00\\
  &0.10&0.93&0.94&0.94&0.94&0.94&0.95&0.00\\
  &0.15&0.90&0.91&0.91&0.91&0.91&0.92&0.00\\
\multirow{-5}{*}{Mice.Norm}&0.20&0.86&0.87&0.88&0.88&0.88&0.88&0.00\\

\hline  &0.01&0.99&0.99&0.99&0.99&0.99&1.00&0.00\\
  &0.05&0.96&0.97&0.97&0.97&0.97&0.97&0.00\\
  &0.10&0.93&0.94&0.94&0.94&0.94&0.95&0.00\\
  &0.15&0.90&0.90&0.91&0.91&0.91&0.92&0.00\\
\multirow{-5}{*}{Mice.PMM}&0.20&0.87&0.87&0.88&0.88&0.88&0.89&0.00\\

\hline  &0.01&0.99&0.99&0.99&0.99&0.99&1.00&0.00\\
  &0.05&0.96&0.97&0.97&0.97&0.97&0.97&0.00\\
  &0.10&0.93&0.94&0.94&0.94&0.94&0.95&0.00\\
  &0.15&0.90&0.90&0.91&0.91&0.91&0.92&0.00\\
\multirow{-5}{*}{Mice.RF}&0.20&0.87&0.87&0.88&0.88&0.88&0.89&0.00\\

\hline Naive &0.01&0.99&0.99&0.99&0.99&0.99&1.00&0.00\\
  Imputation&0.05&0.96&0.97&0.97&0.97&0.97&0.97&0.00\\
  &0.10&0.93&0.93&0.94&0.94&0.94&0.94&0.00\\
  &0.15&0.89&0.90&0.90&0.90&0.91&0.91&0.00\\
 &0.20&0.86&0.87&0.87&0.87&0.88&0.88&0.01\\

\hline  &0.01&0.99&1.00&1.00&1.00&1.00&1.00&0.00\\
  &0.05&0.98&0.98&0.98&0.98&0.98&0.98&0.00\\
  &0.10&0.95&0.96&0.96&0.96&0.96&0.97&0.00\\
  &0.15&0.93&0.94&0.94&0.94&0.94&0.95&0.00\\
\multirow{-5}{*}{MissRanger}&0.20&0.91&0.92&0.92&0.92&0.92&0.93&0.00\\

\hline 
\end{tabular} 
\end{minipage} 
\end{table}

\newpage 
\begin{table}[H] 
\caption{minimum ($q_0$), first quartile ($q_{0.25}$),
               median ($q_{0.5}$), mean ($\bar{x}$), third quartile ($q_{0.75}$),
               maximum ($q_{1}$) and standard deviation (sd) of the observed values of Cram\'{e}rs $V$ for the variable \texttt{beatm} for the different
               imputation methods and missing rates (r)
               creating missing values for the hospital data (DRG, Cram\'{e}rs $V$, \texttt{beatm}, MCAR)}\label{Tab:drg.vals.cramersv.beatm }\begin{minipage}[t]{\textwidth} 

\centering 
\begin{tabular}[t]{|l|c|c|c|c|c|c|c|c|}\hline
 method&r&$q_{0}$& $q_{0.25}$&$q_{0.5}$&$\bar{x}$&$q_{0.75}$& $q_{1}$&sd\\ \hline 
\hline  &0.01&0.92&0.97&0.97&0.97&0.98&0.99&0.01\\
  &0.05&0.84&0.88&0.89&0.89&0.90&0.93&0.02\\
  &0.10&0.80&0.82&0.83&0.83&0.85&0.88&0.02\\
  &0.15&0.74&0.78&0.79&0.79&0.80&0.84&0.02\\
\multirow{-5}{*}{Amelia}&0.20&0.71&0.74&0.75&0.75&0.77&0.79&0.02\\

\hline  &0.01&0.94&0.99&0.99&0.99&1.00&1.00&0.01\\
  &0.05&0.92&0.96&0.97&0.96&0.97&0.98&0.01\\
  &0.10&0.88&0.92&0.93&0.93&0.94&0.96&0.02\\
  &0.15&0.84&0.88&0.89&0.89&0.91&0.93&0.02\\
\multirow{-5}{*}{Mice.Norm}&0.20&0.79&0.84&0.86&0.86&0.87&0.90&0.02\\

\hline  &0.01&0.94&0.99&0.99&0.99&1.00&1.00&0.01\\
  &0.05&0.92&0.96&0.97&0.96&0.97&0.98&0.01\\
  &0.10&0.88&0.92&0.93&0.93&0.94&0.96&0.01\\
  &0.15&0.85&0.88&0.89&0.89&0.91&0.93&0.02\\
\multirow{-5}{*}{Mice.PMM}&0.20&0.79&0.84&0.86&0.86&0.87&0.90&0.02\\

\hline  &0.01&0.94&0.99&0.99&0.99&1.00&1.00&0.01\\
  &0.05&0.92&0.96&0.97&0.96&0.97&0.98&0.01\\
  &0.10&0.88&0.92&0.93&0.93&0.94&0.96&0.02\\
  &0.15&0.85&0.88&0.90&0.89&0.91&0.93&0.02\\
\multirow{-5}{*}{Mice.RF}&0.20&0.79&0.84&0.86&0.86&0.87&0.91&0.02\\

\hline Naive &0.01&0.94&0.99&1.00&0.99&1.00&1.00&0.01\\
  Imputation&0.05&0.93&0.97&0.98&0.97&0.98&0.99&0.01\\
  &0.10&0.90&0.94&0.95&0.95&0.96&0.98&0.02\\
  &0.15&0.88&0.91&0.92&0.92&0.94&0.96&0.02\\
 &0.20&0.83&0.88&0.90&0.90&0.91&0.94&0.02\\

\hline  &0.01&0.94&1.00&1.00&1.00&1.00&1.00&0.01\\
  &0.05&0.93&0.97&0.98&0.98&0.99&0.99&0.01\\
  &0.10&0.91&0.95&0.96&0.96&0.97&0.98&0.01\\
  &0.15&0.88&0.92&0.93&0.93&0.95&0.97&0.02\\
\multirow{-5}{*}{MissRanger}&0.20&0.83&0.89&0.91&0.91&0.92&0.95&0.02\\

\hline 
\end{tabular} 
\end{minipage} 
\end{table}

\newpage 
\begin{table}[H] 
\caption{minimum ($q_0$), first quartile ($q_{0.25}$),
               median ($q_{0.5}$), mean ($\bar{x}$), third quartile ($q_{0.75}$),
               maximum ($q_{1}$) and standard deviation (sd) of the observed values of Cram\'{e}rs $V$ for the variable \texttt{entl\_grd} for the different
               imputation methods and missing rates (r)
               creating missing values for the hospital data (DRG, Cram\'{e}rs $V$, \texttt{entl\_grd}, MCAR)}\label{Tab:drg.vals.cramersv.entlgrd }\begin{minipage}[t]{\textwidth} 

\centering 
\begin{tabular}[t]{|l|c|c|c|c|c|c|c|c|}\hline
 method&r&$q_{0}$& $q_{0.25}$&$q_{0.5}$&$\bar{x}$&$q_{0.75}$& $q_{1}$&sd\\ \hline 
\hline  &0.01&0.97&0.98&0.98&0.98&0.98&0.99&0.00\\
  &0.05&0.89&0.90&0.91&0.91&0.91&0.93&0.01\\
  &0.10&0.80&0.82&0.83&0.83&0.84&0.86&0.01\\
  &0.15&0.74&0.76&0.76&0.76&0.77&0.80&0.01\\
\multirow{-5}{*}{Amelia}&0.20&0.68&0.69&0.70&0.70&0.70&0.75&0.01\\

\hline  &0.01&0.98&0.99&0.99&0.99&0.99&1.00&0.00\\
  &0.05&0.95&0.96&0.96&0.96&0.97&0.97&0.00\\
  &0.10&0.90&0.92&0.92&0.92&0.93&0.94&0.01\\
  &0.15&0.87&0.88&0.88&0.88&0.89&0.91&0.01\\
\multirow{-5}{*}{Mice.Norm}&0.20&0.82&0.83&0.84&0.84&0.85&0.86&0.01\\

\hline  &0.01&0.99&0.99&0.99&0.99&0.99&1.00&0.00\\
  &0.05&0.95&0.96&0.96&0.96&0.96&0.97&0.00\\
  &0.10&0.90&0.92&0.92&0.92&0.92&0.94&0.01\\
  &0.15&0.87&0.88&0.88&0.88&0.89&0.90&0.01\\
\multirow{-5}{*}{Mice.PMM}&0.20&0.82&0.83&0.84&0.84&0.84&0.86&0.01\\

\hline  &0.01&0.98&0.99&0.99&0.99&0.99&1.00&0.00\\
  &0.05&0.95&0.96&0.96&0.96&0.96&0.97&0.00\\
  &0.10&0.90&0.92&0.92&0.92&0.92&0.94&0.01\\
  &0.15&0.86&0.88&0.88&0.88&0.89&0.91&0.01\\
\multirow{-5}{*}{Mice.RF}&0.20&0.82&0.84&0.84&0.84&0.85&0.86&0.01\\

\hline Naive &0.01&0.99&0.99&1.00&1.00&1.00&1.00&0.00\\
  Imputation&0.05&0.96&0.97&0.97&0.97&0.98&0.98&0.00\\
  &0.10&0.93&0.94&0.95&0.95&0.95&0.96&0.01\\
  &0.15&0.90&0.92&0.92&0.92&0.93&0.94&0.01\\
 &0.20&0.87&0.89&0.89&0.89&0.90&0.91&0.01\\

\hline  &0.01&0.99&0.99&1.00&1.00&1.00&1.00&0.00\\
  &0.05&0.96&0.97&0.97&0.97&0.98&0.98&0.00\\
  &0.10&0.93&0.94&0.95&0.95&0.95&0.96&0.01\\
  &0.15&0.91&0.92&0.92&0.92&0.93&0.94&0.01\\
\multirow{-5}{*}{MissRanger}&0.20&0.87&0.89&0.89&0.89&0.90&0.91&0.01\\

\hline 
\end{tabular} 
\end{minipage} 
\end{table}

\newpage 
\begin{table}[H] 
\caption{minimum ($q_0$), first quartile ($q_{0.25}$),
               median ($q_{0.5}$), mean ($\bar{x}$), third quartile ($q_{0.75}$),
               maximum ($q_{1}$) and standard deviation (sd) of the observed values of Cram\'{e}rs $V$ for the variable \texttt{fab\_max} for the different
               imputation methods and missing rates (r)
               creating missing values for the hospital data (DRG, Cram\'{e}rs $V$, \texttt{fab\_max}, MCAR)}\label{Tab:drg.vals.cramersv.fabmax }\begin{minipage}[t]{\textwidth} 

\centering 
\begin{tabular}[t]{|l|c|c|c|c|c|c|c|c|}\hline
 method&r&$q_{0}$& $q_{0.25}$&$q_{0.5}$&$\bar{x}$&$q_{0.75}$& $q_{1}$&sd\\ \hline 
\hline  &0.01&0.98&0.98&0.98&0.98&0.99&0.99&0.00\\
  &0.05&0.92&0.92&0.93&0.93&0.93&0.94&0.00\\
  &0.10&0.85&0.86&0.87&0.87&0.87&0.88&0.01\\
  &0.15&0.79&0.80&0.81&0.81&0.82&0.83&0.01\\
\multirow{-5}{*}{Amelia}&0.20&0.73&0.75&0.76&0.76&0.77&0.78&0.01\\

\hline  &0.01&0.99&0.99&0.99&0.99&0.99&1.00&0.00\\
  &0.05&0.95&0.96&0.96&0.96&0.96&0.97&0.00\\
  &0.10&0.91&0.92&0.92&0.92&0.93&0.93&0.00\\
  &0.15&0.87&0.88&0.88&0.88&0.89&0.90&0.01\\
\multirow{-5}{*}{Mice.Norm}&0.20&0.83&0.84&0.85&0.84&0.85&0.86&0.01\\

\hline  &0.01&0.99&0.99&0.99&0.99&0.99&1.00&0.00\\
  &0.05&0.95&0.96&0.96&0.96&0.96&0.97&0.00\\
  &0.10&0.91&0.92&0.92&0.92&0.93&0.93&0.00\\
  &0.15&0.87&0.88&0.88&0.88&0.89&0.89&0.00\\
\multirow{-5}{*}{Mice.PMM}&0.20&0.83&0.84&0.84&0.84&0.85&0.86&0.01\\

\hline  &0.01&0.99&0.99&0.99&0.99&0.99&0.99&0.00\\
  &0.05&0.95&0.96&0.96&0.96&0.96&0.97&0.00\\
  &0.10&0.91&0.92&0.92&0.92&0.93&0.93&0.00\\
  &0.15&0.87&0.88&0.88&0.88&0.89&0.90&0.01\\
\multirow{-5}{*}{Mice.RF}&0.20&0.83&0.84&0.84&0.84&0.85&0.86&0.01\\

\hline Naive &0.01&0.99&0.99&0.99&0.99&1.00&1.00&0.00\\
  Imputation&0.05&0.97&0.97&0.97&0.97&0.98&0.98&0.00\\
  &0.10&0.93&0.94&0.95&0.95&0.95&0.95&0.00\\
  &0.15&0.90&0.92&0.92&0.92&0.92&0.93&0.00\\
 &0.20&0.87&0.89&0.89&0.89&0.90&0.90&0.01\\

\hline  &0.01&0.99&0.99&1.00&1.00&1.00&1.00&0.00\\
  &0.05&0.97&0.98&0.98&0.98&0.98&0.98&0.00\\
  &0.10&0.94&0.95&0.96&0.96&0.96&0.97&0.00\\
  &0.15&0.92&0.93&0.93&0.93&0.94&0.94&0.00\\
\multirow{-5}{*}{MissRanger}&0.20&0.90&0.91&0.91&0.91&0.91&0.92&0.01\\

\hline 
\end{tabular} 
\end{minipage} 
\end{table}

\newpage 
\begin{table}[H] 
\caption{minimum ($q_0$), first quartile ($q_{0.25}$),
               median ($q_{0.5}$), mean ($\bar{x}$), third quartile ($q_{0.75}$),
               maximum ($q_{1}$) and standard deviation (sd) of the observed values of Cram\'{e}rs $V$ for the variable \texttt{sex} for the different
               imputation methods and missing rates (r)
               creating missing values for the hospital data (DRG, Cram\'{e}rs $V$, \texttt{sex}, MCAR)}\label{Tab:drg.vals.cramersv.sex }\begin{minipage}[t]{\textwidth} 

\centering 
\begin{tabular}[t]{|l|c|c|c|c|c|c|c|c|}\hline
 method&r&$q_{0}$& $q_{0.25}$&$q_{0.5}$&$\bar{x}$&$q_{0.75}$& $q_{1}$&sd\\ \hline 
\hline  &0.01&0.99&0.99&0.99&0.99&0.99&0.99&0.00\\
  &0.05&0.95&0.95&0.96&0.96&0.96&0.96&0.00\\
  &0.10&0.90&0.91&0.91&0.91&0.91&0.92&0.00\\
  &0.15&0.86&0.86&0.87&0.87&0.87&0.88&0.00\\
\multirow{-5}{*}{Amelia}&0.20&0.81&0.82&0.82&0.82&0.83&0.83&0.00\\

\hline  &0.01&0.99&0.99&0.99&0.99&0.99&0.99&0.00\\
  &0.05&0.95&0.95&0.95&0.95&0.96&0.96&0.00\\
  &0.10&0.90&0.91&0.91&0.91&0.91&0.91&0.00\\
  &0.15&0.85&0.86&0.86&0.86&0.86&0.87&0.00\\
\multirow{-5}{*}{Mice.Norm}&0.20&0.81&0.81&0.81&0.81&0.82&0.82&0.00\\

\hline  &0.01&0.99&0.99&0.99&0.99&0.99&0.99&0.00\\
  &0.05&0.95&0.95&0.95&0.95&0.96&0.96&0.00\\
  &0.10&0.90&0.91&0.91&0.91&0.91&0.91&0.00\\
  &0.15&0.85&0.86&0.86&0.86&0.86&0.87&0.00\\
\multirow{-5}{*}{Mice.PMM}&0.20&0.80&0.81&0.81&0.81&0.82&0.82&0.00\\

\hline  &0.01&0.99&0.99&0.99&0.99&0.99&0.99&0.00\\
  &0.05&0.95&0.95&0.95&0.95&0.96&0.96&0.00\\
  &0.10&0.90&0.91&0.91&0.91&0.91&0.92&0.00\\
  &0.15&0.85&0.86&0.86&0.86&0.86&0.87&0.00\\
\multirow{-5}{*}{Mice.RF}&0.20&0.80&0.81&0.81&0.81&0.82&0.82&0.00\\

\hline Naive &0.01&0.99&0.99&0.99&0.99&0.99&0.99&0.00\\
  Imputation&0.05&0.95&0.95&0.95&0.95&0.96&0.96&0.00\\
  &0.10&0.90&0.91&0.91&0.91&0.91&0.92&0.00\\
  &0.15&0.86&0.87&0.87&0.87&0.87&0.88&0.00\\
 &0.20&0.82&0.83&0.83&0.83&0.84&0.84&0.00\\

\hline  &0.01&0.99&0.99&0.99&0.99&0.99&1.00&0.00\\
  &0.05&0.96&0.96&0.96&0.96&0.96&0.97&0.00\\
  &0.10&0.92&0.92&0.92&0.92&0.93&0.93&0.00\\
  &0.15&0.87&0.88&0.89&0.88&0.89&0.90&0.00\\
\multirow{-5}{*}{MissRanger}&0.20&0.83&0.84&0.85&0.85&0.85&0.87&0.01\\

\hline 
\end{tabular} 
\end{minipage} 
\end{table}

\newpage 
\begin{table}[H] 
\caption{minimum ($q_0$), first quartile ($q_{0.25}$),
               median ($q_{0.5}$), mean ($\bar{x}$), third quartile ($q_{0.75}$),
               maximum ($q_{1}$) and standard deviation (sd) of the observed values of Cram\'{e}rs $V$ for the variable \texttt{typ\_alter} for the different
               imputation methods and missing rates (r)
               creating missing values for the hospital data (DRG, Cram\'{e}rs $V$, \texttt{typ\_alter}, MCAR)}\label{Tab:drg.vals.cramersv.typalter }\begin{minipage}[t]{\textwidth} 

\centering 
\begin{tabular}[t]{|l|c|c|c|c|c|c|c|c|}\hline
 method&r&$q_{0}$& $q_{0.25}$&$q_{0.5}$&$\bar{x}$&$q_{0.75}$& $q_{1}$&sd\\ \hline 
\hline  &0.01&0.99&0.99&0.99&0.99&0.99&0.99&0.00\\
  &0.05&0.95&0.96&0.96&0.96&0.96&0.96&0.00\\
  &0.10&0.91&0.91&0.91&0.91&0.92&0.92&0.00\\
  &0.15&0.86&0.87&0.87&0.87&0.87&0.88&0.00\\
\multirow{-5}{*}{Amelia}&0.20&0.82&0.83&0.83&0.83&0.83&0.84&0.00\\

\hline  &0.01&0.99&0.99&0.99&0.99&0.99&0.99&0.00\\
  &0.05&0.96&0.96&0.96&0.96&0.96&0.97&0.00\\
  &0.10&0.91&0.92&0.92&0.92&0.92&0.93&0.00\\
  &0.15&0.87&0.88&0.88&0.88&0.88&0.89&0.00\\
\multirow{-5}{*}{Mice.Norm}&0.20&0.83&0.84&0.84&0.84&0.84&0.85&0.00\\

\hline  &0.01&0.99&0.99&0.99&0.99&0.99&0.99&0.00\\
  &0.05&0.96&0.96&0.96&0.96&0.96&0.97&0.00\\
  &0.10&0.92&0.92&0.92&0.92&0.92&0.93&0.00\\
  &0.15&0.87&0.88&0.88&0.88&0.88&0.89&0.00\\
\multirow{-5}{*}{Mice.PMM}&0.20&0.83&0.84&0.84&0.84&0.84&0.85&0.00\\

\hline  &0.01&0.99&0.99&0.99&0.99&0.99&0.99&0.00\\
  &0.05&0.96&0.96&0.96&0.96&0.96&0.96&0.00\\
  &0.10&0.91&0.92&0.92&0.92&0.92&0.93&0.00\\
  &0.15&0.87&0.88&0.88&0.88&0.88&0.89&0.00\\
\multirow{-5}{*}{Mice.RF}&0.20&0.83&0.84&0.84&0.84&0.85&0.85&0.00\\

\hline Naive &0.01&0.99&0.99&0.99&0.99&0.99&0.99&0.00\\
  Imputation&0.05&0.96&0.96&0.97&0.97&0.97&0.97&0.00\\
  &0.10&0.93&0.93&0.93&0.93&0.94&0.94&0.00\\
  &0.15&0.90&0.90&0.91&0.91&0.91&0.91&0.00\\
 &0.20&0.87&0.88&0.88&0.88&0.88&0.89&0.00\\

\hline  &0.01&0.99&0.99&0.99&0.99&0.99&0.99&0.00\\
  &0.05&0.96&0.97&0.97&0.97&0.97&0.97&0.00\\
  &0.10&0.93&0.93&0.94&0.94&0.94&0.94&0.00\\
  &0.15&0.90&0.90&0.90&0.90&0.90&0.91&0.00\\
\multirow{-5}{*}{MissRanger}&0.20&0.86&0.87&0.87&0.87&0.87&0.88&0.00\\

\hline 
\end{tabular} 
\end{minipage} 
\end{table}

\newpage 
\begin{table}[H] 
\caption{minimum ($q_0$), first quartile ($q_{0.25}$),
               median ($q_{0.5}$), mean ($\bar{x}$), third quartile ($q_{0.75}$),
               maximum ($q_{1}$) and standard deviation (sd) of the observed values of Cram\'{e}rs $V$ for the variable \texttt{typ\_op} for the different
               imputation methods and missing rates (r)
               creating missing values for the hospital data (DRG, Cram\'{e}rs $V$, \texttt{typ\_op}, MCAR)}\label{Tab:drg.vals.cramersv.typop }\begin{minipage}[t]{\textwidth} 

\centering 
\begin{tabular}[t]{|l|c|c|c|c|c|c|c|c|}\hline
 method&r&$q_{0}$& $q_{0.25}$&$q_{0.5}$&$\bar{x}$&$q_{0.75}$& $q_{1}$&sd\\ \hline 
\hline  &0.01&1.00&1.00&1.00&1.00&1.00&1.00&0.00\\
  &0.05&0.99&0.99&1.00&1.00&1.00&1.00&0.00\\
  &0.10&0.99&0.99&0.99&0.99&0.99&0.99&0.00\\
  &0.15&0.98&0.98&0.98&0.98&0.98&0.99&0.00\\
\multirow{-5}{*}{Amelia}&0.20&0.97&0.98&0.98&0.98&0.98&0.98&0.00\\

\hline  &0.01&1.00&1.00&1.00&1.00&1.00&1.00&0.00\\
  &0.05&0.99&0.99&0.99&0.99&0.99&1.00&0.00\\
  &0.10&0.98&0.98&0.99&0.99&0.99&0.99&0.00\\
  &0.15&0.97&0.98&0.98&0.98&0.98&0.98&0.00\\
\multirow{-5}{*}{Mice.Norm}&0.20&0.96&0.97&0.97&0.97&0.97&0.98&0.00\\

\hline  &0.01&1.00&1.00&1.00&1.00&1.00&1.00&0.00\\
  &0.05&0.99&0.99&0.99&0.99&0.99&0.99&0.00\\
  &0.10&0.98&0.98&0.99&0.99&0.99&0.99&0.00\\
  &0.15&0.97&0.98&0.98&0.98&0.98&0.98&0.00\\
\multirow{-5}{*}{Mice.PMM}&0.20&0.96&0.97&0.97&0.97&0.97&0.98&0.00\\

\hline  &0.01&1.00&1.00&1.00&1.00&1.00&1.00&0.00\\
  &0.05&0.99&0.99&0.99&0.99&0.99&0.99&0.00\\
  &0.10&0.98&0.98&0.99&0.99&0.99&0.99&0.00\\
  &0.15&0.97&0.98&0.98&0.98&0.98&0.98&0.00\\
\multirow{-5}{*}{Mice.RF}&0.20&0.96&0.97&0.97&0.97&0.97&0.98&0.00\\

\hline Naive &0.01&0.99&0.99&0.99&0.99&0.99&1.00&0.00\\
  Imputation&0.05&0.95&0.96&0.96&0.96&0.96&0.97&0.00\\
  &0.10&0.91&0.92&0.92&0.92&0.92&0.93&0.00\\
  &0.15&0.87&0.88&0.88&0.88&0.89&0.89&0.00\\
 &0.20&0.84&0.85&0.85&0.85&0.85&0.86&0.00\\

\hline  &0.01&1.00&1.00&1.00&1.00&1.00&1.00&0.00\\
  &0.05&1.00&1.00&1.00&1.00&1.00&1.00&0.00\\
  &0.10&1.00&1.00&1.00&1.00&1.00&1.00&0.00\\
  &0.15&1.00&1.00&1.00&1.00&1.00&1.00&0.00\\
\multirow{-5}{*}{MissRanger}&0.20&1.00&1.00&1.00&1.00&1.00&1.00&0.00\\

\hline 
\end{tabular} 
\end{minipage} 
\end{table}

\newpage 
\begin{table}[H] 
\caption{minimum ($q_0$), first quartile ($q_{0.25}$),
               median ($q_{0.5}$), mean ($\bar{x}$), third quartile ($q_{0.75}$),
               maximum ($q_{1}$) and standard deviation (sd) of the observed values of Cram\'{e}rs $V$ for the variable \texttt{typ\_vwd} for the different
               imputation methods and missing rates (r)
               creating missing values for the hospital data (DRG, Cram\'{e}rs $V$, \texttt{typ\_vwd}, MCAR)}\label{Tab:drg.vals.cramersv.typvwd }\begin{minipage}[t]{\textwidth} 

\centering 
\begin{tabular}[t]{|l|c|c|c|c|c|c|c|c|}\hline
 method&r&$q_{0}$& $q_{0.25}$&$q_{0.5}$&$\bar{x}$&$q_{0.75}$& $q_{1}$&sd\\ \hline 
\hline  &0.01&0.99&0.99&0.99&0.99&0.99&0.99&0.00\\
  &0.05&0.94&0.95&0.95&0.95&0.95&0.95&0.00\\
  &0.10&0.90&0.90&0.90&0.90&0.90&0.91&0.00\\
  &0.15&0.85&0.86&0.86&0.86&0.86&0.87&0.00\\
\multirow{-5}{*}{Amelia}&0.20&0.81&0.82&0.82&0.82&0.83&0.84&0.00\\

\hline  &0.01&0.99&1.00&1.00&1.00&1.00&1.00&0.00\\
  &0.05&0.97&0.98&0.98&0.98&0.98&0.98&0.00\\
  &0.10&0.95&0.95&0.95&0.95&0.96&0.96&0.00\\
  &0.15&0.92&0.93&0.93&0.93&0.93&0.94&0.00\\
\multirow{-5}{*}{Mice.Norm}&0.20&0.90&0.90&0.91&0.91&0.91&0.92&0.00\\

\hline  &0.01&0.99&1.00&1.00&1.00&1.00&1.00&0.00\\
  &0.05&0.97&0.98&0.98&0.98&0.98&0.98&0.00\\
  &0.10&0.95&0.95&0.95&0.95&0.96&0.96&0.00\\
  &0.15&0.92&0.93&0.93&0.93&0.93&0.94&0.00\\
\multirow{-5}{*}{Mice.PMM}&0.20&0.90&0.90&0.91&0.91&0.91&0.92&0.00\\

\hline  &0.01&0.99&1.00&1.00&1.00&1.00&1.00&0.00\\
  &0.05&0.97&0.98&0.98&0.98&0.98&0.98&0.00\\
  &0.10&0.95&0.95&0.96&0.96&0.96&0.96&0.00\\
  &0.15&0.92&0.93&0.93&0.93&0.93&0.94&0.00\\
\multirow{-5}{*}{Mice.RF}&0.20&0.89&0.91&0.91&0.91&0.91&0.92&0.00\\

\hline Naive &0.01&0.99&0.99&0.99&0.99&0.99&1.00&0.00\\
  Imputation&0.05&0.96&0.97&0.97&0.97&0.97&0.97&0.00\\
  &0.10&0.93&0.94&0.94&0.94&0.94&0.94&0.00\\
  &0.15&0.90&0.91&0.91&0.91&0.91&0.92&0.00\\
 &0.20&0.87&0.88&0.88&0.88&0.88&0.89&0.00\\

\hline  &0.01&1.00&1.00&1.00&1.00&1.00&1.00&0.00\\
  &0.05&0.99&0.99&0.99&0.99&0.99&1.00&0.00\\
  &0.10&0.98&0.99&0.99&0.99&0.99&0.99&0.00\\
  &0.15&0.97&0.98&0.98&0.98&0.98&0.99&0.00\\
\multirow{-5}{*}{MissRanger}&0.20&0.97&0.97&0.97&0.97&0.98&0.98&0.00\\

\hline 
\end{tabular} 
\end{minipage} 
\end{table}

\newpage 

%% file: table_fivepoint_KolmogorowSmirnow_drg.txt
\begin{table}[H] 
\caption{minimum ($q_0$), first quartile ($q_{0.25}$),
               median ($q_{0.5}$), mean ($\bar{x}$), third quartile ($q_{0.75}$),
               maximum ($q_{1}$) and standard deviation (sd) of the observed values of the realization of the Kolmogorow-Smirnow-statistic for the variable \texttt{beatm} for the different
               imputation methods and missing rates (r)
               creating missing values for the hospital data (DRG, Kolmogorow-Smirnow, \texttt{beatm}, MCAR)}\label{Tab:drg.vals.ksmirn.beatm }\begin{minipage}[t]{\textwidth} 

\centering 
\begin{tabular}[t]{|l|c|c|c|c|c|c|c|c|}\hline
 method&r&$q_{0}$& $q_{0.25}$&$q_{0.5}$&$\bar{x}$&$q_{0.75}$& $q_{1}$&sd\\ \hline 
\hline  &0.01&0.00&0.00&0.00&0.00&0.00&0.00&0.00\\
  &0.05&0.00&0.00&0.00&0.00&0.00&0.00&0.00\\
  &0.10&0.00&0.01&0.01&0.01&0.01&0.01&0.00\\
  &0.15&0.01&0.01&0.01&0.01&0.01&0.01&0.00\\
\multirow{-5}{*}{Amelia}&0.20&0.01&0.01&0.01&0.01&0.01&0.02&0.00\\

\hline  &0.01&0.00&0.00&0.00&0.00&0.00&0.00&0.00\\
  &0.05&0.00&0.00&0.00&0.00&0.00&0.00&0.00\\
  &0.10&0.00&0.00&0.00&0.00&0.00&0.00&0.00\\
  &0.15&0.00&0.00&0.00&0.00&0.00&0.00&0.00\\
\multirow{-5}{*}{Mice.Norm}&0.20&0.00&0.00&0.00&0.00&0.00&0.01&0.00\\

\hline  &0.01&0.00&0.00&0.00&0.00&0.00&0.00&0.00\\
  &0.05&0.00&0.00&0.00&0.00&0.00&0.00&0.00\\
  &0.10&0.00&0.00&0.00&0.00&0.00&0.00&0.00\\
  &0.15&0.00&0.00&0.00&0.00&0.00&0.00&0.00\\
\multirow{-5}{*}{Mice.PMM}&0.20&0.00&0.00&0.00&0.00&0.00&0.01&0.00\\

\hline  &0.01&0.00&0.00&0.00&0.00&0.00&0.00&0.00\\
  &0.05&0.00&0.00&0.00&0.00&0.00&0.00&0.00\\
  &0.10&0.00&0.00&0.00&0.00&0.00&0.00&0.00\\
  &0.15&0.00&0.00&0.00&0.00&0.00&0.00&0.00\\
\multirow{-5}{*}{Mice.RF}&0.20&0.00&0.00&0.00&0.00&0.00&0.01&0.00\\

\hline Naive &0.01&0.00&0.00&0.00&0.00&0.00&0.00&0.00\\
  Imputation&0.05&0.00&0.00&0.00&0.00&0.00&0.00&0.00\\
  &0.10&0.00&0.00&0.00&0.00&0.00&0.00&0.00\\
  &0.15&0.00&0.00&0.00&0.00&0.00&0.00&0.00\\
 &0.20&0.00&0.00&0.00&0.00&0.00&0.01&0.00\\

\hline  &0.01&0.00&0.00&0.00&0.00&0.00&0.00&0.00\\
  &0.05&0.00&0.00&0.00&0.00&0.00&0.00&0.00\\
  &0.10&0.00&0.00&0.00&0.00&0.00&0.00&0.00\\
  &0.15&0.00&0.00&0.00&0.00&0.00&0.00&0.00\\
\multirow{-5}{*}{MissRanger}&0.20&0.00&0.00&0.00&0.00&0.00&0.01&0.00\\

\hline 
\end{tabular} 
\end{minipage} 
\end{table}

\newpage 
\begin{table}[H] 
\caption{minimum ($q_0$), first quartile ($q_{0.25}$),
               median ($q_{0.5}$), mean ($\bar{x}$), third quartile ($q_{0.75}$),
               maximum ($q_{1}$) and standard deviation (sd) of the observed values of the realization of the Kolmogorow-Smirnow-statistic for the variable \texttt{cm\_vol} for the different
               imputation methods and missing rates (r)
               creating missing values for the hospital data (DRG, Kolmogorow-Smirnow, \texttt{cm\_vol}, MCAR)}\label{Tab:drg.vals.ksmirn.cmvol }\begin{minipage}[t]{\textwidth} 

\centering 
\begin{tabular}[t]{|l|c|c|c|c|c|c|c|c|}\hline
 method&r&$q_{0}$& $q_{0.25}$&$q_{0.5}$&$\bar{x}$&$q_{0.75}$& $q_{1}$&sd\\ \hline 
\hline  &0.01&0.00&0.00&0.00&0.00&0.00&0.00&0.00\\
  &0.05&0.00&0.01&0.01&0.01&0.01&0.01&0.00\\
  &0.10&0.01&0.01&0.01&0.01&0.01&0.02&0.00\\
  &0.15&0.01&0.02&0.02&0.02&0.02&0.02&0.00\\
\multirow{-5}{*}{Amelia}&0.20&0.02&0.02&0.02&0.02&0.03&0.03&0.00\\

\hline  &0.01&0.00&0.00&0.00&0.00&0.00&0.00&0.00\\
  &0.05&0.00&0.00&0.00&0.00&0.00&0.00&0.00\\
  &0.10&0.00&0.00&0.00&0.00&0.00&0.01&0.00\\
  &0.15&0.00&0.00&0.00&0.00&0.01&0.01&0.00\\
\multirow{-5}{*}{Mice.Norm}&0.20&0.00&0.01&0.01&0.01&0.01&0.01&0.00\\

\hline  &0.01&0.00&0.00&0.00&0.00&0.00&0.00&0.00\\
  &0.05&0.01&0.01&0.01&0.01&0.01&0.01&0.00\\
  &0.10&0.01&0.01&0.02&0.02&0.02&0.02&0.00\\
  &0.15&0.02&0.02&0.02&0.02&0.02&0.03&0.00\\
\multirow{-5}{*}{Mice.PMM}&0.20&0.02&0.03&0.03&0.03&0.03&0.04&0.00\\

\hline  &0.01&0.00&0.00&0.00&0.00&0.00&0.00&0.00\\
  &0.05&0.00&0.00&0.00&0.00&0.00&0.00&0.00\\
  &0.10&0.00&0.00&0.00&0.00&0.00&0.01&0.00\\
  &0.15&0.00&0.00&0.00&0.00&0.00&0.01&0.00\\
\multirow{-5}{*}{Mice.RF}&0.20&0.00&0.00&0.01&0.01&0.01&0.01&0.00\\

\hline Naive &0.01&0.00&0.01&0.01&0.01&0.01&0.01&0.00\\
  Imputation&0.05&0.03&0.03&0.03&0.03&0.03&0.03&0.00\\
  &0.10&0.06&0.06&0.06&0.06&0.06&0.07&0.00\\
  &0.15&0.09&0.09&0.10&0.10&0.10&0.10&0.00\\
 &0.20&0.12&0.13&0.13&0.13&0.13&0.13&0.00\\

\hline  &0.01&0.00&0.00&0.00&0.00&0.00&0.00&0.00\\
  &0.05&0.01&0.01&0.01&0.01&0.01&0.02&0.00\\
  &0.10&0.02&0.03&0.03&0.03&0.03&0.03&0.00\\
  &0.15&0.03&0.04&0.04&0.04&0.04&0.05&0.00\\
\multirow{-5}{*}{MissRanger}&0.20&0.04&0.05&0.05&0.05&0.06&0.07&0.01\\

\hline 
\end{tabular} 
\end{minipage} 
\end{table}

\newpage 
\begin{table}[H] 
\caption{minimum ($q_0$), first quartile ($q_{0.25}$),
               median ($q_{0.5}$), mean ($\bar{x}$), third quartile ($q_{0.75}$),
               maximum ($q_{1}$) and standard deviation (sd) of the observed values of the realization of the Kolmogorow-Smirnow-statistic for the variable \texttt{typ\_alter} for the different
               imputation methods and missing rates (r)
               creating missing values for the hospital data (DRG, Kolmogorow-Smirnow, \texttt{typ\_alter}, MCAR)}\label{Tab:drg.vals.ksmirn.typalter }\begin{minipage}[t]{\textwidth} 

\centering 
\begin{tabular}[t]{|l|c|c|c|c|c|c|c|c|}\hline
 method&r&$q_{0}$& $q_{0.25}$&$q_{0.5}$&$\bar{x}$&$q_{0.75}$& $q_{1}$&sd\\ \hline 
\hline  &0.01&0.00&0.00&0.00&0.00&0.00&0.00&0.00\\
  &0.05&0.00&0.00&0.00&0.00&0.00&0.00&0.00\\
  &0.10&0.00&0.00&0.01&0.01&0.01&0.01&0.00\\
  &0.15&0.01&0.01&0.01&0.01&0.01&0.01&0.00\\
\multirow{-5}{*}{Amelia}&0.20&0.01&0.01&0.01&0.01&0.01&0.01&0.00\\

\hline  &0.01&0.00&0.00&0.00&0.00&0.00&0.00&0.00\\
  &0.05&0.00&0.00&0.00&0.00&0.00&0.01&0.00\\
  &0.10&0.00&0.01&0.01&0.01&0.01&0.01&0.00\\
  &0.15&0.01&0.01&0.01&0.01&0.01&0.02&0.00\\
\multirow{-5}{*}{Mice.Norm}&0.20&0.01&0.01&0.01&0.01&0.02&0.02&0.00\\

\hline  &0.01&0.00&0.00&0.00&0.00&0.00&0.00&0.00\\
  &0.05&0.00&0.00&0.00&0.00&0.00&0.01&0.00\\
  &0.10&0.00&0.01&0.01&0.01&0.01&0.01&0.00\\
  &0.15&0.01&0.01&0.01&0.01&0.01&0.01&0.00\\
\multirow{-5}{*}{Mice.PMM}&0.20&0.01&0.01&0.01&0.01&0.02&0.02&0.00\\

\hline  &0.01&0.00&0.00&0.00&0.00&0.00&0.00&0.00\\
  &0.05&0.00&0.00&0.00&0.00&0.00&0.01&0.00\\
  &0.10&0.00&0.01&0.01&0.01&0.01&0.01&0.00\\
  &0.15&0.00&0.01&0.01&0.01&0.01&0.02&0.00\\
\multirow{-5}{*}{Mice.RF}&0.20&0.01&0.01&0.01&0.01&0.01&0.02&0.00\\

\hline Naive &0.01&0.00&0.01&0.01&0.01&0.01&0.01&0.00\\
  Imputation&0.05&0.03&0.03&0.03&0.03&0.03&0.04&0.00\\
  &0.10&0.06&0.06&0.06&0.06&0.07&0.07&0.00\\
  &0.15&0.09&0.09&0.10&0.10&0.10&0.11&0.00\\
 &0.20&0.12&0.13&0.13&0.13&0.13&0.14&0.00\\

\hline  &0.01&0.00&0.00&0.00&0.00&0.00&0.00&0.00\\
  &0.05&0.00&0.00&0.01&0.01&0.01&0.01&0.00\\
  &0.10&0.01&0.01&0.01&0.01&0.01&0.02&0.00\\
  &0.15&0.01&0.01&0.02&0.02&0.02&0.02&0.00\\
\multirow{-5}{*}{MissRanger}&0.20&0.01&0.02&0.02&0.02&0.02&0.03&0.00\\

\hline 
\end{tabular} 
\end{minipage} 
\end{table}

\newpage 
\begin{table}[H] 
\caption{minimum ($q_0$), first quartile ($q_{0.25}$),
               median ($q_{0.5}$), mean ($\bar{x}$), third quartile ($q_{0.75}$),
               maximum ($q_{1}$) and standard deviation (sd) of the observed values of the realization of the Kolmogorow-Smirnow-statistic for the variable \texttt{typ\_vwd} for the different
               imputation methods and missing rates (r)
               creating missing values for the hospital data (DRG, Kolmogorow-Smirnow, \texttt{typ\_vwd}, MCAR)}\label{Tab:drg.vals.ksmirn.typvwd }\begin{minipage}[t]{\textwidth} 

\centering 
\begin{tabular}[t]{|l|c|c|c|c|c|c|c|c|}\hline
 method&r&$q_{0}$& $q_{0.25}$&$q_{0.5}$&$\bar{x}$&$q_{0.75}$& $q_{1}$&sd\\ \hline 
\hline  &0.01&0.00&0.00&0.00&0.00&0.00&0.00&0.00\\
  &0.05&0.01&0.01&0.01&0.01&0.01&0.01&0.00\\
  &0.10&0.01&0.01&0.02&0.02&0.02&0.02&0.00\\
  &0.15&0.02&0.02&0.02&0.02&0.02&0.03&0.00\\
\multirow{-5}{*}{Amelia}&0.20&0.03&0.03&0.03&0.03&0.03&0.03&0.00\\

\hline  &0.01&0.00&0.00&0.00&0.00&0.00&0.00&0.00\\
  &0.05&0.00&0.00&0.00&0.00&0.00&0.00&0.00\\
  &0.10&0.00&0.01&0.01&0.01&0.01&0.01&0.00\\
  &0.15&0.01&0.01&0.01&0.01&0.01&0.01&0.00\\
\multirow{-5}{*}{Mice.Norm}&0.20&0.01&0.01&0.01&0.01&0.01&0.02&0.00\\

\hline  &0.01&0.00&0.00&0.00&0.00&0.00&0.00&0.00\\
  &0.05&0.00&0.00&0.00&0.00&0.00&0.00&0.00\\
  &0.10&0.00&0.01&0.01&0.01&0.01&0.01&0.00\\
  &0.15&0.01&0.01&0.01&0.01&0.01&0.01&0.00\\
\multirow{-5}{*}{Mice.PMM}&0.20&0.01&0.01&0.01&0.01&0.01&0.02&0.00\\

\hline  &0.01&0.00&0.00&0.00&0.00&0.00&0.00&0.00\\
  &0.05&0.00&0.00&0.00&0.00&0.00&0.00&0.00\\
  &0.10&0.00&0.01&0.01&0.01&0.01&0.01&0.00\\
  &0.15&0.01&0.01&0.01&0.01&0.01&0.01&0.00\\
\multirow{-5}{*}{Mice.RF}&0.20&0.01&0.01&0.01&0.01&0.01&0.02&0.00\\

\hline Naive &0.01&0.00&0.01&0.01&0.01&0.01&0.01&0.00\\
  Imputation&0.05&0.03&0.03&0.03&0.03&0.03&0.03&0.00\\
  &0.10&0.05&0.06&0.06&0.06&0.06&0.07&0.00\\
  &0.15&0.08&0.09&0.09&0.09&0.09&0.10&0.00\\
 &0.20&0.11&0.12&0.12&0.12&0.12&0.13&0.00\\

\hline  &0.01&0.00&0.00&0.00&0.00&0.00&0.00&0.00\\
  &0.05&0.00&0.00&0.00&0.00&0.00&0.00&0.00\\
  &0.10&0.00&0.00&0.00&0.00&0.00&0.00&0.00\\
  &0.15&0.00&0.00&0.00&0.00&0.00&0.00&0.00\\
\multirow{-5}{*}{MissRanger}&0.20&0.00&0.00&0.00&0.00&0.00&0.01&0.00\\

\hline 
\end{tabular} 
\end{minipage} 
\end{table}

\newpage 

%% file: table_fivepoint_CramerVonMises_drg.txt
\begin{table}[H] 
\caption{minimum ($q_0$), first quartile ($q_{0.25}$),
               median ($q_{0.5}$), mean ($\bar{x}$), third quartile ($q_{0.75}$),
               maximum ($q_{1}$) and standard deviation (sd) of the observed values of the realization of the Cramer-von-Mises-statistic for the variable \texttt{beatm} for the different
               imputation methods and missing rates (r)
               creating missing values for the hospital data (DRG, Cramer-von-Mises, \texttt{beatm}, MCAR)}\label{Tab:drg.vals.cramervonmises.beatm }\begin{minipage}[t]{\textwidth} 

\centering 
\begin{tabular}[t]{|l|c|c|c|c|c|c|c|c|}\hline
 method&r&$q_{0}$& $q_{0.25}$&$q_{0.5}$&$\bar{x}$&$q_{0.75}$& $q_{1}$&sd\\ \hline 
\hline  &0.01&0.00&0.00&0.00&0.00&0.00&0.01&0.00\\
  &0.05&0.01&0.04&0.05&0.06&0.07&0.09&0.02\\
  &0.10&0.10&0.18&0.21&0.21&0.23&0.37&0.05\\
  &0.15&0.26&0.39&0.48&0.48&0.54&0.81&0.11\\
\multirow{-5}{*}{Amelia}&0.20&0.48&0.73&0.85&0.86&0.96&1.54&0.19\\

\hline  &0.01&0.00&0.00&0.00&0.00&0.00&0.00&0.00\\
  &0.05&0.00&0.00&0.00&0.00&0.00&0.01&0.00\\
  &0.10&0.00&0.00&0.01&0.01&0.01&0.03&0.01\\
  &0.15&0.00&0.01&0.02&0.02&0.02&0.05&0.01\\
\multirow{-5}{*}{Mice.Norm}&0.20&0.01&0.02&0.03&0.03&0.04&0.14&0.02\\

\hline  &0.01&0.00&0.00&0.00&0.00&0.00&0.00&0.00\\
  &0.05&0.00&0.00&0.00&0.00&0.00&0.01&0.00\\
  &0.10&0.00&0.00&0.01&0.01&0.01&0.03&0.01\\
  &0.15&0.00&0.01&0.02&0.02&0.03&0.05&0.01\\
\multirow{-5}{*}{Mice.PMM}&0.20&0.01&0.02&0.03&0.03&0.04&0.14&0.02\\

\hline  &0.01&0.00&0.00&0.00&0.00&0.00&0.00&0.00\\
  &0.05&0.00&0.00&0.00&0.00&0.00&0.01&0.00\\
  &0.10&0.00&0.00&0.01&0.01&0.01&0.03&0.01\\
  &0.15&0.00&0.01&0.02&0.02&0.02&0.05&0.01\\
\multirow{-5}{*}{Mice.RF}&0.20&0.00&0.02&0.03&0.04&0.04&0.14&0.02\\

\hline Naive &0.01&0.00&0.00&0.00&0.00&0.00&0.00&0.00\\
  Imputation&0.05&0.00&0.00&0.01&0.01&0.01&0.02&0.00\\
  &0.10&0.01&0.02&0.02&0.03&0.03&0.06&0.01\\
  &0.15&0.02&0.04&0.05&0.05&0.06&0.10&0.02\\
 &0.20&0.05&0.08&0.10&0.10&0.12&0.27&0.03\\

\hline  &0.01&0.00&0.00&0.00&0.00&0.00&0.00&0.00\\
  &0.05&0.00&0.00&0.00&0.00&0.00&0.01&0.00\\
  &0.10&0.00&0.01&0.01&0.02&0.02&0.04&0.01\\
  &0.15&0.01&0.02&0.03&0.03&0.04&0.07&0.01\\
\multirow{-5}{*}{MissRanger}&0.20&0.02&0.04&0.06&0.06&0.07&0.17&0.02\\

\hline 
\end{tabular} 
\end{minipage} 
\end{table}

\newpage 
\begin{table}[H] 
\caption{minimum ($q_0$), first quartile ($q_{0.25}$),
               median ($q_{0.5}$), mean ($\bar{x}$), third quartile ($q_{0.75}$),
               maximum ($q_{1}$) and standard deviation (sd) of the observed values of the realization of the Cramer-von-Mises-statistic for the variable \texttt{cm\_vol} for the different
               imputation methods and missing rates (r)
               creating missing values for the hospital data (DRG, Cramer-von-Mises, \texttt{cm\_vol}, MCAR)}\label{Tab:drg.vals.cramervonmises.cmvol }\begin{minipage}[t]{\textwidth} 

\centering 
\begin{tabular}[t]{|l|c|c|c|c|c|c|c|c|}\hline
 method&r&$q_{0}$& $q_{0.25}$&$q_{0.5}$&$\bar{x}$&$q_{0.75}$& $q_{1}$&sd\\ \hline 
\hline  &0.01&0.00&0.00&0.00&0.00&0.00&0.01&0.00\\
  &0.05&0.03&0.05&0.05&0.05&0.06&0.10&0.01\\
  &0.10&0.11&0.18&0.21&0.21&0.24&0.34&0.04\\
  &0.15&0.22&0.40&0.46&0.46&0.51&0.64&0.08\\
\multirow{-5}{*}{Amelia}&0.20&0.55&0.71&0.78&0.80&0.86&1.22&0.14\\

\hline  &0.01&0.00&0.00&0.00&0.00&0.00&0.01&0.00\\
  &0.05&0.00&0.01&0.01&0.01&0.01&0.02&0.00\\
  &0.10&0.01&0.01&0.01&0.02&0.02&0.05&0.01\\
  &0.15&0.00&0.02&0.02&0.03&0.03&0.07&0.01\\
\multirow{-5}{*}{Mice.Norm}&0.20&0.01&0.03&0.03&0.04&0.05&0.12&0.02\\

\hline  &0.01&0.00&0.00&0.00&0.00&0.01&0.01&0.00\\
  &0.05&0.05&0.07&0.08&0.08&0.09&0.15&0.02\\
  &0.10&0.20&0.29&0.31&0.32&0.35&0.44&0.05\\
  &0.15&0.47&0.65&0.73&0.72&0.80&1.02&0.11\\
\multirow{-5}{*}{Mice.PMM}&0.20&0.83&1.17&1.25&1.26&1.34&1.72&0.18\\

\hline  &0.01&0.00&0.00&0.00&0.00&0.00&0.01&0.00\\
  &0.05&0.00&0.00&0.01&0.01&0.01&0.02&0.00\\
  &0.10&0.00&0.01&0.01&0.01&0.02&0.05&0.01\\
  &0.15&0.01&0.01&0.02&0.02&0.03&0.07&0.01\\
\multirow{-5}{*}{Mice.RF}&0.20&0.01&0.02&0.03&0.04&0.04&0.10&0.02\\

\hline Naive &0.01&0.03&0.04&0.05&0.05&0.06&0.08&0.01\\
  Imputation&0.05&0.98&1.20&1.26&1.28&1.37&1.58&0.12\\
  &0.10&4.41&4.93&5.16&5.17&5.37&5.96&0.32\\
  &0.15&10.48&11.25&11.69&11.72&12.15&13.12&0.60\\
 &0.20&18.94&20.45&20.92&20.96&21.45&22.61&0.77\\

\hline  &0.01&0.00&0.01&0.01&0.01&0.01&0.03&0.00\\
  &0.05&0.17&0.23&0.26&0.27&0.30&0.38&0.05\\
  &0.10&0.76&1.00&1.09&1.09&1.19&1.49&0.15\\
  &0.15&1.78&2.25&2.46&2.45&2.68&3.19&0.31\\
\multirow{-5}{*}{MissRanger}&0.20&2.53&3.73&4.24&4.23&4.74&6.29&0.65\\

\hline 
\end{tabular} 
\end{minipage} 
\end{table}

\newpage 
\begin{table}[H] 
\caption{minimum ($q_0$), first quartile ($q_{0.25}$),
               median ($q_{0.5}$), mean ($\bar{x}$), third quartile ($q_{0.75}$),
               maximum ($q_{1}$) and standard deviation (sd) of the observed values of the realization of the Cramer-von-Mises-statistic for the variable \texttt{typ\_alter} for the different
               imputation methods and missing rates (r)
               creating missing values for the hospital data (DRG, Cramer-von-Mises, \texttt{typ\_alter}, MCAR)}\label{Tab:drg.vals.cramervonmises.typalter }\begin{minipage}[t]{\textwidth} 

\centering 
\begin{tabular}[t]{|l|c|c|c|c|c|c|c|c|}\hline
 method&r&$q_{0}$& $q_{0.25}$&$q_{0.5}$&$\bar{x}$&$q_{0.75}$& $q_{1}$&sd\\ \hline 
\hline  &0.01&0.00&0.00&0.00&0.00&0.00&0.00&0.00\\
  &0.05&0.00&0.01&0.01&0.01&0.02&0.03&0.00\\
  &0.10&0.01&0.04&0.04&0.04&0.05&0.08&0.01\\
  &0.15&0.05&0.08&0.10&0.10&0.11&0.20&0.02\\
\multirow{-5}{*}{Amelia}&0.20&0.09&0.14&0.16&0.17&0.19&0.27&0.04\\

\hline  &0.01&0.00&0.00&0.00&0.00&0.00&0.01&0.00\\
  &0.05&0.01&0.02&0.02&0.02&0.03&0.06&0.01\\
  &0.10&0.02&0.07&0.09&0.09&0.10&0.18&0.03\\
  &0.15&0.05&0.14&0.17&0.19&0.23&0.41&0.07\\
\multirow{-5}{*}{Mice.Norm}&0.20&0.12&0.25&0.35&0.34&0.41&0.58&0.11\\

\hline  &0.01&0.00&0.00&0.00&0.00&0.00&0.01&0.00\\
  &0.05&0.01&0.02&0.02&0.02&0.03&0.06&0.01\\
  &0.10&0.01&0.06&0.08&0.08&0.11&0.17&0.03\\
  &0.15&0.04&0.13&0.18&0.18&0.23&0.39&0.06\\
\multirow{-5}{*}{Mice.PMM}&0.20&0.13&0.26&0.34&0.34&0.43&0.63&0.11\\

\hline  &0.01&0.00&0.00&0.00&0.00&0.00&0.00&0.00\\
  &0.05&0.00&0.01&0.02&0.02&0.03&0.06&0.01\\
  &0.10&0.03&0.06&0.09&0.08&0.10&0.19&0.03\\
  &0.15&0.04&0.14&0.18&0.19&0.24&0.49&0.07\\
\multirow{-5}{*}{Mice.RF}&0.20&0.16&0.27&0.34&0.35&0.41&0.60&0.10\\

\hline Naive &0.01&0.03&0.05&0.06&0.06&0.07&0.09&0.01\\
  Imputation&0.05&1.12&1.35&1.44&1.45&1.57&1.85&0.15\\
  &0.10&4.86&5.44&5.70&5.71&5.99&6.61&0.40\\
  &0.15&10.62&12.05&12.67&12.54&13.00&14.93&0.76\\
 &0.20&19.10&20.99&21.66&21.74&22.39&24.94&1.16\\

\hline  &0.01&0.00&0.00&0.00&0.00&0.00&0.01&0.00\\
  &0.05&0.01&0.02&0.04&0.04&0.05&0.09&0.02\\
  &0.10&0.06&0.11&0.13&0.14&0.17&0.24&0.04\\
  &0.15&0.16&0.24&0.32&0.31&0.36&0.60&0.09\\
\multirow{-5}{*}{MissRanger}&0.20&0.22&0.46&0.57&0.56&0.65&1.07&0.17\\

\hline 
\end{tabular} 
\end{minipage} 
\end{table}

\newpage 
\begin{table}[H] 
\caption{minimum ($q_0$), first quartile ($q_{0.25}$),
               median ($q_{0.5}$), mean ($\bar{x}$), third quartile ($q_{0.75}$),
               maximum ($q_{1}$) and standard deviation (sd) of the observed values of the realization of the Cramer-von-Mises-statistic for the variable \texttt{typ\_vwd} for the different
               imputation methods and missing rates (r)
               creating missing values for the hospital data (DRG, Cramer-von-Mises, \texttt{typ\_vwd}, MCAR)}\label{Tab:drg.vals.cramervonmises.typvwd }\begin{minipage}[t]{\textwidth} 

\centering 
\begin{tabular}[t]{|l|c|c|c|c|c|c|c|c|}\hline
 method&r&$q_{0}$& $q_{0.25}$&$q_{0.5}$&$\bar{x}$&$q_{0.75}$& $q_{1}$&sd\\ \hline 
\hline  &0.01&0.00&0.00&0.00&0.00&0.00&0.01&0.00\\
  &0.05&0.02&0.03&0.04&0.04&0.04&0.06&0.01\\
  &0.10&0.09&0.13&0.14&0.14&0.16&0.23&0.03\\
  &0.15&0.25&0.28&0.31&0.31&0.33&0.42&0.04\\
\multirow{-5}{*}{Amelia}&0.20&0.46&0.52&0.56&0.56&0.59&0.68&0.05\\

\hline  &0.01&0.00&0.00&0.00&0.00&0.00&0.00&0.00\\
  &0.05&0.01&0.02&0.03&0.03&0.03&0.06&0.01\\
  &0.10&0.06&0.09&0.11&0.11&0.13&0.21&0.03\\
  &0.15&0.17&0.23&0.27&0.27&0.30&0.49&0.06\\
\multirow{-5}{*}{Mice.Norm}&0.20&0.32&0.44&0.49&0.51&0.57&0.79&0.10\\

\hline  &0.01&0.00&0.00&0.00&0.00&0.00&0.00&0.00\\
  &0.05&0.01&0.02&0.03&0.03&0.03&0.05&0.01\\
  &0.10&0.07&0.09&0.11&0.11&0.13&0.17&0.03\\
  &0.15&0.16&0.23&0.26&0.27&0.30&0.43&0.06\\
\multirow{-5}{*}{Mice.PMM}&0.20&0.29&0.42&0.50&0.51&0.57&0.73&0.10\\

\hline  &0.01&0.00&0.00&0.00&0.00&0.00&0.00&0.00\\
  &0.05&0.01&0.02&0.03&0.03&0.03&0.05&0.01\\
  &0.10&0.06&0.10&0.11&0.12&0.13&0.21&0.03\\
  &0.15&0.17&0.23&0.27&0.27&0.30&0.42&0.05\\
\multirow{-5}{*}{Mice.RF}&0.20&0.20&0.42&0.51&0.50&0.56&0.72&0.10\\

\hline Naive &0.01&0.04&0.07&0.08&0.08&0.09&0.15&0.02\\
  Imputation&0.05&1.51&1.95&2.11&2.11&2.23&2.87&0.24\\
  &0.10&7.00&8.28&8.62&8.71&9.14&10.83&0.64\\
  &0.15&16.97&19.11&20.01&19.97&20.85&22.96&1.27\\
 &0.20&30.97&35.16&36.54&36.68&37.98&43.84&2.10\\

\hline  &0.01&0.00&0.00&0.00&0.00&0.00&0.00&0.00\\
  &0.05&0.00&0.00&0.00&0.00&0.00&0.00&0.00\\
  &0.10&0.00&0.00&0.00&0.00&0.00&0.01&0.00\\
  &0.15&0.00&0.00&0.01&0.01&0.01&0.01&0.00\\
\multirow{-5}{*}{MissRanger}&0.20&0.00&0.01&0.01&0.01&0.02&0.02&0.00\\

\hline 
\end{tabular} 
\end{minipage} 
\end{table}

\newpage 

%% file: table_fivepoint_KullbackLeibler_drg.txt
\begin{table}[H] 
\caption{minimum ($q_0$), first quartile ($q_{0.25}$),
               median ($q_{0.5}$), mean ($\bar{x}$), third quartile ($q_{0.75}$),
               maximum ($q_{1}$) and standard deviation (sd) of the observed values of the Kullback-Leibler divergence for the variable \texttt{cm\_vol} for the different
               imputation methods and missing rates (r)
               creating missing values for the hospital data (DRG, Kullback-Leibler, \texttt{cm\_vol}, MCAR)}\label{Tab:drg.vals.kullbackleibler.cmvol }\begin{minipage}[t]{\textwidth} 

\centering 
\begin{tabular}[t]{|l|c|c|c|c|c|c|c|c|}\hline
 method&r&$q_{0}$& $q_{0.25}$&$q_{0.5}$&$\bar{x}$&$q_{0.75}$& $q_{1}$&sd\\ \hline 
\hline  &0.01&0.00&0.00&0.00&0.00&0.00&0.00&0.00\\
  &0.05&0.00&0.00&0.00&0.00&0.00&0.00&0.00\\
  &0.10&0.00&0.00&0.00&0.00&0.00&0.00&0.00\\
  &0.15&0.00&0.00&0.00&0.00&0.00&0.00&0.00\\
\multirow{-5}{*}{Amelia}&0.20&0.00&0.00&0.00&0.00&0.00&0.00&0.00\\

\hline  &0.01&0.00&0.00&0.00&0.00&0.00&0.00&0.00\\
  &0.05&0.00&0.00&0.00&0.00&0.00&0.00&0.00\\
  &0.10&0.00&0.00&0.00&0.00&0.00&0.00&0.00\\
  &0.15&0.00&0.00&0.00&0.00&0.00&0.00&0.00\\
\multirow{-5}{*}{Mice.Norm}&0.20&0.00&0.00&0.00&0.00&0.00&0.00&0.00\\

\hline  &0.01&0.00&0.00&0.00&0.00&0.00&0.00&0.00\\
  &0.05&0.00&0.00&0.00&0.00&0.00&0.00&0.00\\
  &0.10&0.00&0.00&0.00&0.00&0.00&0.00&0.00\\
  &0.15&0.00&0.00&0.00&0.00&0.00&0.00&0.00\\
\multirow{-5}{*}{Mice.PMM}&0.20&0.00&0.00&0.00&0.00&0.00&0.00&0.00\\

\hline  &0.01&0.00&0.00&0.00&0.00&0.00&0.00&0.00\\
  &0.05&0.00&0.00&0.00&0.00&0.00&0.00&0.00\\
  &0.10&0.00&0.00&0.00&0.00&0.00&0.00&0.00\\
  &0.15&0.00&0.00&0.00&0.00&0.00&0.00&0.00\\
\multirow{-5}{*}{Mice.RF}&0.20&0.00&0.00&0.00&0.00&0.00&0.00&0.00\\

\hline Naive &0.01&0.00&0.00&0.00&0.00&0.00&0.00&0.00\\
  Imputation&0.05&0.00&0.00&0.00&0.00&0.00&0.00&0.00\\
  &0.10&0.00&0.00&0.00&0.00&0.00&0.00&0.00\\
  &0.15&0.00&0.00&0.00&0.00&0.00&0.00&0.00\\
 &0.20&0.00&0.00&0.00&0.00&0.00&0.00&0.00\\

\hline  &0.01&0.00&0.00&0.00&0.00&0.00&0.00&0.00\\
  &0.05&0.00&0.00&0.00&0.00&0.00&0.00&0.00\\
  &0.10&0.00&0.00&0.00&0.00&0.00&0.00&0.00\\
  &0.15&0.00&0.00&0.00&0.00&0.00&0.00&0.00\\
\multirow{-5}{*}{MissRanger}&0.20&0.00&0.00&0.00&0.00&0.00&0.00&0.00\\

\hline 
\end{tabular} 
\end{minipage} 
\end{table}

\newpage 

%% file: table_fivepoint_Mallows_drg.txt
\begin{table}[H] 
\caption{minimum ($q_0$), first quartile ($q_{0.25}$),
               median ($q_{0.5}$), mean ($\bar{x}$), third quartile ($q_{0.75}$),
               maximum ($q_{1}$) and standard deviation (sd) of the observed values of Mallows $L^2$-distance $\left(\text{with factor} \frac{1}{1e+18}\right)$ for the variable \texttt{cm\_vol} for the different
               imputation methods and missing rates (r)
               creating missing values for the hospital data (DRG, Mallows, \texttt{cm\_vol}, MCAR)}\label{Tab:drg.vals.mallows.cmvol }\begin{minipage}[t]{\textwidth} 

\centering 
\begin{tabular}[t]{|l|c|c|c|c|c|c|c|c|}\hline
 method&r&$q_{0}$& $q_{0.25}$&$q_{0.5}$&$\bar{x}$&$q_{0.75}$& $q_{1}$&sd\\ \hline 
\hline  &0.01&0.03&0.06&0.08&0.08&0.10&0.16&0.03\\
  &0.05&0.26&0.40&0.45&0.45&0.51&0.65&0.08\\
  &0.10&0.77&0.92&1.02&1.02&1.09&1.38&0.13\\
  &0.15&1.35&1.56&1.68&1.68&1.77&2.21&0.17\\
\multirow{-5}{*}{Amelia}&0.20&1.87&2.29&2.43&2.45&2.61&3.25&0.22\\

\hline  &0.01&<0.01&<0.01&<0.01&<0.01&<0.01&0.01&<0.01\\
  &0.05&0.01&0.01&0.02&0.02&0.02&0.06&0.01\\
  &0.10&0.02&0.03&0.03&0.04&0.04&0.12&0.02\\
  &0.15&0.01&0.04&0.05&0.06&0.07&0.24&0.04\\
\multirow{-5}{*}{Mice.Norm}&0.20&0.03&0.06&0.08&0.09&0.10&0.26&0.04\\

\hline  &0.01&0.10&0.16&0.19&0.21&0.24&0.41&0.06\\
  &0.05&0.63&1.00&1.08&1.08&1.16&1.52&0.14\\
  &0.10&1.92&2.32&2.45&2.43&2.57&3.00&0.22\\
  &0.15&3.24&3.70&3.86&3.90&4.13&4.70&0.28\\
\multirow{-5}{*}{Mice.PMM}&0.20&4.80&5.29&5.57&5.57&5.81&6.31&0.36\\

\hline  &0.01&<0.01&<0.01&<0.01&<0.01&<0.01&0.01&<0.01\\
  &0.05&0.01&0.01&0.01&0.02&0.02&0.05&0.01\\
  &0.10&0.01&0.02&0.03&0.03&0.04&0.11&0.02\\
  &0.15&0.02&0.03&0.05&0.05&0.06&0.23&0.03\\
\multirow{-5}{*}{Mice.RF}&0.20&0.03&0.05&0.06&0.08&0.10&0.21&0.04\\

\hline Naive &0.01&0.04&0.06&0.07&0.07&0.07&0.10&0.01\\
  Imputation&0.05&1.15&1.38&1.49&1.49&1.61&1.81&0.16\\
  &0.10&4.50&5.35&5.69&5.68&5.90&6.96&0.42\\
  &0.15&11.05&12.00&12.44&12.46&12.88&14.60&0.73\\
 &0.20&19.74&20.93&21.73&21.75&22.40&24.28&1.04\\

\hline  &0.01&0.01&0.01&0.02&0.02&0.02&0.03&0.01\\
  &0.05&0.18&0.26&0.30&0.30&0.34&0.41&0.06\\
  &0.10&0.82&1.04&1.13&1.15&1.26&1.52&0.15\\
  &0.15&1.74&2.30&2.50&2.50&2.67&3.38&0.30\\
\multirow{-5}{*}{MissRanger}&0.20&2.90&3.92&4.26&4.31&4.64&5.76&0.56\\

\hline 
\end{tabular} 
\end{minipage} 
\end{table}

\newpage 